# UNIVERSITÀ DEGLI STUDI DI TRIESTE
## Dipartimento di Fisica

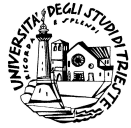

XXII CICLO DEL DOTTORATO DI RICERCA IN FISICA
(I CICLO DELLE SCUOLE DI DOTTORATO)

# The Swift-XRT Survey of Groups and Clusters of Galaxies

DOTTORANDO
Andrea Bignamini
email: andrea.bignamini@gmail.com
        bignamini@oats.inaf.it

COORDINATORE DEL COLLEGIO DEI DOCENTI
Chiar.mo Prof. Gaetano Senatore
Università degli Studi di Trieste
Firma:_______________________

RELATORE
Dr. Paolo Tozzi
INAF - Osservatorio Astronomico di Trieste
Firma:_______________________

SUPERVISORE
Chiar.mo Prof. Stefano Borgani
Università degli Studi di Trieste
Firma:_______________________

Anno Accademico 2008/2009

*Force quit and move to trash*
*I was right there at the top of the bottom*
*On the edge of the known Universe where I wanted to be*
*I had driven to the scene of the accident*
*And I sat there waiting for me*
*Restart and re-boot yourself*
*You're free to go.*

U2 – Unknown Caller

# CONTENTS

# LIST OF FIGURES

## LIST OF TABLES

## ACRONYMS

1AXG    First ASCA X-Ray Gas imaging spectrometer sources

2dFGRS    2dF Galaxy Redshift Survey

2MASX    2 Micron All Sky survey eXtended objects

ABELL     Abell clusters of galaxies

AGN       Active Galactic Nucleus

ARF       Ancillary Response File

BAT       Burst Alert Telescope

BCG       Brightest Cluster Galaxy

C-stat    Cash Statistic

CC        Cool Core

CCD       Charge Coupled Device

CDFS      Chandra Deep Field South

CFHT      Canada France Hawaii Telescope

CIAO      Chandra Interactive Analysis of Observations

CIZA      Clusters In the Zone of Avoidance

CMB       Cosmic Microwave Background

CTI       Charge Transfer Inefficiency

CTIO      Cerro Tololo Inter-American Observatory

Dec       Declination

DSS       Digital Sky Survey

ECF       Energy Conversion Factor

ENR       Electroformed Nickel-cobalt Replicas

FoV       Field of View

GRB       Gamma Ray Burst

GTI       Good Time Interval

HEW       Half Energy Width

HST       Hubble Space Telescope

ICM       Intra Cluster Medium

| | |
|---|---|
| ISM | Inter Stellar Medium |
| MaxBCG | Maximum likelihood redshift Brightest Cluster Galaxy |
| MCMC | Markov Chain-Monte Carlo |
| MHD | Modified Hausdorff Distance |
| NASA | National Aeronautics and Space Administration |
| NED | NASA/IPAC Extragalactic Database |
| PDF | Probability Density Function |
| PHA | Pulse Height Amplitude |
| PI | Pulse Invariant |
| PS | Press & Schechter |
| PSF | Point Spread Function |
| QE | Quantum Efficiency |
| R.A. | Right Ascension |
| RASS | ROSAT All Sky Survey |
| RCS | Red-sequence Cluster Survey |
| RDCS | ROSAT Deep Cluster Survey |
| RMF | Redistribution Matrix File |
| RMS | Root Mean Square |
| RXS | ROSAT X-ray Source |
| SDSS | Sloan Digital Sky Survey |
| SDSS-GC1 | Sloan Digital Sky Survey Galaxy Cluster |
| SiC | Silicon-Carbide |
| SNR | Signal to Noise Ratio |
| SXCS | Swift-XRT Cluster Survey |
| SZ | Sunyaev-Zel'dovich |

TDF      Two Dimensional Fitting

TNG      Telescopio Nazionale Galileo

UVOT     UltraViolet/Optical Telescope

VLT      Very Large Telescope

WFXT     Wide Field X-ray Telescope

WGA      White+Giommi+Angelini ROSAT X-Ray sources list

XRB      X-Ray Background

XRT      Swift X-Ray Telescope

ZwCl     Zwicky's catalog of galaxies and Clusters of galaxies

# INTRODUCTION

My Ph.D. Thesis is devoted to the study of groups and clusters of galaxies in the X-ray band. This field has been very active in the last ten years, thanks to the data gathered from the *Chandra* and *XMM-Newton* satellites. Clusters of galaxies are prominent X-ray sources thanks to thermal bremsstrahlung emission from the diffuse Intra Cluster Medium (ICM) heated to $10^7$-$10^8$ Kelvin, which provides about 15% of their total mass. The analysis of the X-ray emission from groups and clusters allows to study the large scale structure of the Universe, to constrain the cosmological parameters, and to investigate the interaction between the ICM and the cluster galaxies.

My scientific work is mainly focused on the realization of a new X-ray survey of galaxy clusters, the Swift-XRT Cluster Survey (SXCS), obtained from the previously unexplored archive of the X-Ray Telescope (XRT) on board of the Swift satellite. The goal is not only to build a new catalogue, but also to characterize the thermodynamical and chemical properties of the brightest groups and clusters in the survey catalogue. Moreover, given the overall characteristics of the survey, I also expect to detect some clusters at redshift $z > 1$, which will have a strong impact in the study of the large scale structure of the Universe and the cosmological parameters.

Along the line of investigating the properties of high redshift clusters, I also analysed *Chandra* data for distant, optically selected clusters, included in the Red-sequence Cluster Survey (RCS). As a complement to the SXCS survey, I intend to assess the evolutionary stage of these objects, in order to point out the differences with respect to the X-ray selected clusters.

During my work I also contributed substantially to the image simulator code of a new proposed X-ray mission submitted to the NASA Astro 2010 Decadal Survey: the Wide Field X-ray Telescope (WFXT). This work represents an important part of the scientific case of WFXT, since, for first time in the simulations I included realistic populations of all the source types contributing to the extragalactic X-ray sky, namely groups and clusters of galaxies, active galactic nuclei, and star-forming galaxies. Thanks to this work, the scientific cases of WFXT can now be tested on solid ground.

My Thesis is divided into six Chapters. The Chapter 1 is a concise review about the cluster of galaxies and the physics of the ICM, while the Chapter 6 summarizes all the main results. The original work of my PhD project is presented in four main topics, which constitute the remaining Chapters and are briefly summarized below.

## CHAPTER 2: ANALYSIS OF SWIFT-XRT DATA AND SURVEY CATALOGUE

In collaboration with the Swift group of the Osservatorio Astronomico di Brera, I performed the reduction of the Swift-XRT archive data and updated the telescope calibration. Then, I investigated in details the XRT characteristics, in particular the effective area, the instrumental background, and the spatial resolution. A proper characterization of the spatial resolution is crucial in order to distinguish the point sources, mostly Active Galactic Nuclei (AGNs) and star-forming galaxy, from the extended sources (galaxy clusters and groups).

To identify the sources in the XRT images, I used the wavdetect algorithm within the CIAO software package developed for *Chandra*. However, this is not sufficient to identify extended sources. In order to perform this fundamental step, I developed a new Two Dimensional Fitting (TDF) algorithm based on Markov Chain-Monte Carlo (MCMC) and maximum likelihood criterion, to measure the extension of each detected source. As a matter of fact, in astronomical images, a point source has a well known size depending on the spatial resolution of the instrument, while extended sources have a larger extension. To calibrate both the algorithms I realized a new X-ray image simulator adapted to the technical XRT properties. After a detailed analysis I found that the source-detection algorithm, wavdetect and the TDF algorithm are efficient in identifying clusters of galaxies in X-ray images.

From the entire Swift-XRT data archive, I firmly detected about 250 clusters, which constitute the SXCS catalogue. The total number of detected sources as function of the flux limit is in perfect agreement with previous results by other authors (e.g. Rosati et al., 1998), indicating that the criteria used to define the SXCS catalogue produce a fair and complete sample of cluster of galaxies. Within the whole SXCS catalogue I selected a subsample of about 50 high redshift candidate, for which we are planning optical follow-up to measure the redshift. We expect about $\sim 10$ clusters at $z > 1$ among them, thus doubling the number of known X-ray clusters at $z > 1$. The survey catalogue will be published in Moretti, Bignamini et al. (2010).

### CHAPTER 3: A COMPLETE SAMPLE OF X-RAY GROUPS AND CLUSTERS OF GALAXIES FROM THE SXCS

To perform a detailed X-ray spectral analysis of a cluster of galaxy and therefore to measure the thermodynamical and physical properties of the ICM, you need to extract a spectrum with at least a few hundreds of net counts. So I selected all the sources in the SXCS catalogue with more than 200 net counts in the soft band, getting 32 objects that constitute the bright subsample. Most of these objects were not previously detected. For the clusters in the bright sample I performed a detailed spectral analysis to measure average ICM temperature, iron abundance, redshift, bolometric luminosity, total mass and gas mass.

The measure of the redshift, in particular, was obtained for the first time for a complete sample of X-ray selected clusters from X-ray data alone. I confirmed the validity of the derived redshifts by comparing them with spectroscopic redshifts obtained with a proposal at the Telescopio Nazionale Galileo (TNG), and with the few photometric redshifts available in the literature. I derived scaling relations (luminosity-temperature, mass-temperature, etc.) for the bright sample, finding agreement with previous studies on X-ray selected samples (e.g. Arnaud et al., 2007; Branchesi et al., 2007c; Sun et al., 2009). The distribution of iron abundance and its dependence on the cosmic epoch is also in agreement with that found in local and distant cluster samples (e.g. Balestra et al., 2007). In this way I show that, given the quality of XRT data, it is possible to build a sample of clusters of galaxies entirely characterized on the basis of the X-ray data alone. These results will be published in Bignamini et al. (2010).

### CHAPTER 4: X-RAY PROPERTIES OF HIGH-Z OPTICALLY SELECTED CLUSTERS

A useful complement to the study of X-ray selected clusters of galaxies, is the X-ray analysis of optically selected cluster samples. I analysed the X-ray properties of 11 optically selected clusters of galaxies in the Red-sequence Cluster Survey (RCS) at redshift $0.6 < z < 1.2$ observed with *Chandra* (Yee & Gladders, 2001). Thanks to the deep and high quality *Chandra* data, I investigated the thermodynamical properties of the ICM through spectral analysis and the distribution of AGNs in the surroundings. I detected extended X-ray emission for the majority of the clusters, except for three, for which I obtained only a marginal detection at $\sim 3\sigma$. I found that the normalization of the luminosity-temperature relation of RCS clusters is a factor $\sim 2$ lower than that of X-ray selected clusters, possibly indicating an incomplete virialization. I confirmed that the iron abundance in the detected objects is consistent with that of X-ray selected clusters at the same redshift. Moreover, I analysed the fraction of Cool Cores in the RCS sample, finding that it is lower than that found in X-ray selected samples. For the cluster RCS0224-0002 I computed the total mass obtained from the X-ray temperature of the ICM, and I show that it is in agreement with the one derived from strong lensing observed with the Hubble Space Telescope (HST). Moreover, I found an excess of AGNs towards the center of the clusters, showing the presence of AGN activity in the central galaxies of high redshift clusters as expected in several feedback models. These results are published in Rzepecki et al. (2007), Bignamini et al. (2008) and Santos et al. (2008).

### CHAPTER 5: PLANNING A FUTURE MISSION: THE WIDE FIELD X-RAY TELESCOPE

The Wide Field X-ray Telescope (WFXT) is a new proposed X-ray mission submitted to the NASA Astro 2010 Decadal Survey that will address many outstanding cosmological and astrophysical objectives, such as the formation and evolution of clusters of galaxies and associated implications on cosmology and fundamental physics, properties and evolution of AGNs, and cosmic star formation history. I contributed substantially to the realization of realistic X-ray simulated images to help in the choice of the optimal design in order to achieve the mission goals.

The simulated images take into account all the sources that contribute to the X-ray sky: AGNs and star forming galaxies (extrapolated from observations and theoretical models), groups and clusters of galaxies, and the X-ray Galactic foreground. Groups and clusters of galaxies are sampled from the observed X-ray luminosity function and added to the images. In order to preserve the complexity of these extended sources, I applied the cloning technique to the detailed *Chandra* images of local clusters, as described in the paper Santos et al. (2008).

The image simulator accounts for all the instrumental features: Point Spread Function (PSF) profile, instrumental background, vignetting, effective area, and response matrix. The sim-ulations were run for different choices of instrumental design in order to compare different configurations and to help in the choice of the most effective one. The image simulations represent a crucial aspect of the scientific case of WFXT. This work is presented in the white paper by Giacconi et al. (2009) submitted to the NASA Decadal Survey Committee.

# CLUSTER OF GALAXIES AND THE PHYSICS OF THE INTRA CLUSTER MEDIUM

In this Chapter I describe the main properties of clusters of galaxies, with a strong focus on their X-ray characteristics and the related physics.

## 1.1 CLUSTERS AS ASTROPHYSICAL TOOLS

Clusters of galaxies are the largest self gravitating structures in the Universe. They typically contain hundreds to thousands of galaxies, spread over a region of a few Mpc. Their total masses vary from $10^{13} M_\odot$ up to $10^{15} M_\odot$. They were first studied in detail by Wolf (1906), where clusters of galaxies were identified as large galaxy concentrations in the projected galaxy distribution. A great advance in the systematic study of the properties of clusters occurred when Abell compiled an extensive, statistically complete catalog of rich clusters of galaxies (Abell, 1958). For the last quarter century, this catalog has been the most important resource in the study of galaxy clusters.

However, despite of the name the galaxies are not the dominant componet in a cluster of galaxies. The Intra Cluster Medium (ICM) is a very hot gas present in the space between galaxies. This plasma is heated to temperatures of between roughly $10^7$-$10^8$ K and consists mainly of ionised hydrogen and helium, containing most of the baryonic material in the cluster. The ICM strongly emits X-ray radiation and its total mass provide about 15% of the total cluster mass. The total cluster mass is dominated by an unseen and non-baryonic component, the dark matter, whose presence can be inferred from gravitational effects on visible matter.

Today we have a clear cosmic scenario with galaxy clusters as an integral part of the large-scale structure of the Universe. They are the largest matter aggregates which have collapsed under their own gravity and are closely approaching a dynamical equilibrium. Quantitatively, the mass composition of a cluster is roughly subdivided as follow: 80% dark matter, 15% hot baryons in the ICM, and 5% cool baryons in stars and galaxies.

Two features of clusters make them uniquely useful tracers of cosmic evolution. First, clusters are the biggest objects whose masses can be reliably measured because they are the largest objects to have undergone gravitational relaxation and entered into virial equilibrium. Mass measurements of nearby clusters can therefore be used to trace the large scale structure in the Universe on scales of $10^{14}$-$10^{15} M_\odot$, and comparisons of the present-day cluster mass distribution with the mass distribution at earlier times can be used to measure the rate of structure formation, placing important constraints on cosmological models. Second, clusters are essentially "closed boxes", despite the enormous energy input associated with supernovae and active galactic nuclei, because the gravitational potential wells of clusters are deep enough to retain all the diffuse baryons. The baryonic component of clusters therefore contains a wealth of information about the processes associated with galaxy formation, including the efficiency with which baryons are converted into stars and the effects of the resulting feedback processes on galaxy formation. In this sense clusters of galaxies are unique laboratories for studying the baryon history along the cosmic time.

In the following section I expound the main physical properties of galaxy clusters and the observables I use in the rest of my Thesis.

## 1.2 OPTICAL PROPERTIES

CLUSTER GALAXY POPULATION    Early-type galaxies are the dominant population in the center of all galaxy cluster. In a color-magnitude diagram early-type galaxies are arranged according to a well-defined relation, the so called red-sequence. Baum (1959) first discovered color-magnitude relation in cluster member early-type galaxies, observing that less luminous elliptical galaxies are bluer with respect to the more massive ones. The dispersion, the color and the slope of the red-sequence are used to study the evolution of high redshift clusters (Bower et al., 1992; Aragón-Salamanca et al., 1993; Stanford et al., 1998; Gladders et al., 1998; Mei et al., 2006). Yee, Gladders & López-Cruz (1999) showed how it is possible to use early-type galaxies red-sequence to detect high redshift clusters and to measure at the same time their photometric redshifts. The Red-sequence Cluster Survey (RCS) is a survey of optically selected clusters identified with this technique (Gladders & Yee, 2000). In Chapter 4 I discuss the X-ray properties of optically selected clusters of galaxies in the RCS observed with the *Chandra* satellite, at redshifts $0.6 < z < 1.2$.

GALAXY VELOCITY DISPERSION    Once a cluster has been optically identified as a galaxy overdensity in an optical or infrared image, obtaining the radial velocities $v_r$ of the cluster galaxies from their redshifts is crucial. The knowledge of the redshift distribution helps in identifying the true cluster members and then mitigating projection effects, and in measuring the cluster's mass. Because the velocity distribution of a relaxed cluster's galaxies is expected to be gaussian in velocity space, galaxies with velocities falling well outside the best-fitting Gaussian evelope are unlikely to be cluster members and are generally discarded. Fitting the velocity distribution with a Gaussian $\exp[-(v_r - \bar{v}_r)^2/2\sigma_{1D}^2]$ provides a one-dimensional velocity dispersion $\sigma_{1D}$ for the cluster galaxies, which depends on the gravitational potential well of the cluster and it is connected to the total cluster mass through the virial theorem. If the velocity distribution of a cluster candidate is far from Gaussian, then it is probably not a real cluster but rather a chance superposition. Quantitatively, the velocity dispersion of cluster galaxies is of the order of $\sim 10^3$km/s.

Zwicky (1933, 1937) measured a cluster's velocity dispersion for the first time, finding $\sigma_{1D} \sim 700$km/s for the Coma cluster. He correctly concluded that, given its radius, the Coma cluster requires a large amount of unseen matter to bind its fast moving galaxies, providing the first clear evidence of Dark Matter presence in the Universe.

Detailed information on the spatial distribution of galaxy velocities is of great help in measuring the masses of large, nearby clusters but similar information is very difficult to obtain for distant clusters, for which different mass measuring techniques are needed.

GRAVITATIONAL LENSING  Gravitational lensing is one of the most attractive methods to directly study the mass distribution in the Universe on different scales, regardless of its componend and dynamical state (Peacock & Schneider 2006). As a matter of fact, this phenomenon only depends on the gravitational potential well, giving the opportunity to study directly the matter distribution in clusters. Interesting review about lensing can be found in Tyson et al. (1990), Kaiser & Squires (1993), Hoekstra et al. (1998), Mellier (1999), Bartelmann & Schneider (2001). The effective space-time curvature is ruled by the effective refraction index $n \equiv 1 - 2\phi/c^2$, where $\phi$ is the gravitational potential due to the mass distribution. The effective refraction index determines the deflection angle of the light passing throughout the "gravitational lens". This leads to the distortion of the images of background galaxies stretched tangentially to the gradient of the gravitational potential, and to the formation of multiple images and giant luminous arcs (Schneider et al 1992). Observations of these features allow to reconstruct the distribution of the mass responsible for the deflection, and in particular provide accurate estimates of the total mass, irrespective of its dynamical status. However, this technique should be applied carefully, because gravitational lensing are plagued by projection effects, since all the mass concentrated along the line of sight contribute to the distortion of background galaxies images. In Section 4.6 we present the mass reconstruction of the cluster RCS0224-0002 at $z = 0.773$ from strong lensing features, and we compare it with the mass obtained from X-ray observations.

## 1.3  X-RAY PROPERTIES

In 1966, X-ray emission was detected from the region around the galaxy M87 in the center of the Virgo cluster (Byram et al., 1966; Bradt et al., 1967). In fact, M87 was the first object outside of our galaxy to be identified as a source of astronomical X-ray emission. Five years later, X-ray sources were also detected in the directions of the Coma and Perseus clusters (Fritz et al., 1971; Gursky et al., 1971a, b; Meekins et al., 1971). Since these are three of the nearest rich clusters, it was suggested that clusters of galaxies might generally be X-ray sources (Cavaliere et al., 1971).

The launch of the Uhuru X-ray astronomy satellite (Giacconi et al., 1972) indicated that many clusters are bright X-ray sources, with luminosities typically in the range of $10^{43}$-$10^{45}$ erg/s. The X-ray sources associated with clusters were found to be spatially extended;

their sizes were comparable to the size of the galaxy distribution in the clusters (Kellogg et al., 1972; Forman et al., 1972). Unlike other bright X-ray sources but consistent with their spatial extents, cluster X-ray sources did not vary temporally in their brightness (Elvis, 1976). Although several emission mechanisms were proposed, the X-ray spectra of clusters were most consistent with thermal bremsstrahlung from hot gas.

### 1.3.1 *Thermal Bremsstrahlung*

Felten et al. (1966) first suggested that the X-ray emission from clusters was due to hot diffuse gas, the Intra Cluster Medium (ICM), that is in thermal equilibrium within the gravitational potential well. Since the gas shares the same dynamics of the member galaxies, it would have a temperature such that the typical atomic velocity is similar to the velocity of the galaxies in the cluster. That is

$$k_B T \simeq \mu m_p \sigma_{1D}^2 \simeq \left( \frac{\sigma_{1D}}{10^3 \text{km/s}} \right)^2 \text{keV},$$ (1.1)

where $m_p$ is the proton mass, $\mu$ is the mean molecular weight ($\mu = 0.6$ for a primordial composition with 76% of hydrogen), and $\sigma_{1D}$ is the line-of-sight velocity dispersion of galaxies in the cluster. At such temperatures, the primary emission process for a gas composed mainly of hydrogen is the thermal bremsstrahlung emission. The emissivity at a frequency $\nu$ of a ion on charge $Z$ in a plasma with an electron temperature $T$ is given by (Ribicky & Lighmann, 1979)

$$\epsilon_\nu = \frac{2^5 \pi e^6}{3mc^3} \left( \frac{2\pi}{3 k_B m} \right)^{1/2} T^{-1/2} z^2 n_e n_i e^{-h\nu/k_B T} g_{ff}(Z, T, \nu),$$ (1.2)

where $n_i$ and $n_e$ are the number density of ions and electrons, respectively. The emissivity is defined as the emitted energy per unit time, frequency $\nu$ and volume $V$,

$$\epsilon_\nu \equiv \frac{dL}{dV d\nu}.$$ (1.3)

The Gaunt factor $g_{ff}(Z, T, \nu)$ corrects for quantum mechanics effects and for the effect of distant collisions, and is a slowly varying function of frequency and temperature (Karzas & Latter 1961; Kellogg et al. 1975). Integrating Equation 1.2 with respect to the energy and the volume, we obtain the total X-ray luminosity

$$L_X = \int_V \left( \frac{\rho_{gas}}{\mu m_p} \right)^2 \Lambda(T) dV,$$ (1.4)

where we assumed $n_i \sim n_e$ and $\Lambda(T)$ is the cooling function, which takes into account all the other terms of Equation 1.2, included metal lines emission (see 1.3.2), and in first order approximation $\Lambda(T) \propto T^{1/2}$. The total X-ray luminosity falls typically in the range of $10^{43}$-$10^{45}$ erg/s and allows to detect galaxy clusters as extended X-ray sources even at high redshift.

For relaxed clusters the ICM gas is assumed to be in hydrostatic equilibrium. In the further assumption of spherical symmetry we have

$$\frac{dp}{dr} = -\frac{GM(<r)\rho_{gas}(r)}{r^2}, \tag{1.5}$$

where $p$ is the pressure, $\rho_{gas}(R)$ is the gas density as function of the radius $R$ and $M(R)$ is the total mass within $R$. Using the equation of state of ideal gas $p = \rho_{gas}k_B T/m_p$, we can write

$$M(<r) = -\frac{k_B Tr}{G\mu m_p} \left( \frac{d\log\rho_{gas}}{d\log r} + \frac{d\log T}{d\log r} \right). \tag{1.6}$$

This equation shows how the quantities $\rho_{gas}$ and $T$, which can be measured with X-ray observations, are directly related to the total cluster mass. A common description of the gas density profile is the $\beta$-model

$$\rho_{gas}(r) = \rho_{gas,0} \left[ 1 + \left( \frac{r}{r_c} \right)^2 \right]^{-3\beta_m/2}, \tag{1.7}$$

which was introduced by Cavaliere & Fusco-Femiano (1976) to describe an isothermal gas in hydrostatic equilibrium within the gravitational potential well where the dark matter density profile is given by a King profile (Sarazin 1988). In Equation 1.7, $r_c$ is the core radius, $\rho_{gas,0} \equiv \rho_{gas}(r=0)$, and $\beta_m$ is the ratio of ther kinetic energy of galaxies and the thermal energy of the gas,

$$\beta_m \equiv \frac{\mu m_p \sigma_{1D}^2}{k_B T}. \tag{1.8}$$

Since the X-ray luminosity is proportional to the square of the gas density (see Equation 1.4), the surface brightness, $S(r)$, of an ICM whose gas density is described by Equation 1.7 is still a King profile with exponent $\beta = 3\beta_m - 0.5$,

$$S(r) \propto \left[ 1 + \left( \frac{r}{r_c} \right)^2 \right]^{-\beta} \propto \left[ 1 + \left( \frac{r}{r_c} \right)^2 \right]^{-3\beta_m+0.5}. \tag{1.9}$$

These theoretical concepts are widely employed in this thesis. In particular, in Chapters 3 and 4 the thermal bremsstrahlung emission model, Equations 1.2 and 1.4, is used to fit the X-ray spectra of galaxy clusters to measure ICM average temperature and to derive total X-ray luminosity; in Chapter 2 the surface brightness of the X-ray sources are fitted with a King profile, Equation 1.9, to distinguish between extended X-ray sources (namely galaxy clusters) and X-ray point sources (Active Galactic Nuclei (AGNs) and star forming galaxies).

### 1.3.2 *Metal Emission Lines*

In 1976, X-ray line emission from iron was detected from the Perseus cluster of galaxies (Mitchell et al., 1976), and shortly thereafter from Coma and Virgo as well (Serlemitsos et al., 1977). The emission mechanism for this line is thermal, and its detection confirmed the thermal interpretation of cluster X-ray sources. Altogether, the X-ray spectrum of a galaxy cluster is a continuum thermal bremsstrahlung with metal emission lines.

Emission lines of heavy elements are another important feature of the X-ray spectra of clusters of galaxies. For clusters with temperature greater than 3.0 keV the contribution to the luminosity is not relevant, since at these temperatures most of the atoms are completely ionized. Instead, line emission is more important for low temperature clusters, in particular below 2.0 keV.

Since most of the elements in the ICM are completely ionized, only emission lines corresponding to K-shell or L-shell transitions of highly ionized heavy elements (e. g. iron, oxygen and silicon) with one ore two orbiting electrons can be observed. For high redshift clusters, because of the lower Signal to Noise Ratio (SNR) of the spectra, the only detectable emission lines are the iron K-shell ones at $\sim 6.7 - 6.9$ keV. The presence of emission lines in an X-ray spectrum of a cluster is an important aspect, since it allows to study the metal production in clusters and their diffusion in the ICM (Rasmussen & Ponman 2007).

For galaxy clusters analyzed in Chapters Chapters 3 and 4, the metal content of the ICM is derived from the X-ray spectrum emission lines; in Chapter 3 metal emission lines are also used to constrain the redshift directly from the X-ray spectrum for a sample of a previously undetected galaxy clusters.

### 1.4 SUNYAEV-ZEL'DOVICH EFFECT

The hot gas in galaxy clusters can also be detected in the microwave, by observing the induced distortion by the ICM on the Cosmic Microwave Background (CMB). The CMB spectrum is nearly a perfect black body spectum with a temperature of about 2.7 K (Mather et al 1990). Weymann (1965, 1966) investigated the possibility of distortions of the residual black body due to hot, fully ionized gas, and Sunyaev & Zel'dovich (1970, 1972) predicted that ICM electrons would indeed produce such a distortion into the CMB spectrum. This phenomenon is called Sunyaev-Zel'dovich (SZ) effect.

Two decades after this prediction there were only a few marginal detections (Birkinshaw et al., 1991), but many clusters were detected at high significance in the ensuing decade (Birkinshaw, 1999; Carlstrom et al., 2000). With multiple new and highly capable SZ instruments coming on line in the next few years, another quantum leap in this area is poised to happen, enabling wide-field cosmological studies of clusters to extend through much of the observable universe (Carlstrom et al., 2002). A number of recent reviews elucidate the details of the S-Z effect (e.g., Birkinshaw, 1999; Carlstrom et al., 2002). Here I summarize only a few fundamentals.

To lowest order, the shape of the distorted spectrum depends on a single parameter proportional to the product of the probability that a photon passing through the cluster will suffer Compton scatter and the typical amount of energy gained by a scattered photon gains:

$$y = \int \frac{k_B T}{m_e c^2} n_e \sigma_T \, dl \ , \tag{1.10}$$

where $\sigma_T$ is the Thomson cross-section and the integral is over a line of sight through the cluster. Because the optical depth of the cluster is small, the change in microwave intensity at any frequency is linearly proportional to $y \ll 1$, with reduced intensity at long wavelengths and enhanced intensity at short wavelengths. A cluster's motion with respect to the microwave background produces additional distortion, known as the kinetic SZ effect.

Cosmological applications of the thermal S-Z effect in clusters benefit greatly from the fact that the effect is independent of distance, unlike optical and X-ray surface brightness. Thus, a dedicated S-Z cluster survey may efficiently find clusters out to arbitrarily high redshifts. Because not all these clusters will be well resolved, the measure will be a volume average of the distortion parameter:

$$Y = \int y \, dA \propto \int n_e T \, dV \ , \tag{1.11}$$

where the first integral is over a cluster's projected surface area and the second is over its volume. The $Y$ parameter therefore tells us the total thermal energy of the electrons, from which one easily derives the total gas mass times its mass-weighted temperature within a given region of space.

The impressive power of the S-Z effect for finding distant clusters also has a significant drawback, namely sky confusion owing to projection effects. Along any line of sight through the entire observable universe, the probability of passing within the virial radius of a cluster or group of galaxies is of order unity (e.g., Voit et al., 2001). Because a cluster's S-Z distortion does not fade with distance, many of the objects in a highly sensitive S-Z survey will therefore significantly overlap.

## 1.5 CLUSTER SURVEYS

Determining the evolution of the space density of clusters requires counting the number of clusters of a given mass per unit volume at different redshifts. Therefore, three essential tools are required for a cosmological test: first, an efficient method to find clusters over a wide redshift range; second, an observable estimator of the cluster mass; third, a method to compute the selection function or equivalently the survey volume within which clusters are found.

MASS FUNCTION   The shape and evolution of the cluster mass function, as derived from X-ray survey of Clusters of Galaxies, has been used for many years as a method to constrain cosmological parameters (see Oukbir & Blanchard 1992, Eke et al. 1998, Bahcall et al. 1999, Borgani et al. 2001, Henry 2004). At the same time, X-ray observations of clusters of galaxies over a significant range of redshifts have been used to investigate the chemical and thermo-dynamical evolution of the X-ray emitting Intra Cluster Medium (ICM, see Ettori et al. 2004; Balestra et al. 2007; Maughan et al. 2008), whose properties depend on the interaction between the diffuse hot baryons and the nuclear or star forming activity in the cluster galaxies. These two approaches can be pursued at the same time, provided that the X-ray data have a good spatial and spectral resolution, and that the cluster catalogs are drawn from samples with a well defined completeness. In this respect, X-ray surveys of clusters of galaxies represent a key tool for cosmology and large scale structure physics. In order to build statistically complete cluster catalogues, a wide and deep coverage of the X-ray sky is mandatory.

Press & Schechter (1974) presented an analytical formula to describe the mass function of virialized haloes as function of the cosmic epoch,

$$
\begin{aligned}
\frac{dn(M,z)}{dM} &= \frac{2}{V_M}\frac{\partial p_{>\delta_c}(M,z)}{\partial M} \\
&= \sqrt{\frac{2}{\pi}}\frac{\bar{\rho}}{M^2}\frac{\delta_c}{\sigma_M(z)}\left|\frac{d\log\sigma_M(z)}{d\log M}\right|\exp\left(-\frac{\delta_c^2}{2\sigma_M(z)^2}\right).
\end{aligned}
$$ (1.12)

This equation shows the reason for which the mass function of galaxy clusters is a powerful probe of cosmological models. Cosmological parameters enter in eqz. 26 through the mass variance $\sigma_M$, which depends on the power spectrum and on the cosmological density parameters, thorugh the linear perturbation growth factor, and, to a lesser degree, thorugh the critical density contrast $\delta_c$. Taking this expression in the limit of massive objects (i.e. rich galaxy clusters), the shape of the mass function is dominated by the exponential tail. This implies that the mass function becomes exponentially sensitive to the choice of the cosmological parameters. In other words, a reliable observational determination of the mass function of rich clusters would allow us to place tight constraints on cosmological parameters.

### 1.5.1 *Optical Survey*

Optical survey of galaxy clusters are historically the main method to compile large sample of clusters. Simply, the detection of galaxy clusters in optical images is based on the detection of over-densities in the projected galaxies distribution. Clusters are characterized by their "richness", i.e. the membership of a catalogue of clusters is determined by the criteria used to define a "rich cluster". These criteria must specify the threshold in the surface number density enanchement and the linear or angular scale of the enhancement. As clustering exists on a very wide range of angular and intesity scales (Peebles, 1974), it is not possible to give a unique and unambiguous definition of a rich cluster.

Abell (1958) provided the firs extensive, statistically complete sample of galaxy clusters. Abell's criteria require: that the cluster contains at least 50 galaxies brighter than a magnitude limit; that these galaxies are contained within a $3h_{50}^{-1}$Mpc radius. The Abell catalogue contains 2712 clusters, of which 1682 satisfy all these criteria. The other 1030 were discovered during the search and were included to provide a more extensive list. The Abell catalogue gives estimates of the cluster center position, distance, and the richness of the clusters.

For the Zwicky et al. (1968) catalogue the criteria were as follows: the size of the cluster was determined by the isopleth where the galaxy surface density fell to twice the local background density; the isopleth has to contain at least 50 galaxies brighter than a magnitude limit. No distance limits were specified. Even if the study of the galaxy surface density distribution provides a more objective criterion for cluster detection, the completeness of the catalogue is strongly redshift dependent. As a matter of fact, a rich cluster at higher redshift has a lower over-density in the galaxy surface density distribution, with respect to the same cluster at lower redshift.

The main issue in optical survey is due to the projection effects in the selection of the clusters candidates. Infact, projection effects in galaxies distribution can complicate the detection of over-density associated with clusters. Filamentary structures along the line of sight can mimic a moderately rich cluster when projected onto the plane of the sky. In addition, the background galaxy distribution, against which two dimensional over-densities are selected, is far from uniform. Therefore, because of contamination of foreground galaxies and projection effects, searching for high redshift cluster is difficult and undergoes a lot of systematic effects difficult to quantify.

Beside projection effect problem, a common feature of all these methods of cluster identification is that they classify clusters according to richness, which generally have a loose relation with the cluster mass. This represents a serious limitation for any cosmological application, which requires the observable, on which the cluster selection is based, to be a reliable proxy of the cluster mass. This also makes difficult to determine the exact survey volume that is essential for cosmological tests.

RED-SEQUENCE CLUSTER SURVEY  The Red-sequence Cluster Survey (RCS) is a new optical survey of cluster of galaxies designed specifically to unveil a fairly large number of clusters out to $z \sim 1.4$. This identification technique uese the color-magnitude relation of early-type galaxies described in Section 1.2. The red-sequence represents the reddest galaxies of a set of galaxies at the same redshift, and is easilly detectable in the color-magnitude diagram since early-type galaxies dominate the bright end of luminosity function in clusters. Therefore, selecting galaxies around the red-sequence, it is possible to exclude most of background and foreground galaxies, which are redder and bluer respectively. So, the criterion to find cluster candidates is based on identifying galaxy over-densities in the four-dimension space given by color, magnitude and sky coordinate.

The RCS is a 100 square degree imaging survey obtained using mosaic CCD cameras on 4 m-class telescopes. The survey area is equally divided between the northern hemisphere, observed with the CFHT telescope, and the southern hemisphere, observed with the CTIO. To optimize the sensitivity to high redshift cluster detection, passband filter $z'$ (920 nm) and $R_C$ (650 nm) are used. This choice assures that the wavelength of the break ($\sim$ 400 nm) in the optical spectrum of galaxies falls between the two passbands for galaxies at $0.6 < z < 1.4$. Exposure times are chosen to detect galaxies down to magnitude 25.2 in $R_C$ band, and 23.6 in $z'$.

Afterwards, images in the two bands are elaborated with the algorithm PPP (Yee 1991) for galaxy detection and photometric analysis. Finally, the detection of cluster candidates is performed searching for galaxy over-densities in the four-dimension space given by color, magnitude and sky coordinate.

Therefore, galaxy clusters in RCS are detected through bright early-type galaxies concentrations. No other dynamical information guarantees that such systems are gravitationally bound. In this respect, observations of the X-ray properties of these objects can add crucial information to understand their nature and their importance in the cosmological context.

In Chapter 4 a detailed analysis of the X-ray properties of a subsample of 11 RCS clusters is described. We intended to assess the evolutionary stage of optically selected high-z clusters of galaxies, performing a spectral analysis of the diffuse emission from their ICM. We also investigated the distribution of AGN in their surroundings.

### 1.5.2 X-ray Survey

X-ray surveys offer an efficient means of constructing samples of galaxy clusters out to cosmologically interesting redshifts. First, the X-ray selection has the advantage of revealing physically-bound systems, because diffuse emission from a hot ICM is the direct manifestation of the existence of a potential well within which the gas is in dynamical equilibrium with the galaxies and the dark matter. Second, the X-ray luminosity is well correlated with the cluster mass Third, the X-ray emissivity is proportional to the square of the gas density , hence cluster emission is more concentrared than the optical bidimensional galaxy distribution. In combination with the relatively low surface density of X-ray sources, this property makes clusters high contrast objects in the X-ray sky, and alleviates problems due to the projection effects that affect optical selection. Fourth, an inherent fundamental advantage of X-ray selection is the ability to define flux-limited samples with well-understood selection functions. This leads to a simple evaluation of the survey volume and therefore to a straightforward computation of space densities.

To date, there are no X–ray surveys which are not based on previous ROSAT surveys and evantual optical follow up to derive redshift and thus luminosities. The most recent constraints on cosmological parameters from clusters, are based on the *Chandra* follow up of 400 deg$^2$ ROSAT serendipitous survey and of the All-Sky Survey (Vikhlinin et al. 2009a; 2009b; Manz et al. 2009). Brand new X–ray surveys of clusters of galaxies in the *Chandra* and *XMM-Newton* era are based on the compilation of serendipitous medium and deep–exposure extragalactic pointings not associated to previously known X–ray clusters (Boschin 2002; Barkhouse et al. 2006) or a program like XMM–Newton LSS (Pierre et al. 2008). The impact of the ongoing *Chandra* and *XMM-Newton* surveys is limited, despite the good imaging and spectroscopic quality. In the case of *Chandra*, the process of assembling a wide and deep survey is extremely slow due to the small Field of View and the low collecting area. In the case of *XMM-Newton* the identification of extended sources, in particular at medium and high redshift, is hampered by the size of the Point Spread Function, whose Half Energy Width is $15''$ at the aimpoint, and its degradation as a function of the off-axis angle. In addition, the relatively high particle background makes more difficult the detection of low surface brightness sources. In conclusion, both telescopes are not efficient for obtaining a wide area X-ray survey of clusters of galaxies.

**SKY COVERAGE AND FLUX LIMIT** Once the survey flux limit and the sky coverage are defined one can compute the maximum search volume, $V_{max}$, within which a cluster of a given luminosity is found in that survey:

$$V_{max} = \int_0^{z_{max}} S[f(L,z)] \left(\frac{d_L(z)}{1+z}\right)^2 \frac{c\,dz}{H(z)}. \tag{1.13}$$

Here $S(f)$ is the survey sky coverage, which depends on the flux $f = L/(4\pi d_L^2)$, $d_L(z)$ is the luminosity distance, and $H(z)$ is the Hubble constant at $z$. $z_{max}$ is defined as the maximum redshift out to which the flux of an object of luminosity $L$ lies above the flux limit.

## 1.6 CLUSTER SCALING RELATION

The simplest model to explain the physics of the ICM is based on the assumption that gravity only determines the thermodynamical properties of the hot diffuse gas (Kaiser, 1986). Since gravity does not have a preferred scale, we expect clusters of different sizes to be the scaled version of each other as long as gravity only determines the ICM evolution. This is the reason why the ICM model based on the effect of gravity only is said to be self-similar.

If we define $M_{\Delta_c}$ as the mass contained within the radius $R_{\Delta_c}$, which describes the sphere within which the cluster overdensity is $\Delta_c$ times the critical density, then

$$M_{\Delta_c} \propto \rho_c(z)\Delta_c R_{\Delta_c}^3. \tag{1.14}$$

Here $\rho_c(z)$ is the critical density of the universe which scales with redshift as

$$\rho_c(z) = \rho_{c,0}E^2(z), \tag{1.15}$$

where

$$\rho_{c,0} = \frac{3H_z^2}{8\pi G}, \tag{1.16}$$

$$E(z) = [(1+z)^4\Omega_R + (1+z)^3\Omega_m + (1+z)^2(1-\Omega_0) + (1+z)^{3(1+w)}\Omega_{DE}]^{1/2}, \tag{1.17}$$

$G$ is the gravitational constant, and $H_0$ is the Hubble constant. In Equation 1.17, $\Omega_R$ is the density parameter contributed by relativistic matter, $\Omega_m$ by non-relativistic matter, and $\Omega_{DE}$ by Dark Energy with equation of state $p = w\rho c^2$ (if the Dark Energy term is provided by cosmological constant $w = -1$).

On the other hand, the cluster size $R_{\Delta_c}$ scales with $z$ and $M_{\Delta_c}$ as

$$R_{\Delta_c} \propto M_{\Delta_c}^{1/3} E^{-2/3}(z). \tag{1.18}$$

Therefore, assuming hydrostatic equilibrium, the cluster mass scales with the temperature $T$ as

$$M_{\Delta_c} \propto T^{3/2} E^{-1}(z). \tag{1.19}$$

Further assuming that the gas distribution traces the dark matter distribution, $\rho_{\text{gas}}(r) \propto \rho_{DM}(r)$, the X-ray luminosity for pure thermal bremsstrahlung emission in Equation 1.4 scales as

$$L_X \propto M_{\Delta_c} \rho_c T^{1/2} \propto T^2 E(z). \tag{1.20}$$

As for the CMB intensity decrement due to the thermal SZ effect we have

$$\Delta S \propto \int y(\theta) d\Omega \propto d_A^{-2} \int T n_e d^3 r \propto d_A^{-2} T^{5/2} E^{-1}(z), \tag{1.21}$$

where $y$ is the Comptonization parameter, $d_A$ is the angular size distance and $n_e$ is the electron number density. We can also write $\Delta S$ in a different way to get the explicit dependence on $y_0$:

$$\Delta S \propto y_0 d_A^{-2} \int d\Omega \propto y_0 d_A^{-2} M^{2/3} E^{-4/3}(z) \propto y_0 d_A^{-2} T E^{-2}(z). \tag{1.22}$$

In this way, we obtain the following scalings for the central value of the Comptonization parameter:

$$y_0 \propto T^{3/2} E(z) \propto L_X^{3/4} E^{1/4}(z). \tag{1.23}$$

Eqs.(1.19), (1.20) and (1.23) are unique predictions for the scaling relations among ICM physical quantities and, in principle, they provide a way to relate the cluster masses to observables at different redshifts. As we shall discuss in the following, deviations with respect to these relations witness the presence of more complex physical processes, beyond gravitational dynamics only, which affect the thermodynamical properties of the diffuse baryons and, therefore, the relation between observables and cluster masses.

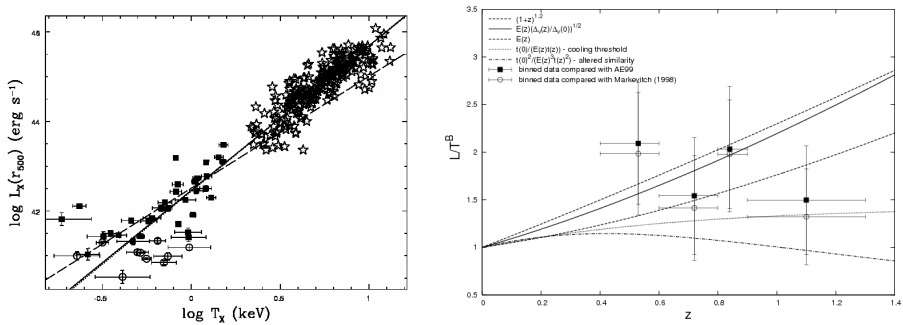

Figure 1: Left panel: the $L_X - T_X$ relation for nearby clusters and groups (from Osmond & Ponman, 2004). The star symbols are the for the sample of clusters, with temperature measured from ASCA, while the filled squares and open circles are for a sample of groups, also with ASCA temperatures. Right panel: the evolution of the $L_X - T_X$ relation, normalized to the local relation Maughan et al. (from 2006), using *Chandra* temperatures of clusters at $z > 0.4$.

### 1.6.1 *Phenomenological scaling relations*

The relation between X–ray luminosity and temperature of nearby clusters is considered as one of the most robust observational facts against the self–similar model of the ICM. A number of observational determinations now exist, pointing toward a relation $L_X \propto T^\alpha$, with $\alpha \simeq 2.5$-3 Xue & Wu (e. g. 2000), possibly flattening towards the self–similar scaling only for the very hot systems with $T \gtrsim 10$ keV (Allen & Fabian, 1998). While in general the scatter around the best–fitting relation is non negligible, it has been shown to be significantly reduced after excising the contribution to the luminosity from the cluster cooling regions Markevitch (1998) or by removing from the sample clusters with evidence of cooling flows Arnaud & Evrard (1999). As for the behaviour of this relation at the scale of groups, $T \lesssim 1$ keV, the emerging picture now is that it lies on the extension of the $L_X$–$T$ relation of clusters, with no evidence for a steepening Mulchaey & Zabludoff (1998), although with a significant increase of the scatter Osmond & Ponman (2004), possibly caused by a larger diversity of the groups population when compared to the cluster population. This result is reported in the left panel of Figure 1 (from Osmond & Ponman, 2004), which shows the $L_X - T_X$ relation for a set of clusters with measured ASCA temperatures and for a set of groups.

As for the evolution of the $L_X$–$T$ relation, a number of analyses have been performed, using *Chandra* (Holden et al., 2002; Vikhlinin et al., 2002; Ettori et al., 2004; Maughan et al., 2006) and *XMM-Newton* (Kotov & Vikhlinin, 2005; Lumb et al., 2004) data. Although some differences exist between the results obtained from different authors, such differences are most likely due to the convention adopted for the radii within which luminosity and temperature are estimated. In general, the emerging picture is that clusters at high redshift are relatively brighter, at fixed temperature. The resulting evolution for a cosmology with $\Omega_m = 0.3$ and $\Omega_\Lambda = 0.7$ is consistent with the predictions of the self–similar scaling, although the slope of the high–$z$ $L_X$–$T$ relation is steeper than predicted by self–similar scaling, in agreement with results for nearby clusters. The left panel of Figure 1 shows the evolution of the $L_X$–$T$ relation from Maughan et al. (2006), where *Chandra* and *XMM-Newton* observations of 11 clusters with redshift $0.6 < z < 1.0$ were analyzed. The vertical axis reports the quantity $L_X/T^B$, where $B$ is the slope of the local relation. Quite apparently, distant clusters are systematically brighter relatively to the local ones. However, the uncertainties are still large enough not to allow the determination of a precise redshift dependence of the $L_X$–$T$ normalization.

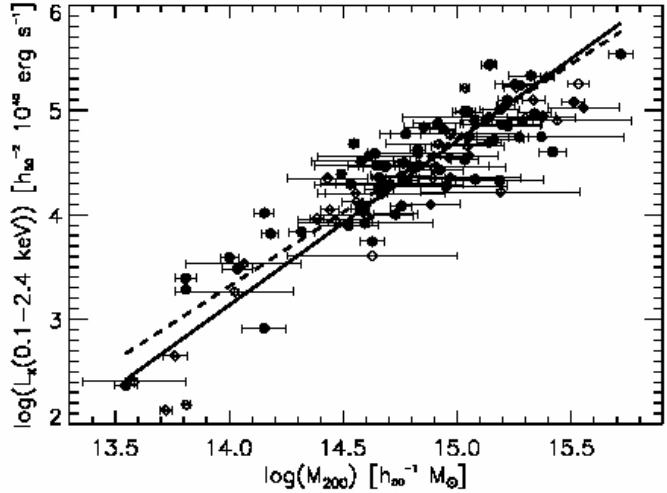

Figure 2: The $L_X - M$ relation for nearby clusters (from Reiprich & Böhringer (2002)). *X*–ray luminosities are from the RASS, while masses are estimated using ASCA temperatures and assuming hydrostatic equilibrium for isothermal gas.

As for the relation between X–ray luminosity and mass, its first calibration has been presented in Reiprich & Böhringer (2002), for a sample of bright clusters extracted from the ROSAT All Sky Survey (RASS). In their analysis, these authors derived masses by using temperatures derived from ASCA observations and applying the equation of hydrostatic equilibrium for an isothermal $\beta$-model. The resulting $L_X - M$ relation is shown in Figure 2. From the one hand, this relation demonstrates that a well defined relation between X–ray luminosity and mass indeed exist, although with some scatter, thus confirming that $L_X$ can indeed be used as a proxy of the cluster mass. From the other hand, the slope of the relation is found to be steeper than the self-similar scaling, thus consistent with the observed $L_X - T_X$ relation.

# ANALYSIS OF SWIFT-XRT DATA AND SURVEY CATALOGUE

In this Chapter I introduce the new Swift-XRT Cluster Survey (SXCS). This survey is obtained from the archive of the Swift X-Ray Telescope (XRT), which contains hundreds of serendipitous and long exposure X-ray pointings. First, I show the main characteristics of the instrument and of the survey itself. Then I describe the new detection algorithm I developed for this survey, and the new X-ray image simulator algorithm I developed to test the capabilities of the detection algorithm itself. Finally, I show the catalogue of extended sources and its properties. Within the whole SXCS catalogue I selected the brightest object, for which in Chapter 3 I will present the detailed spectral analysis. Furthermore, in Chapter 5 I will use a modified version of the image simulator described in this Chapter adapted to the technical characteristics of WFXT. This work is done in collaboration with A. Moretti, G. Tagliaferri and S. Campana of the Osservatorio Astronomico di Brera, and the survey catalogue willl be published in Moretti, et al. (2010).

## 2.1 THE SWIFT MISSION

The Swift mission (Burrows et al., 2005) is devoted to the study of the Gamma Ray Bursts (GRBs) Gehrels et al., 2009). GRBs are events that occur approximately once per day and are randomly distributed in the sky. Swift is equipped with three telescope: the Burst Alert Telescope (BAT), a $\gamma$-ray telescope to detect the initial position of the GRBs, the X-Ray Telescope (XRT) and the UltraViolet/Optical Telescope (UVOT) to study the long afterglow of the GRB. So in the XRT archive hundreds of serendipitous and long exposure X-ray pointings are available to build an unbiased survey, since the targets are not previously known X-ray sources.

XRT is a small instrument with a low collecting area (about one fifth of that of *Chandra* at 1.0 keV) with two characteristics that make it a sort of ideal prototype for future X-ray cluster surveys: a low background, about one tenth of that of *Chandra* (Pagani et al., 2007), and a constant PSF across the Field of View (FoV) (Moretti et al 2005). Since XRT and the Swift mission itself were not conceived to identify and to study extended sources and cluster of galaxies, some adjustments are necessary to achieve a more refined data reduction, which is essential for an accurate spatial and spetral analysis of the extended sources.

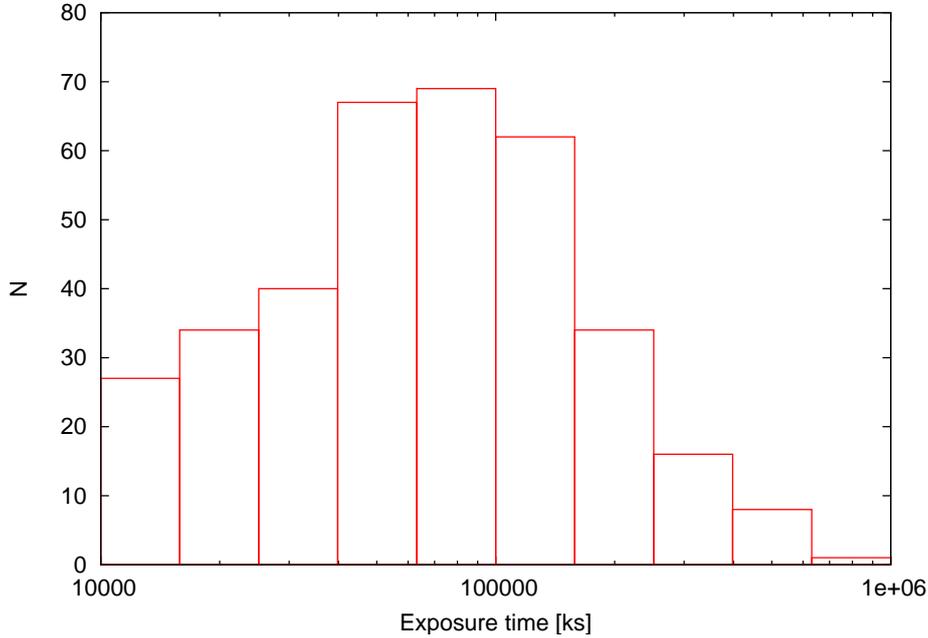

Figure 3: Distribution of the exposure times for the 390 fields of the SXCS.

## 2.2 SURVEY OVERVIEW

The Swift-XRT Cluster Survey (SXCS) is based on the 390 archived XRT fields at the date of February 2009. The distribution of the exposure times of the XRT fields, spanning from $10^4$ s to $10^6$ s, is shown in Figure 3.

Galactic fields are not excluded *a priori*. In fact, as can be seen in Figure 4 from the distribution of Galactic hydrogen column density, $N_H$, there are field with large $N_H$ values, although more than 90% of fields have $N_H < 0.5 \times 10^{22} \mathrm{cm}^{-2}$. However, the two fields of GRB060116 and GRB060421 are excluded from the survey, since their lines of sight pass accidentally throughout the Orion Nebula and the Cave Nebula, respectively. The X-ray emission of the two nebulae yields to a extremely high X-ray foreground that makes impossible the detection of extragalactic sources.

The obvious limit of the SXCS survey is the relatively high flux limits due to the small effective area. Nevertheless, the total solid angle is about 60 deg$^2$, and the expected sky coverage at $10^{-14}$ erg s$^{-1}$ cm$^{-2}$ is more than 20 deg$^2$ (see Section 2.9). This makes the SXCS competitive with present day X-ray clusters surveys with *Chandra* and *XMM-Newton*.

The realization of a survey of groups and clusters of galaxies consists of two basic steps: the source detection and the separation of the detected sources between point sources and extended sources. The source detection is performed with `wavdetect` (see Section 2.5) in the 0.5-2.0 keV band, where galaxy clusters have an higher SNR and where XRT has the maximum effective area. The separation between point sources and extended sources is performed with a newly developed Two Dimensional Fitting (TDF) algorithm that fits directly 2D surface brightness of the detected sources (see Section 2.6).

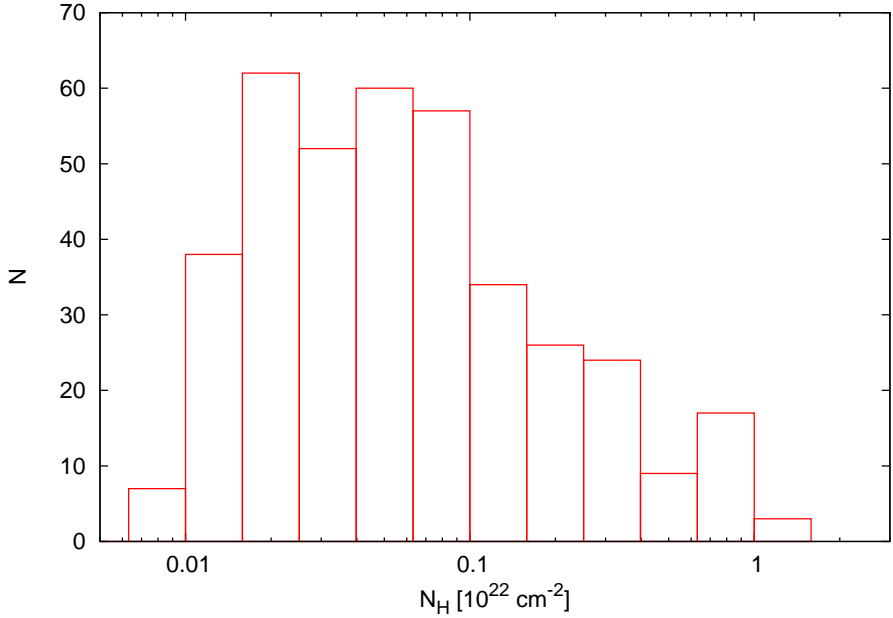

Figure 4: Distribution of the values of the Galactic hydrogen column density, $N_H$, for the 390 fields of the SXCS.

However, before apllying the "detection machine" on real data, it is fundamental to investigate the main characteristics of XRT that mostly affect the results and the efficiency of extended sources detection.

## 2.3 SWIFT-XRT CHARACTERISTICS

XRT uses grazing incidence Wolter I mirror (originally built for Jet-X) to focus X-ray onto a CCD detector similar to the EPIC MOS detector flown on *XMM-Newton*. The main XRT characteristics are listed in Table 1 and a complete description of the instruments is given in Burrows et al. (2003) and Hill et al. (2004).

| | |
|---|---|
| Telescope: | Wolter I 3.5m focal length |
| Detector: | E2V CCD-22 |
| Pixel Size: | $40\mu m \times 40\mu m$ |
| Pixel Scale: | $2.36''$ per pixel |
| Field of View: | $23.6 \times 23.6$ arcmin |
| PSF: | $18''$ HPD at 1.5 keV |
| | $22''$ HPD at 8.1 keV |
| Position accuracy: | $3''$ |
| Energy Range: | 0.2-10.0 keV |
| Energy Resolution: | 140 eV at 5.9 keV |
| Effective Area: | 135 cm$^2$ at 1.5 keV |
| | 20 cm$^2$ at 8.1 keV |
| Sensitivity: | $2 \times 10^{-14}$ erg cm$^{-2}$ s$^{-1}$ at $10^4$ s |

Table 1: List of main XRT characteristics.

The dimension of the CCD on the XRT is $600 \times 600$ pixels and it is equipped with four calibration sources located at each corner of the detector. The energy of the line emitted by the sources are 5.9 keV and 6.4 keV. The location and the radius of the calibration sources in detector coordinates are in Table 2.

| Source | Position | Radius |
|---|---|---|
| Cal 0 | $(35, 570)$ | 47 |
| Cal 1 | $(573, 561)$ | 48 |
| Cal 2 | $(36, 27)$ | 47 |
| Cal 3 | $(576, 20)$ | 44 |

Table 2: Position and radius of the XRT calibration sources. Quantities are in pixels and the position is expressed as (*x coordinate*, *y coordinate*).

XRT can operate in four different science modes: Image Long and Short, Low rate and Piled-up Photodiode, Windowed Timing, and Photon Counting. An exaustive description of all these modes is far beyond my purpose, therefore I will describe briefly only the Photon Counting mode, in which XRT images used in this work are taken. The Photon Counting mode retains full imaging and spectroscopic resolution with a time resolution of 2.5 s. The pixels are processed on board in a $5 \times 5$ matrix to eliminate most of the cosmic rays and chip defects. For each valid event, the $3 \times 3$ matrix is telemetred. On the ground a single Pulse Height Amplitude (PHA) value is reconstructed and GRADE is assigned.

During operations for most of the time the standard window setting is reduced to $480 \times 480$ pixels, the so-called w2 mode, excluding the calibration sources. However, as shown in Section 2.3.4 this is not enough to exclude completely events from calibration sources, since residual events from calibration sources remain and during the reading of the CCD the charge is transferred also to the pixels belonging to the same column.

Table 3: The typical values of the parameter of the PSF King profile.

| Energy [keV] | $\theta$ [arcmin] | $r_c$ [arcsec] | $\beta$ | XMM $r_c$ [arcsec] | XMM $\beta$ |
|---|---|---|---|---|---|
| 0.5 | 0 | 5.6 | 1.5 | 4.1 | 1.4 |
| 0.5 | 7 | 5.0 | 1.4 | | |
| 4.0 | 0 | 6.1 | 1.6 | 5.1 | 1.5 |
| 4.0 | 7 | 5.0 | 1.6 | | |

To exclude completely the contribution from calibration sources, a further image resize is performed during the data reduction adopted in this work (see Section 2.4), excluding all events with column number lower than 90 or greater than 510 (see Section 2.3.4).

A proper characterization of the telescope properties, such as PSF profile, effective area, or background, is crucial in order to detect sources and to distinguish point sources (mostly AGNs and star-forming galaxies) from extended sources (galaxy clusters and groups). In the following I analyse the main XRT characteristics, which are studied in details directly on a set of ten real images without obvious extended sources or extremely bright point sources.

### 2.3.1 *Point Spread Function*

In an imaging system, each source within the FoV has its own observed image in the image plane. What distinguishes extended sources from point sources is the "size" of the source observed images. The definition of "size" is not unique (for example it can be defined as the value which includes a given fraction of the total flux of the source), but it still depends on the telescope Point Spread Function (PSF). The PSF describes the response of an imaging system to a point object or point source. In more general terms the observed image of each source is the result of the convolution between the PSF and the original image of the source. Obviously, since by definition a Dirac $\delta$ is assumed as the original image of point sources, the PSF itself is the observed image of point sources. Instead, for extended sources the larger is the real image, the larger is the observed one with respect to the PSF. So an effective separation between point sources and extended sources must rely on a precise determination and characterization of the PSF as function of the energy and the off-axis angle.

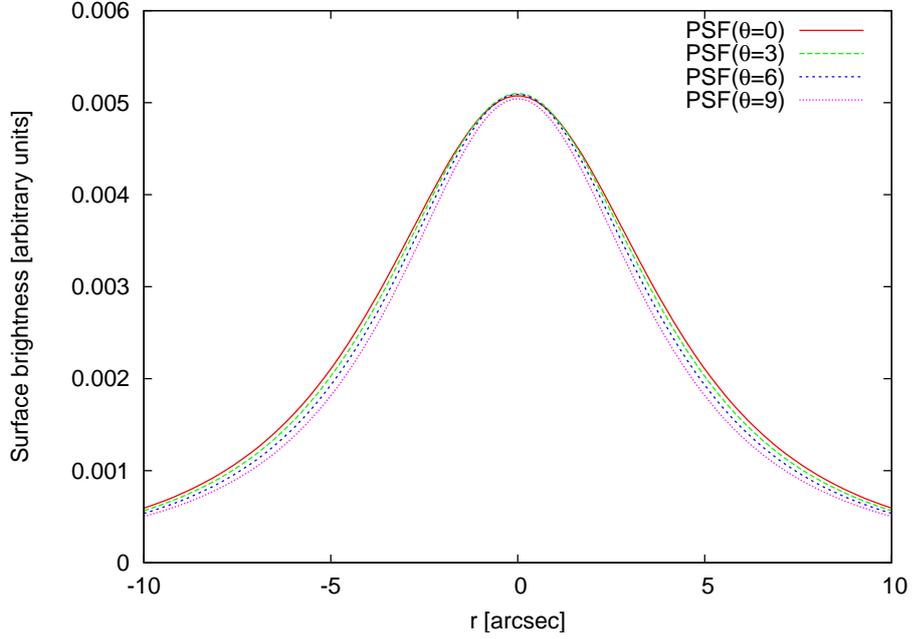

Figure 5: XRT PSF profile at different off-axis. The profiles refers to the PSF at 0.5 keV for different off-axis angle $\theta = 0'$, $3'$, $6'$, $9'$.

The theoretical model of the XRT PSF was analyzed by Moretti et al. (2005, 2006). The PSF is almost constant across the Field of View, a characteristic which is very important for surveys, since it preserves an high sky coverage at low fluxes for a large fraction of the total solid angle. The PSF can be modeled with a King profile:

$$PSF(r) \propto \left[ 1 + \left( \frac{r}{r_c} \right)^2 \right]^{-\beta}. \tag{2.1}$$

This model has two free parameters (plus the normalization): the core radius $r_c$ and the slope $\beta$. These parameters are function of the energy $E$ and the off-axis angle $\theta$:

$$r_c(E, \theta) = a_1 + b_1 \times \theta + c_1 \times E + d_1 \times E \times \theta \tag{2.2}$$

$$\beta(E, \theta) = a_2 + b_2 \times \theta + c_2 \times E + d_2 \times E \times \theta. \tag{2.3}$$

The values of the coefficients are available in the XRT calibration database[1], whereas in Table 3 the typical values of $r_c$ and $\beta$ at different energies and off-axis angles are listed, in comparison with the analogous *XMM-Newton* ones. The XRT PSF is more or less comparable with the *XMM-Newton* PSF, but it has also the valuable property to be almost independent on the off-axis angle. Figure 5 shows that the PSF at different off-axis is almost identical.

---

[1] http://swift.gsfc.nasa.gov/docs/swift/analysis

A constant PSF is crucial in the detection of extended sources, for two main reasons. First, the criteria to distinguish between point sources and extended sources can be defined independently on the off-axis angle, since all point sources have the same "size" at all off-axis. Second, this property is very favourable in case of multiple pointings, possibly not perfectly overlapped. Indeed, this is the case of XRT fields, which in general are obtained as merging of several pointings. In this situation the resulting PSF in the total image is the superimposition of different PSFs in single pointing images, weighted for the exposure time of each pointing. A PSF that strongly depends on the off-axis angle would be hard to model in case of a multiple pointings image, and it can be very different from the PSF modeled in a single pointing image. However, as the XRT PSF is almost constant on the whole chip, this is a marginal effect, unlike *Chandra* or *XMM-Newton* where the off-axis PSF is distorted and extended with respect to the on-axis PSF.

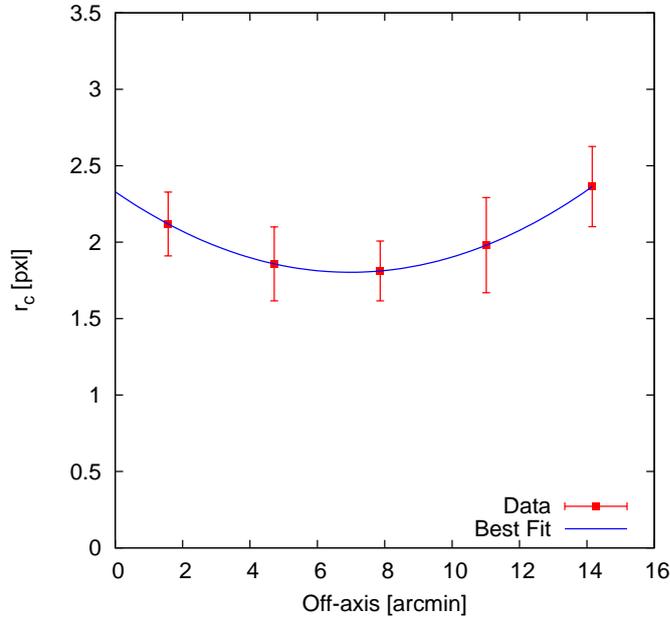

Figure 6: Measured core radius, $r_c$, of the XRT PSF profile as function of the off-axis angle. Error bars refers to the RMS in each bin.

Since the source detection will be performed in the soft band (0.5-2.0 keV), where galaxy clusters have an higher SNR, a further analysis of the XRT PSF in this energy band has been done. In particular, the PSF is modelled directly on real soft band images without obvious extended sources and without not-alligned multiple pointings.

All point sources in the images are identified with `wavdetect`, as described in Section 2.5, and their surface brightness is fitted with a single King profile, using the algorithm described in Section 2.6. Since $\beta$ does not change so much in the range 0.5-2.0 keV, it is fixed to 1.5 for all sources[2].

The Figure 6 shows the measured $r_c$ for all the detected point sources as function of the off-axis angle. This behaviour is fitted with a polymonial law:

$$r_c = 1.6777 \times 10^{-5} \times \theta^2 - 0.005947 \times \theta + 2.33, \tag{2.4}$$

where both the off-axis angle, $\theta$, and the core radius, $r_c$, are in pixels. With this prescription $r_c = 2.33$ on-axis, it reaches the minimum $r_c \simeq 1.8$ at $\theta \simeq 170 \simeq 7$arcmin, and at the image edges $r_c \simeq 2.4$.

### 2.3.2 *Effective Area and Redistribution Matrix*

Reflectivity and vignetting, among other effects, cause the geometric area of a telescope to be reduced to a smaller "effective area". This effective area is the area that must be used when calculating the physical properties of sources in the sky (e.g. flux, surface brightness), and it has units of [cm²]. The XRT effective area is shown in Figure 7.

The vignetting effect is caused by the reduced reflection efficiency and lower geometric collecting area for X-ray photons that are offset from the optical axis of the telescope. The XRT vignetting can be modeled as a function of distance from the center of the chip as:

$$V = 1 - c \times \theta^2, \tag{2.5}$$

where $\theta$ is the off-axis angle in arcminutes and $c$ is a function of the energy $E$ of the events of the form:

$$c = A_0 \times A_1^E + A_2, \tag{2.6}$$

where $A_0 = 0.000124799$, $A_1 = 1.55219$, $A_2 = 0.00185628$, and the energy is in [keV]. More details about the vignetting can be found in the calibration database of XRT[3].

The Redistribution Matrix File (RMF) is a map from energy space into detector Pulse Height Amplitude (PHA) space. In practice, it is a matrix to convert the integrated charge of an event recorded in a detector, the PHA, into the energy associated to the event. Since detectors are not perfect, this involves a spreading of the observed counts due to the detector resolution, which is expressed as a matrix multiplication. In high resolution instruments (e. g. diffraction gratings) the matrix is almost diagonal. In proportional counters the matrix elements are non-zero over a large area. CCD detectors, such as the XRT detector, are an intermediate case, with most of the response being almost diagonal, but with escape peaks and low energy tails adding significant contributions. On-Axis RMF is provided by the XRT calibration database. However the dependence of the RMF on the chip position has not been analyzed yet, and neither an analytical formula for the RMF as function of the off-axis does exist. So the on-axis RMF is used for all the spectral fits.

---

2 Moreover, as shown in Section 2.8.2, there is a degeneracy between $\beta$ and $r_c$, so it is convenient to fix one of the free parameter.

3 http://swift.gsfc.nasa.gov/docs/swift/analysis

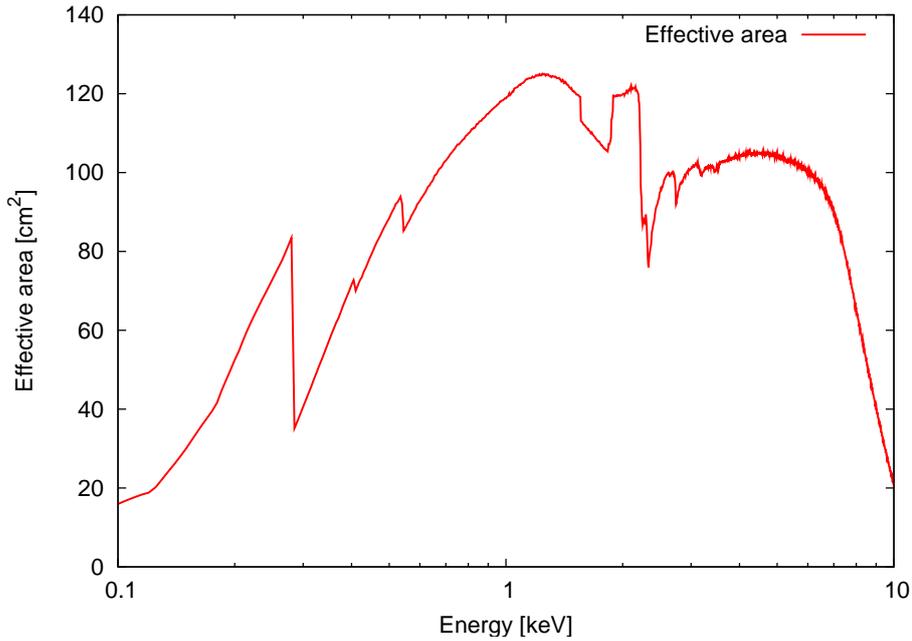

Figure 7: XRT effective area as function of energy.

The Ancillary Response File (ARF) contains the combined "effective area" and the Quantum Efficiency (QE) as a function of energy. The effective area is in [cm$^2$] and the QE is [counts/photon]; they are multiplied together to create the ARF, resulting in [cm$^2$ counts/photon]. A specific ARF is builded for each source. In the software package of XRT is not available a routine to make ARF for extended sources, so I wrote a routine that can compensate for this lack. The ARF is calculated from exposure maps and non-vignetted exposure maps (see Section 2.4) taking into account the following: mirror effective area (i.e. on-axis ARF provided by the calibration database), vignetting correction, and source profile correction. Details about this "homemade" routine are in Appendix Section A.3. RMF and ARF are used in the next Section to derive ECF and in Chapter 3 for source spectral analysis.

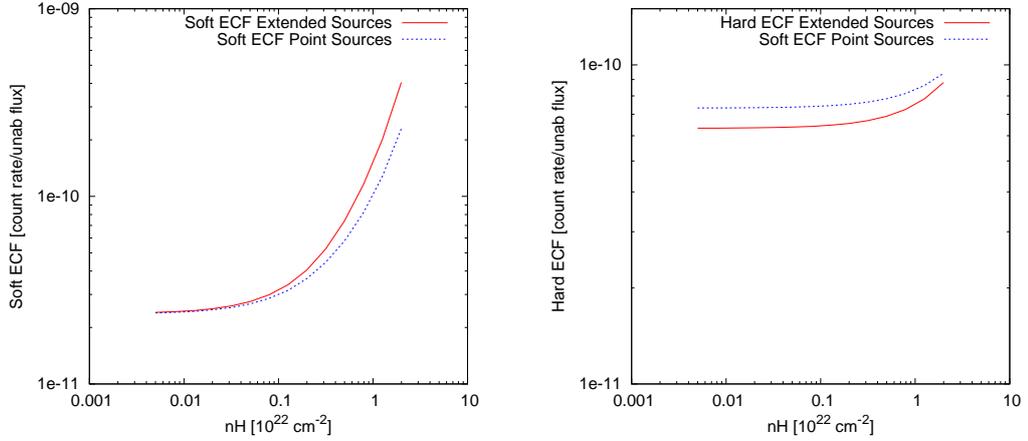

Figure 8: Soft and hard XRT ECFs for extended sources and point sources. Left panel (right panel), average soft (hard) ECFs for point sources in blue dashed line and extended sources in red solid line as function of the Galactic $N_H$.

### 2.3.3 *Energy Conversion Factors*

By definition, the Energy Conversion Factor (ECF) is the ratio between the unabsorbed flux in an energy range and the expected count rate in the same energy range, and it is needed to convert the measured count rate into flux. The unabsorbed flux is the flux from a source spectrum as if there were no absorption due to the Galaxy Inter Stellar Medium (ISM). So the ECF depends on the source spectrum and on the Galactic $N_H$. Extragalactic X-ray sources can be divided into point sources (mostly AGNs and star forming galaxies) and extended sources (diffuse emission from ICM of clusters). I used X-ray thermal emission spectra to compute ECFs for extended source and power law spectra for point soruce. I computed the ECFs in the soft and hard bands, using on-axis ARF and RMF. The plots in Figure 8 show the dependence of average ECFs on the Galactic $N_H$.

Soft ECFs have a strong dependence on the Galactic $N_H$, while hard ECFs depend much less on $N_H$, increasing only for high $N_H$ values. In the soft band the ECF of AGNs is very similar to the average ECF of thermal bremsstrahlung sources, especially for $N_H < 0.5 \times 10^{22} \text{cm}^{-2}$, so for the large majority of SXCS fields (see Figure 4). Instead, in the hard band the ECF of AGN is on average 10% larger than that of thermal bremsstrahlung sources. In Appendix Section A.4 more details about XRT ECFs and their dependence on spectral parameters are discussed.

### 2.3.4 *XRT Background*

A detailed knowledge of the background, its spatial variation and its spectrum can be very important for source detection and spectral analysis. The operational temperature of the CCD is the main factor that influences the XRT background level. The XRT background varies by more than a factor of 10 from the coldest temperature, $\sim -75°C$ which is achieved by the passive cooling, to the highest teperature at which XRT can collect scientifically useful data, $\sim -50°C$. Because of these variability, for spectral analysis of galaxy clusters in Chapter 3, the background is always sampled from the same observation.

The background is contributed by two main components: a part is due to the internal or instrumental background and another part is due to unresolved astrophysical X-ray sources (e.g. cosmic X-ray background and Galactic X-ray background). On average the instrumental background is responsible for 30% of the total background (Pagani et al., 2007). The XRT background is estimated directly on real images without obvious extended sources or extremely bright point sources. All other point sources in the images are identified with `wavdetect`, as described in Section 2.5, and masked by removing a region centered at the source positions with radius 3 times the PSF. This choice is made empirically and ensures that nearly all the source flux is excluded. Figure 9 shows the background spectrum extracted from real images after removing all the detected sources.

The spectral lines at 5.9 keV and 6.5 keV are due, respectively, to Mn $K_\alpha$ and Mn $K_\beta$ emission from the $^{55}$Fe corner calibration sources, that are not completely masked out in the standard screening process. The other spectral features correspond to the Ni $K_\alpha$ and $K_\beta$ lines at 7.48 keV and 8.26 keV and are due to high-energy particles interacting with the camera body. To avoid these background emission lines, during data reduction used in this work events with column number lower than 90 or greater than 510 are excluded, and the high energy cutoff for spectral analysis is set to 7.0 keV. In Figure 10 the spectrum of the background extracted from the center of the image is compared with the one extracted at the edge of the images, showing that excluding the outer columns the intrumental emission lines below 7.0 keV are removed.

On average the XRT background is very low, about one tenth of that of *Chandra* (see also Pagani et al., 2007); the background count rate in the soft band (0.5-2.0 keV) and in the hard band (2.0-7.0 keV) is $1.7 \times 10^{-4}$ cts s$^{-1}$ arcmin$^{-2}$ ($\sim 0.06$ cts s$^{-1}$ per field) and $2.0 \times 10^{-4}$ cts s$^{-1}$ arcmin$^{-2}$ ($\sim 0.07$ cts $^{-1}$ per field) respectively.

## 2.4 DATA REDUCTION

To create screened event files, the starting point is the Level 1 data as stored in the archive. Most of the procedure is the standard one, here I summarize only the main points. For detailed description see *"The SWIFT-XRT Data Reduction Guide"* document[4]. The standard processing involves identification and removal of bad pixels or bad columns, hot and flickering pixels, coordinates transformation, time tagging of events, reconstruction of events, computation of PHA values, and elimination of the piledup frames and partially exposed pixels. Then the events are screened by applying conditions for specified parameters. The screening uses the GTIs obtained by setting conditions on instrumental-specific housekeeping parameters, and on attitude and orbit related quantities. Additional selections are applied on the GRADE and the STATUS columns.

---

4 http://heasarc.gsfc.nasa.gov/docs/swift/analysis/

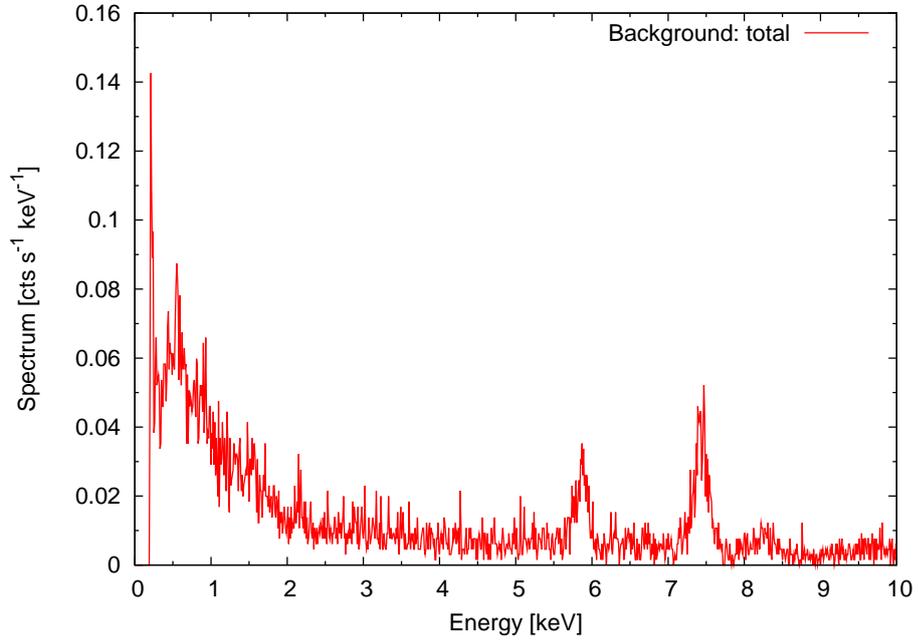

Figure 9: Spectrum of the XRT background extracted form real images after masking all the detected sources.

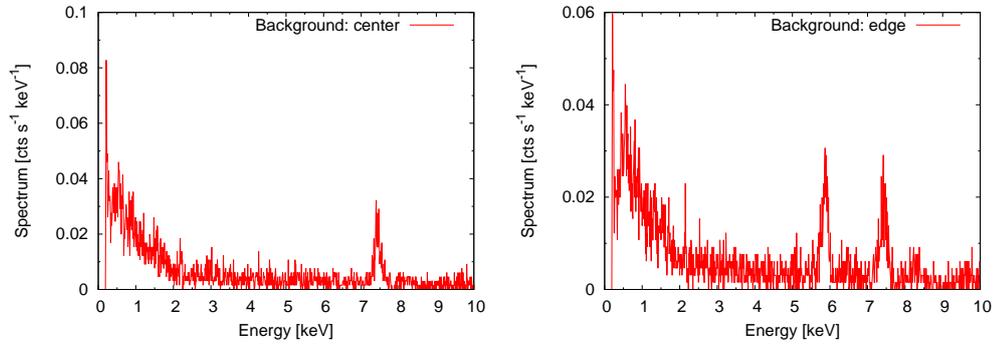

Figure 10: Spectrum of the XRT background extracted at the center (left panel) and at the edge (right panel) of the image.

Besides the standard data reduction, events with column number lower than 90 or greater than 510 are excluded, to avoid completely the contribution from calibration sources (see Section 2.3). Moreover, events are selected in the energy range 0.5-7.0 keV. The cut at low energies is imposed by the rapidly increasing background, while removing high energy events is necessary to avoid residual instrumental emission lines (see Section 2.3.4).

Standard exposure map at 1.5 keV are created. Exposure maps are expressed in seconds and take into account most of all the vignetting. Also exposure maps without vignetting correction[5] are created, since they are necessary to create specific ARFs (see Section 2.3.2).

## 2.5 SOURCE DETECTION

The `wavdetect` algorithm within the CIAO software developed for *Chandra* was used to identify sources in the 390 XRT fields of SXCS. This algorithm exploits Mexican-Hat wavelet for source detection, correlating the image with wavelets of different scales and then searching the results for significant correlations. For a description of the theory and operation of `wavdetect`, see Freeman et al. (2002).

The algorithm was run on the images obtained in the soft (0.5-2.0 keV) band. The soft band is optimal to identify extended ICM emission, because it is in this band that XRT has the highest effective area and clusters have higher SNR. In this way the sensitivity in the detection of extended source powered by thermal bremsstrahlung is maximized. The best input parameters are chosen empirically. In particular the wavelet radii, the parameter `scales` expressed in pixels, are chosen as power $2^{n/2}$ with the integer $n \in [0:13]$. The threshold parameter, `sigthresh`, is set to $10^{-6}$ significantly larger than one over the number of pixels in the image. The `wavdetect` algorithm is used only for source detection, while the analysis of source extention and therefore the separation between point sources and extended sources is performed by fitting the surface brightness of detected sources with the new two-dimensional fitting anglorithm described in Section 2.6.

---

5 Attention: exposure maps are pixel by pixel maps that only take into account the effective exposure time at the fixed energy 1.5. Specific ARF at source position are needed to perform spectral analysis.

## 2.6 THE TWO DIMENSIONAL FITTING ALGORITHM

After the source detection a method is need to assign a "size" to each source in order to distinguish between point sources and extended sources, and also a method to do the photometry and measure the number of counts.

Since XRT has a large PSF Half Energy Width (HEW) (e. g. compared with the *Chandra* PSF), sometimes `wavdetect` identifies two sources with centroids closer than the HEW. In this situation the surface brightness profiles of the two sources are then superimposed and a simple criterion to assign a "size" based on the fit of the profile of the integrated brightness (e. g. fitting the surface brightness profile in concentrical rings) may be inaccurate. Furthermore, measuring the net counts of detected sources with simple aperture photometry may introduce significant errors. The situation can be more complicated for extended sources, which have by definition an extension larger than the PSF. Sometimes it happens that within an extended source is identified a point source due to a background Active Galactic Nucleus (AGN).

A possible solution to this problems is to fit at the same time the 2D surface brightness of all sources present in an image. However, to accomplish this aim you need to fit about 20-50 sources per field and even for simple 2D surface brightness profiles, such as a King profile or a Gaussian profile, the number of free parameters can be larger than 100, and it increases linearly with the number of sources. So, to explore the multi-dimensional space of the parameters I have developed a new Two Dimensional Fitting (TDF) algorithm based on MCMC. This algorithm exploits "simulated annealing" technique, based on MCMC and maximum likelihood criterion. The theory behind this algorithm is described in Appendix Section A.1. The aim is to fit simultaneously the background and the surface brightness profile of all the detected sources in the image, in order to assign to each detected source a "size" from which discriminate between point sources and extended sources, and to measure the number of net counts of all detected sources.

TDF has as input parameters only the positions and the surface brightness models to use for each sourcs. The background is assumed constant on all the image. Both the constant background and the source profiles are corrected for vignetting. This algorithm is versatile allowing to chose different 2D function to model the surface brightenss profile of the sources. However, for the case of XRT, it is more convenient use King profile both for point sources and extended sources.

At the state of the art the main disadvantage of this algorithm is the machine time which is approximately proportional to the square of the number of sources to fit. In Section 2.8.2 I test widely this new algorithm on realistic simulated XRT images, to understand its capability in recovering the simulated source input parameters and how it can be exploited to distinguish between point sources and extended sources.

To test the efficiency of the detection algorithm and the capability for the separation between point and extended sources, I developed an X-ray image simulator to perform a set of simulations which realistically reproduce XRT pointed observations in the soft (0.5-2.0 keV) band. This image simulator is a very flexible algorithm, which can be easily adapted to the technical specifications of other X-ray telescopes. In the Chapter 5 I adapted a modified version of this image simulator for the scientific case of WFXT.

I realized a set of simulations that include point sources, whose flux distribution is consistent with that of the Chandra Deep Field South (CDFS) (Rosati et al., 2002), and extended sources, whose surface brightness is modeled as simple King profile and flux distribution is derived from the ROSAT Deep Cluster Survey (RDCS) (Rosati et al., 1998). However, I do not attempt a more realistic modelization of the shape of low-surface brightness extended sources, since at this level it is more interesting and important to evaluate the detection efficiency for sources with different sizes, and how the chosen criteria are able to recover the right source size to distinguish between point and extended sources. These simulated images also take into account the X-ray background, which is directly derived from real XRT images (see Section 2.3.4), vignetting correction and PSF effect.

The main properties of the simulations are described in the following Sections. The following list is a concise summary of the properties of the simulations used to test the "detection machine":

- 400 simulated images;

- image size $480 \times 480$ pixel (like in the *w2* mode of XRT);

- exposure times between 50ks and 300ks;

- energy band 0.5-2.0 keV;

- lowest flux of the simulated sources $10^{-16}$ erg cm$^{-2}$ s$^{-1}$;

- PSF modeled with a King profile with fixed $\beta$ and $r_c$ as function of the off-axis angle;

- background 0.06 cts s$^{-1}$ per field;

- ECF equal to $2.33 \times 10^{-11}$ erg cm$^{-2}$ cts$^{-1}$ for all sources.

### 2.7.1 *Technical Issues*

The images are realized using C and IDL codes. The image setup is chosen in order that it can be as more realistic as possible and somehow representative of XRT fields. I made and analysed 400 simulated images with different exposure times. According to Figure 3, four exposure times are used to mimic the survey exposure time distribution: 50ks, 100ks, 200ks and 300ks. For each exposure there are 100 simulated images. The images are simulated in the energy band chosen for the source detection, 0.5-2.0 keV, since in this band galaxy clusters have an higher SNR. The image size is $480 \times 480$ pixels, because for most of the XRT fields the standard window setting is $480 \times 480$ pixels, the so-called w2 mode, excluding the calibration sources (see Section 2.3).

Even if most of the XRT fields are made of several pointing, the configuration with one single pointing per image is chosen. Since the XRT PSF is almost constant on the whole chip, considering images made of multiple pointings would be a further complication with marginal effect, whose analysis is beyond the simulation purpose.

The PSF is modeled with a King profile as described in Section 2.3.1 (see also Moretti et al. 2005). The parameter $\beta = 1.5$ is fixed and is independent on the off-axis angle, while $r_c$ changes with the off-axis according to the Equation 2.4. With this prescription $r_c = 2.33$ on-axis, it reaches the minimum $r_c \simeq 1.8$ at $\theta \simeq 7$ arcmin, and at the image edges $r_c \simeq 2.4$.

### 2.7.2 *Physical Issues*

In these simulated images the simulated photon counts in each pixel are obtained from a mixture of three components: point sources, extend sources and background.

POINT SOURCES   The total number of point sources are simulated according to the soft LogN-LogS as measured in the CDFS (Rosati et al., 2002). The distribution of fluxes of the point sources is obtained through a Monte Carlo sampling of the differential number counts distribution as measured in deep *Chandra* observations. The differential number counts are modeled with a double power law:

$$\frac{dN(>S)}{dS} = \begin{cases} k_1\,S^{-\alpha_1-1} & \text{if } S < S_B \\ k_2\,S^{-\alpha_2-1} & \text{if } S > S_B \end{cases} \tag{2.7}$$

where $S$ is expressed in $10^{-15}$ erg cm$^{-2}$ s$^{-1}$. The best fit values for the CDFS LogN-LogS in the soft band are:

- $S_B = 1.43219668 \times 10^{-14}$ erg cm$^{-2}$ s$^{-1}$

- $\alpha_1 = 0.67$

- $\alpha_2 = 1.70$

- $k_1 = 458.3$ deg$^{-2}$

- $k_2 = 7100.0$ deg$^{-2}$

Point sources are sampled down to the flux $10^{-17}$ erg cm$^{-2}$ s$^{-1}$, well below the detection limit. On average $\sim 1500$ point sources are simulated in each field. The count rate for each source is computed from the source flux, assuming an average ECF=$2.33 \times 10^{-11}$ erg cm$^{-2}$ cts$^{-1}$ obtained for a power law spectrum with $\gamma = 1.4$ (see Section 2.3.3).

The sources are distributed randomly in the image, i. e. without any clustering prescription. The source profiles are modeled according to the PSF profile, with $\beta = 1.5$ and $r_c$ computed with the Equation 2.4 at the source position. To the total net counts of the sources Poissonian noise is added as well.

EXTENDED SOURCES    For extended sources, neraly the same procedure as for point sources is adopted. The total number of extended sources per field and the distribution of fluxes are obtained through Monte Carlo sampling of the soft differential number counts distribution as measured in the RDCS (Rosati et al., 1998). The differential number counts are modeled with a single power law:

$$\frac{dN(> S)}{dS} = k \, S^{-\alpha - 1} \tag{2.8}$$

where $S$ is expressed in $10^{-15}$ erg cm$^{-2}$ s$^{-1}$. The best fit values for the RDCS LogN-LogS in the soft band is:

- $\alpha = 1.11$

- $k = 171.92$ deg$^{-2}$

Extended sources are sampled down to the flux $10^{-16}$ erg cm$^{-2}$ s$^{-1}$, well below the detection limit. On average $\sim 200$ extended sources are simulated in each field.

The plots in Section 2.3.3 show that the ECF for a thermal bremsstrahlung spectrum in the soft band strongly depends on the Galactic $N_H$, increasing rapidly for $N_H > 0.1 \times 10^{22}$cm$^{-2}$, but it has a less severe dependence on temperature, redshift and metallicity. Since most of the fields in the survey have $N_H < 0.1$ (see Figure 4) for simplicity an average ECF=$2.33 \times 10^{-11}$erg cm$^{-2}$ cts$^{-1}$ for all the sources is assumed.

A very precise modelization of the shape of the extended sources requires a huge effort handling different physical issues (e. g. luminosity and temperature function, redshift distribution, etc.), and it is far beyond the aim of this analysis.

For simplicity, the surface brightness profile is modeled with single King profile with fixed exponent $\beta = 1.5$[6] and core radius, $r_c$, variyng uniformly in the range $[5 : 16]$ pixels, roughly correspondig to $[10'' : 30'']$. Extended sources are distributed randomly in the image, without any clustering prescription or correlation between point sources and extended sources. Poissonian noise is added to the total net counts of the sources.

---

6 Attention: a surface brightness derived from a $\beta$-model with $\beta_m = 0.67$ means a King profile with exponent $\beta = 3\beta_m - 0.5 = 1.5$ (see Equation 1.9).

BACKGROUND    In X-ray images the background is contributed by several components: the flux of unresolved extragalactic sources below the detection limit; instrumental background; galactic foreground, which is dominant in Galactic fields and it is subject to strong fluctuation. These effect are very different from one another. The instrumental background depends on the telescope operating mode and the instrumental performance, it is almost constant and it is not subject to vignetting correction. The Galactic foreground depends only on the field coordinates and on the Galactic $N_H$, while the shorter is the exposure time, the larger is the total flux of unresolved extragalactic sources below the detection limit.

According to background calibration on real images as described in Section 2.3.4, beyond unresolved sources simulated down to the flux $10^{-16}$ erg cm$^{-2}$ s$^{-1}$, an extra background of 0.06 cts s$^{-1}$ per field is added by hand to take into account the contribution of the other background components. This extra background is assumed to be constant on the whole image, and it is added Poissonianly after vignetting correction, since only instrumental background is not affected by vignetting correction and it very low (Pagani et al., 2007).

## 2.8    SIMULATION ANALYSIS

Simulated images were analysed in detail to test the efficiency of source detection and fitting algorithms. The detection with `wavdetect` was tested on all the 400 simualted fields and the output catalogue was matched with the input catalogue to get the fraction of detected sources and false detection as function of the flux of the sources. The TDF was run first on 4000 single snapshots with one single source plus background to test the efficiency in recovering the input net counts, and then it was run on the output catalogue of `wavdetect` to test the efficiency in separating point sources from extended sources. At the end of this detailed simulation analysis the "detection machine" has been set, ready to perform the detection on real images.

### 2.8.1    Detection Ability Test

The `wavdetect` algorithm was run on the 400 simulated images with exposure times of 50 ks, 100 ks, 200 ks and 300 ks. The adopted input parameter set-up is the one described in Section 2.5. The catalogue of detected sources was matched with the input catalogue. The fraction of detected sources as function of flux and of net counts for the different exposures is shown in Figure 11. On average for all the exposures, even at 50 ks, the fraction of recovered souces is larger than 90% at fluxes larger than $10^{-14}$ erg cm$^{-2}$ s$^{-1}$ or similarly for more than 50 net counts.

The number of false detections with respect to the number of real detection is ∼ 1% and most of them have less than 20 net counts as measured by TDF (see Section 2.8.2).

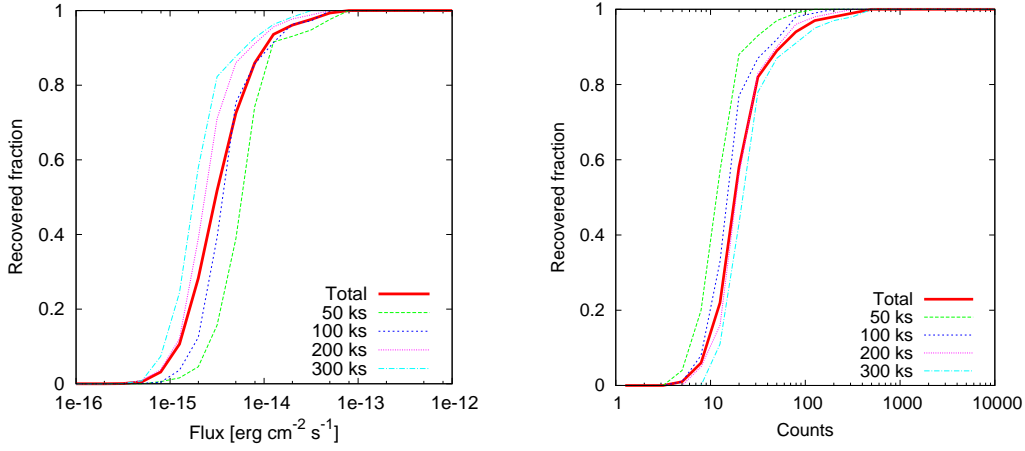

Figure 11: Fraction of sources recovered by `wavdetect` as function of the source flux (left panel) and of the source net counts (right panel).

### 2.8.2 *Two Dimensional Fitting Test*

Since the algorithm described in Section 2.6 is a new fitting algorithm based on surface brightness modelization, I widely tested it on simulated images. The target is to understand how the algorithm works and how much it is reliable in the reconstruction of the profile parameters, above all in case of King profile.

As a preliminray test, I analysed 4000 single snapshots with one single source plus background, to test the capability of TDF in recovering the input parameters of simulated sources. These sources are all simulated with the on-axis PSF profile $r_c = 2.33$ pxl and $\beta = 1.5$. The normalizations and the exposure times are randomlly chosen in the range $[0 : 1000]$ counts and $[10 : 500]$ ks.

The algorithm was run fixing the position of each source and fitting the surface brightness with a King profile,

$$ K(r) = N \frac{\beta - 1}{\pi r_c^2} \left[ 1 + \left( \frac{r}{r_c} \right)^2 \right]^{-\beta}, \tag{2.9} $$

leaving free all the three parameters, $r_c$, $\beta$ and $N$. The multiplicative term $(\beta - 1)/(\pi r_c^2)$ makes the normalization $N$ equal to the number of net counts.

In Figures 12, 13 and 14 I show the comparison between the input and output parameters of the fit: in Figure 12 the correlation between the measured $r_c$ and $\beta$; in Figure 13 the com-parison between the input and output normalization; in Figure 14 the correlation between the measured $\beta$ and the ratio between output and input normalization.

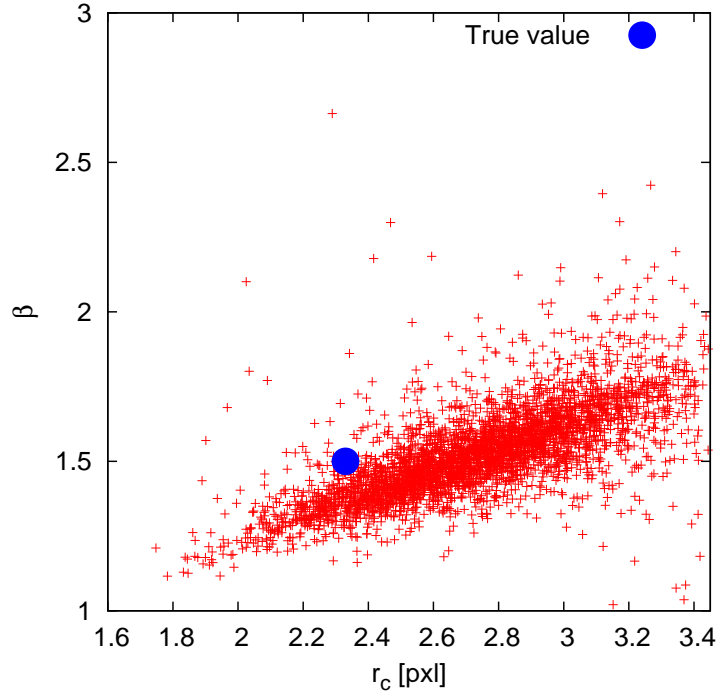

Figure 12: Best fit $r_c$ *versus* best fit $\beta$ in XRT simulations. Each red cross represents a simulated point source, while the big blue point shows the true value, $\beta = 1.5$ and $r_c = 2.33$.

As one can see there is strong correlation between the fitting parameters, in particular between $r_c$ and $\beta$. This is even more strange because the effect is present for all values of normalization even the largest ones. On average the *a posteriori* probability distribution of $\beta$ is centered on the real value, while $r_c$ and $N$ are overestimeted by 20% and 10% respectively. Moreover, in some cases the number of net counts are overestimated by a factor $\sim 2$, but seldom they are underestimated.

I interpret this issue in the following way. Since on average $r_c$ is overestimated, in Equation 2.9 a larger normalization $N$ is required to well fit the central pixels of the profile that have a larger number of counts, i. e. they have a larger weight. Moreover, as shown in Figure 14 small $\beta$ calls large $N$. Since the measured $\beta$ is mostly in the range $[1:2]$, the multiplicative term $(\beta - 1)$ can eventually vary by a factor $[0:2]$ with respect to the true value, forcing the $N$ to vary of the inverse amount to compensate. These effects are not primarily a limitation of the algorithm, but most of all they are due to the strong degeneration between the King profile parameters and to the presence of the background.

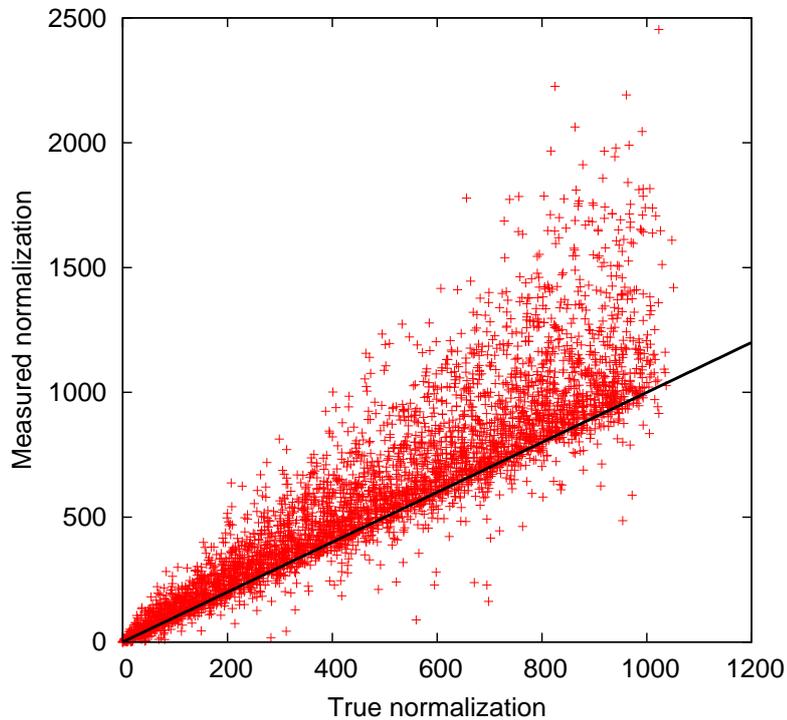

Figure 13: True normalization *versus* best fit normalization in XRT simulations. Each red cross represents a simulated point source, while the black solid line shows the true values.

Indeed, profiles with (opportune combinations of) different values of $r_c$ and $\beta$ differ mostly in the wings rather than in the core. This is evident, for example, in Figure 5, where the XRT PSF profile at different off-axis is almost identical while the $r_c$ and $\beta$ are different. But beyond a certain radius the wings fall below the background (where the background dominates) and so they become "invisible" to any fitting algorithm. In this sense a profile with larger values of $r_c$, $\beta$ and $N$ can be equal to one with smaller values of the parameters, since the part of the profile where they differ are the wings, which are not fitted.

At this point, a furthermore consideration is important. Simulating a source, the parameter $N$ in Equation 2.9 takes on the double meaning of normalization constant and expected number of net counts. This does not apply when the same equation is used in the fitting algorithm, since $N$ retains only the meaning of normalization constant. As in the King profile wings descend very slowly compared to, for example, a Gaussian profile, the amount of flux in the wings can be very relevant. But, since the parameters of the fit are degenerate, a higher value of normalization (in case combined with a $\beta$ close to 1 or a larger $r_c$) could be only a fictitious higher number of net counts due to unfitted "invisible" wings. Figure 13 shows incontrovertibly that $N$ is not suitable to estimate directly the number of net counts. So another criterion is needed to reconstruct the number of net counts.

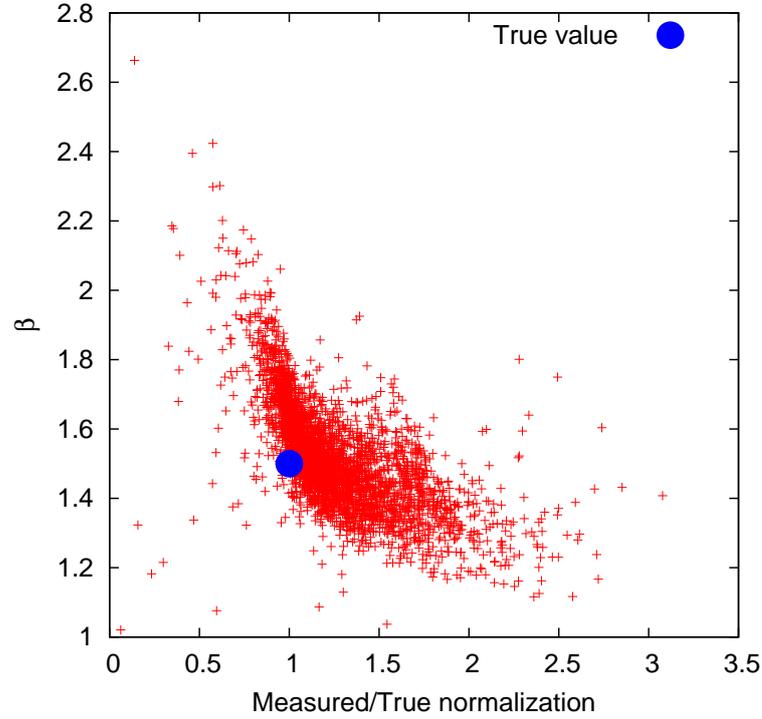

Figure 14: Best fit $\beta$ *versus* the ratio between best fit normalization over true normalization in XRT simulations. Each red cross represents a simulated point source, while the big blue point shows the true value, $\beta = 1.5$.

On the basis of these considerations, I introduce the radius $r_{B2}$ and I will probe that, at least in simulated images, the integral of Equation 2.9 within $r_{B2}$ is a good estimate of the number of net counts. The radius $r_{B2}$ is defined as the radius where the background is twice the surface brightness of the source. For the King profile (Equation 2.9) it reads:

$$N\frac{\beta-1}{\pi r_c^2}\left[1+\left(\frac{r_{B2}}{r_c}\right)^2\right]^{-\beta} = \frac{B}{2}, \tag{2.10}$$

then

$$r_{B2} = r_c\left[\left(\frac{2N(\beta-1)}{\pi r_c^2 B}\right)^{1/\beta} - 1\right]^{1/2}, \tag{2.11}$$

where $B$ is the number of background counts per pixel. The integral of Equation 2.9 within $r_{B2}$ reads:

$$\int_0^{r_{B2}} NK(r')dr' = N\left[1 - \left(\frac{2N(\beta-1)}{\pi r_c^2 B}\right)^{\frac{1-\beta}{\beta}}\right].\qquad(2.12)$$

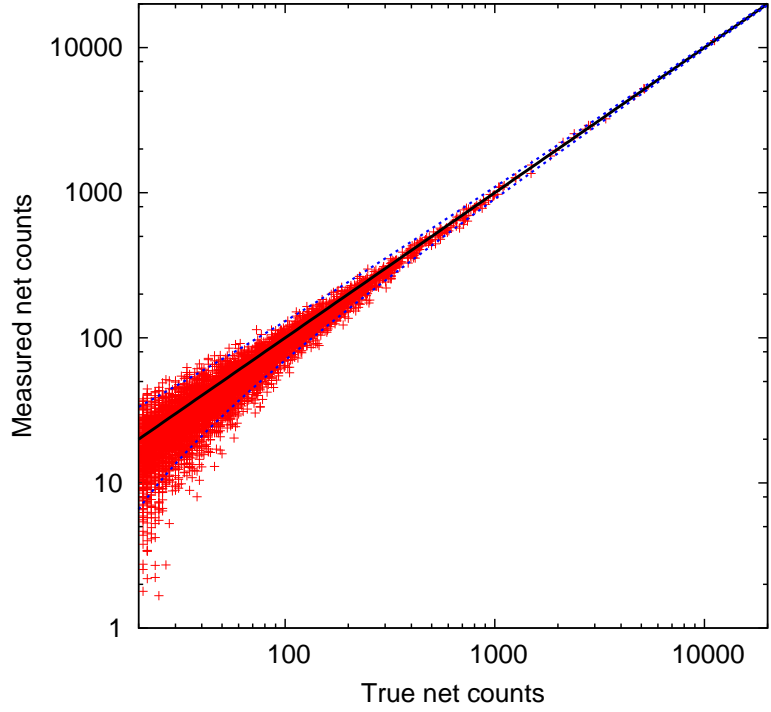

Figure 15: Net counts recovered from the best fit *versus* true net counts. Each red cross represents a simulated point source, while the black solid line shows the true value and the two dashed blue lines enclose the $3\sigma$ Poissonian noise region.

I assume that the integral in Equation 2.12 represents the measured number of net counts given by the fitting algorithm. In this sense $N$ is only a first order approximation of the number of net counts, and the Equation 2.12 goes in the right direction to suppress $N$ when $r_c$ is overextimated[7] or $\beta$ is close to 1. I stress again that the normalization itself is not interesting, but net counts are crucial since from them is derived the source flux.

These new prescriptions are applied to the analysis of the 400 simulated fields. In particular the TDF was run fitting all the sources in the catalogue obtained with wavdetect. For all the sources $\beta$ was fixed to 1.5 to avoid degeneracy between parameters and to reduce considerably the number of free parameters, and thus to reduce machine time.

The recovered number of net counts using the $r_{B2}$ criterion and the Equation 2.12 is shown in Figure 15. The agreement between input and recovered counts is excellent within $3\sigma$ Poissonian noise for all values of net counts. Regarding the false detected source in the wavdetect catalogue, most of them have less than 20 net counts as measured by TDF.

### 2.8.3 *Separation between Point Sources and Extended Sources*

Extended sources and point sources are discriminated by their "size". The point sources have by definition the size of the PSF, while extended sources have a larger size. To select a catalogue of extended sources, the issue is to locate in the parameter space where are the point sources, and then to choice a selection criterion to identify point sources and exclude them from the catalogue.

After several tests, once again the best choice turned out to be the radius $r_{B2}$, as shown in Figure 16 where $r_{B2}$ is compared with the quantity $N/(Br_c^2)$, being $N$ the normalization and $B$ the background counts per square arcsec. Even if there is superposition between the region occupied by point sources and the one by extended sources, however it is possible to separate point and extended sources using an analytical equation derived from Equation 2.11. In Figure 16 the blue solid line is the "theoretical" relation of $r_{B2}$ for point sources (see Equation 2.11), while the dashed black line is an analytical relation, chosen empirically modifying Equation 2.11, that bounds at $7\sigma$ the point sources distribution (for the derivation and the forumla of the black line see in Appendix Section A.2). With this $7\sigma$ criterion the contamination is virtually null, and moreover, the majority of extended sources below the black line are detected with less than 20 net counts.

---

7 Attention: the exponent $\frac{1-\beta}{\beta}$ is always negative since by definition $\beta > 1$.

Actually, the choice of the $7\sigma$ threshold is conservative, since in simulation also a $3\sigma$ threshold can exclude the majority of point sources. However, the difference between point sources and extended sources in real images is expected to be less pronounced than as idealized in these simulations.

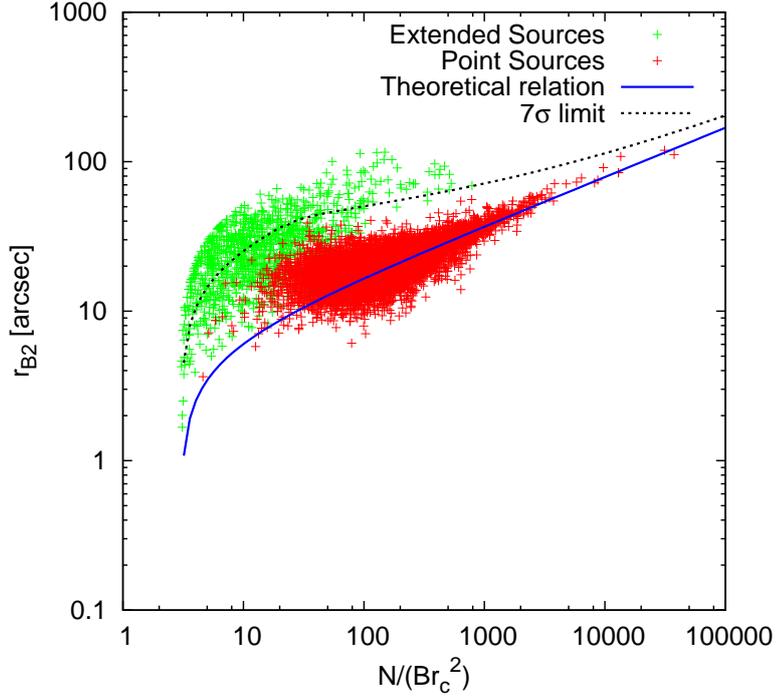

Figure 16: Relation between $r_{B2}$ and the best fit parameters $N/(Br_c^2)$ in XRT simulations. $N$ is the normalization of the surface brightness profile, $B$ the background counts per square arcsec, and $r_c$ is the core radius. Red points are point sources, green points extended sources. The blue solid line is the "theoretical" relation expected for point sources, the black dashed line is the $7\sigma$ limit of the $r_{B2}$ distribution of point sources.

### 2.8.4 *Final Setup of the Detection Machine*

I briefly summarize the final set-up of the "detection machine" chosen after this detailed analysis of the simulations and the step-by-step analysis that will be applied to real images.

1. The `wavdetect` algorithm is run on the soft 0.5-2.0 keV images. With the input parameters described in Section 2.5, the expected fraction of recovered sources is larger than 90% for fluxes larger than $10^{-14}$ erg cm$^{-2}$ s$^{-1}$ for all the exposure times of the SXCS fields.

2. The TDF algorithm is run on the soft 0.5-2.0 keV images fixing the position of the sources to those detected by `wavdetect`. The background is assumed constant and corrected for the vignetting, and every source is fitted with a single King profile with fixed $\beta = 1.5$, which is representative both of the PSF profile of point sources and of the X-ray surface brightness profile of galaxy clusters with gas density profile described by a $\beta$-model with $\beta_m = 2/3$.

3. The net counts of each source are computed as the integral of the best fit King profile within the radius $r_{B2}$ (see Equation 2.12). All the sources with less than 20 net counts are removed form the catalogue to exclude most of false detection.

4. In the output catalogue extended sources are selected according to the $7\sigma$ criterion described in Section 2.8.3.

## 2.9 EXTENDED SOURCE CATALOGUE

As calibrated on simulated images, `wavdetect` was run with the input parameters described in Section 2.5. The total number of identified sources of all the fields is $10^4$ sources. All the detected sources were checked visually looking for cases where `wavdetect` eventually fails the detection. In general, `wavdetect` works fine also in real images as in simulated images, but it fails the detection close to Galactic stars. In fact, very bright stars (approximately with magnitude $V < 8$) produce pile-up events[8] that are removed during the data reduction and they appear as "holes" in the image surrounded by luminous "rings". `wavdetect` fails the detection at the "hole" position identifying about ten sources along the luminous "ring". All these false detected sources are removed from the `wavdetect` catalogue. However, since are only six bright stars within the SXCS fields, the final catalogue is practically unaffected by these corrections.

Then TDF was run on the soft 0.5-2.0 keV images of all the fields to measure the soft net counts of each source and to separate between point sources and extended sources. All the sources in the `wavdetect` catalogue were fitted with a single King profile fixing the source position and $\beta = 1.5$. So the free parameters in each image fit are the normalization and the $r_c$ of each source and the image background. The results of the fits are shown in Figure 17, which is the analogous for the real fields of Figure 16. The lines in Figure 17 are the ones calibrated on the simulations; in particular, the blue solid line is the "theoretical" relation of $r_{B2}$ for point sources, while the dashed black line is the $7\sigma$ of the point sources distribution.

---

8 For a very bright source, there is a non-negligible probability for two or more photons to arrive in the same pixel during an integration time. The detector will be unable to distinguish the two events. This phenomena is called "pile-up".

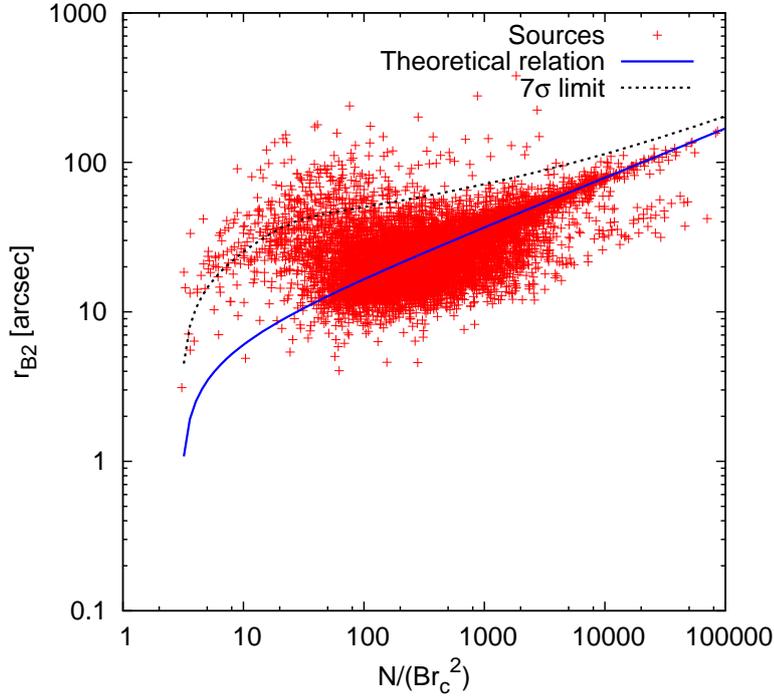

Figure 17: Relation between $r_{B2}$ and the best fit parameters $N/(Br_c^2)$ in real XRT fields. $N$ is the normalization of the surface brightness profile, $B$ the background counts per square arcsec, and $r_c$ is the core radius. Each red point represents a detected source. The blue solid line is the "theoretical" relation expected for point sources, the black dashed line is the $7\sigma$ limit of the $r_{B2}$ distribution of point sources calibrated in XRT simulations.

The net counts of each source were measured integrating the best fit surface brightness up to the radius $r_{B2}$ according to the Equation 2.12. Afterwards, the TDF was also run on the hard 2.0-7.0 keV images of all the fields to measure exclusively the hard net counts of each source. Then soft and hard net counts were converted to flux through specific ECFs. As shown in Section 2.3.3 both for point sources and extended sources the ECF depends most of all on the Galactic neutral hydrogen column density, $N_H$. Since each field has its own $N_H$ derived from radio data (Dickey & Lockman, 1990), four specific ECFs, for each energy band (soft and hard) and for each source type (point and extended), were computed from the average ECFs shown in Figure 8.

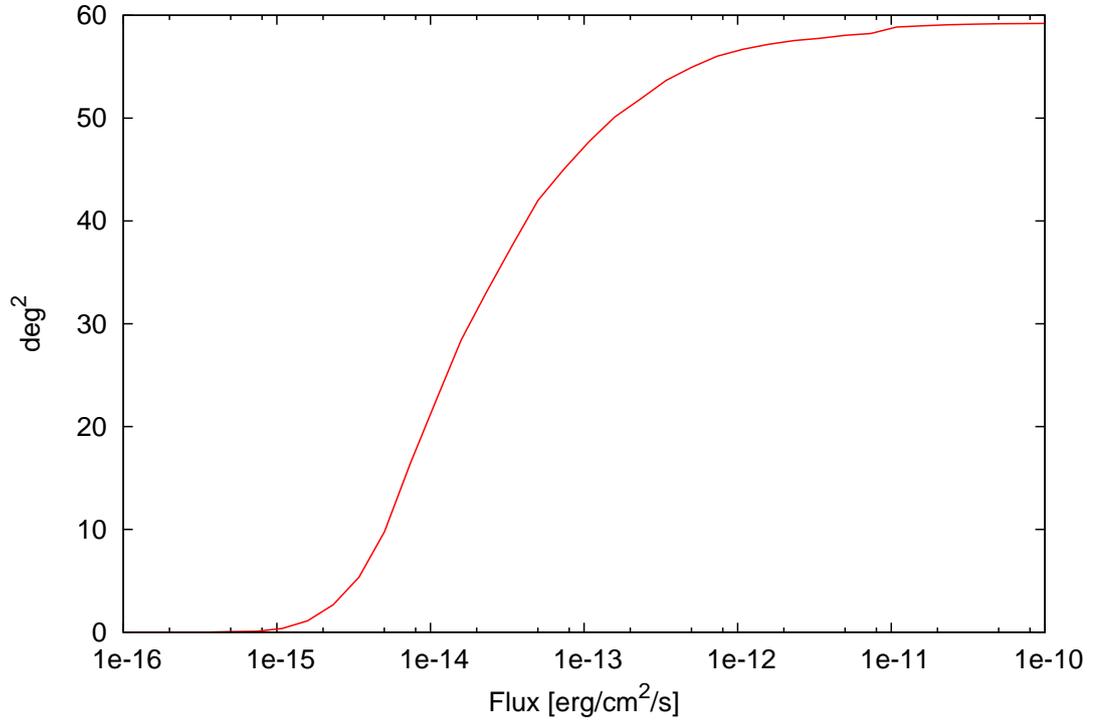

Figure 18: Soft sky coverage of the whole SXCS survey.

Finally, the catalogue of extended sources of the SXCS was compiled selecting all sources above the $7\sigma$ threshold and with at least 20 net counts in the soft band. The extended source catalogue obtained in this way consists of 243 sources. In the catalogue, which will be published in Moretti et al. (2010), each source is archived with the following parameters:

- source name based on the source position;

- R.A. and Dec;

- GRB field;

- exposure time of the field, and exposure corrected at the source position;

- best fit $r_c$ and image background rate;

- $r_{B2}$, soft and hard net counts, SNR, and hardness ratio;

- Galactic $N_H$ of the field, soft and hard ECFs, and fluxes.

SKY COVERAGE    The sky-coverage at a given flux is defined as the area of the survey where a source with that flux can be detected. The sky-coverage depends most of all on the exposure time of all the fields, vignetting correction and PSF correction. However, since the XRT PSF is almost constant on the whole image the correction due to PSF degradation at high off-axis is marginal. The SXCS sky-coverage is shown in Figure 18 and it is simply computed integrating for each flux the solid angle of the survey in which an extended source with that flux could be detected with at least 20 net counts. To convert flux into counts the specific ECFs computed from the $N_H$ of each field were used, like in Section 2.9 (see also Figure 8 left panel). Figure 19 shows that SXCS is comparable in terms of sky coverage with other present day X-ray cluster surveys, most of which based on ROSAT, *Chandra* and *XMM-Newton*.

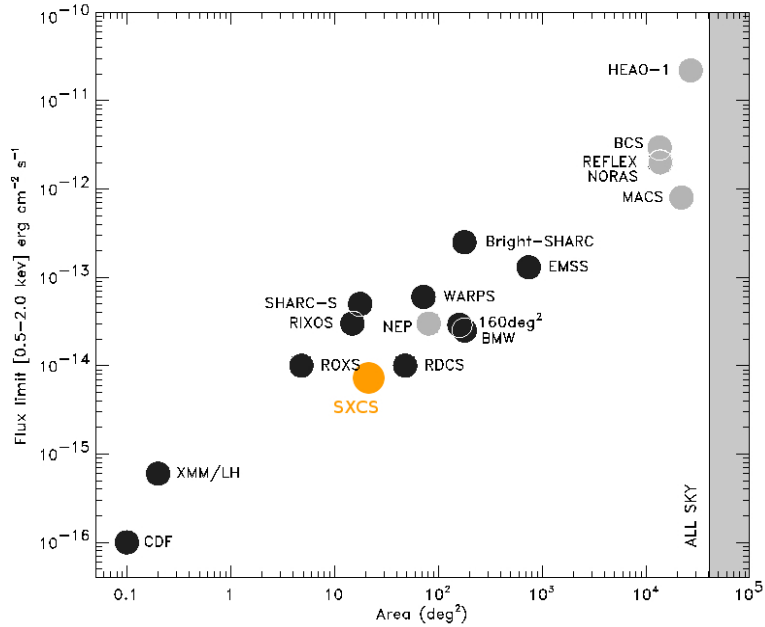

Figure 19: SXCS flux limit *versus* solid angle (orange point) compared with other present day X-ray cluster surveys.

LOGN-LOGS    The LogN-LogS is defined as the logarithm of the number of sources per unit of solid angle with flux greater than a flux limit, $\text{Log}(N(>S))$, as a function of the flux limit, $S$. The number density $N(>S)$ can be computed as:

$$N(>S) = \sum_i \frac{1}{\omega(S_i)},$$

(2.13)

where $\omega(S_i)$ is the value of the sky-coverage evaluated at the flux of the source $S_i$ and the sum is over all the sources with flux $S_i > S$. The resulting LogN-LogS of SXCS is shown in Figure 20, and it is consistent with the LogN-LogS measured with the RDCS by Rosati et al. (1998).

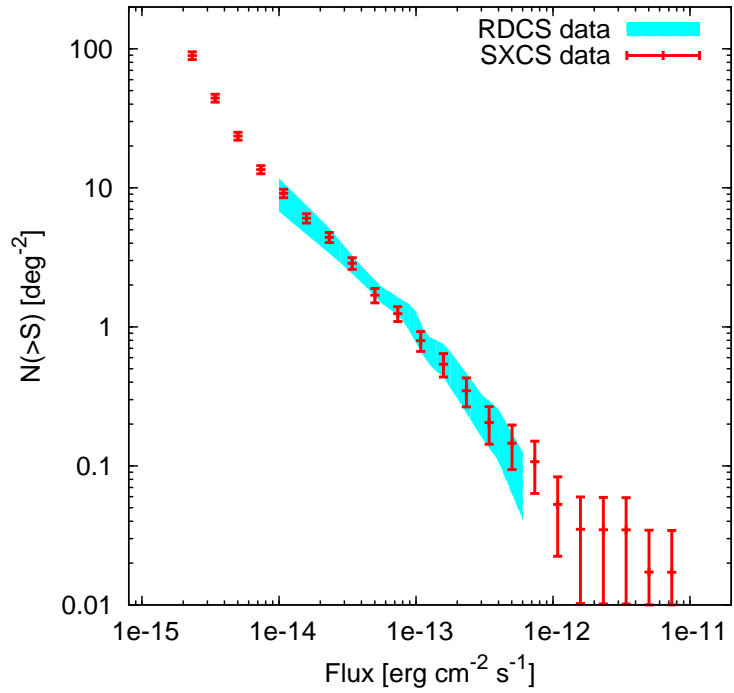

Figure 20: Soft LogN-LogS of SXCS. The red points are the data of the SXCS, which are compared with the RDCS data by Rosati et al. (1998) (shaded region).

## 2.10 CONCLUSIONS

In this Chapter I focused on the details about the realization of a new X-ray survey of galaxy clusters, the Swift-XRT Cluster Survey (SXCS). This survey is obtrined from the Swift X-Ray Telescope (XRT) archive. Beyond the detailed investigation of the XRT characteristics and the analysis of the XRT data, I developed two new algorithm: the Two Dimensional Fitting (TDF) algorithm and an X-ray image simulator. The former is a fitting algorithm that models the surface brightness profile of all the sources in an X-ray images and the background at the same time; the latter creates realistic X-ray simulated images including populations of all the source types contributing to the extragalactic X-ray sky. Both these algorithms are created specifically for XRT, but they are also flexible and can be used for other present and future X-ray telescopes.

The final catalogue of the SXCS consists of 243 extended sources detected with high significance. The number of sources detected as function of the flux limit is in very good agreement with previous studies, suggesting that this sample is complete and the flux measurement is accurate.

Among the sources in the SXCS catalogue, I selected the brightest ones requiring a minimum of 200 net counts to create a bright sample. The idea is to analyze and entirely characterize these sources on the basis of the X-ray data alone. This is possible given the high number of net counts. The results about the analysis of the bright sample are described in Chapter 3. Moreover, among all the sources in the SXCS catalogue I also selected a subsample of about 50 sources as possible high redshift candidates. In collaboration with P. Rosati, J. Santos and Y. Hang, we plan to do optical follow-up with Magellan and VLT for these candidates. This subsample was selected by choosing the sources whose optical counterpart is not visible in the Digital Sky Survey (DSS) or in Sloan Digital Sky Survey (SDSS), in which galaxy clusters are typically identified up to redshift 0.5. We expect to found about 10-15 clusters with redshift greater than one, given the SXCS sky-coverage. The future perspective is to extend the optical follow up to the whole SXCS catalogue to measure the redshift of as many sources as possible and eventually perform cosmological tests based on the entire cluster sample.

# A COMPLETE SAMPLE OF X-RAY GROUPS AND CLUSTERS OF GALAXIES FROM THE SXCS

In this Chapter I present the detailed X-ray properties of the brightest sources in the SXCS catalogue described in the previous Chapter. A cluster is included in the bright sample if it is detected with more than 200 net counts in the soft (0.5-2.0 keV) band. With this prescription 32 extended sources are selected. With such a high number of net counts it is possible to perform a high-quality spectral analysis for all the sources in the bright sample. The goal is to measure temperature, total luminosity, metal abundance and redshift, to get the total and the gas mass, and to compare the properties of this sample with the ones of other X-ray selected sample. This study will provide the first X-ray flux limited sample of clusters of galaxies fully characterized and usable for cosmology without additional optical follow-up. The results presented in this Chapter will be published in Bignamini et al. (2010).

## 3.1 BRIGHT SAMPLE PROPERTIES

The sources in the bright sample were extracted from the SXCS catalogue requiring a minimum of 200 net counts in the soft (0.5-2.0 keV) band. As described in Chapter 2, the surface brightness of all the sources was fitted with a single King profile and were selected as extended all the sources whose radius $r_{B2}$[1] is at more than $7\sigma$ from the average $r_{B2}$ of point sources, for a given SNR. With such a conservative choice, the expected contamination from spurious sources is virtually null. Therefore the bright sample can be considered complete and free of contamination.

As a check against any selection bias or any possible correlation with the Gamma Ray Burst (GRB) position, in Figure 21, left panel, the positions of the sources in the bright sample is plotted with respect to the GRB positions set to (0,0). We do not find any hint of an increasing surface density of extended sources towards the GRB position, as shown also in the right panel of Figure 21. Therefore, we can safely assume that our field selection is unbiased with respect to X-ray clusters (see also Berger et al., 2007).

The list of the 32 extended sources in the bright sample is shown in Table 4. We checked for counterparts in known X-ray or optical surveys with the NASA/IPAC Extragalactic Database (NED). In particular, the X-ray centroid of the bright sample sources was matched with the objects found in the NASA/IPAC Extragalactic Database (NED) database, searching for galaxy clusters and galaxies within a distance from the X-ray centroid of $1'$ and $10''$, respectively. For galaxies we also required that they are archived with optical spectroscopic redshift. In this way, given the proximity with the center of the X-ray sources, archived clusters can be tentatively identified as the detected clusters while the archived galaxies as cluster members. With these prescription the bright sample is subdivided as follows: 18 sources are new object; 5 are already known as galaxy clusters; 9 are already known but not identified as clusters, of which 5 as ROSAT X-ray sources and 4 as galaxies[2]. In addition, 14 clusters have SDSS data available, with at least photometric redshift for all the galaxies in the field. Moreover, we observed with the Telescopio Nazionale Galileo (TNG) 3 clusters in our sample, confirming the existence of the clusters themself and successfully measuring the spectroscopic redshift[3] of their member galaxies. In summary we gathered 12 optical (spectroscopic or photometric) redshifts, which are listed in the last column of Table 4.

---

1 The definition of $r_{B2}$ is given in Equation 2.11.

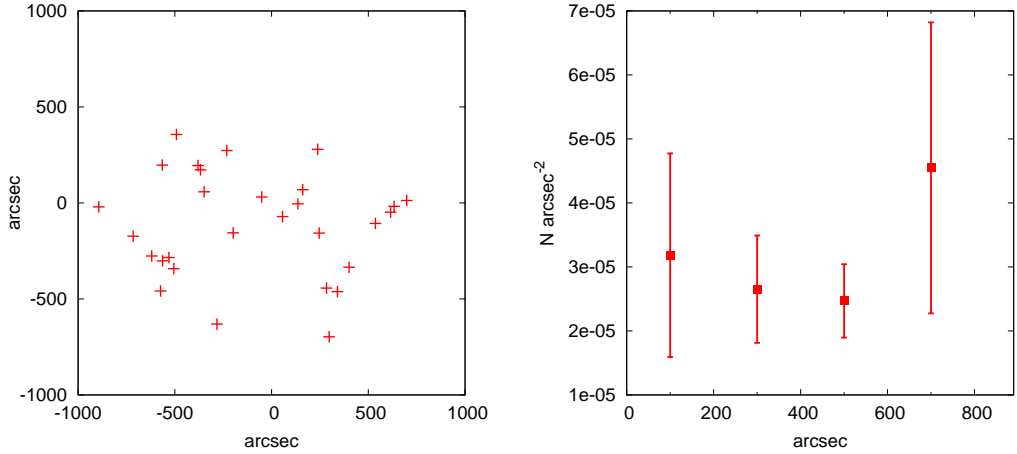

Figure 21: Left panel: positions of the sources in the bright sample with respect to the GRB position. Right panel: surface density of the sources in the bright sample as a function of the distance from the GRB.

In Table 5 we list the basic properties of the X-ray sources in the bright sample. The photometry of our extended sources refers to the extraction radius $r_{ext} = r_{B2}$ defined as the radius where the fitted surface brightness falls below half of the measured background of the image. As we said in advance, the bright sample is obtained by selecting all the sources with at least 200 net counts in the soft band. Therefore, the sky-coverage of the bright sample is obtained directly from the sharp limit of 200 net detected counts in the soft band applied

---

2 Indeed, 2MASXJ21451552-1959406 with spectroscopic redshift $z = 0.058700$ is about $1''$ from the X-ray centrod of SWJ2145-1959. However, it is clearly a foreground galaxy, since the cluster redshift as measured from X-ray spectral analysis is $z = 0.335^{0.013}_{0.014}$ and its K-$\alpha$ iron emission line is unmistakable.

3 Thanks to D. Fugazza.

| Name | GRB | R.A. | Dec | Catalogue | Distance | Published $z$ |
|------|-----|------|-----|-----------|----------|---------------|
| SWJ0217-5014 | GRB050406 | 34.270653 | -50.236660 | 1AXG | 0.787 | - |
| SWJ0927+3010 | GRB050505* | 141.874588 | 30.179701 | 1AXG | 0.314 | 0.35[7] |
| SWJ0926+3013 | GRB050505* | 141.708237 | 30.229328 | - | - | - |
| SWJ0927+3013 | GRB050505* | 141.831726 | 30.228964 | MaxBCG | 0.186 | 0.30[1,7] |
| SWJ0239-2505 | GRB050603 | 39.850075 | -25.084248 | 2dFGRS | 0.084 | 0.174[2] |
| SWJ2322+0548 | GRB050803 | 350.702240 | 5.803986 | - | - | - |
| SWJ1737+4618 | GRB050814 | 264.339539 | 46.309334 | - | - | - |
| SWJ2323-3130 | GRB051001 | 350.939148 | -31.512819 | - | - | - |
| SWJ1330+4200 | GRB051008* | 202.731781 | 42.004669 | 2MASX | 0.061 | 0.017 |
| SWJ0847+1331 | GRB051016B* | 131.955215 | 13.528370 | - | - | - |
| SWJ0821+3200 | GRB051227* | 125.307526 | 32.002674 | - | - | - |
| SWJ1949+4616 | GRB060105 | 297.330414 | 46.272861 | - | - | - |
| SWJ1406+2743 | GRB060204B* | 211.654831 | 27.730812 | WGA | 1.045 | - |
| SWJ1406+2735 | GRB060204B* | 211.664337 | 27.597130 | MaxBCG | 0.161 | 0.25[1,7] |
| SWJ1145+5953 | GRB060319* | 176.469711 | 59.887722 | 2MASX | 0.080 | 0.1463 |
| SWJ1551+4451 | GRB060904A* | 237.823303 | 44.857132 | - | - | - |
| SWJ0352-0043 | GRB060904B* | 58.247223 | -0.727059 | SDSS | 0.015 | 0.335 |
| SWJ0629+4607 | GRB061028 | 97.311806 | 46.120747 | CIZA ZwCl | 0.148 | 0.13[4] |
| SWJ0948-1316 | GRB061121 | 147.070740 | -13.278816 | - | - | - |
| SWJ2015+1529 | GRB061122 | 303.848236 | 15.497838 | - | - | - |
| SWJ0003-5255 | GRB070110 | 0.808732 | -52.920033 | - | - | - |
| SWJ0003-5253 | GRB070110 | 0.849941 | -52.898319 | - | - | - |
| SWJ1935+0214 | GRB070917 | 293.840546 | 2.235848 | - | - | - |
| SWJ1059+5348 | GRB071018* | 164.941711 | 53.802010 | SDSS-GCl | 0.098 | 0.072[5] |
| SWJ2336-3136 | GRB071028B | 354.066040 | -31.604403 | ABELL | 0.358 | 0.0625[6] |
| SWJ0024-5803 | GRB071031 | 6.157714 | -58.064888 | - | - | - |
| SWJ1432+3617 | GRB080319B* | 218.097107 | 36.297855 | - | - | - |
| SWJ0232-7117 | GRB080411 | 38.172527 | -71.298302 | - | - | - |
| SWJ0233-7116 | GRB080411 | 38.262535 | -71.276505 | - | - | - |
| SWJ2144-1956 | GRB080413B | 326.041351 | -19.933523 | RXS | 0.537 | - |
| SWJ2145-1959 | GRB080413B | 326.314331 | -19.994783 | - | - | - |
| SWJ0150+6125 | GRB081024A* | 27.589357 | 61.418316 | WGA | 0.180 | - |

Table 4: List of the sources in the bright sample. The first column is the source ID in the XRT source catalog (Moretti et al. 2010); the second column refers to the corresponding GRB field (fields marked with "*" have SDSS data). Third and fourth columns show the R.A. and Dec position of the X-ray centroid. The fifth column shows the catalog of the counterpart found in NED with the criteria described in the text. The sixth column shows the distance from the X-ray centroid in arcmin, and the last column the published redshift. Notes. [1] Koester et al. (2007); [2] Colless et al. (2003); [3] Gal et al. (2003); [4] Kocevski et al. (2007); [5] von der Linden et al. (2007); [6] Schwope et al. (2000); [7] TNG follow-up.

| Name | Expmap [s] | Net Counts | SNR | Flux [$10^{-13}$ erg/cm$^2$/s] | $N_H$ [$10^{22}$cm$^{-2}$] | Background [$10^{-3}$ cts/arcsec$^2$] | $r_{ext}$ [arcsec] |
|---|---|---|---|---|---|---|---|
| SWJ0217-5014 | 133717 | 543±30 | 18.0 | 1.02±0.06 | 0.018 | 5.18 | 116.6 |
| SWJ0927+3010 | 85805 | 767±34 | 23.2 | 2.23±0.10 | 0.017 | 7.47 | 116.2 |
| SWJ0926+3013 | 164969 | 289±22 | 13.2 | 0.44±0.03 | 0.017 | 7.47 | 65.0 |
| SWJ0927+3013 | 157316 | 370±31 | 13.8 | 0.59±0.05 | 0.017 | 7.47 | 90.3 |
| SWJ0239-2505 | 75421 | 341±27 | 13.5 | 1.15±0.09 | 0.022 | 9.63 | 129.8 |
| SWJ2322+0548 | 207450 | 1622±55 | 32.9 | 2.15±0.07 | 0.050 | 9.39 | 118.8 |
| SWJ1737+4618 | 137248 | 448±31 | 15.4 | 0.83±0.06 | 0.023 | 6.52 | 106.4 |
| SWJ2323-3130 | 102561 | 204±17 | 11.6 | 0.49±0.04 | 0.011 | 3.76 | 68.9 |
| SWJ1330+4200 | 168584 | 732±39 | 22.2 | 1.06±0.06 | 0.010 | 6.70 | 97.8 |
| SWJ0847+1331 | 74419 | 4294±78 | 59.3 | 15.04±0.27 | 0.032 | 8.16 | 223.7 |
| SWJ0821+3200 | 119159 | 697±40 | 20.6 | 1.53±0.09 | 0.034 | 5.72 | 119.3 |
| SWJ1949+4616 | 117420 | 404±26 | 14.6 | 1.23±0.08 | 0.148 | 6.05 | 103.7 |
| SWJ1406+2743 | 80171 | 1039±42 | 27.5 | 3.23±0.13 | 0.017 | 3.81 | 142.5 |
| SWJ1406+2735 | 78534 | 591±37 | 18.7 | 1.87±0.12 | 0.017 | 3.81 | 148.4 |
| SWJ1145+5953 | 479992 | 499±37 | 15.6 | 0.26±0.02 | 0.015 | 19.48 | 74.8 |
| SWJ1551+4451 | 147518 | 226±20 | 11.2 | 0.38±0.03 | 0.013 | 5.06 | 78.2 |
| SWJ0352-0043 | 106949 | 1516±47 | 34.0 | 4.65±0.15 | 0.114 | 4.07 | 144.1 |
| SWJ0629+4606 | 7220 | 803±29 | 25.3 | 38.09±1.38 | 0.131 | 1.55 | 379.6 |
| SWJ0948-1316 | 160740 | 266±26 | 12.2 | 0.44±0.04 | 0.040 | 10.01 | 73.7 |
| SWJ2015+1529 | 169258 | 1484±56 | 28.9 | 2.95±0.11 | 0.125 | 10.42 | 134.6 |
| SWJ0003-5255 | 316209 | 721±37 | 19.1 | 0.57±0.03 | 0.016 | 18.99 | 77.8 |
| SWJ0003-5253 | 303740 | 694±35 | 18.0 | 0.57±0.03 | 0.016 | 18.99 | 84.8 |
| SWJ1935+0214 | 90758 | 1133±41 | 29.4 | 4.71±0.17 | 0.169 | 5.42 | 109.4 |
| SWJ1059+5348 | 22491 | 248±25 | 11.9 | 2.69±0.27 | 0.008 | 1.26 | 174.4 |
| SWJ2336-3136 | 43924 | 1492±48 | 31.2 | 8.35±0.27 | 0.012 | 3.44 | 201.0 |
| SWJ0024-5803 | 76010 | 298±27 | 13.6 | 0.97±0.09 | 0.012 | 3.84 | 94.1 |
| SWJ1432+3617 | 333299 | 454±32 | 16.9 | 0.33±0.02 | 0.011 | 13.65 | 64.8 |
| SWJ0232-7117 | 324085 | 699±44 | 21.1 | 0.61±0.04 | 0.058 | 14.59 | 68.8 |
| SWJ0233-7116 | 324059 | 842±37 | 21.2 | 0.73±0.03 | 0.058 | 14.59 | 92.7 |
| SWJ2144-1956 | 78298 | 1784±51 | 37.5 | 5.92±0.17 | 0.031 | 5.62 | 129.4 |
| SWJ2145-1959 | 64304 | 1457±71 | 27.6 | 5.88±0.29 | 0.031 | 5.62 | 238.3 |
| SWJ0150+6125 | 13989 | 991±41 | 27.2 | 79.73±3.34 | 0.771 | 0.93 | 278.0 |

Table 5: X-ray properties of the sources in the bright sample. The second column refers to the effective exposure of the field at the position of the source in seconds. The third column shows the net counts and the $1\sigma$ error in the soft band (0.5-2.0 keV) computed from simple aperture photometry. In fourth column the SNR is shown. The fluxes in the fifth column refer to the soft band as well and are in units of [$10^{-13}$ erg/cm$^2$/s], and they are corrected by local Galactic absorption, shown in the sixth column in units of [$10^{22}$cm$^{-2}$]. The seventh column shows the measured background in units of [$10^{-3}$ cts/arcsec$^2$]. The last column shows $r_{ext}$ in arcsec, which is the radius where the surface brightness is half of the background.

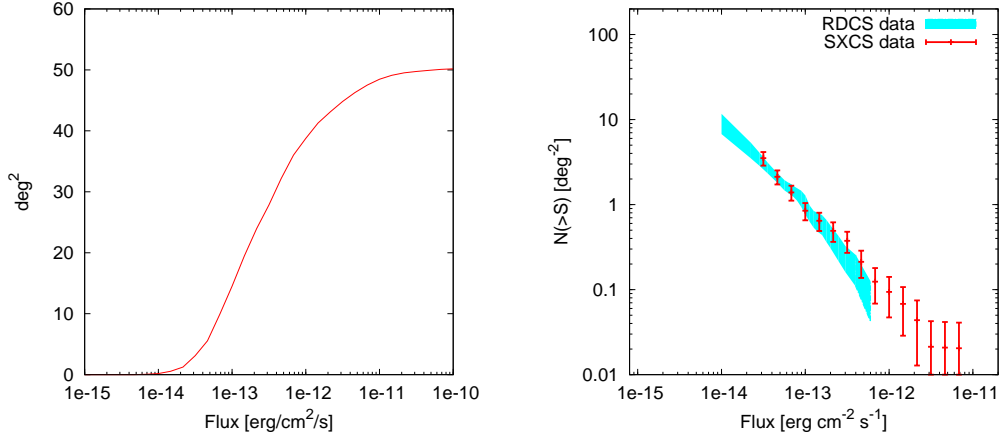

Figure 22: Sky-coverage and LogN-LogS for the SXCS bright sample. In left panel, the sky-coverage is obtained assuming a detection limit of 200 net counts and specific ECFs computed from the Galactic hydrogen column density $N_H$ of each field is used. In right panel, cumulative number counts derived from the bright sample (red square), compared to the LogN-LogS derived from the RDCS (shaded region Rosati et al., 1998).

to each field, after accounting for the vignetting. The sky-coverage of the bright sample is shown in left panel of Figure 22.

The flux of each source in the soft band is measured after the spectral fitting (see next Section). No correction has been made to correct for the expected flux at radii larger than $r_{ext}$. The resulting LogN-LogS is shown in right panel of Figure 22, and it is consistent with the LogN-LogS measured with the RDCS (Rosati et al., 1998) shown with shaded region.

## 3.2 SPECTRAL ANALYSIS

The spectra were extracted within the radius $r_{ext}$ after removal of the point sources. We sampled the background from the same observation, in a region typically three times larger by size than the extraction region of the source spectra. The net counts were obtained with simple aperture photometry, by subtracting the total number of events in the background extraction region scaled by the area ratio from the number of events in the cluster region. The net counts error was obtained from the Poissonian error of the numbers of counts. Calibration files, RMF and ARF, were built for each extraction region, as described in Chapter 2.

The background subtracted spectra were analyzed with *XSpec* v.12.3.0 (Arnaud, 1996). We used the C-stat as the statistical criterion to find the best fit models (see Cash, 1979; Bevington & Robinson, 2002; Arnaud, 2004). All our sources are fitted with a single temperature `mekal` model (Kaastra, 1992; Liedahl et al., 1995) with four free parameters: temperature, metallicity, redshift and normalization. The ratio between the elements were fixed to the solar values as in Grevesse & Sauval (1998). We modeled the Galactic absorption with `tbabs` (Wilms, Allen & McCray, 2000) fixing the Galactic neutral Hydrogen column density to the values measured by Dickey & Lockman (1990).

The fits were performed over the energy range 0.5-7.0 keV. We removed lower energy photons in order to avoid uncertainties due to rapidly increasing XRT background below 0.5 keV. The cut at high energies, instead, is imposed by the rapidly decreasing SNR ratio, due to the combination of the lower effective area of XRT, and of the exponential cut-off to the thermal spectra. With this choice we also avoid the strong calibration lines present in the XRT background (see Section 2.3).

Single temperatures were measured with a typical $1\sigma$ error of 10%. The presence of emission lines, mostly the K-shell iron emission lines at 6.7-6.9 keV, allow us to measure at the same time the iron abundance and the redshift. For groups, whose temperature is significantly below 3 keV (e.g. SWJ1330+4200, SWJ1145+5953, SWJ0948-1316, and SWJ0024-5803) redshift and iron abundance are measured thanks to the L-shell emission lines at $\sim$ 1.0-2.0 keV.

Taking advantage of the good spectral resolution of XRT, we measured the redshift for all the sources in our sample except SWJ1737+4618 and SWJ0233-7116. The spectrum of SWJ1737+4618 has too noisy emission lines and the redshift determination is extremely uncertain. Instead, SWJ0233-7116 does not have any emission line in its spectrum. However, they are not false detection since in both cases the X-ray emission is detected with high significance and both sources are clearly extended.

## 3.3 LUMINOSITY AND MASS DERIVATION

The luminosity and the cluster mass were computed from the surface brightness profile fit. As described in Chapter 2, the surface brightness of all sources was fitted with a King profile

$$S(r) \propto \left[ 1 + \left( \frac{r}{r_c} \right) \right]^{-3\beta_m + 1}, \tag{3.1}$$

with $\beta = 2/3$ for all sources. The electron number density profile associated with this surface brightness profile is the $\beta$-model

$$n_e(r) = n_e(0) \left[ 1 + \left( \frac{r}{r_c} \right)^2 \right]^{-3/2\beta_m}. \tag{3.2}$$

The bolometric luminosities, and the total and gas mass were then evaluated at $R_{\Delta_c} = R_{500}$ that describes the sphere within which the cluster overdensity is $\Delta_c = 500$ with respect to the critical density. Under the assumption of isothermality and hydrostatic equilibrium, we calculated therefore:

$$M_{tot}(<r) = -\frac{kTr}{\mu m_p G}\frac{d\log n_e}{d\log r}, \tag{3.3}$$

$$R_{500} = \left(\frac{3M_{tot}(<R_{500})}{4\pi 500\rho_c(z)}\right)^{1/3}, \tag{3.4}$$

$$M_{gas}(<R_{500}) = \int_0^{R_{500}} \rho_{gas}(r)4\pi r^2 dr. \tag{3.5}$$

In the previous equations $m_p$ is the proton mass, $\mu$ is the mean molecular weight ($\mu = 0.6$ for a primordial composition with 76% of hydrogen), and $G$ is the gravitational constant. The critical density at redshift $z$ is equal to $\rho_c(z) = 3H_z^2/(8\pi G)$ with $H_z = H_0 E(z)$. The temperature $kT$ is the one computed with the spectral fit within $r_{ext}$. The gas density, $\rho_{gas}(r)$ has the same profile of the electron number density profile, and the central value is obtained converting the central surface brightness by a conversion factor derived from the *XSpec* fit. The gas mass fraction is then $f_{gas}(R_{\Delta_c}) = M_{gas}(<R_{\Delta_c})/M_{tot}(<R_{\Delta_c})$.

We computed the X-ray bolometric luminosities within $r_{ext}$ integrating over the entire X-ray band. We computed the total luminosity by accounting for emission outside the extraction radius, by extrapolating the surface brightness profile up to the radius $R_{500}$.

Errors on $R_{500}$ and on the masses were estimated from Monte Carlo simulations. For each output parameter ($R_{500}$, the total mass and the gas mass) a new set of input parameters (temperature, abundance, redshift and $r_c$) was simulated 1000 times assuming a Gaussian distribution with $\sigma$ equal to the measure error of the input parameters. The errors of the output parameters were obtained from the $1\sigma$ confidence level of the resulting probability distribution.

Instead, errors on luminosities were taken from the Poissonian error on the net detected counts combined with the errors on redshift (upon which depends the luminosity distance) with the usual rules of error propagation. The final results are shown in Table 5. We adopted a $\Lambda$CDM cosmology with $\Omega_m = 0.3$, $\Omega_\Lambda = 0.7$ and $H_0 = 70$ km s$^{-1}$ Mpc$^{-1}$ (Spergel et al. 2007).

## 3.4 REDSHIFT DISTRIBUTION

Having a reliable estimate of the cluster redshift is a crucial aspect to study cluster properties. In fact, the ICM temperature is derived from the X-ray spectrum, and the exponential cut off in the thermal bremsstrahlung spectrum is redshifted by $(1 + z)$. The redshift is fundamental to measure the X-ray luminosity from the X-ray flux, and masses from the integral of the X-ray surface brightness profiles within a physical radius.

| Name | $kT$ [keV] | $z$ | $X_{Fe}/X_{\odot}$ | $r_{ext}$ [kpc] | $r_{500}$ [kpc] | $L_{ext}$ [$10^{44}$ erg/s] | $L_{500}$ [$10^{44}$ erg/s] |
|---|---|---|---|---|---|---|---|
| SWJ0217-5014 | $6.3^{+0.8}_{-0.7}$ | $0.516^{+0.008}_{-0.008}$ | $1.08^{+0.38}_{-0.28}$ | $730\pm39$ | $936\pm51$ | $4.21^{+0.33}_{-0.33}$ | $4.53^{+0.35}_{-0.36}$ |
| SWJ0927+3010 | $3.0^{+0.2}_{-0.2}$ | $0.417^{+0.018}_{-0.039}$ | $0.49^{+0.17}_{-0.13}$ | $540\pm52$ | $741\pm72$ | $2.49^{+0.59}_{-1.07}$ | $2.63^{+0.62}_{-1.13}$ |
| SWJ0926+3013 | $2.9^{+0.4}_{-0.2}$ | $0.412^{+0.058}_{-0.029}$ | $1.21^{+0.83}_{-0.44}$ | $392\pm42$ | $658\pm71$ | $0.84^{+0.20}_{-0.12}$ | $0.96^{+0.23}_{-0.14}$ |
| SWJ0927+3013 | $2.2^{+0.3}_{-0.3}$ | $0.252^{+0.043}_{-0.023}$ | $0.72^{+0.57}_{-0.31}$ | $402\pm123$ | $632\pm194$ | $0.29^{+0.08}_{-0.05}$ | $0.34^{+0.10}_{-0.06}$ |
| SWJ0239-2505 | $3.4^{+0.5}_{-0.5}$ | $0.201^{+0.005}_{-0.007}$ | $1.19^{+0.70}_{-0.50}$ | $430\pm24$ | $833\pm46$ | $0.27^{+0.04}_{-0.04}$ | $0.33^{+0.04}_{-0.05}$ |
| SWJ2322+0548 | $2.9^{+0.3}_{-0.1}$ | $0.227^{+0.003}_{-0.002}$ | $1.09^{+0.25}_{-0.16}$ | $432\pm18$ | $771\pm33$ | $0.82^{+0.04}_{-0.04}$ | $0.91^{+0.05}_{-0.04}$ |
| SWJ1737+4618 | $3.5^{+0.5}_{-0.3}$ | $0.280^{+0.030}_{-0.032}$ | $0.81^{+0.49}_{-0.29}$ | $416\pm32$ | $825\pm64$ | $0.33^{+0.12}_{-0.12}$ | $0.41^{+0.15}_{-0.15}$ |
| SWJ2323-3130 | $6.1^{+1.3}_{-0.7}$ | $0.964^{+0.033}_{-0.037}$ | $0.48^{+0.38}_{-0.24}$ | $548\pm47$ | $708\pm61$ | $8.06^{+1.11}_{-1.20}$ | $8.55^{+1.21}_{-1.27}$ |
| SWJ1330+4200 | $0.7^{+0.0}_{-0.0}$ | $0.055^{+0.007}_{-0.010}$ | $0.69^{+0.68}_{-0.16}$ | $120\pm8$ | $420\pm29$ | $0.01^{+0.00}_{-0.00}$ | $0.01^{+0.00}_{-0.00}$ |
| SWJ0847+1331 | $6.4^{+0.3}_{-0.3}$ | $0.346^{+0.004}_{-0.003}$ | $0.43^{+0.09}_{-0.08}$ | $1096\pm91$ | $1072\pm89$ | $19.08^{+0.74}_{-0.60}$ | $19.03^{+0.74}_{-0.60}$ |
| SWJ0821+3200 | $5.0^{+0.7}_{-0.5}$ | $0.706^{+0.031}_{-0.019}$ | $0.44^{+0.26}_{-0.19}$ | $855\pm89$ | $727\pm76$ | $6.76^{+1.07}_{-0.73}$ | $6.38^{+1.02}_{-0.69}$ |
| SWJ1949+4616 | $3.3^{+0.3}_{-0.3}$ | $0.113^{+0.031}_{-0.025}$ | $0.61^{+0.29}_{-0.16}$ | $375\pm39$ | $837\pm88$ | $1.06^{+0.21}_{-0.05}$ | $1.30^{+0.25}_{-0.06}$ |
| SWJ1406+2743 | $9.2^{+1.3}_{-0.8}$ | $0.600^{+0.057}_{-0.008}$ | $0.74^{+0.31}_{-0.20}$ | $674\pm70$ | $1033\pm108$ | $26.98^{+4.11}_{-1.10}$ | $29.84^{+4.54}_{-1.22}$ |
| SWJ1406+2735 | $4.4^{+0.6}_{-0.4}$ | $0.216^{+0.012}_{-0.032}$ | $1.05^{+0.53}_{-0.36}$ | $520\pm117$ | $939\pm212$ | $3.21^{+0.55}_{-1.14}$ | $3.72^{+0.64}_{-1.32}$ |
| SWJ1145+5953 | $1.1^{+0.1}_{-0.2}$ | $0.200^{+0.014}_{-0.045}$ | $0.18^{+0.11}_{-0.06}$ | $246\pm49$ | $477\pm96$ | $0.05^{+0.01}_{-0.03}$ | $0.06^{+0.01}_{-0.03}$ |
| SWJ1551+4451 | $5.1^{+1.0}_{-0.6}$ | $0.401^{+0.007}_{-0.006}$ | $1.83^{+1.20}_{-0.66}$ | $420\pm56$ | $917\pm123$ | $3.13^{+0.34}_{-0.35}$ | $3.84^{+0.41}_{-0.43}$ |
| SWJ0352-0043 | $3.5^{+0.2}_{-0.2}$ | $0.311^{+0.008}_{-0.008}$ | $0.37^{+0.11}_{-0.09}$ | $657\pm60$ | $807\pm74$ | $2.96^{+0.23}_{-0.24}$ | $3.05^{+0.24}_{-0.24}$ |
| SWJ0629+4606 | $6.3^{+0.6}_{-0.4}$ | $0.113^{+0.008}_{-0.016}$ | $0.53^{+0.22}_{-0.16}$ | $393\pm11$ | $1209\pm36$ | $8.23^{+1.34}_{-0.97}$ | $9.86^{+1.61}_{-1.16}$ |
| SWJ0948-1316 | $0.7^{+0.0}_{-0.0}$ | $0.108^{+0.022}_{-0.017}$ | $0.36^{+0.46}_{-0.16}$ | $147\pm49$ | $412\pm139$ | $0.02^{+0.01}_{-0.01}$ | $0.03^{+0.01}_{-0.01}$ |
| SWJ2015+1529 | $3.6^{+0.3}_{-0.2}$ | $0.403^{+0.007}_{-0.008}$ | $0.32^{+0.11}_{-0.08}$ | $726\pm309$ | $743\pm316$ | $4.19^{+0.28}_{-0.30}$ | $4.22^{+0.28}_{-0.30}$ |
| SWJ0003-5255 | $4.0^{+0.3}_{-0.3}$ | $0.281^{+0.007}_{-0.007}$ | $0.98^{+0.37}_{-0.27}$ | $330\pm18$ | $870\pm47$ | $1.10^{+0.11}_{-0.11}$ | $1.47^{+0.14}_{-0.14}$ |
| SWJ0003-5253 | $4.7^{+0.4}_{-0.4}$ | $0.756^{+0.036}_{-0.009}$ | $0.50^{+0.20}_{-0.16}$ | $624\pm38$ | $666\pm40$ | $4.26^{+0.60}_{-0.29}$ | $4.42^{+0.62}_{-0.30}$ |
| SWJ1935+0214 | $7.4^{+0.6}_{-0.4}$ | $0.634^{+0.010}_{-0.008}$ | $0.49^{+0.14}_{-0.10}$ | $749\pm26$ | $969\pm34$ | $27.73^{+1.66}_{-1.46}$ | $28.70^{+1.71}_{-1.52}$ |
| SWJ1059+5348 | $2.9^{+0.4}_{-0.3}$ | $0.086^{+0.010}_{-0.009}$ | $1.52^{+0.88}_{-0.50}$ | $282\pm145$ | $828\pm426$ | $0.15^{+0.05}_{-0.04}$ | $0.20^{+0.06}_{-0.06}$ |
| SWJ2336-3136 | $2.0^{+0.1}_{-0.1}$ | $0.049^{+0.009}_{-0.013}$ | $0.46^{+0.09}_{-0.16}$ | $240\pm14$ | $691\pm42$ | $0.25^{+0.06}_{-0.04}$ | $0.29^{+0.07}_{-0.10}$ |
| SWJ0024-5803 | $1.1^{+0.1}_{-0.1}$ | $0.178^{+0.016}_{-0.016}$ | $0.50^{+0.36}_{-0.16}$ | $282\pm79$ | $491\pm138$ | $0.11^{+0.03}_{-0.03}$ | $0.13^{+0.04}_{-0.03}$ |
| SWJ1432+3617 | $2.7^{+0.4}_{-0.2}$ | $0.527^{+0.038}_{-0.032}$ | $0.67^{+0.45}_{-0.24}$ | $407\pm57$ | $620\pm86$ | $0.69^{+0.14}_{-0.13}$ | $0.76^{+0.16}_{-0.15}$ |
| SWJ0232-7117 | $5.4^{+1.1}_{-0.5}$ | $0.550^{+0.335}_{-0.193}$ | $0.19^{+0.14}_{-0.14}$ | $459\pm136$ | $815\pm242$ | $2.70^{+2.78}_{-1.65}$ | $3.04^{+3.14}_{-1.86}$ |
| SWJ0233-7116 | $5.5^{+0.6}_{-0.5}$ | $0.366^{+0.171}_{-0.154}$ | $0.20^{+0.26}_{-0.15}$ | $470\pm38$ | $971\pm79$ | $1.08^{+1.05}_{-0.96}$ | $1.34^{+1.31}_{-1.18}$ |
| SWJ2144-1956 | $7.4^{+0.6}_{-0.3}$ | $0.610^{+0.008}_{-0.008}$ | $0.50^{+0.13}_{-0.10}$ | $878\pm39$ | $980\pm43$ | $27.38^{+1.36}_{-1.35}$ | $27.76^{+1.38}_{-1.37}$ |
| SWJ2145-1959 | $4.1^{+0.4}_{-0.3}$ | $0.335^{+0.013}_{-0.020}$ | $0.49^{+0.19}_{-0.15}$ | $519\pm247$ | $862\pm409$ | $1.89^{+0.59}_{-0.61}$ | $2.57^{+0.80}_{-0.82}$ |
| SWJ0150+6125 | $9.6^{+1.3}_{-0.8}$ | $0.344^{+0.020}_{-0.005}$ | $0.80^{+0.23}_{-0.19}$ | $919\pm393$ | $1298\pm556$ | $43.27^{+6.45}_{-2.52}$ | $46.67^{+6.96}_{-2.72}$ |

Table 6: Results of the spectral fits of all the sources in the bright sample. Error bars refer to $1\sigma$ confidence levels. $r_{ext}$ is the extraction radius in kpc at the redshift measured from the X-ray data, while $r_{500}$ is computed with Equation 3.4. $L_{ext}$ is the bolometric luminosity within the extraction radius $r_{ext}$, while $L_{500}$ is the value extrapolated to $r_{500}$. Errors on luminosities are obtained from errors on net counts and redshift.

| Name | $M_{500}$ [$10^{13} M_\odot$] | $M_{gas,500}$ [$10^{13} M_\odot$] | $f_{gas,500}$ |
|---|---|---|---|
| SWJ0217-5014 | 41.78±6.76 | 2.53±0.62 | 0.060±0.005 |
| SWJ0927+3010 | 16.15±0.59 | 1.33±0.21 | 0.082±0.010 |
| SWJ0926+3013 | 13.83±1.33 | 0.69±0.13 | 0.050±0.005 |
| SWJ0927+3013 | 9.88±1.41 | 0.43±0.10 | 0.043±0.004 |
| SWJ0239-2505 | 20.46±4.41 | 0.79±0.16 | 0.040±0.009 |
| SWJ2322+0548 | 16.47±1.41 | 0.70±0.15 | 0.042±0.006 |
| SWJ1737+4618 | 20.82±2.70 | 0.58±0.13 | 0.028±0.003 |
| SWJ2323-3130 | 31.36±6.86 | 2.80±0.78 | 0.088±0.005 |
| SWJ1330+4200 | 2.25±0.10 | 0.02±0.00 | 0.010±0.002 |
| SWJ0847+1331 | 50.42±3.33 | 7.43±1.56 | 0.146±0.021 |
| SWJ0821+3200 | 24.30±3.49 | 3.07±0.68 | 0.125±0.010 |
| SWJ1949+4616 | 20.25±2.01 | 1.24±0.26 | 0.061±0.007 |
| SWJ1406+2743 | 69.00±10.87 | 8.64±2.08 | 0.124±0.010 |
| SWJ1406+2735 | 29.65±5.06 | 3.94±0.98 | 0.131±0.010 |
| SWJ1145+5953 | 3.81±0.58 | 0.15±0.04 | 0.039±0.003 |
| SWJ1551+4451 | 34.19±8.15 | 2.09±0.49 | 0.063±0.016 |
| SWJ0352-0043 | 20.74±1.50 | 1.47±0.32 | 0.070±0.010 |
| SWJ0629+4606 | 56.59±6.81 | 5.91±1.36 | 0.103±0.012 |
| SWJ0948-1316 | 2.23±0.19 | 0.06±0.01 | 0.025±0.004 |
| SWJ2015+1529 | 17.91±1.10 | 2.11±0.39 | 0.117±0.014 |
| SWJ0003-5255 | 25.18±1.87 | 1.85±0.33 | 0.073±0.008 |
| SWJ0003-5253 | 19.67±1.52 | 2.27±0.41 | 0.114±0.012 |
| SWJ1935+0214 | 52.36±4.92 | 6.06±1.36 | 0.114±0.015 |
| SWJ1059+5348 | 17.80±3.03 | 0.29±0.08 | 0.016±0.002 |
| SWJ2336-3136 | 9.99±0.63 | 0.26±0.06 | 0.026±0.004 |
| SWJ0024-5803 | 4.04±0.33 | 0.16±0.04 | 0.039±0.006 |
| SWJ1432+3617 | 12.23±1.45 | 0.74±0.16 | 0.060±0.006 |
| SWJ0232-7117 | 30.00±2.95 | 2.52±0.34 | 0.084±0.005 |
| SWJ0233-7116 | 38.20±1.69 | 1.34±0.17 | 0.035±0.003 |
| SWJ2144-1956 | 53.40±3.39 | 6.36±1.23 | 0.118±0.015 |
| SWJ2145-1959 | 22.97±1.91 | 1.81±0.35 | 0.078±0.009 |
| SWJ0150+6125 | 89.99±14.85 | 11.53±2.95 | 0.126±0.012 |

Table 7: Total mass and gas mass neasured for the SXCS bright sample. The total mass is obtained with Equation 3.3, while the gas mass is obtained with Equation 3.5 integrating the gas distribution up to $r_{500}$. The gas fraction in the last column is obtained simply by the ratio between the gas mass and the total mass. Error bars refer to $1\sigma$ confidence levels.

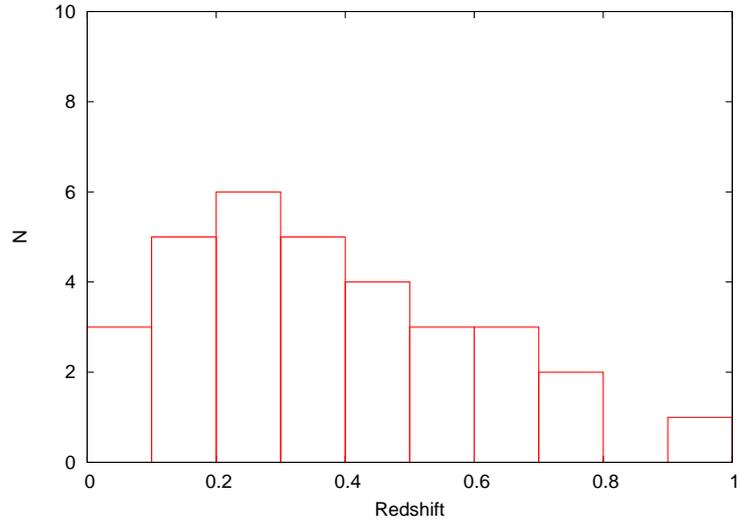

Figure 23: Redshift distribution for the clusters in the bright sample, according to the redshift measured by fitting X-ray spectra.

Taking advantage of the good spectral resolution of XRT, we measured the redshift for all the sources in our sample except for SWJ1737+4618 and SWJ0233-7116. Most of the measured redshift are based on the detection of the K-shell iron emission lines at 6.7-6.9 keV, but in some cases, for temperatures below 3 keV, L-shell lines are dominant (the spectra are shown in the Appendix B). The resulting redshift distribution is shown in Figure 23. Most of our clusters are at $z < 0.5$, but there are also 9 objects with redshift at $0.5 < z < 1.0$.

A check on the reliability of the X-ray measured redshfts can be done comparing them with optical redshifts available in literature or obtained from optical follow up. In particular, the X-ray centroid of the bright sample sources was matched with the objects found in the NED database, searching for clusters with archived photometric or spectroscopic redshift and galaxies with optical spectroscopic redshift within a distance from the X-ray centroid of $1'$ and $10''$ respectively. In this way, given the proximity with the center of the X-ray sources, archived clusters can be tentatively identified as the detected clusters while the archived galaxies as cluster members. Moreover, we observed with the Telescopio Nazionale Galileo (TNG) 3 clusters in our sample and we successfully measured the spectroscopic redshift of their member galaxies. In summary, there are: 2 clusters with photometric redshift, from MaxBCG catalogue; 3 clusters with spectroscopic redshift, one from ZwCl catalogue, one from ABELL, and one from SDSS-GCl; 4 galaxies with spectroscopic redshift, two from 2MASX catalogue, one from 2dFGRS, and one from SDSS; 3 clusters with spectroscopic redshift from the TNG. As shown in Figure 24, we found a satisfactory agreement from the comparison of the X-ray derived redshift and the redshift obtained from the optical.

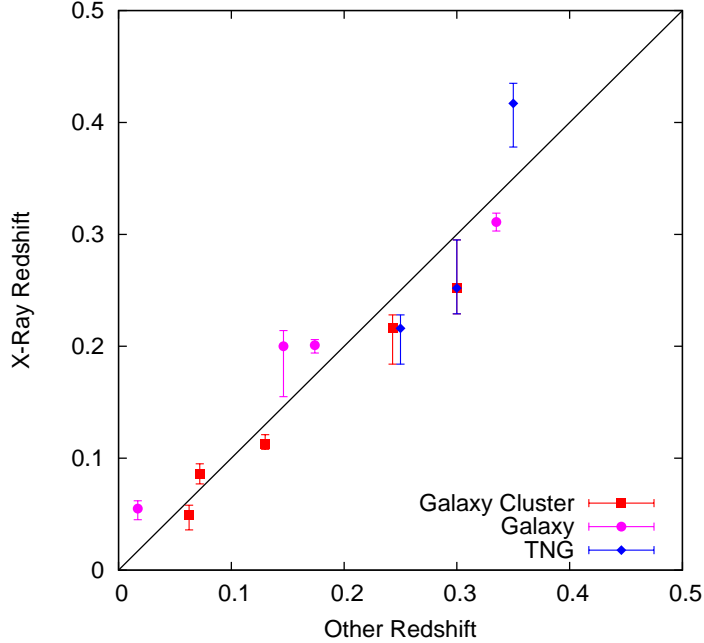

Figure 24: Comparison between the redshift derived from X-ray spectrum and the optical redshifts obtained from the NED catalogue or from the TNG follow-up.

The reliability of the measured redshift can also be tested against the distribution of photometric redshifts in all the cases, fourteen, where SDSS data are available. The redshift measured by X-ray spectral analysis almost always matches a peak in the distribution of photometric redshifts from the SDSS in the region of the cluster. The redshift histograms are shown in Appendix B.

From this simple check, we conclude that the X-ray redshift is reliable and in agreement with redshift form the optical. However, as one can see from Figure 24, the redshift errors are always very small with respect to the distance form the optical redshift. This seems to point towards a systematic underestimation of the error of X-ray derived redshifts. As a matter of fact, on closer inspection it seems to be a statistical issue due to the intrinsic nature of cluster X-ray spectra.

The $1\sigma$ error is computed form the $\Delta$C-stat, which depends on the degrees of freedom in the spectral fit (Avni, 1978; Cash 1981). But the error computed in this way gets the usual meaning of enclosing the 68% of the probability distribution only when the marginalized posterior probability distribution has a Gaussian-like shape, otherwise it still represents a "measure" of the error but its quantification and interpretation can be measleaded. Posterior probability distribution of the temperature and of the metallicity are almost Gaussian, but this is not always the case of redshift, whose marginalized C-stat has a Gaussian-like shape only for clusters with emission lines detected with high SNR and for a narrow redshift range close to the C-stat minimum (see the marginalized C-stat in Appendix B).

I suppose that this behaviour is due to the combination of two effects. First, X-ray spectra are represendad as counts divided into energy bins, therefore we have Poissonian error in each bin. Poissonian fluctuations of the spectrum can amplify or reduce the detection of real lines, thus modifying artificially the redshift error. Second, emission lines in thermal bremsstrahlung spectra of clusters are not single emission lines, but complexes of lines. Also the strongest emission lines, the K-shell iron lines, is a superimposition of lines with energies between 6.7 and 6.9 keV. This often produces in the marginalized C-stat of the redshift multiple relative minimum that can be confused, making the redshift posterior probability distribution non-Gaussian.

So to have a reliable estimate of the redshift error measured from X-ray spectra, it is needed a more robust statistical method that can take into account these systematics. This is a crucial issue for an X-ray mission like the future WFXT, whose intent is not only to detect a lot of new clusters, but also to measure directly from X-ray data the cluster properties, among which redshift has a key relevance.

### 3.5 THERMODYNAMICAL PROPERTIES OF THE ICM

We present and discuss the results on the thermodynamics of the ICM in our cluster sample. In Figure 25 we show the distribution of the measured temperatures, centered around 4.5 keV. Most of the clusters ($\sim$ 80%) have temperatures below 6 keV, however there are also a few objects with very high temperatures, two of tem (SWJ1406+2743 and SWJ1935+0214) with temperatures above 10 keV.

In Figure 26 left panel we show the correlation between iron abundance and temperature. The data suggest constant $X_{Fe} \sim 0.5 X_{Fe\odot}$ at temperature higher than 5 keV, whereas at lower temperature we found a much larger scatter and, on average, an higher iron abundance, and, at the level of groups, below 2 keV, again lower values of $X_{Fe}$. This is in line with findings by Baumgartner et al. (2003) for local clusters and by Balestra et al. (2007) at higher redshift. Finally, the correlation between iron abundance and redshift is shown in Figure 26 right panel. We compare our data with the results of Balestra et al. (2007) and Maughan et al. (2008). We found general agreement, with previous results, however the large errors on our best fit values hamper us from contsraining the $X_{Fe}$ evolution on the basis of the XRT data alone.

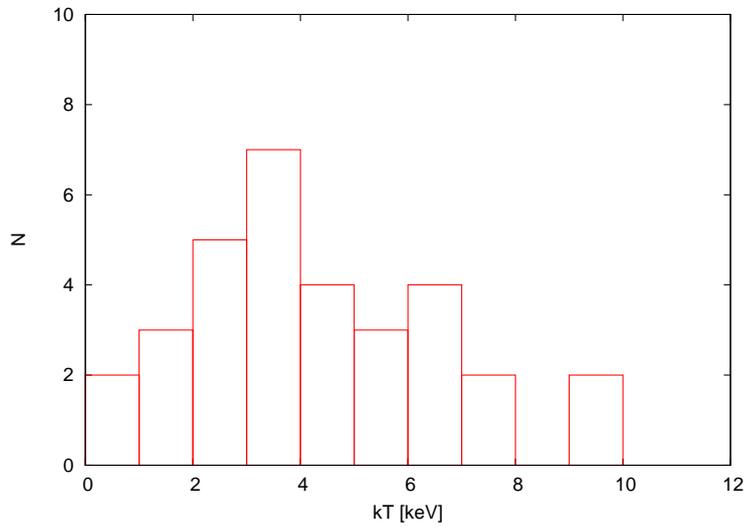

Figure 25: Temperature distribution for the clusters in the bright sample.

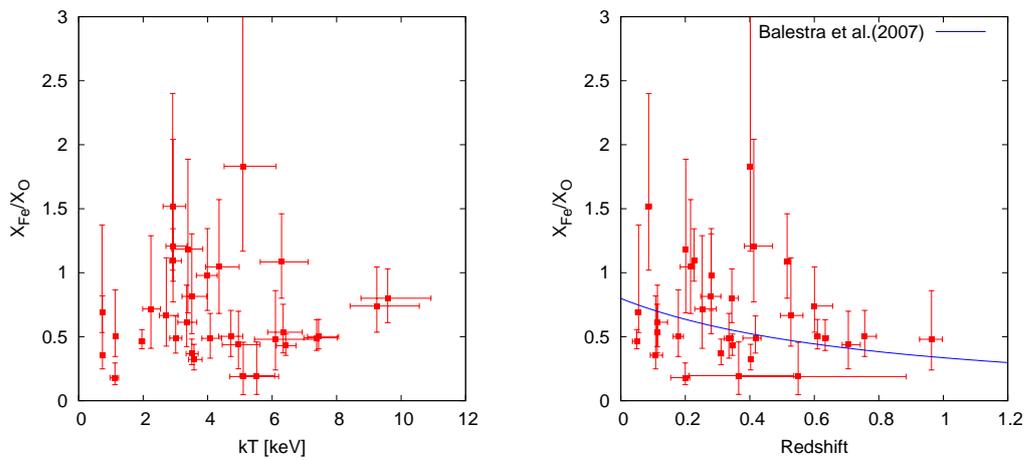

Figure 26: Left panel, iron abundance *versus* the temperature for the bright sample. Right panel, iron abundance *versus* X-ray redshift for the bright sample, compared with the best fit, blue solid line, obtained by Balestra et al. (2007). Error bars refer to $1\sigma$ confidence level.

## 3.6 SCALING RELATIONS

We present the scaling relations between total mass, gas mass and temperature of our sample compared with the results obtained by other authors studying X-ray selected clusters. From this analysis we excluded SWJ1737+4618 and SWJ0233-7116, for which the redshift is almost undetermined, and SWJ0239-2505 and SWJ1059+5348, since they are on the edge of the image and their X-ray emission is cut, missing a considerable part of their luminosity.

In Figure 27 left panel, we compare the bolometric luminosity within $R_{500}$ and the temperature measured for our clusters to the best fit of the $L_X - T_X$ relation as obtained by Branchesi et al. (2007) for a sample of 39 X-ray selected clusters with redshift $0.25 < z < 1.3$. The agreement between our data and the best fit of Branchesi et al. (2007) is very good, both at low and high temperatures.

In Figure 27 right panel, we compare the total mass within $R_{500}$ and the temperature for our sample to the best fits obtained by Arnaud et al. (2007) for a sample of 10 relaxed nearby clusters, $z < 0.15$, observed with *XMM-Newton*, and by Sun et al. (2009) for a sample of 43 nearby relaxed groups, $kT < 2.7$ and $z < 0.12$, observed with *Chandra*. Our data are in agreement with the best fit of the other authors, in particular at temperature above 2.0 keV. At lower temperature there is a small departure from the best fits, but the displacement is smaller if we consider only the data of Sun et al. (2009), blue solid line, since their sample consists only of groups.

In Figure 28 we investigate the gas mass in relation to the total mass and to the temperature, comparing our results with the ones obtained by Arnaud et al. (2007) and Sun et al. (2009). In Figure 28 left panel we show the $M - Y_X$ and in Figure 28 right panel the gas fraction within $R_{500}$ as function of the temperature. While the two reference best fit of $M - Y_X$ relation are almost identical, our data show a flatter slope and are in agreement with the previous results only for $Y_X > 10^{14}$ keV $M_\odot$. Moreover, the gas fraction is always smaller than the best fit from Sun et al. (2009), especially at low temperatures. Therefore combining these results with the fact that the $M - T$ of our sample is in agreement with previous works, we can conclude that possibly we underestimate the gas mass.

This effect can be ascribed to the different ways the gas mass is computed. First, Sun et al. (2009) adopted the 3D temperature profile as used in Vikhlinin et al. (2006) to compute $R_{500}$ and to extrapolate the temperature at that radius. Instead, we used one single temperature computed within $r_{ext}$, which is not directly related to $R_{500}$. As pointed out by Ettori et al. (2009), the gas fraction within $R_{500}$ obtained assuming the 3D temperature profile of Vikhlinin et al. (2006) is on average 16% larger than the gas fraction obtained estimating recursively $R_{500}$ with one single temperature.

Moreover, Sun et al. (2009) uses a more complex gas density profile with respect to the simple $\beta$-model, to better model the inner part of the gas distribution. This model combines the profile used by Vikhlinin et al. (2006) and profile proposed by Ettori et al. (2000) for cool core. Instead, we fitted the gas density profile with a single $\beta$-model, likely not properly following the gas profile in the center and then missing of a significant fraction of the gas mass. However, fitting the temperature profile or modeling the gas distribution with a more accurate function is impractical and counterproductive in the case of XRT, since the large PSF spreads out both the surface brightness and the temperature. Conversely, the data of Sun et al. (2009) are observed with *Chandra* that has a much better PSF, allowing this more refined analysis. Finally, there are considerable differences between the samples. Arnaud et al. (2007) and Sun et al. (2009) used samples of nearby, $z < 0.15$, and relaxed groups and clusers, while in our sample we have clusters up to redshift $\sim 1.0$ without any prescription on the relaxation state.

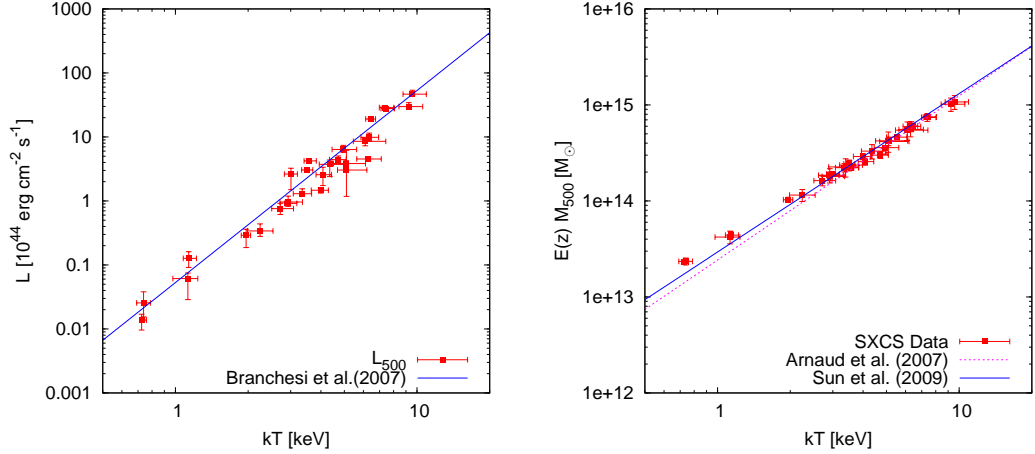

Figure 27: Left panel, $L_X - T_X$ relation for the SXCS bright sample compared with the best fit, blue solid line, for *Chandra* data analyzed by Branchesi et al. (2007c). Right panel, $M - T$ relation for the SXCS bright sample compared with the best fits, magenta dotted line and blue solid line, of Arnaud et al. (2007) and Sun et al. (2009) respectively. Error bars refer to $1\sigma$ confidence level.

## 3.7 CONCLUSIONS

In this Chapter I presented a detailed spectral analysis for the X-ray clusters of galaxies in the bright subsample of the catalogue of the Swift-XRT Cluster Survey (SXCS). For the first time, given the high signal to noise of the spectra and the well defined selection function, it was possible to provide a complete characterization of the X-ray properties of these objects from the X-ray data alone. For all the 32 clusters the measured temperature have a typical error $< 10\%$. The redshift was measured for all sources, except for two which showed no clear emission lines in their spectra. Furthermore, measuring the redshift solely by the X-ray spectrum results in an underestimation of the redshift error. The evolution of the iron abundance of the ICM in agreement with other X-ray slected samples. Scaling relations are consistent with those of other X-ray selected samples of clusters of galaxies as well.

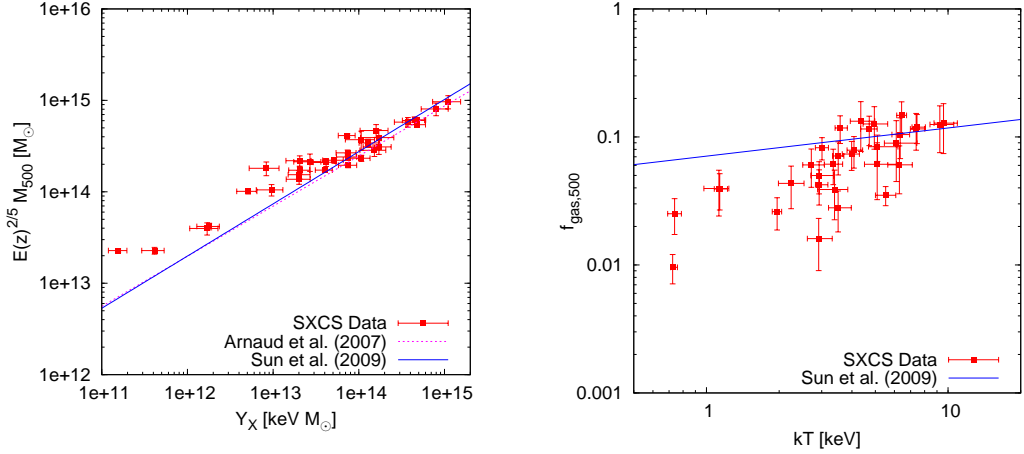

Figure 28: Left panel, $M - Y_X$ relation for SXCS bright sample compared with the best fits, magenta dotted line and blue solid line, of Arnaud et al. (2007) and Sun et al. (2009) respectively. Right panel, Gas fraction as function of the temperature for the SXCS bright sample compared with the best fit, blue solid line, of Sun et al. (2009). Error bars refer to $1\sigma$ confidence level.

These results show that the complete characterization of cluster of galaxies (i.e. detection, spectral analysis and derivation of cluter properties) only through the X-ray data is possible at least in the case of XRT and for sources with a suitable number of net counts. Previous X-ray clusters survey needed optical follow-up to derive redshift and thus luminosity, gas mass and possibly temperatures in order to get mass proxies. Instead, this sample rely on X-ray data only. Future missions dedicated to X-ray survey, e.g. WFXT, with large collecting area will be able to assemble large sample of X-ray clusters fully characterized on the basis of the X-ray data, allowing us to precision cosmology without time expensive follow-up in other wavebands. In the next future, we plan to extend the sample to lower fluxes, and to apply the classical cosmological tests to derive contraints on cosmological parameters.

s

# 4

## X-RAY PROPERTIES OF HIGH-Z OPTICALLY SELECTED CLUSTERS

The study of the thermodynamical and chemical properties of the Intra Cluster Medium (ICM) in high redshift clusters of galaxies is a powerful tool to investigate the formation and evolution of large scale structures. In this Chapter I discuss the X-ray properties of clusters of galaxies optically selected in the Red-sequence Cluster Survey (RCS) observed with the *Chandra* satellite, at redshifts $0.6 < z < 1.2$. I intend to assess the evolutionary stage of optically selected high-z clusters of galaxies, performing a spectral analysis of the diffuse emission from their ICM. I also study the Cool Core fraction in this optically selected sample, and I investigate the distribution of AGN in the cluster surroundings. Moreover, for the cluster RCS0224-0002 I compare the mass obtained from X-ray, with the one derived from strong lensing features observed with the Hubble Space Telescope (HST).

The results described in this Chapter are published in Bignamini et al. (2008), Santos et al. (2008), and Rzepecki et al. (2007).

### 4.1 THE RED-SEQUENCE CLUSTER SURVEY

X-ray selection has been shown to be very effective in providing complete samples of clusters of galaxies, with little contamination and a well-defined selection function. On the other hand, optical selection, based in particular on the presence of a the red-sequence, is very effective (and much cheaper in terms of observing strategies) in finding a large number of candidates. Recently, this technique has been pushed to high redshift, as in the case of the Red-sequence Cluster Survey (RCS), which uses the color-magnitude relation of early-type galaxies as cluster finding method (Gladders & Yee, 2000).

The RCS is a 100 square degree imaging survey obtained using mosaic CCD cameras on 4 m-class telescopes, and it is designed specifically to provide a sample of optically selected $0.2 < z < 1.2$ clusters. The red-sequence method is motivated by the observation that all rich clusters have a population of early-type galaxies which follow a well-defined color-magnitude relation (Yee & Gladders, 2001). The red-sequence represents the nominal reddest galaxy population in any group of galaxies at the same redshift, producing a well-defined signature in the color-magnitude diagram. The criterion to find cluster candidates is based on identifying galaxy over-densities in the four-dimension space given by color, magnitude and sky coordinate. For more details about the implementation of this method and the survey design see Gladders & Yee (2000, 2001, 2005), Yee & Gladders (2001) and Yee (1991).

On the other hand, the X-ray properties of optically selected clusters are usually different from that of X-ray selected clusters (see, e.g. Donahue et al., 2002). In particular, lack of X-ray emission in presence of a well-defined red sequence, can be ascribed to incomplete virialization. In this respect, a systematic study of the X-ray properties of an optically selected sample of clusters can provide key information in order to assess both the reliability of an optical sample for cosmological studies and the dynamical and thermodynamical status of the ICM in these clusters.

In this Chapter I focus on the X-ray properties of clusters of galaxies in the redshift range $0.6 < z < 1.2$ in the Red-sequence Cluster Survey, observed with the *Chandra* satellite. High redshift clusters are relevant both for cosmology and for studies of large scale structure formation and evolution. First, we investigate the thermodynamical properties of the ICM, which, given the low number of photons detected for each object, are described with a single-temperature model. Then we also investigate the distribution of X-ray point sources in the field of RCS clusters.

## 4.2 DATA REDUCTION AND ANALYSIS

Data reduction was performed using the CIAO 3.3 software package with the version 3.2.1 of the Calibration Database (CALDB 3.2.1). The list of the *Chandra* oservations of RCS clusters analyzed in this Chapter is shown in Table 8. The sample consists of all the public *Chandra* archived observations of RCS clusters to date (December 2007), for a total of 11 clusters.

| Cluster | $z$ | ObsID | Mode | Detector | Exp. [ks] | $N_H$ [$10^{20}$ cm$^{-2}$] |
|---|---|---|---|---|---|---|
| RCS1419+5326 | 0.620 | 3240 5886 | VFAINT | ACIS-S | 56.2 | 1.18 |
| RCS1107.3-0523 | 0.735 | 5825 5887 | VFAINT | ACIS-S | 93.0 | 4.25 |
| RCS1325+2858 | 0.750 | 3291 4362 | VFAINT | ACIS-S | 61.5 | 1.15 |
| RCS0224-0002 | 0.778 | 3181 4987 | VFAINT | ACIS-S | 100.9 | 2.92 |
| RCS2318.5+0034 | 0.780 | 4938 | VFAINT | ACIS-S | 50.0 | 4.14 |
| RCS1620+2929 | 0.870 | 3241 | VFAINT | ACIS-S | 33.7 | 2.67 |
| RCS2319.9+0038 | 0.900 | 5750 7172/3/4 | VFAINT | ACIS-S | 73.7 | 4.19 |
| RCS0439.6-2905 | 0.960 | 3577 4438 | VFAINT | ACIS-S | 92.0 | 2.64 |
| RCS1417+5305 | 0.968 | 3239 | VFAINT | ACIS-I | 62.2 | 1.23 |
| RCS2156.7-0448 | 1.080 | 5353 5359 | VFAINT | ACIS-S | 70.7 | 4.60 |
| RCS2112.3-6326 | 1.099 | 5885 | VFAINT | ACIS-S | 67.7 | 3.14 |

Table 8: List of the *Chandra* observations of the RCS cluster sample. Redshift are taken from the literature (Hicks et al., 2008; Gilbank et al., 2007). The sixth column shows the effective exposure times after removal of high background intervals. The last column shows the Galactic $N_H$ values measured by Dickey & Lockman (1990) (Table from Bignamini et al., 2008).

We started the data reduction from the *level* 1 event file. All the observations were taken with ACIS-S (except one with ACIS-I) in the very-faint (VFAINT) mode. This allowed us to run the tool `acis_process_events` to reduce significantly the instrumental background, using the values of the pulse heights in the outer 16 pixels of the $5 \times 5$ event island. We also performed a time-dependent gain adjustment with the tool `acis_process_events`. This adjustment is necessary because the effective gains of the detectors are drifting with time as the result of an increasing charge transfer inefficiency. The gain-file is used to compute the ENERGY and PI of an event from the PHA value. We needed to apply the Charge Transfer Inefficiency (CTI) correction only for ObsID 3239 taken with ACIS-I. This procedure is necessary to recover the original spectral resolution that is partially lost because of the increasing of CTI due to soft protons that damaged the ACIS front illuminated chips in the early phase of the *Chandra* mission. We filtered the data selecting events with the standard set of event grades 0, 2, 3, 4, 6. We also removed by hand hot columns and flickering pixels.[1] We also looked for time intervals with high background, by examining the light-curves of each observation. Usually, we remove only a few hundreds of seconds of the observing time. Only in two cases, for ObsID 3577 and ObsID 5750, we needed to remove a few thousands seconds from the nominal exposure.

Spectra were extracted from regions corresponding to radii which maximize the signal-to-noise in the energy range 0.5-6.0 keV. The extraction radii are in the range between 25 and 45 arcsec or between 200 and 350 kpc. After a visual inspection, we removed the contribution of point sources, most of them low-luminosity AGN, embedded in the ICM diffuse emission (not necessarily related to the cluster) as suggested by Branchesi et al. (2007b). For details about the point sources identification and their flux contribution see Section 4.4.

Background subtraction must be done accurately, given the low surface brightness of the sources. However, this procedure is simplified by the fact that the emission of each source is always within a radius of 40 arcsec. Therefore, we extracted the background from the same observation, in a region typically three times larger by size than the extraction region of the source spectra. Indeed, in our case a different procedure involving the creation of synthetic background spectra to match exactly the same position of the source on the detector, would be risky, especially due to fluctuations in the soft band associated with the Galactic emission. Calibration files (RMF and ARF) were built for the extraction regions, while soft and hard monochromatic exposure maps (for energies of 1.5 and 4.5 keV respectively) are used to compute aperture photometry of point sources.

The background subtracted spectra were analyzed with *Xspec* v.12.3.0 (Arnaud, 1996). Since the signal-to-noise ratio for our clusters is low, we used the C-stat as criterion to find the best fit models (Cash, 1979; Bevington & Robinson, 2002; Arnaud, 2004). In our spectral fits there are three free parameters: temperature, metallicity and normalization. We fitted the spectra with a single temperature `mekal` model (Kaastra, 1992; Liedahl et al., 1995) and model the Galactic absorption with `tbabs` (Wilms, Allen & McCray, 2000), fixing the Galactic neutral Hydrogen column density ($N_H$ in Table 8) to the value obtained with radio data (Dickey & Lockman, 1990). Redshifts were fixed to the values measured from the optical spectroscopy (Hicks et al., 2008; Gilbank et al., 2007).

---

[1] We identify the flickering pixels as the pixels with more than two events contiguous in time, where a single time interval is set to 3.3 s.

It has recently been shown that a methylene layer on the *Chandra* mirrors increases the effective area at energies larger than 2 keV (see Marshall et al., 2003) This has a small effect on the total measured fluxes, but it can have a non-negligible effect on the spectral parameters. To correct for this, we included in the fitting model a "positive absorption edge" (*Xspec* model `edge`) at an energy of 2.07 keV and with $\tau = -0.17$ (Vikhlinin et al., 2005). This multiplicative component artificially increases the hard fluxes by $\simeq 3.5\%$, therefore the final hard fluxes and luminosities computed from the fit are corrected downwards by the same amount.

The ratio between the elements are fixed to the solar values as in Anders & Grevesse (1989). These values for solar metallicity have been superseded by the new values of Grevesse & Sauval (1998) and Asplund et al. (2005), who found a 0.676 and 0.60 times lower iron solar abundance respectively. However, we prefer to report Fe abundances in units of solar abundances by Anders & Grevesse (1989), because most of the literature still refers to them. Since our measures of metallicity are not affected by the presence of other metals, since the only detectable emission lines are the H-like and He-like iron $K_\alpha$ line complex at rest-frame energies of 6.7-6.9 keV, the updated values can be obtained simply rescaling the values reported in Table 9 by the factor 1/0.676 or 1/0.60.

The fits were performed over the energy range 0.5-6.0 keV. Net counts were extracted in the same energy range. We removed low energy photons in order to avoid uncertainties in the ACIS calibration at low energies. The cut at high energies, instead, is imposed by the rapidly decreasing signal-to-noise ratio at energies larger than 6.0 keV, due to the combination of the lower effective area of ACIS, and of the exponential cut-off of the high-z thermal spectra.

Finally, we computed the X-ray bolometric luminosities with *Xspec* integrating over the entire X-ray band the analytical function describing the best fit of each spectrum, and adopting a $\Lambda$CDM cosmology with $\Omega_m = 0.3$, $\Omega_\Lambda = 0.7$ and $H_0 = 70$ km s$^{-1}$ Mpc$^{-1}$.

## 4.3 RESULTS ON THE PROPERTIES OF THE ICM

We present and discuss the results of the X-ray spectral analysis of the high-z RCS clusters. Spectra were extracted within a radius chosen in order to maximize the signal-to-noise ratio in the 0.5-6.0 keV band. The extraction radii and the number of net counts detected for each cluster are shown in Table 9. These values were obtained with simple aperture photometry, by subtracting the total number of events in the background extraction region scaled by the area ratio from the number of events in the cluster region. The net counts error was obtained from the Poissonian error of the numbers of counts.

Best fit temperatures, Fe abundances, fluxes and bolometric luminosities were also shown in Table 9; error bars refer to 1$\sigma$ confidence levels. We show that we are able to measure temperatures with a typical $1\sigma$ error bar of 20-30%. Errors on luminosities are taken from the Poissonian error on the net detected counts.

| Cluster | $r_{ext}$ [arcsec] | $r_{ext}$ [kpc] | Net counts | SNR | $kT$ [keV] | $X_{Fe}/X_{Fe\odot}$ | $S_{0.5-2.0}$ [$10^{-14}$ erg/cm$^2$/s] | $S_{2.0-7.0}$ [$10^{-14}$ erg/cm$^2$/s] | $L_X$ [$10^{44}$ erg/s] |
|---|---|---|---|---|---|---|---|---|---|
| RCS1419+5326 | 37.05 | 252 | $2320 \pm 60$ | 38.0 | $5.0^{+0.4}_{-0.4}$ | $0.29^{+0.06}_{-0.11}$ | $10.9 \pm 0.3$ | $11.6 \pm 0.6$ | $4.63 \pm 0.12$ |
| RCS1107.3-0523 | 28.43 | 207 | $710 \pm 40$ | 15.5 | $4.3^{+0.5}_{-0.8}$ | $0.67^{+0.35}_{-0.27}$ | $2.26 \pm 0.12$ | $2.0 \pm 0.2$ | $1.34 \pm 0.08$ |
| RCS1325+2858 | 29.73 | 218 | $90 \pm 30$ | 2.8 | $1.8^{+1.2}_{-1.2}$ | $0.09^{+0.66}_{-0.09}$ | $0.43 \pm 0.08$ | $0.09 \pm 0.07$ | $0.23 \pm 0.07$ |
| RCS0224-0002 | 36.69 | 273 | $740 \pm 50$ | 13.1 | $5.1^{+1.6}_{-0.8}$ | $0.01^{+0.09}_{-0.01}$ | $1.85 \pm 0.12$ | $1.7 \pm 0.2$ | $1.31 \pm 0.10$ |
| RCS2318.5+0034 | 40.56 | 302 | $970 \pm 50$ | 19.1 | $7.3^{+1.3}_{-1.0}$ | $0.35^{+0.20}_{-0.22}$ | $5.4 \pm 0.3$ | $7.5 \pm 0.6$ | $4.51 \pm 0.23$ |
| RCS1620+2929 | 29.45 | 227 | $190 \pm 20$ | 7.5 | $4.6^{+2.1}_{-1.1}$ | $0.33^{+0.60}_{-0.33}$ | $1.51 \pm 0.17$ | $1.3 \pm 0.3$ | $1.35 \pm 0.18$ |
| RCS2319.9+0038 | 45.62 | 356 | $1490 \pm 60$ | 22.5 | $5.3^{+0.7}_{-0.5}$ | $0.60^{+0.22}_{-0.22}$ | $5.8 \pm 0.2$ | $5.9 \pm 0.4$ | $5.97 \pm 0.26$ |
| RCS0439.6-2905 | 24.71 | 196 | $220 \pm 30$ | 5.7 | $1.8^{+0.4}_{-0.3}$ | $0.44^{+0.27}_{-0.27}$ | $0.65 \pm 0.06$ | $0.11 \pm 0.07$ | $0.56 \pm 0.09$ |
| RCS1417+5305 | 19.68 | 156 | $37 \pm 11$ | 2.9 | $1.0-8.0$ | $(0.3)$ | $< 0.21$ | $< 0.16$ | $< 0.29$ |
| RCS2156.7-0448 | 19.68 | 160 | $60 \pm 20$ | 2.4 | $1.0-8.0$ | $(0.3)$ | $< 0.10$ | $< 0.07$ | $< 0.22$ |
| RCS2112.3-6326 | 19.68 | 161 | $47 \pm 20$ | 2.0 | $1.0-8.0$ | $(0.3)$ | $< 0.12$ | $< 0.19$ | $< 0.21$ |

Table 9: Best fit values of the spectra of each cluster. Errors are at $1\sigma$ confidence level. For each cluster is shown: extraction radius in arcsec and kpc; net counts in the 0.5-6.0 keV band; SNR; best fit temperature; best fit iron abundance; fluxes in the soft (0.5-2.0 keV) and hard (2.0-7.0 keV) band; best bolometric luminosity. Bolometric luminosities are computed for $\Omega_m = 0.3$, $\Omega_\Lambda = 0.7$ and $H_0 = 70$ km s$^{-1}$ Mpc$^{-1}$ (Table from Bignamini et al., 2008).

For three clusters (RCS1417+5305, RCS2112.3-6326 and RCS2156.7-0448), we have only marginal detection of the diffuse emission consistent with noise within $\sim 3\sigma$. So in these three cases we did not compute the maximum signal-to-noise region, but simply selected a circular region with radius of $\sim 20$ arcsec centered in the optical coordinates of the clusters. Since we were not able to perform spectral analysis for these clusters, we only provide an upper limit for the bolometric luminosities corresponding to a temperature range of 1.0-8.0 keV and a fixed metallicity values ($0.3X_{\mathrm{Fe}\odot}$).

In order to obtain a more robust measure of the diffuse emission for these three clusters, we merged together the three X-ray images overlapping the optical centers of the clusters. From the merged image (Figure 45) we found $274 \pm 29$ total net counts and a mean bolometric luminosity for each cluster $L_X \sim 0.2 \times 10^{44}$ erg s$^{-1}$. This implies that on average the three clusters show extended emission. We were not able to derive an average temperature from the combined spectrum, due to the low number of net counts.

As one can see in the Table 8 and Table 9 the cluster RCS1417+5305 is the only one observed with ACIS-I and the one with the lowest number of net counts. Despite this, it is not the one with the lowest signal-to-noise ratio. This is mostly due to the lower background of the front illuminated chips of ACIS-I with respect to that of the back illuminated chip of ACIS-S. Still, RCS1417+5305 is formally undetected in the ACIS-I image.

The detailed spectral analysis of each cluster along with the X-ray spectra are shown in Section 4.8.

### 4.3.1 The Luminosity-Temperature Relation

We checked whether the RCS clusters follow the same $L_X - T_X$ relation observed for X-ray selected clusters. In Figure 29 we compare the bolometric luminosity and the temperature measured for RCS clusters to the best fit of the $L_X - T_X$ relation as obtained by Branchesi et al. (2007c).

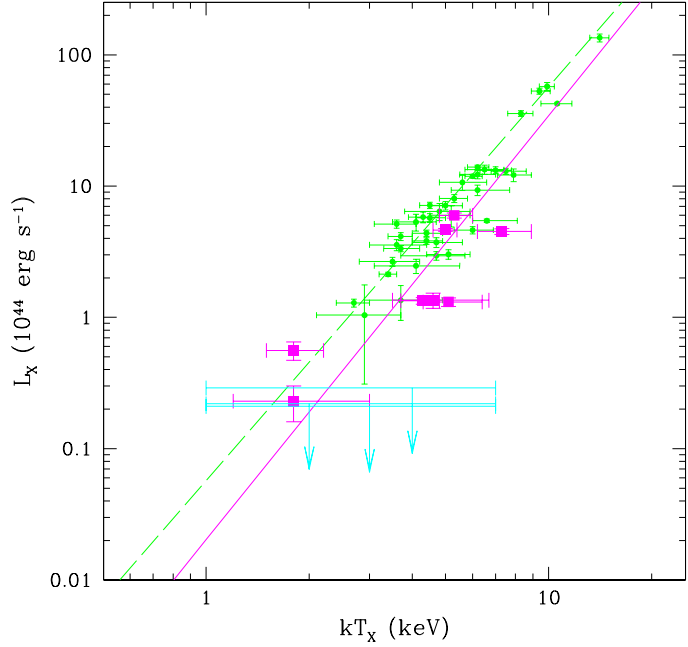

Figure 29: $L_X - T_X$ relation for the RCS sample (magenta squares; upper limits to X-ray luminosity are shown in cyan for RCS clusters without X-ray detection). The solid magenta line is the best fit of the $L_X - T_X$ for our RCS sample. The green points are the X-ray selected sample data by Branchesi et al. (2007c) and the dashed green line is their best fit (Figure from Bignamini et al., 2008).

We compared the slope and normalization of the $L_X - T_X$ relation of RCS clusters with that of X-ray selected high-z clusters. We fitted the $L_X - T_X$ only for the 8 detected clusters with a single power law assuming the following expression:

$$L_{44} = CT_6^\alpha, \tag{4.1}$$

where $L_{44}$ is the bolometric luminosity in units of $10^{44}$ erg s$^{-1}$ and $T_6 = kT/(6 \text{ keV})$. Our best fit is $LogC = 0.82^{+0.08}_{-0.08}$ and $\alpha = 3.2^{+0.7}_{-0.4}$, while best fit of Branchesi et al. (2007c) is $LogC = 1.06 \pm 0.03$ and $\alpha = 3.00^{+0.19}_{-0.18}$. We found that the slope is in agreement within the fairly large uncertainties between the two sample, whereas the RCS normalization is about a factor of $\sim 2$ lower at high temperature and about a factor of $\sim 3$ lower at low temperature. The two fits are inconsistent at a $\sim 4.2\sigma$ level.

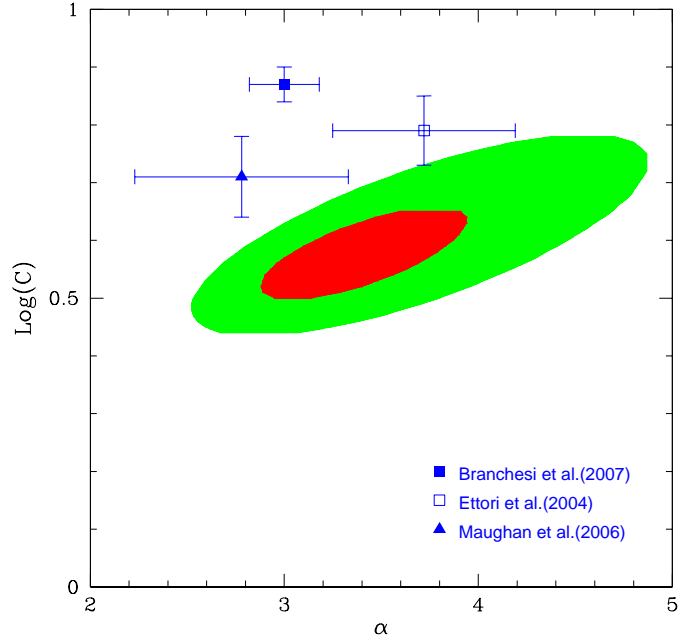

Figure 30: Comparison between the RCS sample and X-ray selected sample $L_X − T_X$ relations, in the normalization-slope, $\log(C)$ *versus* $\alpha$, plane. The red (green) region is $1\sigma$ ($2\sigma$) confidence level for RCS clusters. Blue filled square is the best fit for the sample by Branchesi et al. (2007c), empty square for Ettori et al. (2004) and filled triangle for Maughan et al. (2006). For all these fits the luminosities have been scaled by the cosmological factor $E(z)^{-1}(\Delta_c(z)/\Delta_c(z=0))^{-1/2}$ (Figure from Bignamini et al., 2008).

The three marginally detected RCS clusters, for which we plot in Figure 29 only the upper limits on the luminosity for a wide temperature range, 1.0-8.0 keV, seem to have lower luminosities with respect to the fitted RCS $L_X − T_X$, but only if they have $kT > 3$ keV. Unfortunately, we were not able to estimate their average temperature given the low number of net detected counts.

Branchesi et al. (2007c) repeated the analysis after removing the expected self-similar evolution in the $L_X − T_X$ relation. In order to compare our sample with this analysis and with other authors (namely Ettori et al. 2004 and Maughan et al. 2006), we removed the expected self-similar evolution from the luminosities of our cluster sample, i.e. scaling the luminos-ity by the cosmological factor $E(z)^{-1}(\Delta_c(z)/\Delta_c(z=0))^{-1/2}$. The evolution of the Hubble parameter with redshift reads

$$E(z)^2 = \Omega_m(1+z)^3 + \Omega_\Lambda \tag{4.2}$$

and $\Delta_c(z)$ is the average density of virialized objects in units of the critical density (Eke, Cole & Frenk, 1996). A simple, but accurate expression for $\Delta_c(z)$ is given by Bryan & Norman (1998). Again, we fitted the $L_X - T_X$ only for the 8 detected clusters with a single power law, now assuming the following expression:

$$E(z)^{-1}(\Delta_c(z)/\Delta_c(z=0))^{-1/2}L_{44} = CT_6^\alpha. \tag{4.3}$$

The parameters $LogC$ and $\alpha$, determined by the best fit, are listed in Table 10 compared with those of Branchesi et al. (2007c), Ettori et al. (2004) and Maughan et al. (2006).

| Sample | $\alpha$ | $LogC$ |
|---|---|---|
| RCS | $3.3^{+0.6}_{-0.4}$ | $0.57 \pm 0.08$ |
| Branchesi et al. | $3.00^{+0.19}_{-0.18}$ | $0.87 \pm 0.03$ |
| Ettori et al. | $3.72 \pm 0.47$ | $0.79 \pm 0.06$ |
| Maughan et al. | $2.78 \pm 0.55$ | $0.71 \pm 0.07$ |

Table 10: Slope and normalization of the $L_X - T_X$ relation for RCS and other X-ray selected clusters after removing the expected self-similar evolution (Table from Bignamini et al., 2008).

Also with luminosities scaled by the expected self-similar evolution, we find that the slope is in good agreement with other authors, while always RCS clusters show a factor of $\sim 2$ lower normalization of the $L_X - T_X$ relation at high confidence level with respect to that of X-ray selected clusters at similar redshift (samples inconsistent at $4.0\sigma$ for Branchesi et al. 2007c, $2.6\sigma$ for Ettori et al. 2004 and $1.9\sigma$ for Maughan et al. 2006). Figure 30 summarizes this comparison, showing the $1\sigma$ and the $2\sigma$ confidence level for our fit against the best fit values with corresponding $1\sigma$ error for the X-ray selected samples. The independent analysis by Hicks et al. (2008) found as well that RCS clusters are about a factor of two less luminous for a given temperature than X-ray selected clusters, however their best fit slope is significantly flatter than ours.

Lubin et al. (2004) presented a detailed analysis of two high-z optically selected clusters. The X-ray properties of both clusters are consistent with the high-redshift $L_X - T_X$ relation measured from X-ray selected samples. However, based on the local relations, their X-ray luminosities and temperatures are lower, by a factor of 2-9 with respect to what expected, for their measured velocity dispersion. The exact cause of these results is unclear. The authors claimed that the differences in X-ray properties of these two clusters may result from the fact that clusters at these epochs are still in process of forming.

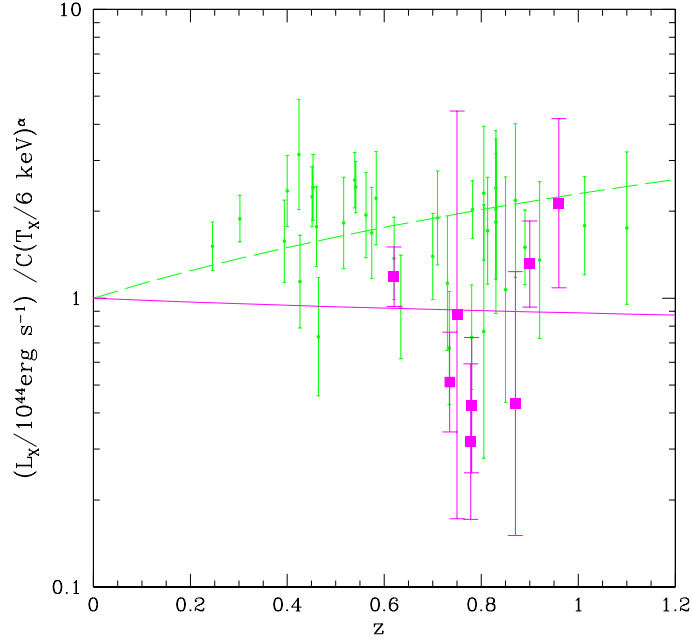

Figure 31: Evolution of the $L_X - T_X$ for the RCS sample. Ratio of the observed luminosity to the expected luminosity form the local relation as in Markevitch (1998) versus redshift. The slope and the normalization of the $L_X - T_X$ relation for local clusters are fixed to be $\alpha = 2.64$ and $log C = 0.80$ respectively. Magenta squares are RCS data and the solid magenta line is the best fit. Small green points are data from Branchesi et al. (2007c) and the dashed green line is their best fit (Figure from Bignamini et al., 2008).

The lower luminosity at fixed temperature of RCS clusters with respect to X-ray selected clusters, can not be entirely ascribed to our different choice of the extraction radius, which results in bolometric luminosity estimated within smaller radius (see below for details about extraction radii and bolometric luminosity measures adopted by other authors). Actually, adopting the $\beta$-model fit by Hicks et al. (2008) for RCS clusters, we estimate on average only $\sim 20\%$ the luminosity loss due to our smaller radius choice.

We also analyzed the evolution with redshift of the $L_X - T_X$ relation for our sample of RCS clusters. In Figure 31, we show the quantity $L_X / C T_X^\alpha$, where $\alpha$ and $C$ are fixed to values for the local $L_X - T_X$ relation as in Markevitch (1998), $\alpha = 2.64$ and $Log C = 0.80$. We fitted the evolution of the observed $L_X/CT_X^\alpha$ as a function of redshift with a single power law, assuming the following expression:

$$L_{44}/CT_6^\alpha = (1+z)^A. \tag{4.4}$$

From the best fit we found $A = -0.2 \pm 0.2$. The slope of the best fit is well consistent with zero, implying no evolution in the $L_X - T_X$. We note that, as pointed out by Santos et al. (2008), only the lowest redshift cluster RCS1419+5326 shows a Cool Core (CC) (see Section 4.5). Therefore we repeated the spectral analysis masking the CC in RCS1419+5326. In this case the temperature obtained by the best fit is $kT = 5.2^{+0.7}_{-0.5}$ keV. We corrected the bolometric luminosity for the removed core emission fitting the radial surface brightness profile with a $\beta$-model and extrapolating the profile in the masked inner region. The corrected bolometric luminosity is $L_X = (3.71 \pm 0.15) \times 10^{44}$ erg s$^{-1}$. Finally, we repeated the fit using these values for RCS1419+5326 and we found $L_X/T_X^\alpha \propto (1+z)^{-0.4\pm0.2}$. In this case we have evidence of slightly negative evolution for the $L_X - T_X$ for RCS clusters, still consistent with no evolution at $2\sigma$ confidence level.

To understand properly the meaning of this result, a rigorous comparison with various studies about X-ray selected clusters can be useful, since the evolution of the $L_X - T_X$ relation for X-ray selected clusters was frequently explored in the last years, in some cases deriving conclusions slightly varying with authors.

Vikhlinin et al. (2002) analysed 22 clusters observed with *Chandra* at redshift between 0.4 and 1.26. Temperatures were measured by fitting a spectrum integrated within a radius of $0.35 - 0.70h_{70}^{-1}$ Mpc with a single temperature mekal model. Bolometric luminosities were extrapolated with a $\beta$-model to a fixed radius of $1.4\ h_{70}^{-1}$ Mpc, excluding the central $70h_{70}^{-1}$ kpc region for clusters with sharply peaked surface brightness profiles. Vikhlinin et al. (2002) claimed a positive evolution, with $A = 1.5 \pm 0.2$.

A similar result was obtained by Lumb et al. (2004) and Kotov & Vikhlinin (2005) who analysed respectively 10 and 8 *XMM-Newton* observed clusters in a smaller and lower redshift range, $0.4 < z < 0.7$. Lumb et al. (2004) adopted a mekal model to extrapolate luminosities within a virial radius, $r_v$, according to the $T - r_v$ relation of Evrard et al. (1996). For each cluster best fit temperatures from spectral fits within $120''$ were used. Lumb et al. (2004) found a positive evolution of the $L_X - T_X$, $A = 1.52^{+0.26}_{-0.27}$. Kotov & Vikhlinin (2005), instead, evaluated bolometric luminosities within r< 1400 kpc, using a mekal model with emission-weighted temperature and correcting the inner 70 kpc emission with a best fit $\beta$-model. Their $L_X - T_X$ shows a positive evolution with $A = 1.8 \pm 0.3$.

Maughan et al. (2006) analysed the evolution with redshift of the $L_X - T_X$, using 11 clusters observed with *Chandra* or *XMM-Newton* at redshift $0.6 < z < 1.0$. They adopted $L_{200}$ luminosities obtained in the following way. First the surface brightness distribution of each cluster was modeled with a two-dimensional $\beta$-model within a detection radius; then the luminosities were extrapolated to $R_{200}$, according to the best fitting surface brightness pro-files. Comparing with the local $L_X - T_X$ of Markevitch (1998), Maughan et al. (2006) found a positive evolution with $A = 0.7 \pm 0.4$

A sample of 28 clusters observed with *Chandra* at redshift between 0.4 and 1.3 was studied by Ettori et al. (2004), who extrapolated bolometric luminosities within $R_{500}$, according to a $\beta$-model fitted within a radius which optimize the signal-to-noise ratio. When they use their whole sample they found a positive evolution $A \sim 0.6$ for the $L_X - T_X$, instead when they use only clusters with redshift greater than 0.6 (16 objects) they found a much weaker evolution with $A \sim 0.1$. The authors explored widely the causes of this relevant differences with other authors, in particular with respect to the similar redshift range sample by Vikhlinin et al. (2002). Basically, beyond differences in fitting procedure, definition of reference radii and temperature estimation, the sample of Ettori et al. (2004) is larger at higher redshift, 16 galaxy clusters at $z > 0.6$ and 4 at $z > 1.0$, whereas Vikhlinin et al. (2002) has 9 and 2 objects respectively, suggesting a different evolution of the $L_X - T_X$ relation according to different redshift range studied.

A similar conclusion was also derived by Branchesi et al. (2007c), who used a sample of 17 clusters from the *Chandra* archive supplemented with additional clusters from Vikhlinin et al. (2002), Ettori et al. (2004) and Maughan et al. (2006), to form a final sample of 39 high redshift, $0.25 < z < 1.3$, clusters. Their luminosities were extrapolated to $R_{500}$ using an isothermal $\beta$-model. For their whole combined sample, Branchesi et al. (2007c) found a positive evolution with $A = 1.20 \pm 0.08$. However, the $\chi^2$ of this fit has a probability lower than 0.1% to be acceptable, suggesting that the evolution of the $L_X - T_X$ on the whole redshift range cannot be described by any power low of the form $\propto (1+z)^A$. In particular, they claimed that a stronger evolution is required at lower redshift, namely up to $z \sim 0.6$, followed by a much weaker evolution at higher redshift, as it is shown by their data over-plotted in Figure 31.

As the careful reader can understand from this concise summary of studies based on high-z X-ray selected clusters, it is evident that the evolution of the $L_X - T_X$ relation is still a matter of debate both for significant differences in measuring luminosities and/or temperature and in the interpretation of the results.

Besides, our claim of no evolution or slightly negative evolution for the $L_X - T_X$ for RCS clusters is bounded by this final remark: we can only compare our $L_X - T_X$ for high-z optically selected clusters with local X-ray based $L_X - T_X$, whereas it would be better to compare our clusters with similarly selected local ones, in order to understand the physical properties evolution of optically selected clusters. However, it is clear that in order to asses the evolution of the $L_X - T_X$ relation for optically selected clusters, we need to collect X-ray data for a larger number of optically selected clusters distributed on a wider redshift range.

### 4.3.2 *The Iron Abundance*

As one can see in Table 9, in the majority of cases the Fe abundances from the spectral analysis of single clusters are consistent with zero within $1\sigma$ and all, except those of RCS1620+2929 and RCS2319.9+0038, are consistent with zero within $2\sigma$. This is mostly due to the low signal-to-noise ratio of the spectra. To obtain a more robust measurement of the average Fe abundance, we fitted at the same time all the spectra using a common value for the iron abundance. In this way we could increase the SNR and measure the mean iron abundance of the sample. This has been already done to measure the evolution of the average Fe abundance at high-z (see Tozzi et al., 2003; Balestra et al., 2007). Like in the single cluster analysis, we fixed the local absorption to the Galactic neutral Hydrogen column density and the redshift to the value measured from the optical spectroscopy. We also fixed the temperatures to the values obtained previously from the single cluster analysis, except for RCS1417+5305, RCS2112.3-6326 and RCS2156.7-0448, whose temperatures vary in the range 1.0-8.0 keV. The combined fit analysis allowed us to measure the average Fe abundance which turns out to be $\langle X_{\mathrm{Fe}} \rangle = 0.37^{+0.09}_{-0.08} X_{\mathrm{Fe}\odot}$. Therefore we detect with high significance the presence of iron at a level consistent with that of X-ray selected clusters at similar redshift.

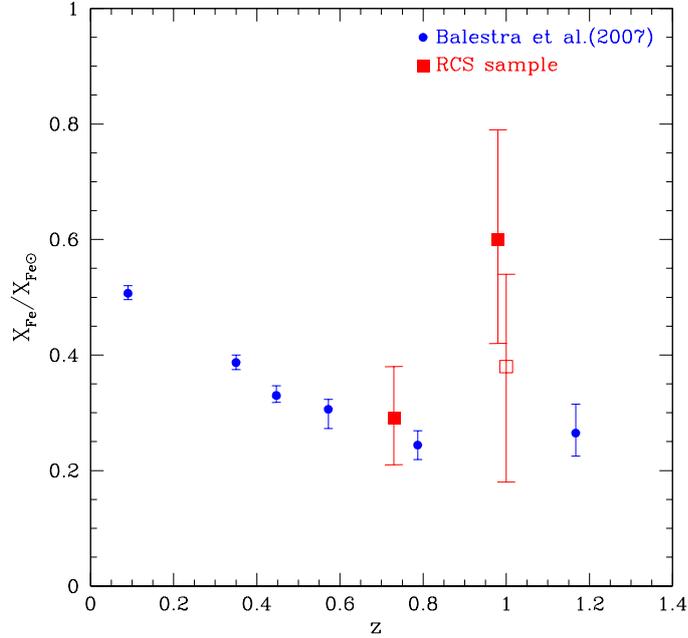

Figure 32: Mean iron abundance of RCS sample, red squares, in two redshift bins compared with the results of Balestra et al. (2007), blue circles. The open square represent the second bin without the high abundance cluster RCS2319.9+0038 (Figure from Bignamini et al., 2008).

We also repeated the same procedure after dividing our sample into two redshift bins. The first bin includes all clusters with redshift lower than 0.80 including 5 object with about 4800 total net counts and $<z> \simeq 0.73$, whereas the second bin includes 6 objects with redshift $z > 0.8$ with about 2000 total net counts and $<z> \simeq 0.98$. The spectral information in this bin is largely dominated by RCS2319.9+0038, which shows an iron abundance slightly larger than the typical value at $z \sim 1$. From the combined fit we found $\langle X_{Fe} \rangle = 0.29^{+0.09}_{-0.08} X_{Fe\odot}$ for the first subsample and $\langle X_{Fe} \rangle = 0.60^{+0.19}_{-0.18} X_{Fe\odot}$ for the second. Excluding RCS2319.9+0038 from the second bin, we found $\langle X_{Fe} \rangle = 0.43^{+0.16}_{-0.20} X_{Fe\odot}$. These values are plotted in Figure 32 and, given the large error bars, are in agreement with the results of Balestra et al. (2007). It is clear that, in order to investigate the typical Fe abundance in the ICM of RCS clusters at high redshift, we need to use a substantially larger sample, observed with medium-deep *Chandra* exposures. However, our results show that also for this sample of optically selected clusters, the ICM was already enriched with iron at a level comparable with that of X-ray selected clusters.

## 4.4 ACTIVE GALACTIC NUCLEI IN THE CLUSTER FIELDS

To study the large scale structure associated with clusters, we computed the number density of active galactic nuclei (AGN) in the regions around RCS clusters, to be compared with the AGN density in the field. An interesting aspect, currently under investigation, is whether the nuclear activity is enhanced in galaxies that belong to high density structures, like clusters or filaments. These effects are difficult to investigate, and preliminary works on deep X-ray fields offered only a tantalizing hint for an enhanced AGN activity (Gilli et al., 2003, 2005; Martel et al., 2006). Recent results from the COSMOS survey are not able to provide statistically significant constraints on a possible enhanced activity associated to large scale structure (Gilli et al., 2008). Still, X-ray data are relevant for this kind of study, thanks to their high efficiency in identifying AGN. In fact, before the current generation of X-ray observatories, like *Chandra*, characterized by an high spatial resolution, the investigation of the AGN distribution in and around clusters was based only on the optical identification of AGN characteristic emission lines in the galaxy spectra (Dressler & Gunn, 1983; Huchra & Burg, 1992; Dressler et al., 1999). This approach misses a large fraction of optically obscured AGN (Martini et al., 2002, 2006), which, on the other hand, can be easily identified by their hard X-ray emission (at least for intrinsic column densities $N_H < 10^{24}$ cm$^{-2}$).

In the last years several studies were performed computing the distribution of X-ray point sources in nearby galaxy clusters. Significant evidence was found of an excess of point sources covering a wide range of redshift and cluster luminosity, around X-ray selected clusters (Henry & Briel, 1991; Cappi et al., 2001; Molnar et al., 2002; Sun & Murray, 2002; Cappelluti et al., 2005). Here we perform a similar study for the fields of the optically selected RCS clusters.

| Cluster | Solid angle [deg²] | Number of sources | | | Flux limit (soft) [erg/cm²/s] | Flux limit (hard) [erg/cm²/s] |
|---|---|---|---|---|---|---|
| | | soft | hard | tot | | |
| RCS1419+5326 | 0.025 | 31 | 19 | 43 | $3.4 \times 10^{-16}$ | $2.8 \times 10^{-15}$ |
| RCS1107.3-0523 | 0.025 | 48 | 33 | 56 | $2.3 \times 10^{-16}$ | $1.9 \times 10^{-15}$ |
| RCS1325+2858 | 0.025 | 35 | 23 | 40 | $3.2 \times 10^{-16}$ | $2.6 \times 10^{-15}$ |
| RCS0224-0002 | 0.024 | 47 | 35 | 70 | $2.1 \times 10^{-16}$ | $1.7 \times 10^{-15}$ |
| RCS2318.5+0034 | 0.021 | 29 | 17 | 40 | $3.6 \times 10^{-16}$ | $2.7 \times 10^{-15}$ |
| RCS1620+2929 | 0.021 | 25 | 18 | 34 | $2.5 \times 10^{-16}$ | $2.0 \times 10^{-15}$ |
| RCS2319.9+0038 | 0.023 | 38 | 26 | 43 | $2.5 \times 10^{-16}$ | $1.9 \times 10^{-15}$ |
| RCS0439.6-2905 | 0.024 | 30 | 26 | 45 | $2.5 \times 10^{-16}$ | $2.1 \times 10^{-15}$ |
| RCS1417+5305 | 0.082 | 114 | 88 | 167 | $1.0 \times 10^{-16}$ | $1.0 \times 10^{-15}$ |
| RCS2156.7-0448 | 0.024 | 34 | 18 | 43 | $2.6 \times 10^{-16}$ | $2.2 \times 10^{-15}$ |
| RCS2112.3-6326 | 0.021 | 38 | 26 | 48 | $3.2 \times 10^{-16}$ | $2.4 \times 10^{-15}$ |

Table 11: Solid angle and number of identified point sources in the soft (0.5-2.0 keV) and hard (2.0-7.0 keV) band, with the corresponding flux limits in each field (Table from Bignamini et al., 2008).

### 4.4.1 *Point Source Identification*

We used two different algorithms to identify point sources in RCS fields. First we ran wavdetect implemented in CIAO (Freeman et al., 2002), and then a modified version of the algorithm *Source-Extractor* (*SExtractor*) (Bertin & Arnouts, 1996). In the second case we performed a background subtraction using a background map obtained by the same X-ray image where any point source candidate has been previously removed. We ran both algorithms on the images obtained in the soft (0.5-2.0 keV), hard (2.0-7.0 keV) and total (0.5-7.0 keV) bands.

Finally, we combined in a single catalogue all the point sources detected with $S/N > 2.1$ as measured from aperture photometry. Here we followed the same procedure used for the Chandra Deep Field South (CDFS) sources as described in Giacconi et al. (2001). We measured the signal-to-noise ratio of all the detected sources in the area of extraction, which is defined as a circle of radius $R_s = 2.4 \times FWHM$ (with a minimum of 5 pixels of radius). The $FWHM$ was modeled as a function of the off-axis angle to reproduce the broadening of the PSF. In each band a detected source has a $S/N \equiv S/\sqrt{S + 2B} > 2.1$ within the extraction area of the image. Here $S$ is the source net counts and $B$ is the background counts found in an annulus with outer radius $R_s + 12''$ and an inner radius of $R_s + 2''$, after masking out other sources, and rescaled to the extraction region. Source counts were measured with simple aperture photometry within $R_s$ in the soft and hard bands separately. Simulations have shown that such aperture photometry leads to an underestimate of the source count rate by approximately 4% (see Tozzi et al., 2001). We corrected such photometric bias before converting count-rates into energy fluxes.

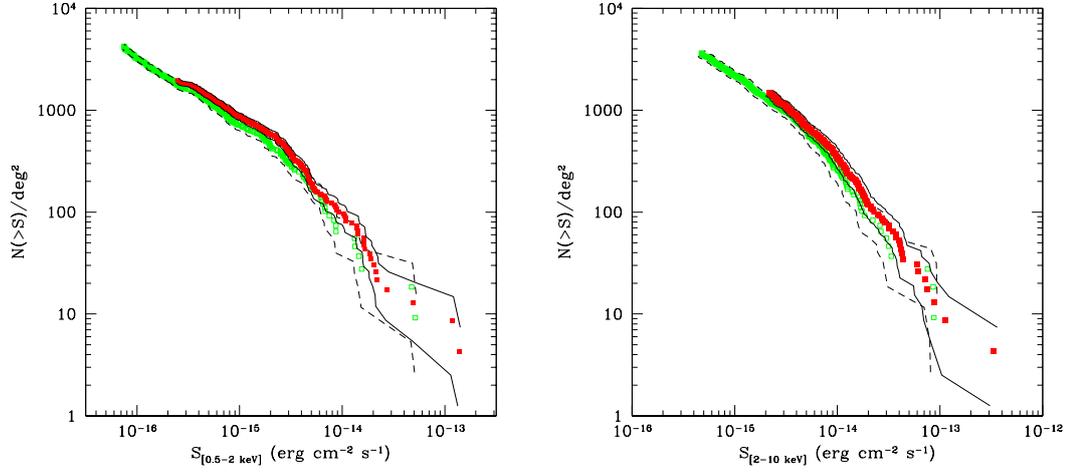

Figure 33: LogN-LogS in the soft band for the observations with the ACIS-S detector. The CDFS data is shown in green and the RCS data is shown in red. The solid (dashed) black line shows the error, $1\sigma$, of the RCS (CDFS) distribution (Figures from Bignamini et al., 2008).

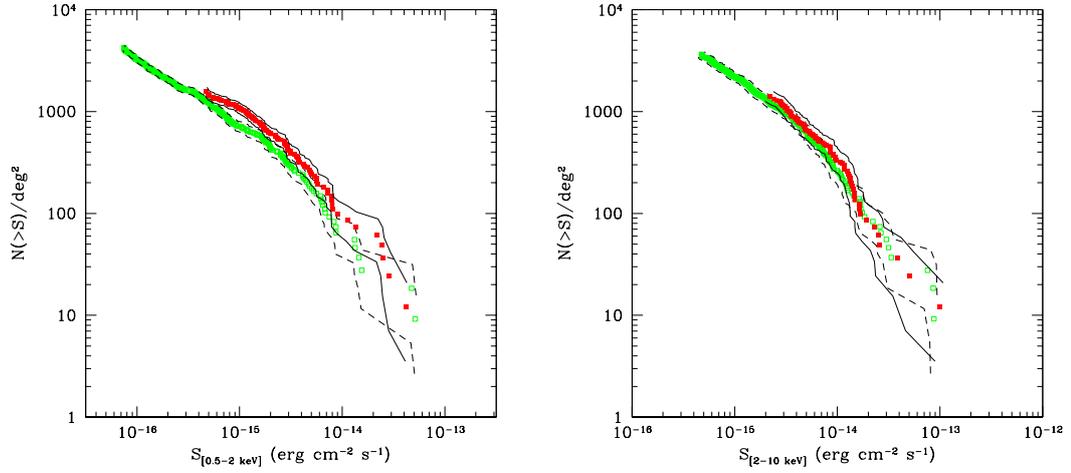

Figure 34: LogN-LogS in the soft band for the observation with the ACIS-I detector. The CDFS data is shown in green and the RCS data is shown in red. The solid (dashed) black line shows the error, $1\sigma$, of the RCS (CDFS) distribution (Figures from Bignamini et al., 2008).

| # | Cluster | CL hardness ratio | PS hardness ratio | $L_X$ [$10^{44}$ erg/s] |
|---|---------|-------------------|-------------------|------------------------|
| 1 | RCS1419+5326 | 0.0 | 0.1 | 0.60 |
| 2 | RCS1107.3-0523 | -0.1 | -0.2 | 0.15 |
| 3 | RCS1107.3-0523 | -0.1 | 0.1 | 0.19 |
| 4 | RCS1325+2858 | -0.7 | 0.9 | 0.09 |
| 5 | RCS0224-0002 | 0.0 | 1.0 | 0.05 |
| 6 | RCS2318.5+0034 | 0.1 | 0.6 | 0.21 |
| 7 | RCS2318.5+0034 | 0.1 | 0.3 | 0.09 |
| 8 | RCS1620+2929 | -0.1 | 0.2 | 1.08 |
| 9 | RCS2319.9+0038 | 0.0 | 0.3 | 0.29 |
| 10 | RCS0439.6-2905 | -0.7 | 0.4 | 0.17 |
| 11 | RCS1417+5305 | -0.2 | 1.0 | 0.14 |
| 12 | RCS2156.7-0448 | -0.2 | 1.0 | 0.31 |

Table 12: Comparison between the hardness ratio of the clusters (CL) and the hardness ratio of the point sources (PS) identified within 150 kpc from the corresponding cluster center. The last column shows the luminosity in the 0.5-10.0 keV band for point sources assumed to be at the same redshift of the cluster (Table from Bignamini et al., 2008).

We removed by means of visual inspection double detections and spurious detections due to occurrences that cannot be handled by the detection algorithm (e. g. sources on the edges of the image or in high background regions). Only in few cases we added by hand obvious sources that were not identified by the algorithm (e.g. point sources missed because too close to a bright source). The final number counts distribution is practically unaffected by these corrections.

We computed the effective sky-coverage at a given flux, which is defined as the area on the sky where a source with a given net count rate can be detected. The computation includes the effect of exposure, vignetting and point spread function variation across the field of view. The count rate to flux Energy Conversion Factors (ECFs) in the 0.5-2.0 keV and in the 2.0-10.0 keV bands were computed using the response matrices. We quoted the fluxes in the canonical 2.0-10.0 keV band, as extrapolated from counts measured from the 2.0-7.0 keV band. The ECFs were computed for $\gamma = 1.4$ at the aimpoint of each field, after including the effect of the Galactic absorbing column (see Table 8).

Before computing the energy flux of each source, the count rates were corrected for vignetting and converted to the count rates that would be measured if the source were in the aimpoint. The correction is simply given by the ratio of the value of the exposure map at the aimpoint to the value of the exposure map at the source position. This is done separately for the soft and the hard band, using the exposure maps computed at energies of 1.5 keV and 4.5 keV. This procedure also accounts for the variations in exposure time across the field of view. In Table 11 we show the solid angle covered by each field, the number of sources detected with our criteria in the soft and hard bands, the total number of sources in the combined catalogue and the flux limits in the soft and hard bands.

### 4.4.2 *AGN Number Counts and Spatial Distribution in RCS Fields*

In order to investigate whether there is any excess of point sources in and around RCS clusters with respect to the field, we computed the point source number density as a function of flux (*LogN-LogS*) and the spatial distribution of point sources.

The *LogN-LogS* is defined as the logarithm of the number of sources per unit of solid angle with flux greater than a given flux, $Log(N(> S))$, as a function of the flux, $S$. The number density $N(> S)$ can be computed as:

$$N(> S) = \sum_i \frac{1}{\omega(S_i)},$$

(4.5)

where $\omega(S_i)$ is the value of the sky-coverage evaluated at the flux of the source $S_i$ and the sum is over all the sources with flux $S_i > S$.

We computed the total *LogN-LogS* in the soft and hard bands by summing all the fields, each one with its own sky-coverage (see Figure 33). In the same Figure we plot the data of the CDFS from Rosati et al. (2002). We treated separately the field of RCS1417+5326 (see Figure 34), since it has been observed with ACIS-I and therefore, having a much larger solid angle, has many more point sources.

Both in Figure 33 and Figure 34 we find a small excess of point sources in RCS fields at all fluxes. In the ACIS-S fields the surface density of point sources exceeds by ∼20% the value expected from the CDFS both in the soft and hard; in the ACIS-I field by ∼ 40% in the soft and ∼ 15% in the hard. The significance of this excess is evaluated separately after fitting the differential number counts with a single slope power law. We found the excess in the normalization to be significant at 2.0 $\sigma$ and at 1.7 $\sigma$ confidence level in the soft and hard band respectively for the ACIS-S fields, and at 2.0 $\sigma$ and at 0.9 $\sigma$ in the soft and hard band respectively for the ACIS-I field.

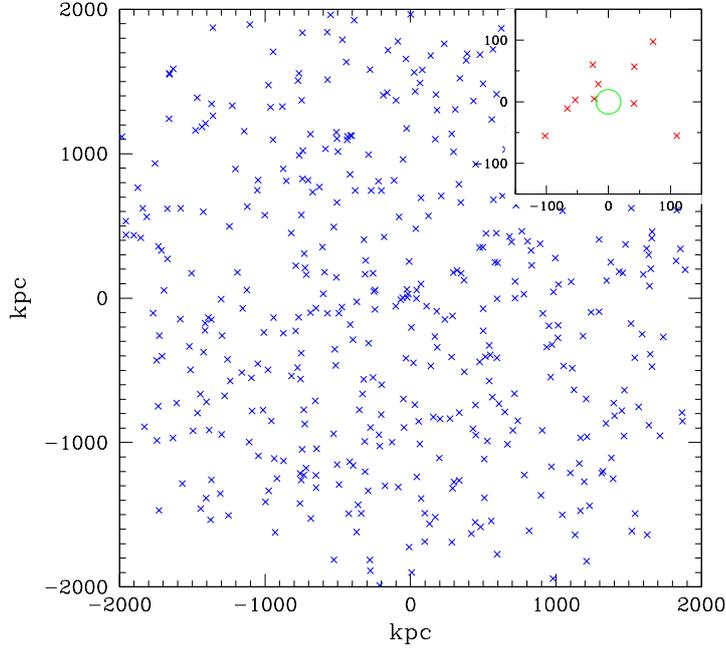

Figure 35: Spatial distribution of point sources in the RCS fields. This image was obtained by stacking the distribution of every single field, matching the centroid of the X-ray emission of the clusters at the coordinates (0,0). We stacked together only point sources with flux higher than the highest flux limit. Distances are rescaled assuming the cluster redshift in each field. Every cross in the image corresponds to a point source. The small panel in the top-right corner is a zoom of the inner region. The small green circle is centered in (0,0) and has a radius of 20 kpc (Figure from Bignamini et al., 2008).

This result is similar to what has been found by other authors. Cappi et al. (2001) found that the 0.5-2.0 keV source surface density (at a flux limit of $1.5 \times 10^{-15}$ erg cm$^{-2}$s$^{-1}$) measured in the area surrounding two clusters at $z \sim 0.5$ exceeds by a factor of $\sim 2$ the value expected in the field *LogN-LogS*, with a significance of $\sim 2\sigma$. Also Cappelluti et al. (2005) found a factor of $\sim 2$ over-density with a significance $> 2\sigma$ in 4 cluster fields studying a sample of 10 high $z$ $(0.24 < z < 1.2)$ clusters. In a similar way Branchesi et al. (2007a) found a $\sim 2\sigma$ excess of sources in the cluster region at the bright end of the *LogN-LogS* for a sample of 18 distant galaxy clusters. In summary, there is a growing evidence that there is an excess of X-ray sources around clusters. The low significance of this evidence is due to the fact that the *LogN-LogS* is dominated by all the sources along the line of sight in the solid angle of each field (for example, at $z \sim 1$ the 8 arcmin size of an ACIS-S field corresponds to 4 Mpc). This result is expected, since the presence of a cluster implies the presence of large scale structure with the consequent excess of galaxies with respect to the field. In order to investigate the genuine enhancement of AGN activity around clusters, we would need to perform an extensive spectroscopic follow up of the identified AGN to select those associated with the cluster and compare their relative density with respect to the field galaxies. This would require an extensive survey of the galaxy population in the cluster and in the field, which is beyond the scope of this work.

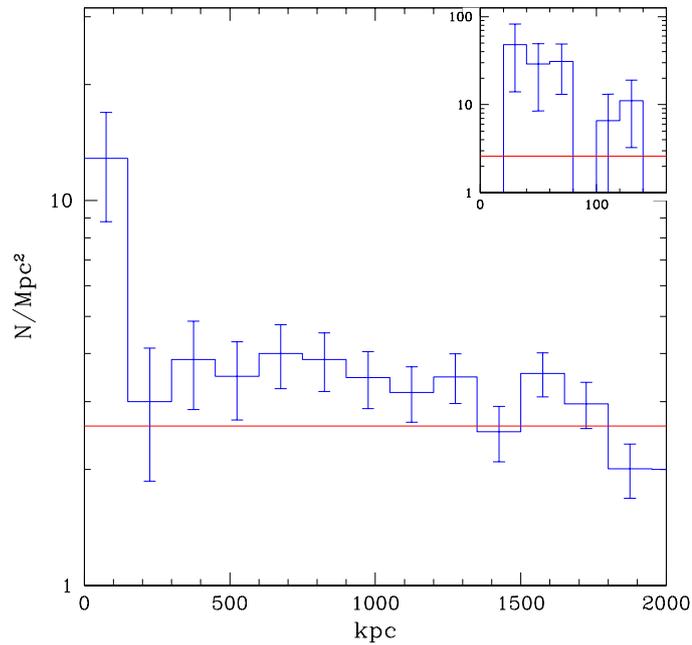

Figure 36: Number density profile of point sources in the fields of the RCS clusters. The profile was obtained from the superposition of the spatial distribution of every single field joining in the clusters position in the center of the image and binning the spatial distribution by concentric rings. The horizontal line shows the mean number density of point sources obtained by the ratio between the total number of point sources in all fields and the total area of all fields. Error bars are at $1\sigma$ level. The small panel in the top-right corner is a zoom of the inner region re-binned with a bin size of 20 kpc (Figure from Bignamini et al., 2008).

Without the spectroscopic information, we can use the spatial distribution of the X-ray sources in order to isolate the AGN actually associated with the cluster. Therefore, we also computed the number density of point sources in RCS fields as a function of the distance from the center of the clusters. Figure 35 shows the spatial distribution of point sources, obtained by stacking the RCS fields overlapping the centroid of the X-ray emission of the clusters. We computed the distances of the point sources from the center of the cluster as if their redshifts were that of the cluster. Since the flux limits slightly vary from one field to another (see Table 11), we stack together only point sources with flux higher than the highest flux limit, which is $3.6 \times 10^{-16}$ erg s$^{-1}$ cm$^{-2}$ in the soft band and $2.8 \times 10^{-15}$ erg s$^{-1}$ cm$^{-2}$ in the hard band.

Altogether there are 12 point sources within 150 kpc ($\sim$ 20 arcsec) from the center of the RCS clusters. Note that 150 kpc is smaller than the extraction radii used for the cluster spectral analysis. Almost every cluster has at least one source within 150 kpc, RCS1107.3-0523 and RCS2318.5+0034 have two sources, while only RCS2112.3-6326 has none. We checked these 12 point sources one by one, to make sure they are not spurious detections due to Poissonian fluctuations in the thermal bremsstrahlung emission of the cluster itself.

When the point source candidate are not clearly resolved we used, as a secondary criterion, the hardness ratio of its emission. Indeed, AGN emission is significantly harder than the thermal bremsstrahlung emission from the ICM. The hardness ratios of these 12 point sources are listed in Table 12. After comparing the hardness ratio of the point sources and the clusters, we removed two sources from the central bin: the first one from RCS1419+5326 and the second one from RCS1107.3-0523. In Figure 36 we plot the number of point sources as a function of the distance from the center, computed at the redshift of the cluster. The excess with respect the mean density in the field is about a factor of $\sim$ 6 and it is significant at $3\sigma$ confidence level in the central bin (corresponding to the innermost 150 kpc).

Both in Figure 35 and Figure 36 we performed a zoom of the inner region, to look more in detail the position of the AGNs with respect the center of the X-ray emission. From the zoomed histogram in top-right corner of Figure 36 we found that no point sources are in the innermost 20 kpc. Unfortunately, given the low number of point sources we have and the low statistic of our small sample it looks difficult to make strong statements. In conclusion, it is clearly evident an excess of point sources towards the center of the X-ray emission of the cluster, which seems not to affect the innermost (20 kpc) central region. On average the nearest point source to the center of the cluster is distant about 70 kpc. Also Hicks et al. (2007) with an independent analysis of RCS0224-0002 found a significant excess of point sources within $R_{200}$.

A similar analysis was performed by Ruderman & Ebeling (2005) with a sample of 51 clusters in the MACS survey. Their source list for all fields is complete to a flux limit of $1.25 \times 10^{-14}$ erg s$^{-1}$ cm$^{-2}$ and in the radial profile of their source surface density they found an evident excess in the central 0.5 Mpc by a factor of $\sim$ 5 significant at the $8.0\sigma$ confidence level. An analogous result was found by Branchesi et al. (2007a) for a sample of 18 distant galaxy clusters.

In the last columns of the Table 12, we show for all the point sources of the central bin (inner 150 kpc) the luminosities that the sources would have at the cluster redshift. In Figure 37 we show the comparison between the total emissions within the extraction radius (ICM plus point sources) and the ICM emission. In the left panel we investigated the contribution of the point sources to the total flux in the soft and hard energy bands separately, by the ratio $(S_{cl} + S_{ps})/S_{cl}$, where $S_{cl}$ and $S_{ps}$ are the flux of the cluster and of the point sources respectively.

As expected the point sources contribution to cluster flux and luminosity is more prominent in the hard band (as one would naturally expect from the power law AGN spectrum harder than a typical thermal spectrum); instead in the soft band for the great majority of cases the point sources contribution to total flux is less than 20%.

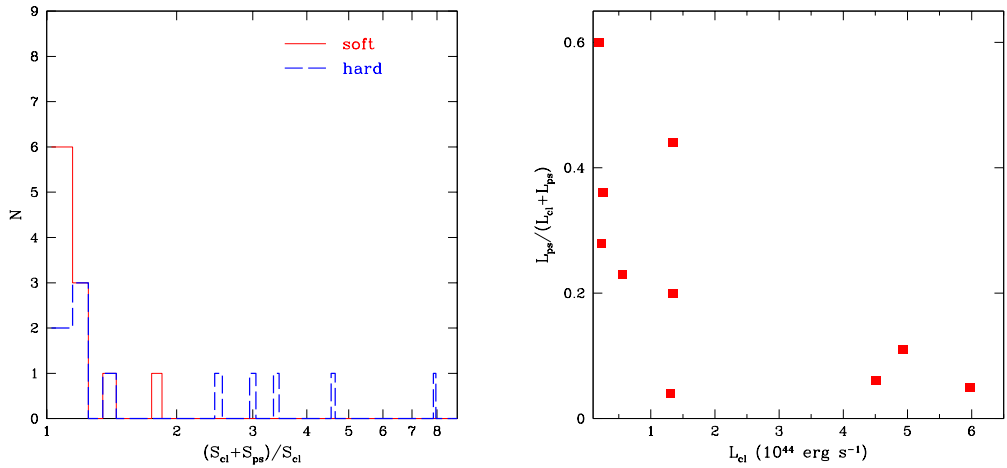

Figure 37: Left panel: histogram of the ratios between the flux of the cluster plus the flux of the point sources embedded and the flux of the extended cluster emission. Number larger than 2 indicates that the emission is dominated by AGN rather than the ICM. The solid red line is for the soft band and the dashed bold blue line is for the hard band. Right panel: ratio between total 0.5-10.0 keV luminosities of point sources and ICM luminosity plus point sources luminosity as a function of the ICM luminosity itself. Lower luminosity clusters tend to be more contaminated by point-source emission (Figures from Bignamini et al., 2008).

In the right panel we plot the quantity $L_{ps}/(L_{cl} + L_{ps})$ *versus* $L_{cl}$, where $L_{cl}$ and $L_{ps}$ are the luminosity in the 0.5-2.0 keV band of the cluster and of the point sources respectively. Large values of this ratio indicate that the emission is dominated by the AGN. The quantity on the $y$ axis can be interpreted as percentage contamination of cluster luminosity due to point sources. As one can see the point sources contamination in greater for clusters with lower luminosity. On the other hand the total luminosity is always dominated by the ICM except in one case.

These considerations are similar to what was obtained by Branchesi et al. (2007b), who conclude that point sources located within the ICM region may affect considerably the estimates of X-ray observable showing that the point source contribution must be removed. However, contamination in the soft band is not severe, and we do not find indication of a possible population of clusters which would have been missed in X-ray due to the presence of bright AGN.

## 4.5 SEARCHING FOR COOL CORES IN RCS CLUSTERS

We analyzed the X-ray surface brightness properties of the RCS clusters. Specifically, we evaluated the presence of Cool Core (CC) in our sample. This analysis is part of a paper (Santos et al., 2008) on the detection of CCs in the distant galaxy cluster population. The purposes of this work are to develop a new method to detect high redshift CC cluster and to measure the CC fraction out to redshift $0.7 < z < 1.4$. This investigation is important in order to understand the nature of CC clusters, whether there is a CC/non-CC bimodal cluster population and eventually the existence of feedback mechanisms which counteract cooling and prevent the formation of a cool core.

The analysis was performed on three different samples: a low-z sample, an high-z sample and the RCS sample described previously in Section 4.2. The high redshift sample (15 clusters) is a statistically complete sample based on the ROSAT Deep Cluster Survey (RDCS) from Rosati et al. 1998. The low redshift sample (11 clusters) was used as a test set to understand our capability in detecting and characterizing CCs.

The most stringent proof for the presence of a cool core is given by temperature decrease in the core with respect to the bulk of the cluster. Unfortunately, the poor photon statistic does not permit a spatially resolved spectroscopy analysis in high-z clusters. Since a central surface brightness excess is a primary indicator (Fabian et al. 1984) of the presence of a CC, we evaluate the core surface brightness in nearby clusters by defining a concentration parameter ($c_{SB}$) as the ratio of the peak over the ambient surface brightness. The $c_{SB}$ parameter was optimized using the low-z sample, varying the radius of the central peak and external radius. The optimal $c_{SB}$ is found for a peak radius of 40 kpc and a cluster bulk radius of 400 kpc:

$$c_{SB} = \frac{SB(r < 40\text{kpc})}{SB(r < 400\text{kpc})}.$$  (4.6)

The efficiency of $c_{SB}$ in separating CC from non-CC is tested in details in Santos et al. (2008) through extensive Monte Carlo simulations of distant galaxy clusters with the "cloning method" (for a detailed description of this technique see Section 5.7.2). According to this analysis we find a strong correlation between the cooling time, $t_{cool}$, and $c_{SB}$, which allow us to reliably relate a physical quantity with a phenomenological parameter. This correlation is described by a power law fit as shown in Figure 38 : $t_{cool} \propto c_{SB}^{-1.10 \pm 0.15}$.

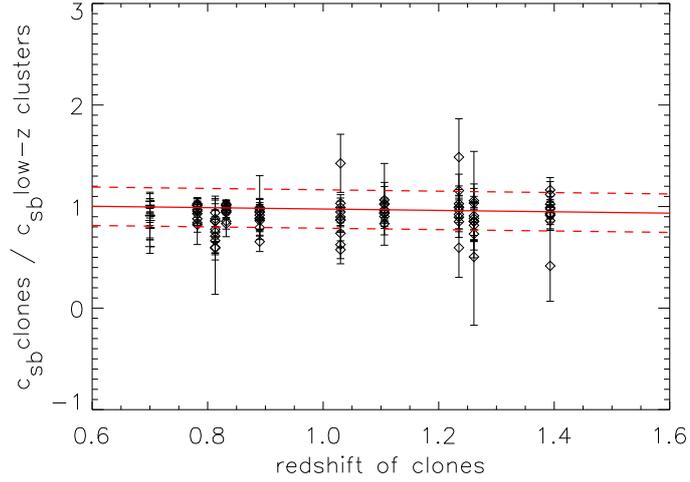

Figure 38: Analysis of 10 realizations of cloning the nearby cluster sample to high-z: we plot the ratio $c_{SB}[clones]/c_{SB}[lowz]$ as a function of cloning redshift. Individual error bars refer to 1-sigma confidence level. The red solid line represents the linear fit to the data and corresponding 1-sigma errors are shown as dashed line (Figure from Santos et al., 2008).

We define three categories of cool cores: non-CC ($c_{SB} < 0.075$), moderate CC ($0.075 < c_{SB} < 0.155$) and strong CC ($c_{SB} > 0.155$). The distribution of the surface brightness concentration and the analysis of the radial profiles show the presence of all the three categories of cool cores in the low-z sample. The same analysis applied to the X-ray selected high-z clusters reveals only two regimes: non-CC and moderate CC.

In the established scenario of hierarchical structure formation, galaxy clusters develop through mergers and by accretion of smaller objects. Numerical simulations (Cohn 2005) predict higher cluster merger rates with increasing redshifts, where the fraction of clusters with recent mass accretion due to major mergers can be doubled at $z = 0.7$, with respect to the local fraction. This framework provides a possible mechanism for preventing the formation of prominent CC at high redshift. However, since high-z clusters are younger, other mechanisms may cause a delay in the formation of cool cores due to some internal energy release, such as AGN activity and star formation processes. The observed absence of pronounced CC at $z > 0.7$ is therefore plausible in the current cosmological framework.

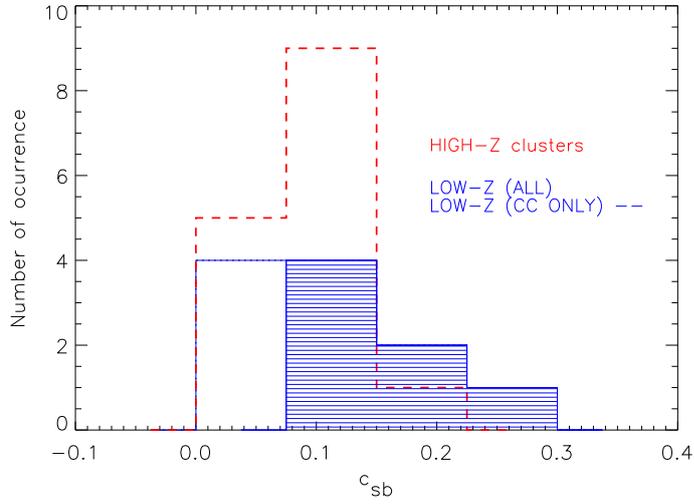

Figure 39: Histograms of $c_{SB}$ for the low-$z$ sample (blue solid line with the known CC clusters indicated in hatched) and for the high-$z$ sample (red dash line) (Figure from Santos et al., 2008).

The same analysis applied to the X-ray selected high redshift sample was applied to the RCS sample. We found that the 3 clusters barely detected in X-rays (see Section 4.3) have such a low surface brightness that a $c_{SB}$ analysis is not possible. The measurement of the surface brightness concentration as described before, yields that the majority (6 of 8) of the RCS clusters with measurable X-ray emission have $c_{SB} < 0.075$, thus falling in the non-CC regime. Of the remaining two: RCS1107.3-0523 has a $c_{SB} = 0.143$ indicating a moderate CC and RCS1419+5326 has $c_{SB} = 0.185$, suggesting a strong CC (see histogram in Figure 40). The $c_{SB}$ distribution of the X-ray distant clusters is overplotted in the same figure (red line) for direct comparison. The outcome of this analysis based on our small samples suggests that high-z optically selected clusters may have a lower fraction of CC with respect to high-z clusters detected in X-rays.

This is what one would expect from such a comparison. In fact, cluster selection based on X-ray flux limited samples will preferentially find high surface brightness clusters, i. e. high luminosity clusters at high redshift. This effect goes in the direction of favoring CC clusters, whose surface brightness is boosted by the bright core, especially at high redshift. On the other hand, in an optically selected sample of clusters, since the detection is based on the galaxies and not on the ICM emission, you would expect to find also low surface brightness object that can be missed in an X-ray flux limited survey. However, larger samples are needed to draw conclusions on possible different physical conditions of the ICM in optically selected clusters, which may have not reached a final state of virialization.

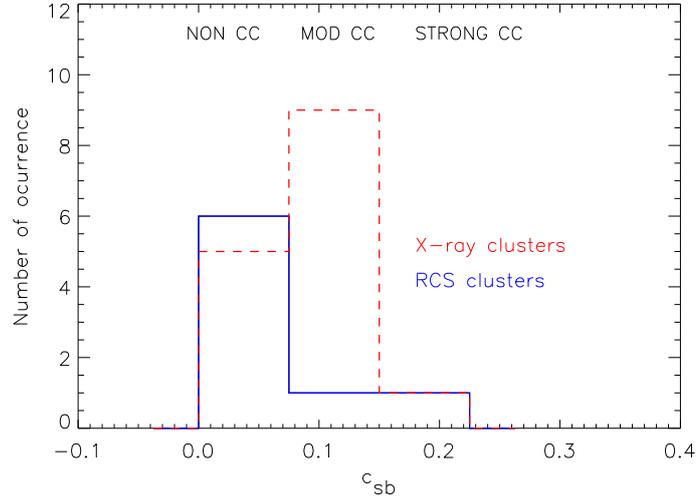

Figure 40: Histogram of $c_{SB}$ for the RCS sample in solid blue line; the high-z X-ray selected sample is overplotted in red dashed line (Figure from Santos et al., 2008).

## 4.6 STRONG LENSING AND X-RAY ANALYSIS OF RCS0224-0002

I briefly present the results published in Rzepecki et al. (2007) about detailed mass reconstruction of the cluster RCS0224-0002 form strong lensing features observed with HST. The mass profile is reconstructed using a novel method to fit extended multiple images based on the Modified Hausdorff Distance (MHD), (Dubuisson & Jain 1994) between observed arcs and the arcs reproduced by the model. The mass derived from the lensing model is in very good agreement with the value obtained from the X-ray observations.

In addition to the *Chandra* observations listed in Table 8, the HST observations of the RCS0224-0002 were taken in two filters, F606W and F814W using the WFPC2 camera. The exposure time for each filter was 1100 seconds. The WFPC2 data reduction was performed by Associations Science Products Pipeline.[2]

---

2 *http://archive.eso.org/archive/hst/wfpc2_asn/wfpc2_products.html*

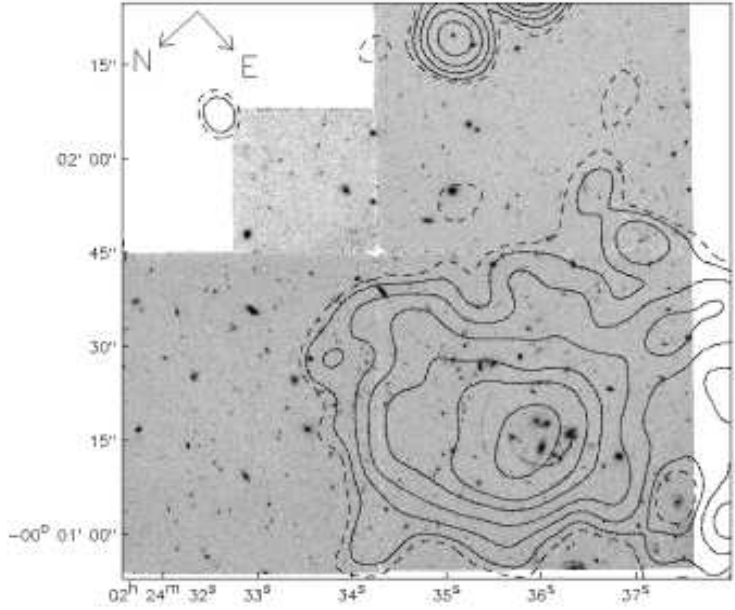

Figure 41: The X-ray emission contours of RCS0224-0002 (smoothed with a Gaussian with $\sigma = 5''$) over-plotted on the F606W WFPC2 HST image (Figure from Rzepecki et al., 2007).

We overlaid the X-ray contours of RCS0224-0002 from the *Chandra* observations in the 0.5-2.0 keV band onto the WFPC2 image in Figure 41. The X-ray emission is not symmetric, with a plume extending North-West, and its peak shifted $\sim 5$ arc seconds north from the two central BCGs. The asymmetry and the displacement may indicate the presence of a merger.

X-RAY EMISSION ANALYSIS  To measure the X-ray temperature, we used an extraction region of $36.7''$ (or 265 kpc), which encompasses most of the X-ray emission by maximizing the SNR. The background subtracted, unfolded spectrum is shown in Figure 43. The data were fitted using the procedure described in Section 4.2.

We measured $742 \pm 35$ total net counts and the spectral analysis gave a best fit temperature of $kT = 5.26^{+1.14}_{-1.07}$ keV (1-sigma error). The unabsorbed flux within the extraction aperture, in the 0.5-2.0 keV band, is $1.84 \times 10^{-14}$erg cm$^{-2}$ s$^{-1}$ and the rest-frame X-ray luminosity $L_X(0.5 - 2\text{keV}) = (0.38 \pm 0.02) \times 10^{44}$ erg s$^{-1}$. The bolometric luminosity returned by the best fit model is $L_{BOL} = (1.28 \pm 0.06) \times 10^{44}$. With these values for X-ray luminosity and temperature, we note that RCS0224-0002, which is an optically selected cluster, lies on the $L_X - T$ relation determined from large samples of X-ray selected clusters (e.g. Rosati, Borgani & Norman, 2002).

We can use the measured cluster temperature to estimate the cluster mass assuming the hydrostatic equilibrium and isothermal distribution of the gas, with a polytropic index $\gamma = 1$. Using the standard $\beta$-model for the gas density profile, $\rho_{\text{gas}}(r) = \rho_0 / [1 + (r/r_c)^2]^{3\beta_m/2}$, the mass within the radius $r$ can be written as (Sarazin, 1988):

$$M(<r) \simeq 1.11 \times 10^{14} \beta_m \gamma \frac{T(r)}{\text{keV}} \frac{r}{h^{-1}\text{Mpc}} \frac{(r/r_c)^2}{1 + (r/r_c)^2} M_\odot. \tag{4.7}$$

A fit to the X-ray surface brightness profile with the corresponding $\beta$-model $S(r) \propto [1 + (r/r_c)^2]^{-3\beta_m+1/2}$ yields a core radius $r_c = (253 \pm 72)$kpc and $\beta_m = 0.97 \pm 0.3$. Therefore the mass within $R_{2000} = 0.4$ Mpc is $(1.7 \pm 1.1) \times 10^{14} M_\odot$.

STRONG LENSING ANALYSIS    RCS0224-0002 has seven prominent luminous arcs and arclets marked as A1, A2, A3, B1, B2, B3, and B4 in Figure 42. Unfortunately, out of those seven arcs, only one arc system A has a confirmed spectroscopic redshift of 4.87 (Gladders et al 2002). The same authors estimated the redshift of system B within the range 1.4 to 2.7 based on the lack of emission lines in their spectra. Since the redshifts of arcs B1, B2, B3, and B4 are not known, an assumption needs to be made of whether all those arcs are images of one source or more sources. Based on very similar color, structure and distance from the center of the cluster we suppose that arcs B1, B2, B3, and B4 are images of one source and we call it system B. This conjecture is supported by the lensing model described below, since by assuming the existence of two separate systems (B1-B2, B3-B4) our model predicts relatively bright multiple images which are not observed. We excluded that the feature D is a radial arc, despite its elongated morphology, since no tangential counter images are visible and because its position and morphology makes this hypothesis unlikely. Our model suggests that feature C is a central demagnified image, which is clearly visible in Figure 42 right panel, where we show the F606W image after subtracting the two cD galaxies. There is also a very faint red arc, labeled E, which was not included in our analysis.

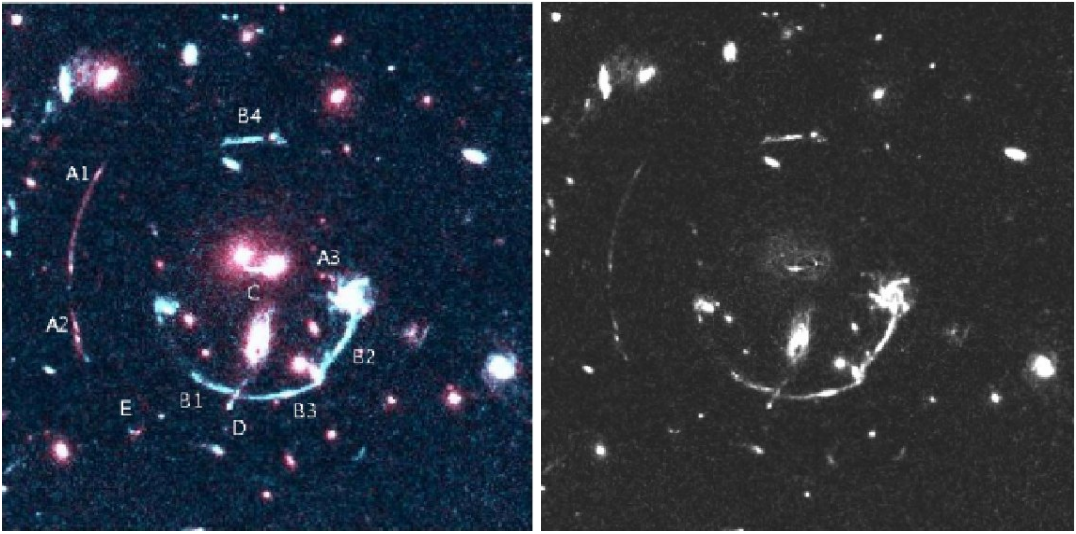

Figure 42: Left panel. The RCS0224-0002 cluster with labeled arcs. Color image composed from F814W and F606W WFPC2 HST images. The image is 40 arcsec across. Right panel. RCS0224-0002 in the F606W filter with subtracted cD galaxies. The central radial feature C is clearly visible (Figure from Rzepecki et al., 2007).

We constructed the mass model of RCS0224-0002 by fitting the position and shapes of the multiple image systems A, B and C (see Figure 42). Based on the light distribution of most luminous red-sequence galaxies, our model consists of several mass components: two Non-singular Isothermal Ellipsoids (NIE1, NIE2) to reproduce global cluster properties; eight Non-singular Isothermal Spheres (NIS1-8) fixed at the position of cluster members refereed to as the substructure; one Non-singular Isothermal Ellipsoids (NIE3) corresponding to the elongated object marked D.

We do not assign any physical meaning to the two distinct smooth components NIE1 & NIE2, since we are interested in the properties of the overall, combined profile. We have also tried to fit the data with only one smooth component (NIE1) and the substructure, however in that case we were not able to fit the arcs system B accurately. The central position of NIE1 was chosen close to the peak of the X-ray emission (i.e. centerend on the ICM distribution), while NIE2 is shifted $\approx 5''$ from NIE1 and it corresponds to the position of the Brightest Cluster Galaxy (BCGs) (i.e. centerend on the dark matter distribution).

Since spectroscopic information is available in the literature only for the two cD galaxies, we identified likely member galaxies in the cluster core from the red-sequence, which is clearly detected in the F606W-F814W color distribution. We identified eight member galaxies, whose position is shown in Figure 42. At the position of each member galaxies is fixed one of the (NIS1-8) of our masss model.

To infer the mass distribution from the position and shapes of the strong lensing features we used a three-step approach: i) fitting minimization on the source plane, ii) fitting minimization on the image plane, based on the MHD, and iii) a refined estimate of the best fit parameters and errors analysis with the MCMC.

In the first step we minimize the distance between the observed position of the sources and the model predicted one in the source plane, while in the second step we minimize in the image plane the distance between the observed image position and the model predicted one for all sources. The first step is computationally very efficient and it provides an approximated solution that is used as the starting point for the second-step image plane analysis. More details about the model adopted and the fitting technique can be found in Rzepecki et al. (2007).

The resulting mass density reproduces all the strong features fairly well. The redshift of the blue arc system B is predicted to be $2.65 \pm 0.08$. The reconstructed mass of the cluster within $R_{2000} = 0.4$Mpc obtained from the model is $(1.9 \pm 0.1) \times 10^{14} M_\odot$, in good agreement with the mass derived above form the X-ray temperature.

## 4.7 CONCLUSIONS

We studied the thermodynamical and chemical X-ray properties of the ICM of a galaxy cluster sample extrated from the Red-sequence Cluster Survey (RCS), in the redshift range $0.6 < z < 1.2$. We detected emission for the majority of the clusters, except for three, for which we have only marginal detection at $\sim 3\sigma$. In general we found that the slope of the $L_X - T_X$ relation of the RCS clusters is in agreement with that of X-ray selected clusters, while the normalization is a factor of 2 lower at high confidence level, in agreement with the result of Hicks et al. (2008) in their independent analysis. Only the three marginally detected RCS clusters seem to have lower luminosities with respect to the RCS $L_X - T_X$, but only if they have $kT > 3$ keV. Unfortunately, for this sample the statistic is too poor to draw any conclusion about possible evolution of the $L_X - T_X$ relation with redshift. Particularly our data are fitted with $L_X / T_X^\alpha \propto (1+z)^{0.2 \pm 0.2}$, consistent with no evolution.

Concerning the iron abundance in RCS clusters, we found that also for this sample of optically selected clusters, the ICM was already enriched with metals at a level comparable with X-ray selected clusters at high redshift (Balestra et al., 2007).

Thanks to the high spatial resolution of the *Chandra* satellite, we also investigated the point source distribution near the region of diffuse cluster emission for RCS with respect to the field. The number counts of point sources as a function of the flux show a significant excess at all fluxes with respect to the data of the CDFS. We quantified this excess between 15% and 40% with a significance of $2\sigma$. This result is in agreement with that found by other authors (Cappi et al., 2001; Branchesi et al., 2007a).

The spatial distribution of point sources in the field of RCS clusters shows a factor of $\sim 6$ over-density at $3\sigma$ confidence level in the inner 150 kpc. The contribution of these point sources to the X-ray emission is limited to few percent in the soft band, showing that the contamination from AGN in the soft band is not severe at least for cluster with $L_X > 10^{44}$ erg s$^{-1}$ up to $z \sim 1$.

Studying the X-ray surface brightness properties of the RCS clusters and evaluating the presence of Cool Core in our sample, we found that high-z optically selected clusters seem to have a lower fraction of CC with respect to high-z clusters detected in the X-ray.

We have performed a strong lensing analysis of the cluster RCS0224-0002. The mass derived from our lensing model is in very good agreement with that obtained from the X-ray temperature, showing that X-ray data are effective in probing the mass of high-z clusters.

## 4.8 PROPERTIES OF INDIVIDUAL CLUSTERS

*RCS1419+5326*

RCS1419+5326 is the lowest redshift RCS cluster ($z = 0.62$) in our sample and was observed with two pointings, ObsID 3240 with nominal exposure time 10 ks, and ObsID 5886 with 50 ks. The extraction radius is $37''$, corresponding to 252 kpc. RCS1419+5326 is detected with the highest number of net counts, $2320 \pm 60$, corresponding to a signal-to-noise ratio $\sim 38$. In the spectrum (Figure 43 top-left panel) there is a clear evidence of the iron $K_\alpha$ line at $\sim 4.14$ keV (observing frame). The resulting Fe abundance is $X_{\text{Fe}} = 0.29^{+0.06}_{-0.11} X_{\text{Fe}\odot}$, with a temperature $kT = 5.0^{+0.4}_{-0.4}$ keV and bolometric luminosity $L_{\text{X}} = 4.63 \pm 0.12 \times 10^{44}$ erg s$^{-1}$. As pointed out by Santos et al. (2008) RCS1419+5326 is a cool-core cluster. Therefore we repeated the spectral analysis masking the cool-core roughly corresponding to the inner 80 kpc. In this case the best fit temperature is $kT = 5.2^{+0.7}_{-0.5}$ keV and the resulting Fe abundance is substantially unchanged $X_{\text{Fe}} = 0.23^{+0.12}_{-0.14}$. We corrected the bolometric luminosity for the removed core emission fitting the radial surface brightness profile with a $\beta$-model

$$S(r) = A \left[ 1 + \left( \frac{r}{r_c} \right)^2 \right]^{-3\beta_m + 0.5}, \tag{4.8}$$

where $A$ is the amplitude at $r = 0$ and $r_c$ is the core radius. The best fit values are $r_c = 17''$ and $\beta_m = 0.78$. Extrapolating the profile in the masked inner region, the corrected bolometric luminosity is $L_{\text{X}} = 3.71 \pm 0.15 \times 10^{44}$ erg s$^{-1}$.

*RCS1107.3+0523*

This cluster at $z = 0.735$ was observed with two pointings, ObsID 5825 and ObsID 5887, both with nominal exposure 50 ks. The extraction radius is $28.4''$, corresponding to 207 kpc. The total number of net counts is $710 \pm 40$. The best fit temperature is $kT = 4.3^{+0.5}_{-0.6}$ keV resulting in a bolometric luminosity of $L_{\text{X}} = 1.34 \pm 0.08 \times 10^{44}$ erg s$^{-1}$. The iron line is well visible at $\sim 3.86$ keV (observing frame). This cluster shows the highest value of Fe abundance in our sample, $X_{\text{Fe}} = 0.67^{+0.35}_{-0.27} X_{\text{Fe}\odot}$

*RCS1325+2858*

This cluster at $z = 0.75$ was observed with two pointings: ObsID 3291 with 31.5 ks and ObsID 4362 with 45 ks. The extraction radius is $29.7''$, corresponding to 218 kpc. We detected $90 \pm 30$ net counts. As one can see from the spectrum (Figure 44 bottom-left panel) the signal is dominated by the background above 2 keV, so we found only an upper limit to the

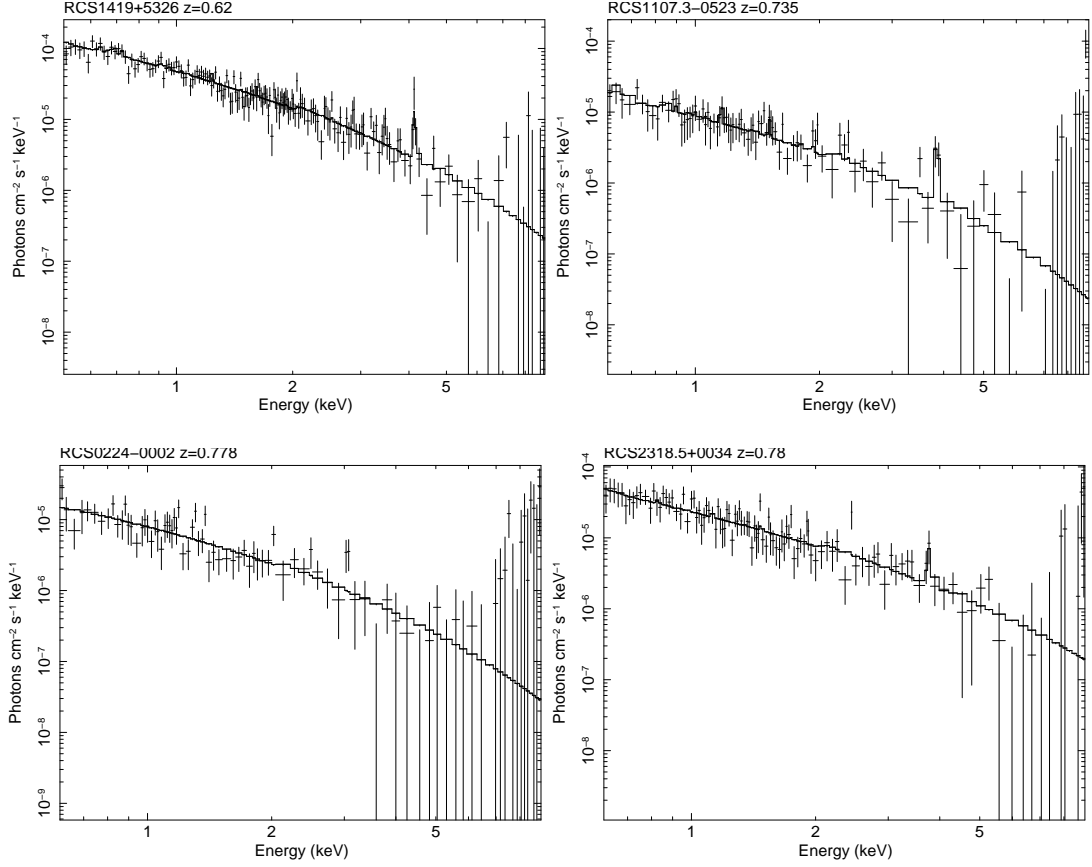

Figure 43: Unfolded spectra of RCS1419+5326, RCS1107.3-0523, RCS0224-0002 and RCS2318.5+0034. The solid lines are the best fit of the model (Figure from Bignamini et al., 2008).

Fe abundance of $X_{Fe} = 0.09^{+0.66}_{-0.09} X_{Fe\odot}$. The best fit temperature is $kT = 1.8^{+1.2}_{-0.6}$ keV and the resulting bolometric luminosity is $L_X = 0.23 \pm 0.07 \times 10^{44}$ erg s$^{-1}$.

*RCS0224-0002*

For this cluster at $z = 0.778$ we analyzed two observations: ObsID 3181 with nominal exposure 15 ks and ObsID 4987 with 100 ks. The extraction radius is 36.7″, corresponding to 273 kpc. The total number of net counts is $740 \pm 50$. The best fit temperature is $kT = 5.1^{+1.3}_{-0.8}$ keV and the bolometric luminosity is $L_X = 1.31 \pm 0.10 \times 10^{44}$ erg s$^{-1}$. In this case we find only an upper limit to the Fe abundance $X_{Fe} < 0.15 X_{Fe\odot}$

*RCS2318.5+0034*

RCS2318.5+0034 ($z = 0.78$) was observed with a single observation, ObsID 4938, with an exposure time of 50 ks. The extraction radius is 40.6″, corresponding to 302 kpc. We detected $970 \pm 50$ net counts for a signal-to-noise ratio of $\sim 19$. The iron $K_\alpha$ line at $\sim 3.76$ keV observing frame is very clear. The resulting Fe abundance is $X_{Fe} = 0.35^{+0.20}_{-0.22} X_{Fe\odot}$, with a best fit temperature of $kT = 7.3^{+1.3}_{-1.0}$ keV and a bolometric luminosity $L_X = 4.51 \pm 0.23 \times 10^{44}$ erg s$^{-1}$.

*RCS1620+2929*

For RCS1620+2929 at $z = 0.87$ we have only one pointing, ObsID 3241, with a nominal exposure time of 35 ks. We detected $190 \pm 20$ net counts, extracted in a region of radius 29.5″, corresponding to 227 kpc. The best fit temperature is $kT = 4.6^{+2.1}_{-1.1}$ keV, the bolometric luminosity is $L_X = 1.35 \pm 0.18 \times 10^{44}$ erg s$^{-1}$ and the Fe abundance is $X_{Fe} = 0.33^{+0.60}_{-0.33} X_{Fe\odot}$.

*RCS2319.9+0038*

RCS2319.9+0038 at $z = 0.9$ was observed with four pointings: ObsID 5750, 7172, 7173 and 7174 with a nominal exposure time of 21 ks, 18 ks, 21 ks and 15 ks, respectively. This is the most distant cluster in our sample for which we have a high number of total net counts, $1490 \pm 60$, for a signal-to-noise ratio of $\sim 22$. It is also the cluster with the largest extraction radius, 45.6″ arcsec (356 kpc), the highest bolometric luminosity, $L_X = 5.97 \pm 0.26 \times 10^{44}$ erg s$^{-1}$. The best fit temperature is $kT = 5.3^{+0.7}_{-0.5}$ keV. The spectrum (Figure 44 bottom-right panel) show a clear $K_\alpha$ line at $\sim 3.53$ keV observing frame. The best fit Fe abundance is $X_{Fe} = 0.60^{+0.22}_{-0.18} X_{Fe\odot}$.

*RCS0439.6-2905*

RCS0439.6-2905 ($z = 0.96$) was observed with two pointings: ObsID 3577 with nominal exposure time of 85 ks and ObsID 4438 with 30 ks. We detected $220 \pm 30$ net counts. The extraction radius is 24.7″, corresponding to 196 kpc. The best fit temperature is $kT = 1.8^{+0.4}_{-0.3}$ keV and the luminosity $L_X = 0.56 \pm 0.09 \times 10^{44}$ erg s$^{-1}$. As one can see from the spectrum (Figure 44 top-right panel) the signal is dominated by the background above 2 keV making impossible to identify the iron $K_\alpha$ line, so we found only an upper limit to the Fe abundance of $X_{Fe} = 0.44 \pm 0.27 X_{Fe\odot}$.

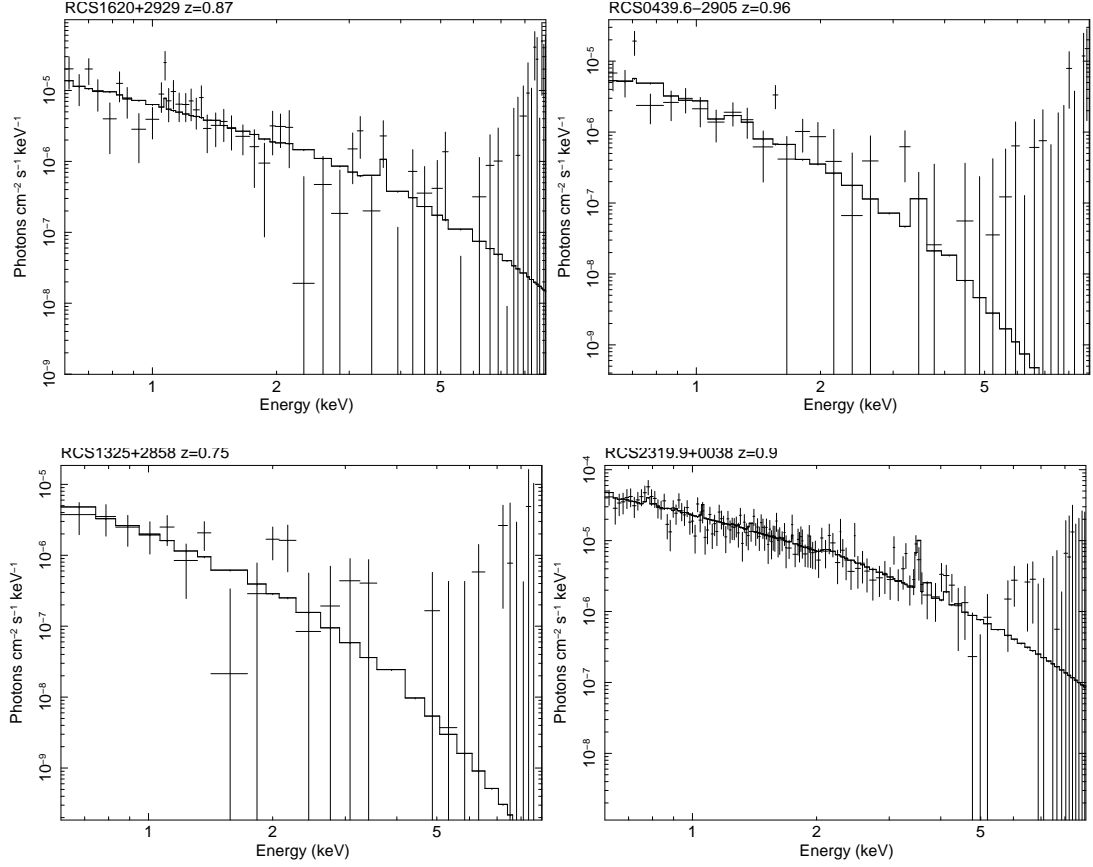

Figure 44: Unfolded spectra of RCS1620+2929, RCS0439.6-2905, RCS1325+2858 and RCS2319.9+0038. The solid lines are the best fit of the model (Figures from Bignamini et al., 2008).

*RCS1417+5305, RCS2112.3-6326, RCS2156.7-0448*

In these three clusters, we were not able to detect the diffuse emission in order to perform a standard spectral analysis. Therefore, we selected a circular region with radius $\sim$ 20 arcsec (20 pixel) centered in the optical coordinates of each cluster. In order to perform a simplified spectral analysis, we constrained the temperature to be in the range 1.0-8.0 keV and the metallicity at $0.3 X_{Fe\odot}$, thawing only the normalization. In this way we obtained a reliable upper limit to the bolometric luminosity.

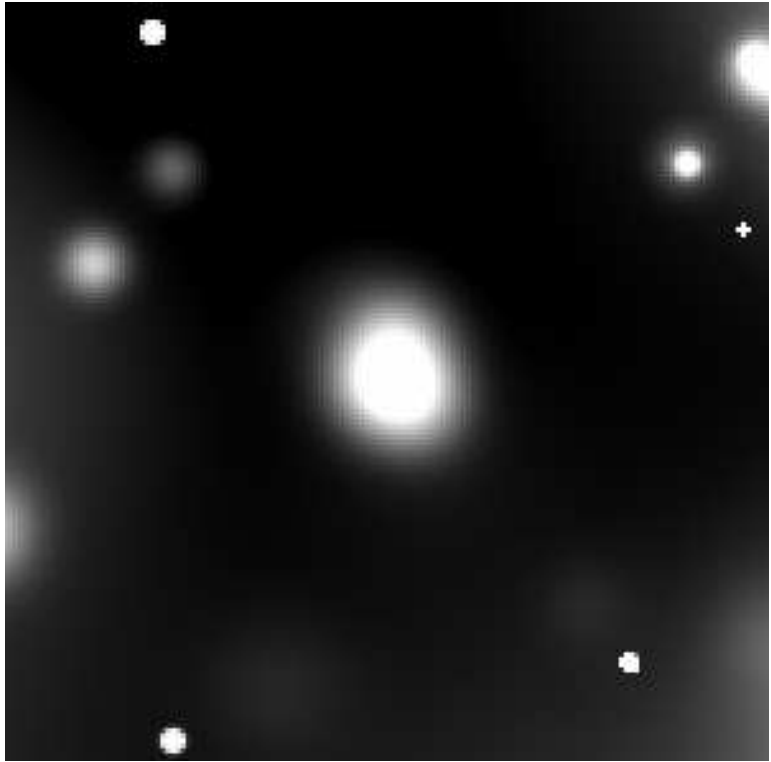

Figure 45: Soft merged image for the three undetected clusters RCS1417+5305, RCS2112.3-6326, RCS2156.7-0448. The size of the image is about $1' \times 1'$ (Figure from Bignamini et al., 2008).

RCS1417+5305 ($z = 0.968$) is the only RCS cluster in our sample observed with ACIS-I (ObsID 3239) with an exposure time of 70 ks. The aperture photometry in the extraction region, corresponding to 156 kpc, gives $37 \pm 11$ net counts, consistent with zero within $3\sigma$. Freezing the temperature in the range 1.0-8.0 keV and the metallicity to $0.3X_{Fe\odot}$, we obtain an upper limit to the bolometric luminosity of $L_X < 0.29 \times 10^{44}$ erg s$^{-1}$.

RCS2112.3-6326 was observed with a single observation, ObsID 5885, with nominal exposure time of 70 ks. This is the most distant cluster in our sample at $z = 1.099$. The extraction radius corresponds to 161 kpc. The total number of net counts is $47 \pm 20$, implying a positive detection at $2\sigma$ only. The upper limit to the bolometric luminosity is $L_X < 0.21 \times 10^{44}$ erg s$^{-1}$.

RCS2156.7-0448, whose redshift is $z = 1.080$, was observed with two pointings: ObsID 5353 with 40 ks and ObsID 5359 with 35 ks. The aperture photometry in the extraction region, corresponding to 160 kpc, gives $60 \pm 20$ net counts, consistent with zero within $3\sigma$. The upper limit to the bolometric luminosity is $L_X < 0.22 \times 10^{44}$ erg s$^{-1}$.

To better constrain the average diffuse emission of these three clusters, we merged together the three images overlapping the optical centers of the clusters. From the merged image we chose the same extraction radius as before ($\sim 20$ arcsec) and we performed the standard spectral analysis, fitting the background subtracted spectrum with three free parameters, with the aim of obtaining a mean value for the thermodynamical properties of these three clusters. Unfortunately, even in this case the fit does not converge. However, from aperture photometry we measured $274 \pm 29$ total net counts, implying a clear detection of diffuse emission from the hot ICM. In the Figure 45 we show the merged image in the soft band, where the diffuse emission contributed by the sum of the three clusters is clearly visible in the center.

Since the spectral fit does not converge, to evaluate also a mean bolometric luminosity for these three clusters from the total net counts, we assumed the typical thermal spectrum for each cluster with the same redshift and exposure map, in order to have the ECFs from net counts to bolometric luminosity. The temperature is fixed at 4 keV and the metallicity at 0.3 $X_{Fe\odot}$. Then we computed an effective ECF from these three single ECFs, weighing them with the net counts of the individual clusters. Multiplying 274 net counts by the effective ECF, we found $\langle L_X \rangle = 0.2 \pm 0.06 \times 10^{44}$ erg s$^{-1}$ as the mean bolometric luminosity for one cluster in agreement with upper limits found from the single cluster analysis.

## PLANNING A FUTURE MISSION: THE WIDE FIELD X-RAY TELESCOPE

In this Chapter I describe the scientific case and the concept of a new proposed X-ray mission submitted to the Astro 2010 NASA Decadal Survey: the Wide Field X-ray Telescope (WFXT). This telescope is planned to cover at each flux three orders of magnitude more area than any previous or planned missions, in order to address many outstanding cosmological and astrophysical topics, such as the formation and the evolution of clusters of galaxies and associated implications on cosmology and fundamental physics, properties and evolutions of AGN, cosmic star formation history, and so on.

Within my Thesis, WFXT can be understood as a deep look into the future in the landscape of X-ray surveys of galaxy clusters, since WFXT will have at the same time up to 2-3 orders of magnitude lower flux limit and larger solid angle covered with respect to The SXCS. In fact, the aim of WFXT is not only to detect a large amount of clusters, but also to measure the redshift for a subsample of the detected object, like I did for the SXCS.

Within the WFXT project, my work constitutes an important part of the scientific case of WFXT. In particular, I contributed substantially to the realization of realistic X-ray simulated images to help in the choice of the optimal design in order to achieve the mission goals. In this Chapter are described all the details about physical and technical issues involved in the realization of the simulated images, and at the end are briefly presented the prelimary results about the ongoing simulation analysis.

### 5.1 THE NEED OF A NEW WIDE ANGLE X-RAY SURVEY

Exploring the high-redshift Universe to the era of galaxy and cluster formation requires an X-ray survey that is both sensitive and extensive, which complements current and planned deep, high-resolution, wide-field surveys in the optical, radio, and infrared bands. To this end, we present the Wide Field X-ray Telescope (WFXT), designed to be three orders of magnitude more effective than previous and planned X-ray missions in carrying out surveys (see Figure 46).

Three co-aligned wide-field X-ray telescopes with a 1 square degree Field of View and a lower than $10''$ (goal of $5''$) angular resolution (HEW) averaged over the full field. With nearly ten times the collecting area of *Chandra* and more than ten times *Chandra*'s FoV, WFXT will perform sensitive deep surveys that will discover and characterize extremely large populations of high redshift AGN and clusters of galaxies. In five years of operation, WFXT will carry out three extragalactic surveys: the first will cover most of the extragalactic sky ($\sim 20000$ deg$^2$) at 100-1000 times the sensitivity, and twenty times better angular resolution of the ROSAT All Sky Survey (RASS); the second survey will map more than 3000 deg$^2$ to deep *Chandra* or *XMM-Newton* sensitivity; and the third will probe more than 100 deg$^2$, or more than 1000 times the area of the Chandra Deep Fields, to the deepest *Chandra* sensitivity. From these surveys, WFXT will generate a legacy dataset of more than $5 \times 10^5$ clusters of galaxies to $z \sim 2$, while characterizing the physics of the intracluster gas for a significant fraction of them, and a sample of more than $10^7$ AGN to $z > 6$, a substantial fraction with X-ray spectra sufficient to distinguish obscured from unobscured quasars.

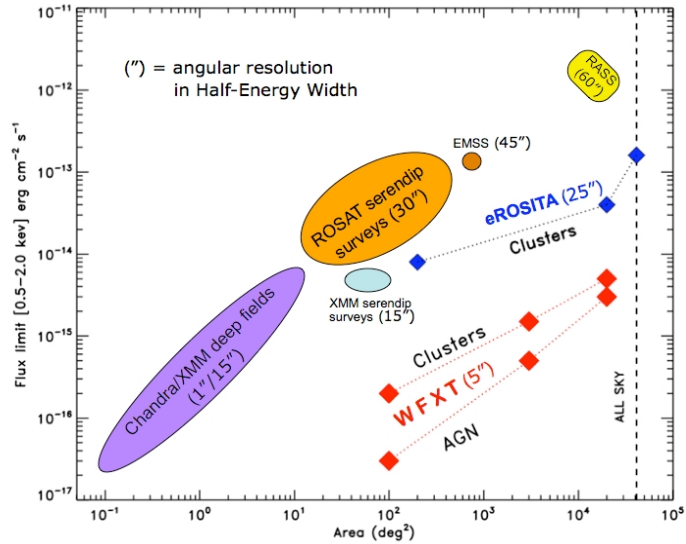

Figure 46: Effective flux limits and sky coverage for past and planned X-ray surveys. WFXT provides an unsurpassed combination of sensitivity and sky coverage.

WFXT will address many outstanding cosmological and astrophysical objectives: formation and evolution of clusters of galaxies and associated implications for cosmology and fundamental physics (e.g. the nature of dark matter, Dark Energy and gravity); black-hole formation and evolution; AGN interaction with ICM and ISM in clusters and galaxies; and the high-energy stellar component and hot-phase interstellar medium of galaxies, including the Milky Way. Unlike previous X-ray astronomy telescopes, WFXT is optimized for performance over its full field of view (rather than mainly on-axis), which makes it an ideal survey instrument.

| Parameter | Requirement | Goal |
|-----------|-------------|------|
| Area (1.0 keV) | 6000 cm$^2$ | 10000 cm$^2$ |
| Area (2.0 keV) | 2000 cm$^2$ | 3000 cm$^2$ |
| Field of View | 1 deg diameter | 1.25 deg diameter |
| Angular Resolution | $< 10''$ | $\leq 5''$ |
| Energy Band | 0.2-4.0 keV | 0.1-7.0 keV |
| Energy Resolution | $\frac{E}{\Delta E} > 10$ | $\frac{E}{\Delta E} > 20$ |
| Time Resolution | $< 3$ s | $< 1$ s |

Table 13: WFXT mission performance requirement (from Murray et al., 2010).

## 5.2 MISSION CONCEPT

In order to carry out the science objectives discussed before, we derived performance requirements for the WFXT mission, which are summarized in Table 13. With these technical requirements a simple mission design was developed. The science payload consists of 3 separate, but identical, X-ray telescope/detector modules. These units mount to a simple spacecraft that provides the usual complement of services (i.e. power, command and data handling, telemetry, attitude control, etc.).

Each of the three identical telescope/detector modules consists of a wide-field X-ray telescope, optical bench, fine attitude sensor, and X-ray detector assembly. The X-ray telescopes are highly nested full shells, using either Electroformed Nickel-cobalt Replicas (ENR) or chemical vapor deposited of Silicon-Carbide (SiC). The optical design is based on a polynomial perturbation of the classical Wolter-I prescription as described by Burrows, Burg & Giacconi (1992). Key to the WFXT mission is the high angular resolution ($< 10''$ HEW) maintained over the full 1 degree field of view. This characteristic, along with its large collecting area, gives WFXT unprecedented sensitivity, so that large surveys can be carried out in just a few years. The energy bandwidth of the detectors is 0.1-7.0 keV.

The detectors are $2 \times 2$ arrays of X-ray CCDs, which have a long history of use as imaging X-ray spectrometers. Our baseline device is a Lincoln Laboratory Charge Coupled Device (CCD) similar to those in operation on *Chandra* (launched in 1999) and *Suzaku* (launched in 2005). These are frame transfer devices, with low noise (2-3 electrons rms) performance resulting in Fano limited energy resolution over the WFXT bandwidth. Based on this preliminary design, the calculated areas at 1.0 keV and 4.0 keV are $\sim 10000$ cm$^2$ and $\sim 3000$ cm$^2$ respectively. WFXT will be placed in a 550 km circular orbit at low ($5°$) inclination, which provides a low particle background, well suited for the mission science.

## 5.3 WFXT SIMULATIONS OVERVIEW

The goal of the simulation tool is to provide WFXT images of the extragalactic sky in different energy bands with realistic populations of all sources that contribute to the X-ray sky.

Starting from a modified version of the X-ray image simulator described in Section 2.7, I developed the bulk of the WFXT image simulator, in close collaboration with R. Gilli (for point sources and AGN modelization), J. Santos (who developed the cloning technique described in Section 5.7.2), P. Rosati, P. Tozzi and Y. Heng (for cluster populations), and G. Pareschi, S. Campana and P. Conconi (for technical issues about PSF, effective area and telescope design).

The simulation algorithm is designed to be flexible and easily adaptable to different specifications still under study, like different PSF profiles or effective areas. Also the mock source input catalogues can be easily updated and changed significantly to meet different needs, e.g. to investigate the impact of a different ratio of Cool Core *versus* non-Cool Core on the detection of high redshift clusters. The simulated images are designed to realistically reproduce the technical characteristics of the telescope fitting the science requirement goals (see Section 5.2).

In the following Sections I present the main details about a set of simulated images realized for the SiC design. Image size is of $4096 \times 4096$ pixel, and the FoV is of $1 \text{deg}^2$, so the resulting pixel scale is $1 \text{pxl} = 0.8789''$. Images are simulated in three standard energy bands and each of these energy bands is associated with a color to create false color images: 0.5-1.0 keV in red, 1.0-2.0 keV in green and 2.0-7.0 keV in blue.

The simulated images are builded for the SiC design, which affects effective area and vignetting (see Section 5.4), and the PSF profile. PSF broadening due to manufactoring errors is chosen so that the average HEW was of $\sim 5''$ fitting the goal requirements (see Section 5.5).

Although these are indeed X-ray images with passband integrated fluxes, each source maintains its own spectral properties. In fact, ECFs are computed one by one for each source according to the different spectra (see Sections 5.6 and 5.7).

Summarizing the simulation algorithm has the following input parameters: exposure time, AGN and galaxies catalogue, groups and clusters of galaxies catalogue, Galactic hydrogen column density $N_H$. Step by step instructions executed by the algorithm are: generation of the expsure maps according to the image exposure time; generation of the point sources and extended source catalogue; simulation of the sources through Poissonian sampling of the source profile, PSF for point sources and cloned template for the extended sources; Poissonian addition of the background. The cosmological parameters used for the simulations are: $H_0$=70 km s$^{-1}$ Mpc$^{-1}$, $\Omega_\Lambda$=0.7 and $\Omega_m$ =0.3.

## 5.4 EFFECTIVE AREA

The reference response function for the WFXT at the aimpoint is based on the Silicon-Carbide (SiC) design. The ARF and RMF have been obtained starting form the *Suzaku* response files. In

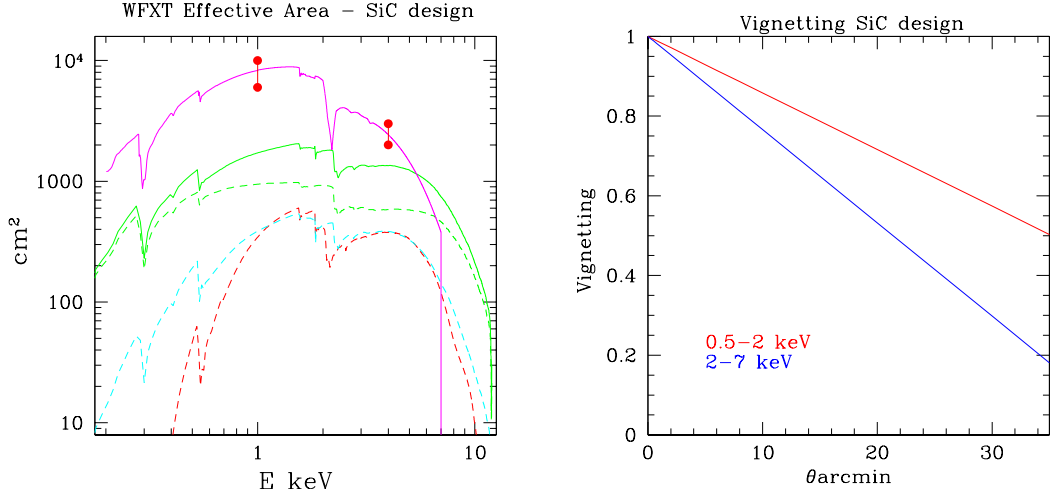

Figure 47: Left panel. Effective area for WFXT Silicon-Carbide design (magenta). Red dots show the requirement and the goal at 1 and 4 keV. The *Chandra* effective area is shown as a red dashed line, while the total XMM-Newton response is shown as a solid green line (green dashed for PN and cyan dashed for MOS). Right panel. Average soft (red) and hard (blue) vignetting for the WFXT Silicon-Carbide design (averaged over the energy band).

Figure 47 left panel, the effective area for the reference SiC design is compared with that of the *Chandra* and *XMM-Newton*.

The vignetting for the SiC design in the soft and hard bands is approximated as a linear radial decreas of the effective area with respet to the aimpoint, after averaging over the corresponding energy bands:

$$V(\theta) = (1 - 0.0142 \times \theta) \text{ in } [0.5 - 2.0] \text{keV} \tag{5.1}$$

$$V(\theta) = (1 - 0.0234 \times \theta) \text{ in } [2.0 - 7.0] \text{keV} \tag{5.2}$$

where $\theta$ is the off-axis angle in arcmin. The average vignetting in the soft and hard band (see Figure 47 right panel) are obtained with arithmetic mean, i.e. assiuming a flat spectrum in each energy band. This is the vignetting approximation that will be used in the simulations to create exposure maps.

## 5.5 SPATIAL RESOLUTION

The spatial resolution is a key aspect of the WFXT mission. The image of each source, point like or extended, is simulated taking into account the shape and the orientation of the Point Spread Function (PSF) at the position of the source, as a function of the off-axis angle $\theta$. The PSF profile is not modeled with an analytic function, but it is directly obtained with ray tracing through telescope optics. The monochromatic PSF snapshots are sampled at 1.0 and 4.0 kev at 3 arcmin step.

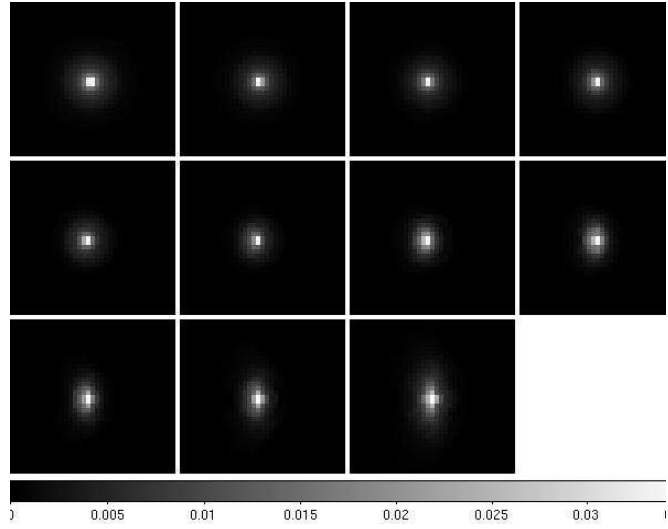

Figure 48: Snapshot images of the WFXT PSF for the SiC design with manufacturing errors in the soft band, aimpoint at the top-left, step $3'$, single image size 1.48 arcmin (after P. Conconi).

Besides the telescope optics, the PSF profile strongly depends on the manufactoring errors, which can be mimicked by smoothing the design PSF profile with a Gaussian kernel. The manufacturing errors tend to give a much flatter PSF across the FoV (see Figure 48). In this sense the HEW of the final PSF can be approximated with the sum under quadrature of the HEW profile from ray tracing and manufacturing errors. The manufacturing errors are chosen so that the average HEW of the PSF satisfies the goal of $5''$. Figure 49, the HEW of the PSF at 1.0 keV and 4.0 keV as a function of the off-axis angle are fitted by the following functions:

$$HEW_{1.0\text{keV}} = 6.6764 - 0.07848 \times \theta - 0.010018 \times \theta^2 + 0.0004192 \times \theta^3; \tag{5.3}$$

$$HEW_{4.0\text{keV}} = 8.364 + 0.06163 \times \theta - 0.022895 \times \theta^2 + 0.000784 \times \theta^3. \tag{5.4}$$

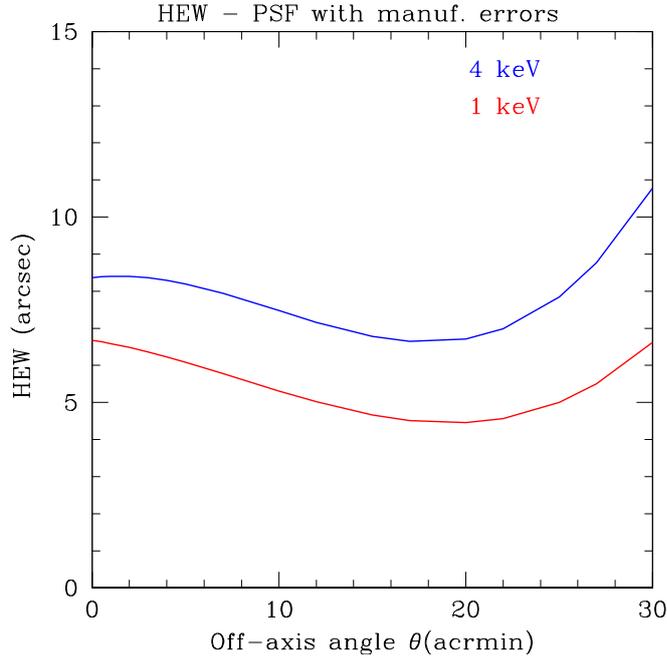

Figure 49: Best fit WFXT HEW for the SiC design as function of the off-axis, including manufacturing errors in the soft (red line, 1.0 keV) and hard (blue line, 4.0 keV) energy bands as a function of the off-axis angle $\theta$.

Each source has its own PSF profile depending on the source off-axis. The image of a point source is simply obtained with Poissonian sampling of the PSF profile normalized by the expected source count rate, while the image of cluster clone is obtained with Poissonian sampling of the clone image convolved with the PSF profile normalized to one.

## 5.6 ENERGY CONVERSION FACTORS

Each source in the input catalogue has its own flux. To covert the flux into count rate, ECF is needed. The ECF is defined as the ratio of the observed energy flux in a given energy band and the observed count rate in the same energy band, and are computed always at the aimpoint. Therefore, the ECF depends on the source spectrum and also on the Galactic absorption. ECFs are computed individually for each simulated source, adopting the true spectral parameters of the source as described in Section 5.7.

The present situation is quite different with respect to how ECFs are employed in the previous Chapters. In fact, in Chapters 2 and 4 ECFs are used to compute the flux of a source from the measured count rate without detailed spectral analysis of the source, therefore average

| Source Type | 0.5-2.0 keV | 2.0-7.0 keV |
|---|---|---|
| AGN Thin, z=1 | $3.08 \times 10^{-13}$ | $2.92 \times 10^{-12}$ |
| AGN Thick, z=1 | $2.39 \times 10^{-13}$ | $2.97 \times 10^{-12}$ |
| AGN Unabsorbed | $2.28 \times 10^{-13}$ | $2.00 \times 10^{-12}$ |
| Galaxies | $2.26 \times 10^{-13}$ | $1.93 \times 10^{-12}$ |
| Cluster kT=5, z=0.5 | $2.30 \times 10^{-13}$ | $1.80 \times 10^{-12}$ |

Table 14: Typical ECF for SiC design at the aimpoint. The units are erg cm$^{-2}$ cts$^{-1}$.

ECFs are employed. Instead in WFXT simulations, for each source the ECFs are computed for the actual source spectrum and they are used to compute the count rate to simulate from the flux of the source. In general, when we will analise the simulations the ECFs used to re-costruct the measured flux from the observed count rate referes to an average spectral shape. Therefore, we should not expect to recover exactly the same input flux of the simulated source, even in the case of a very high number of net detected counts.

ECFs are computed exactly with *XSpec* the three standard energy bands (0.5-1.0 keV, 1.0-2.0 keV and 2.0-7.0 keV) as the ratio between the flux and the corresponding count rate. For each source the specific *XSpec* spectrum model is adopted (see Section 5.7). As explained in Section 2.3.3, ECFs also depend on the Galactic $N_H$, which is an input parameter of WFXT simulated images. To envisage also this effect, an extra multiplicative model tbabs in front of the source spectrum models is used. As an indication, typical ECF values are listed in Table 14 for the SiC design.

## 5.7 SOURCE POPULATIONS

The simulated images take into account all the sources that contribute to the X-ray sky: AGN, normal and starburst galaxies, groups and clusters of galaxies, and residual X-ray background. In the following Sections I will describe how this different components are handled.

### 5.7.1 *Point Sources: Active Galactic Nuclei and Galaxies*

Point sources simulated in WFXT images can be divided into two grops: AGN and galaxies (both normal and starburst galaxies).

AGNs with different column densities[1], $N_H$, have been extracted from the X-Ray Background (XRB) synthesis model of Gilli, Comastri & Hasinger (2007). Sources are randomly extracted from a Probability Density Function (PDF) that takes into account the AGN 0.5-2.0 keV LogN-LogS (see Figure 50) and the redshift distribution. Redshifts and fluxes are assigned to each AGN. It was verified that the simulated sources self-consistently reproduce the model LogN-LogS relations. The AGN population is subdivided into three class according to the intrinsic $N_H$[2]: unobscured AGN for $\mathrm{Log}N_H < 21$, obscured Compton-thin AGN for $21 < \mathrm{Log}N_H < 24$, and obscured Compton-thik AGN for $\mathrm{Log}N_H > 24$. The *XSpec* models used for each AGN class are described in Table 15.

---

[1] Do not confuse the Galactic hydrogen column density with the AGN column density. The former affects every sources and its absorption is local, while the latter is characteristic of AGN and its absorption is redshifted.

| Source | Model |
|---|---|
| Galaxies | `tbabs *zpowerlw` |
| AGNs unabsorbed | `tbabs *(zpowerlw +zwabs *(zhighect *zpowerlw) +zgauss)` |
| AGNs Compton-thin | `tbabs *(zpowerlw +zwabs *(zhighect *(zpowerlw +zgauss)) +zgauss)` |
| AGNs Compton-thick | `tbabs *(zpowerlw +zwabs *(zhighect *(zpowerlw +pexrav)) +zgauss)` |

Table 15: *XSpec* models used for different sources type.

Similarly, the galaxy sample is extracted from the soft LogN-LogS relations by Ranalli et al. (2005). Since the spectrum adopted for the galaxies is a simple power law with photon index $\Gamma = 2$ (see table Table 15), no redshift is assigned to these sources.

AGN and galaxies are extracted down to a 0.5-2.0 keV flux of $2 \times 10^{-18}$ erg cm$^{-2}$ s$^{-1}$. In total about $\sim 30000$ sources per square degree are expected at this flux limit (about 1/3 AGN and 2/3 normal and starburst galaxies). The AGN and the galaxy integrated flux matches the XRB flux. The spectrum model assigned to each source is used to compute source fluxes in the three standard bands (0.5-1.0 keV, 1.0-2.0 keV and 2.0-7.0 keV) from the flux in the 0.5-2.0 keV band, and then specific ECF as described in Section 5.6. The ratio between flux and ECF gives the source count rate in each energy band. Each sources is then simulated through Poissonian sampling of the PSF (see Section 5.5) multiplied by the expected number of net counts to simulate, given by the product between the count rate and the exposure time corrected for vignetting.

### 5.7.2 *Extended Sources: Cloning Galaxy Clusters*

The images of the WFXT simulated clusters are builded from real X-ray images taken with the *Chandra* satellite using a novel cloning technique developed by Santos et al (2008). The list of cluster templates spans a reasonable range of temperatures and morfologies (mostly the CC, non-CC classification), as shown in Table 16. The template list of clusters used for simulations can be easily updated, and the morphology distribution can be changed significantly. As a matter of fact, with this approach we can investigate the impact of a different ratio of CC *versus* non-CC clusters on the detection of high-z clusters and the determination of the cluster evolution and hence cosmology. The *Chandra* images of the templates are shown in Figure 51.

---

2 $N_H$ is measured in unit of cm$^{-2}$

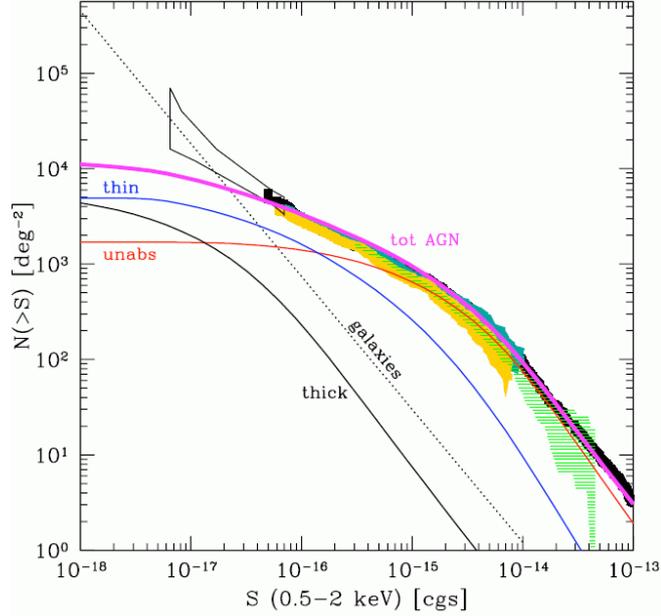

Figure 50: Input LogN-LogS for extragalactic point sources. The contributions of unabsorbed, thin ($N_H \sim 10^{22}$ cm$^{-2}$) and thick ($N_H \sim 10^{24}$ cm$^{-2}$) AGN are highlighted. Normal galaxies becomes dominat at about few $10^{-17}$ erg cm$^{-2}$ s$^{-1}$, just above the expected WFXT detection limit in the Deep Survey (from Gilli et al., 2007).

The cluster population is generated down to $10^{-17}$ erg cm$^{-2}$ s$^{-1}$. The strategy adopted to create the input catalogue involves several steps. First, a catalogue with luminosity and redshift for each source is generated. This can be done with two different approach: a phenomenological method sampling observed X-ray luminosity function (e.g. Rosati et al., 1998; Mullis et al., 2004); a theoretical method based on the theoretical PS-like mass function (e.g. Press & Schechter, 1974; Sheth & Tormen, 1999; Jenkins et al., 2001; Tinker et al., 2008). The former has the advantage of being automatically consistent with the observed LogN-LogS and the observed luminosity function; the latter allows to estimate the accuracy within which the input cosmological parameters used for the input mass function can be recovered. Then, cluster temperatures are extracted from the luminosity according to the observed *L-T* relation assuming also an intrinsic scatter between luminosity and temperature. A random Cool Core (CC) parameter with flat distribution between 0 and 1 is assigned to each source.

| Cluster | $kT$ [keV] | $z$ | class |
|---------|-----------|------|-------|
| RXJ1320 | 1.00 | 0.036 | 0.0 |
| A2717 | 2.4 | 0.047 | 0.0 |
| MKW3S | 3.66 | 0.043 | 0.5 |
| A1621 | 3.6 | 0.085 | 0.0 |
| A3112 | 4.1 | 0.075 | 1.0 |
| A907 | 5.82 | 0.153 | 1.0 |
| A2218 | 6.25 | 0.177 | 0.0 |
| A1835 | 8.1 | 0.253 | 1.0 |
| A2261 | 7.3 | 0.224 | 0.5 |
| A1413 | 7.5 | 0.143 | 0.5 |
| ZWCL3146 | 8.6 | 0.291 | 1.0 |

Table 16: List of cluster templates used in WFXT simulations, ordered from the lowest temperature to the highest. The *class* parameter refers to the CC *versus* non-CC classification, where 0=non-CC, 1=CC, and 0.5 is an intermediate state.

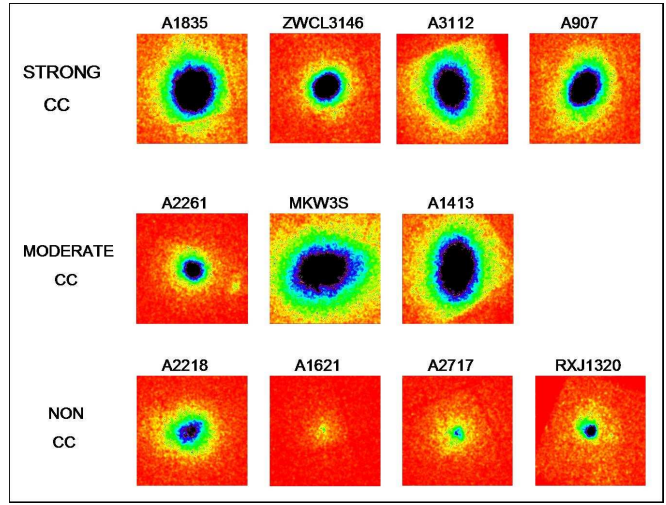

Figure 51: Cluster templates used in the simulations (original *Chandra* images).

Finally, for each input cluster the template which has the closest temperature and the closest CC parameter is selected. The selected template is cloned to the cluster luminosity and redshift according to the cloning technique (Santos et al., 2008) described below. Each cluster is then simulated through Poissonian sampling of the clone image convolved with the PSF profile, taking into account individual ECF and vignetting correction as well.

THE CLONING TECHNIQUE    The cloning technique by Santos et al. (2008) adopted in these simulations applies a pixel-by-pixel K-correction map to high resolution images of local bright groups and clusters of galaxies. We clone low redshift clusters observed with *Chandra* to the redshifts matching those in the input catalogue for each template. The procedure comprises essentially two steps: the flux decrease and the spatial rescaling which the template cluster undergoes.

The cluster flux decrease is due to both the physical luminosity distance, $D_L$, and the redshifted energy band. In the adopted cosmology $D_L$ is given by (see for instance, Carroll, Press & Turner 1992, Perlmutter et al. 1997):

$$D_L = \frac{c(1+z)}{H_0} \int_0^z h(z)^{-1} dz.$$ (5.5)

The X-ray flux is thus given by

$$F_x = \frac{L_x}{4\pi D_L^2},$$ (5.6)

where $L_X$ is the X-ray luminosity in the observed frame. Therefore, when cloning to a higher redshift, we must compute the ratio of the squared luminosity distances.

As we are using images in a given energy band, a bandpass correction, K-correction, plays an important role in the flux rescale. If $f_\nu$ is the unabsorbed spectral flux density, we can express the K-correction for an energy band $[E_1, E_2]$ as:

$$K_{corr} = \frac{f_\nu(z_1, \frac{E_1}{1+z_1}, \frac{E_2}{1+z_2}) \, \text{ECF}(\frac{E_1}{1+z_1}, \frac{E_2}{1+z_2}, z_1, T, Z, N_H)}{f_\nu(z_2, \frac{E_1}{1+z_1}, \frac{E_2}{1+z_2}) \, \text{ECF}(\frac{E_1}{1+z_1}, \frac{E_2}{1+z_2}, z_2, T, Z, N_H)}$$ (5.7)

where $z_1$ and $z_2$ are the starting and final redshifts respectively. The ECFs are computed one by one for each clone according to the appropriate energy range. The ECFs depend on the redshift, temperature, $T$, and metallicity, $Z$, of the cluster, and the Galactic hydrogen column density, $N_H$, of the simulation.

Applying single cluster temperatures, K-corrections are computed with *XSpec*. The thermal bremsstrahlung spectrum is modeled with the optically thin mekal model spectra (Kaastra et al. 1992), setting the metallicity to 0.3 solar and the temperature to the ICM temperature of the template. The Galactic absorption is modeled with the model tbabs.

The spatial rescale which a cluster undergoes when cloned to a higher redshift corresponds to the ratio of the angular distances, $D_A = D_L/(1+z)^2$. Since the premise of this method is non-evolution of the clusters, apart from the redshift resizing, no additional size scaling is applied.

| Energy band | Particle background | Galactic foreground |
|---|---|---|
| 0.5-1.0 keV | 0.05 cts s$^{-1}$ | 21.23 cts s$^{-1}$ |
| 1.0-2.0 keV | 0.11 cts s$^{-1}$ | 0.71 cts s$^{-1}$ |
| 2.0-7.0 keV | 0.53 cts s$^{-1}$ | 0.00 cts s$^{-1}$ |

Table 17: Particle background and Galactic foreground count rates per deg$^2$, corresponding to the WFXT FoV.

### 5.7.3 *Background Components*

The residual background in the WFXT images is the sum of several components. One is the instrumental background, the other contributions are from astrophysical sources. The Galactic foreground is the dominant source of background and it is truly diffuse with fluctuations of the order of a few percent. The extragalactic unresolved background depends on the flux limit where the extragalactic sources are resolved, and therefore it is a function of the field exposures. The extragalactic background is already accounted for in the simulation by the contribution of faint sources well below the detection limit of WFXT. The different components are discussed in details, and the expected count rates in different energy bands are shown in Table 17.

INSTRUMENTAL BACKGROUND   The dark earth data of *Suzaku* are used to estimate the instrumental background of WFXT, since the CCDs will probably be similar. The instrumental background is assumed to have a flat spectrum, which is equal to $\sim 3 \times 10^{-5}$ cts s$^{-1}$ arcmin$^{-2}$ keV$^{-1}$. The instrumental background counts does not suffer vignetting correction since they are not produced by photons traveling throughout the telescope optics, but they are intrinsic of the instrument.

GALACTIC FOREGROUND   We model the Galactic foreground component with a two temperatures mekal model, according to McCammon et al. (2002). The *XSpec* model used is mekal + wabs ( mekal ). The spectrum of the Galactic foreground is shown in Figure 52.

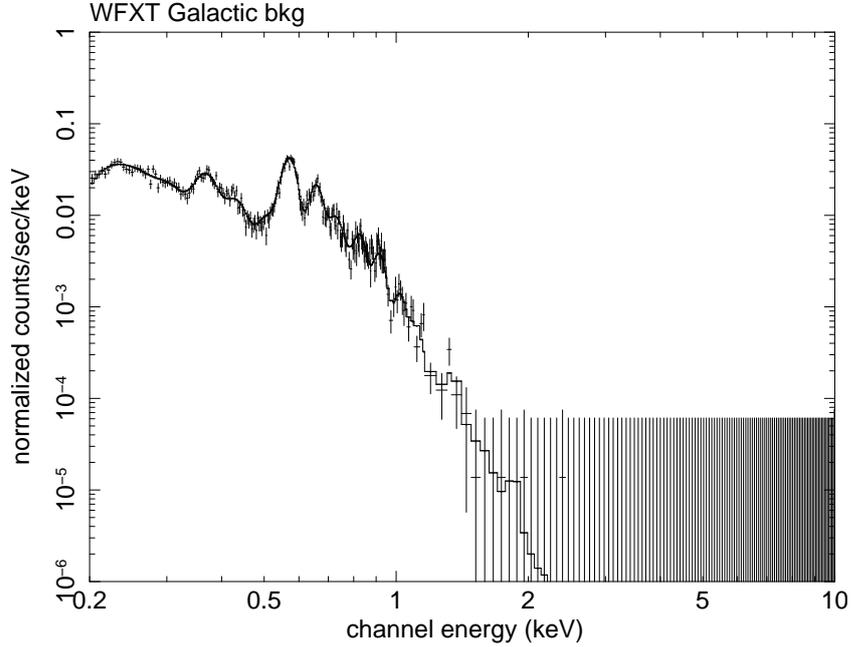

Figure 52: Galactic foreground seen by WFXT from an arcmin$^2$ for an exposure of $10^6$ sec.

## 5.8 WFXT SIMULATED IMAGES

I show some examples of simulated images of WFXT. The filed shown in Figures 53, 54 and 55 was simulated with the three planned exposure times to appreciate the depth associated with each exposure.

Moreover, to make a direct comparison between *XMM-Newton* and WFXT, we simulated a realistic COSMOS field. We used as input catalogue the COSMOS catalogue with 17614 on about 1 deg$^2$.

The XMM-COSMOS (Cappelluti et al., 2007) is a deep X-ray survey over the full 2 deg$^2$ of the COSMOS area. It consists of 55 *XMM-Newton* pointings for a total exposure of $\sim 1.5$ Ms with an average vignetting-corrected depth of 40 ks across the field of view and a sky coverage of 2.13 deg$^2$. The catalogue contains a total of 1887 unique sources detected in at least one band. The survey has a flux limit of $\sim 1.7 \times 10^{-15}$ erg cm$^{-2}$ s$^{-1}$, $\sim 9.3 \times 10^{-15}$ erg cm$^{-2}$ s$^{-1}$ and $\sim 1.3 \times 10^{-14}$ erg cm$^{-2}$ s$^{-1}$ over 90% of the area (1.92 deg$^2$) in the 0.5-2.0 keV, 2.0-10.0 keV and 5.0-10.0 keV energy band, respectively. The 13.2 ks WFXT image of the COSMOS central 1 deg$^2$, compared with that of *XMM-Newton* is shown in Figure 56.

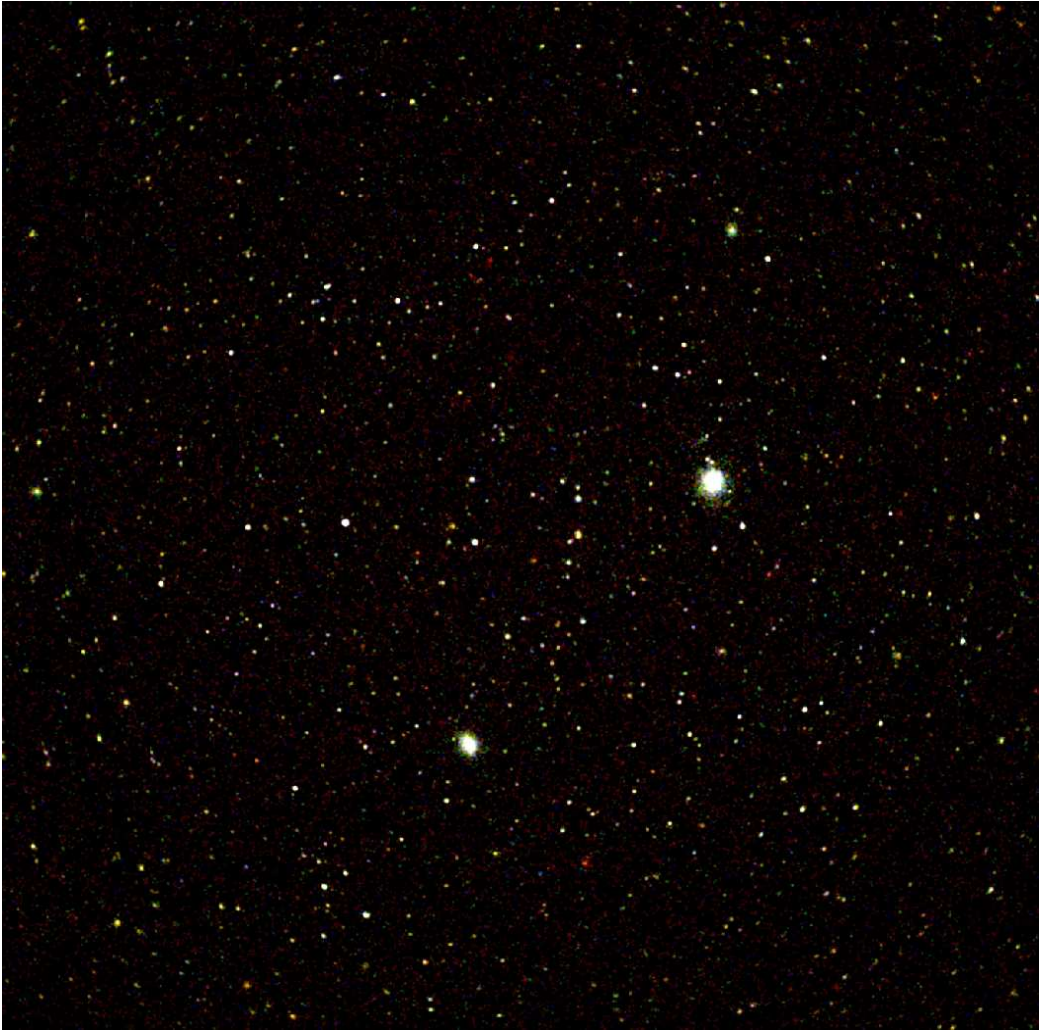

Figure 53: Color image of a 2 ks exposure with WFXT (Wide survey). This field has been choosen in order to have a few hot clusters for displaying purpose.

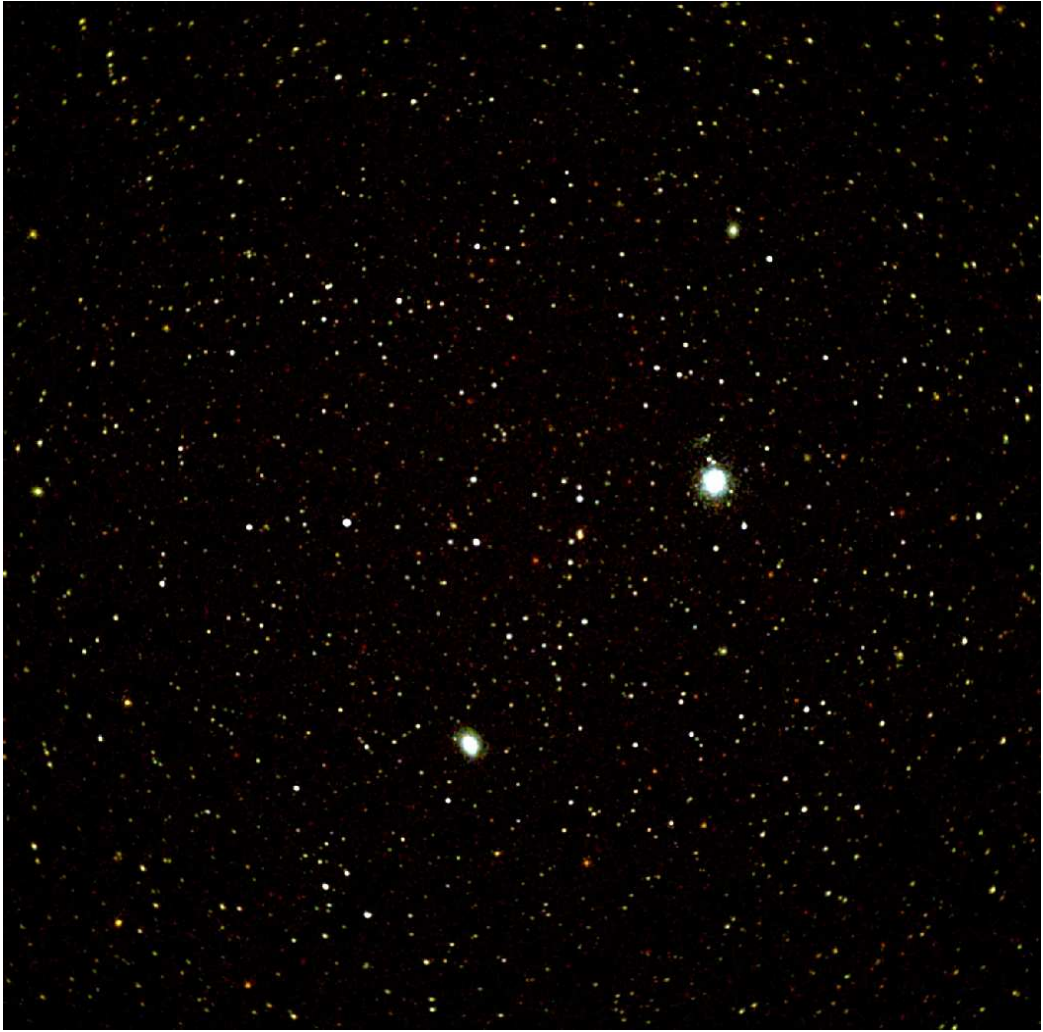

Figure 54: Color image of a 13.2 ks exposure with WFXT (Medium survey). This field has been choosen in order to have a few hot clusters for displaying purpose.

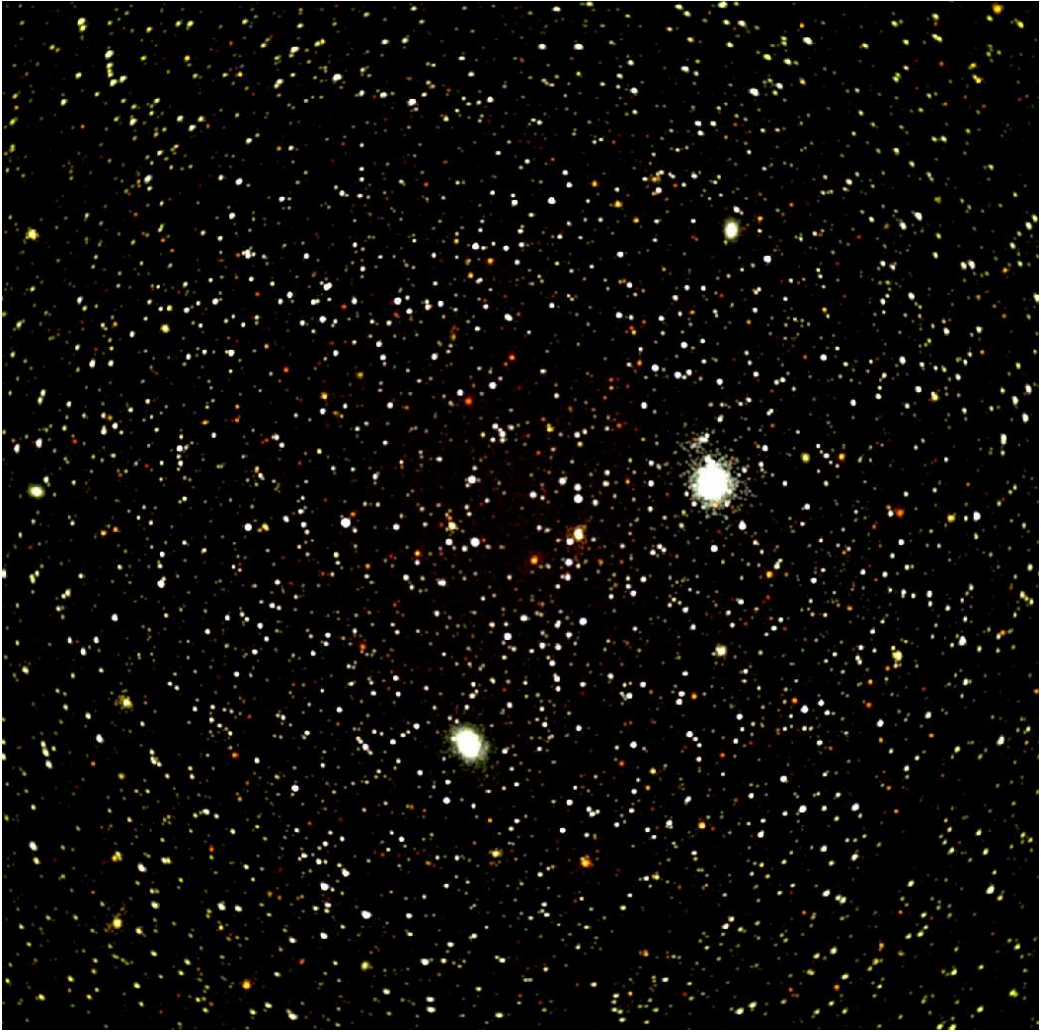

Figure 55: Color image of a 400 ks exposure with WFXT (Deep survey). This field has been choosen in order to have a few hot clusters for displaying purpose.

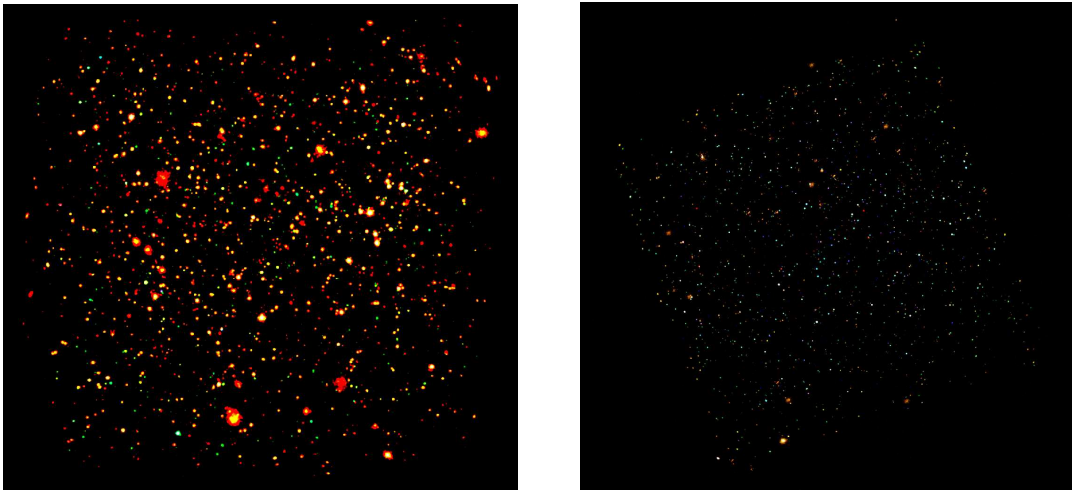

Figure 56: Color image of the COSMOS field observed by *XMM-Newton* (right panel) for a total exposure of 1.5 Ms, compared with the same field observed for 13.2 ks with wfxt (left panel).

## 5.9 PRELIMINARY ANALYSIS OF WFXT SIMULATIONS

A quick and dirty procedure to analyze the simulated images was set up based on the tool `wavdetect`. The detected source catalogue was matched with the input source catalogue, by a simple criterion based on the closest position, searching for matched sources, undetected sources and spurious sources. With these simple products a first indication of the efficiency of the wfxt surveys can be provided. At this level point sources and extended sources were investigated separately using the input information, since an effettive algorithm to separate in wfxt images extended sources from point sources in the detected catalogue does not exist yet.

The `wavdetect` photometry is good for point sources, but it shows a clear offset for extended sources (see Figure 57). The histogram of the number of undetected sources as a function of the input counts gives a good idea of the detection limit. From Figure 58 it is evident that point sources start to be missed below 20 net counts, and the skycoverage is about 10% of the total FoV for about 10 counts. For extended sources, the current detection limit is below 100 counts.

The histogram distribution of the spurious sources gives a good idea of the contamination of the detected catalogue, and it is shown in Figure 59. Clearly spurious sources may be identified as pointlike or extended. As shown in the Figure, above 20 net counts, the number of spurious sources is extremely low.

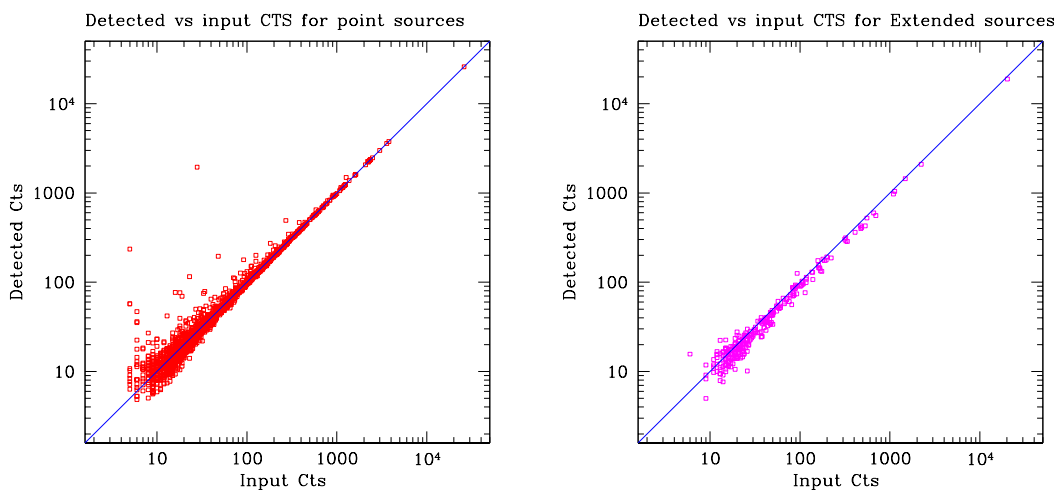

Figure 57: Detected *versus* input counts in the soft band for point sources (left panel, red squares), and for extended sources (right panel, magenta squares), from the medium exposure simulation.

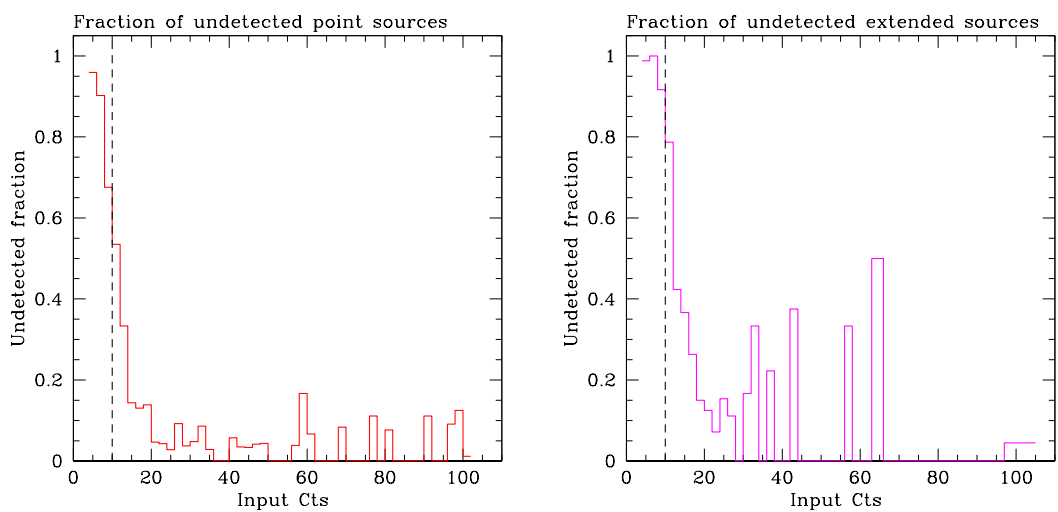

Figure 58: Histogram distribution of the input counts of the undetected sources in the soft band(point sources, left panel, and extended sources, right panel) from the medium exposure simulation.

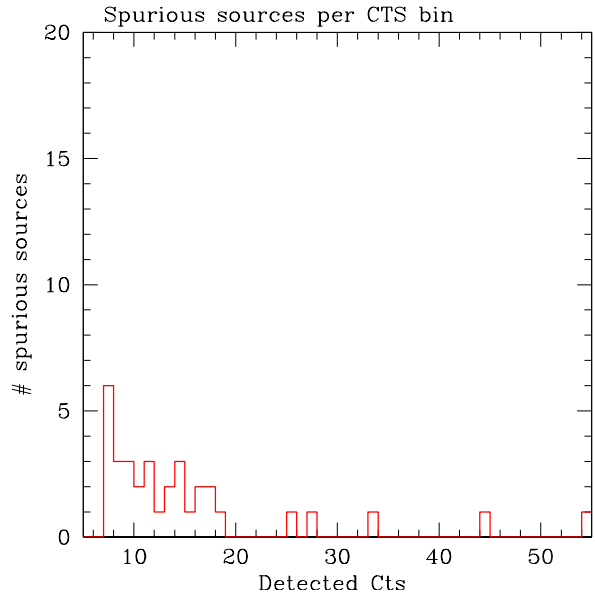

Figure 59: Histogram distribution of the input counts of the spurious sources in the soft band from the medium exposure simulation. This histogram include fake sources that are detected as point-like or extended.

Given the flat PSF, the skycoverage is quite sharp, as shown in Figure 60. The LogN-logS is well recovered for point sources, while for extended sources, at present, it shows the already noticed bias in the recovered number counts (see Figure 61). Overall, this quick and dirty analysis can be used as a rule of thumb to predict the outcome of the WFXT surveys.

## 5.10 FUTURE OF WFXT SIMULATIONS

In this preliminary phase of the mission proposal, simulations are a crucial step not only to support the scientific case of WFXT, but also to test how different technical aspects (e.g. angular resolution, effecttive aree, etc.) influence the overall mission performance.

A proper detection algorithm is needed and a suitable criterion to distinguish between point sources and extended sources must be realized. The procedure described in Chapter 2 for Swift-XRT using the TDF algorithm is impractical. The main obstacle is the TDF machine time, which is proportional to the square of detected sources. As shown in Chapter 2, TDF is very effective in separating point sources from extended sources and in recostructing the number of net counts. However, I wrote TDF for the specific case of XRT images, where on average the number of detected sources per field is between 20 and 50, and where the 15″ PSF frequently makes profiles of different sources superimposed.

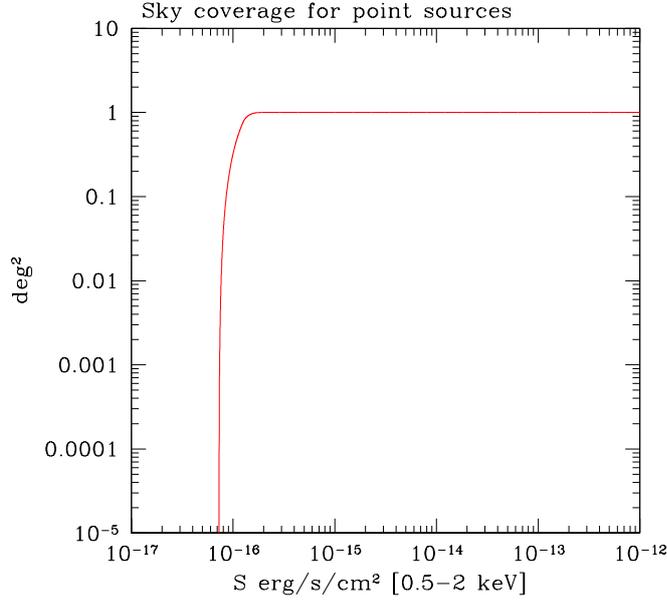

Figure 60: Simple sky-coverage in the soft band for point sources, from the medium exposure simulation.

In the case of WFXT the huge FoV, combined with a smaller PSF and a better detection capability, makes absolutely unnecessary the choice to fit together all the sources in the field, increasing enormously the machine time without any appreciable benefit. A different strategy is needed to reduce the computing time and to adapt TDF to the specific case of WFXT images. In particular this can be done in two different ways. First, improve the efficiency with which the MCMC steps are selected to reduce the number of steps necessary to reach the likelihood minimum. Second, divide the whole image into subimages to be fitted individually, significantly reducing the machine time[3]. The first method can be implemented modifying the "neighborhood function" in the simulated annealing algorithm chosing a "nieghbor state" closer to the minimum avoiding to explore randomly the whole parameter space or alternately decreasing more rapidly the "temperature" of the likelihood. However, both solutions are delicate points: for the first one must be avoided a too rapid convergence that can trap the likelihood around a relative minimum and not close the absolute minimum; for the second method instead is needed a careful treatment of the sources that fall on the edges of the subimages, whose surface bightness is distributed over more than one subimage.

---

3 If $N$ is the number of detected sources in a field and $M$ is the number of subimages $N^2 > M \times \left(\frac{N}{M}\right)^2$

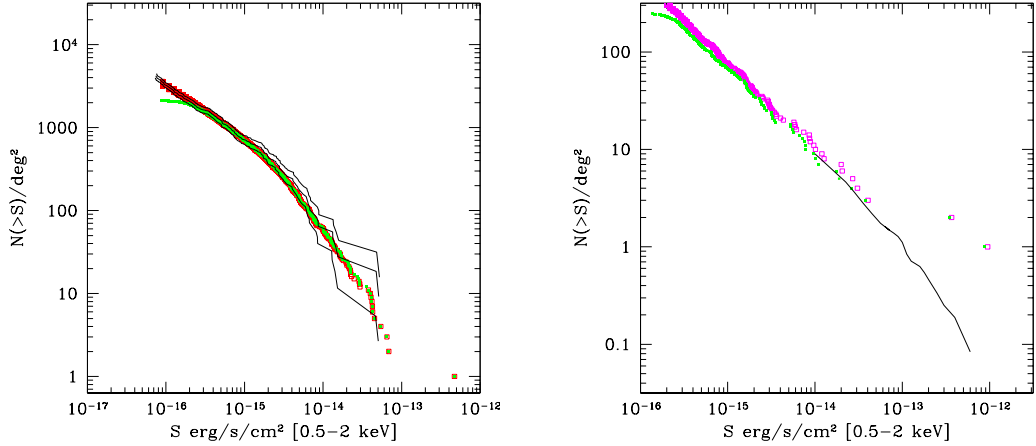

Figure 61: Soft band LogN-LogS (point sources, left panel, and extended sources, right panel) from the medium exposure simulation.

As a secondary aspect, the simulations can be improved for the point of view of source population. For instance, a minor contribution to the X-ray sky is given by late-type stars, which are not included in the simulations at present. Beside this, since WFXT is not only aimed at detecting $\sim 10^6$ clusters out to their formation redshift, but it will characterize the physical properties for a good fraction of these clusters, it is important not only focus on simulated images and source detection, but also on spectral simulations of the sources. For example, as shown in Chapter 3, the redshift measurement starting from the X-ray spectrum is a delicate issue in particular as regards the determination of the error. Therefore, also spectral simulation are essential to understand down to which number of net counts spectral properties of the ICM, first of all the redshift, can be measured satisfactorily.

Beside spectral resolution also spatial resolution is essential for the characterization of detected objects, for instance for the detection of high-z CC clusters. As shown in Section 4.5 the technique introduced by Santos et al. (2008) is effective in detecting high-z CC clusters, but it needs to measure the surface brightness within a peak radius of 40 kpc and a bulk radius of 400 kpc. To give just an example 5″, i. e. the the goal HEW of the PSF, corresponds to $\sim 40$ kpc at redshift 1.0, thus the spatial resolution cannot be neglected not only for the detection of the objects, but also for the characterization of the detected objects.

This is exactly the challenge of the WFXT simulated images: comparing different telescope designs help to understand the balance and the possible compromise to achieve all the goals of the mission.

# 6

## CONCLUSIONS

In this final Chapter I draw the conclusions about the four main topics presented in my Thesis. My PhD was devoted to the study of groups and clusters of galaxies in the X-ray band. In particular, I focused on the realization of a new X-ray selected cluster survey, the Swift-XRT Cluster Survey (SXCS); I entirely characterized on the basis of the X-ray data alone the brightest subsample of the SXCS catalogue; I studied the X-ray properties of optically selected clusters comparing them with X-ray selected clusters; finally, I applied the concepts learned during the realization of SXCS to the scientific case of a new proposed X-ray mission, the Wide Field X-ray Telescope (WFXT).

In Chapter 2 I focused on the details about the realization of the new SXCS survey. This survey is obtained from the Swift X-Ray Telescope (XRT) archive. Beyond the detailed investigation of the XRT characteristics and the analysis of the XRT data, I developed two new algorithms: the Two Dimensional Fitting (TDF) algorithm and an X-ray image simulator. The latter is a fitting algorithm that models the surface brightness profile of all the sources in an X-ray images and the background at the same time; the former create realistic X-ray simulated images including populations of all the source types contributing to the extragalactic X-ray sky. Both these algorithms are created specifically for XRT, but they are also flexible and can be used for other present and future X-ray telescopes.

The final catalogue of the SXCS consists of 243 extended sources detected with high significance. The number of sources detected as function of the flux limit is in very good agreement with previous studies, suggesting that this sample is complete and the flux measurement is accurate.

Among the sources in the SXCS catalogue, we selected the brightest ones requiring a minimum of 200 net counts to create a bright sample. The idea is to analyze and entirely characterize these sources on the basis of the X-ray data alone. This was possible given the high number of net counts. The results about the analysis of the bright sample are described in Chapter 3. Moreover, among all the sources in the SXCS catalogue we also selected a subsample of about 50 sources as possible high redshift candidates. In collaboration with P. Rosati, J. Santos and Y. Hang, we plan to do optical follow-up with Magellan and VLT for these candidates. This subsample was selected by choosing the sources whose optical counterpart is not visible in the DSS or in SDSS, in which galaxy clusters are typically identified up to redshift 0.5. We expect to found about 10-15 clusters with redshift greater than one, given the SXCS sky-coverage. The future perspective is to extend the optical follow up to the whole SXCS catalogue to measure the redshift of as many sources as possible and eventually perform cosmological tests based on the entire cluster sample.

In Chapter 3 I presented a detailed spectral analysis for the X-ray clusters of galaxies in the bright subsample of the catalogue of the Swift-XRT Cluster Survey (sxcs). For the first time, given the high signal to noise of the spectra and the well defined selection function, it was possible to provide a complete characterization of the X-ray properties of these objects from the X-ray data alone. For all the 32 clusters the measured temperature have a typical error $< 10\%$. The redshift was measured for all sources, except for two which showed no clear emission lines in their spectra. Furthermore, measuring the redshift solely by the X-ray spectrum appears to underestimate the redshift error. The evolution of the iron abundance of the ICM in agreement with other X-ray slected samples. Scaling relations are consistent with those of other X-ray selected samples of clusters of galaxies as well.

These results show that the complete characterization of cluster of galaxies (i. e. detection, spectral analysis and derivation of cluter properties) only through the X-ray data is possible at least in the case of XRT and for sources with a suitable number of net counts. Previous X-ray clusters survey needed optical follow-up to derive redshift and thus luminosity, gas mass and possibly temperatures in order to get mass proxies. Instead, this sample rely on X-ray data only. Future missions dedicated to X-ray survey, e. g. WFXT, with large collecting area will be able to assemble large sample of X-ray clusters fully characterized on the basis of the X-ray data, allowing us to precision cosmology without time expensive follow-up in other wavebands. In the next future, we plan to extend the sample to lower fluxes, and to apply the classical cosmological tests to derive contraints on cosmological parameter.

In Chapter 4 we studied the thermodynamical and chemical X-ray properties of the ICM of a galaxy cluster sample of the Red-Sequence Cluster Survey, in the redshift range $0.6 < z < 1.2$. We detected emission for the majority of the clusters, except for three, for which we have only marginal detection at $\sim 3\sigma$. In general we found that the slope of the $L_X - T_X$ relation of the RCS clusters is in agreement with that of X-ray selected clusters, while the normalization is a factor of 2 lower at high confidence level, in agreement with the result of Hicks et al. (2008) in their independent analysis. Only the three marginally detected RCS clusters seem to have lower luminosities with respect to the RCS $L_X - T_X$, but only if they have $kT > 3$ keV. Unfortunately, for this sample the statistic is too poor to draw any conclusion about possible evolution of the $L_X - T_X$ relation with redshift. Particularly our data are fitted with $L_X/T_X^\alpha \propto (1 + z)^{0.2 \pm 0.2}$, consistent with no evolution.

Concerning the iron abundance in RCS clusters we found that also for this sample of optically selected clusters, the ICM was already enriched with metals at a level comparable with X-ray selected clusters at high redshift (Balestra et al., 2007).

Thanks to the high spatial resolution of the *Chandra* satellite, we also investigate the point source distribution near the region of diffuse cluster emission for RCS with respect to the field. The number counts of point sources as a function of the flux show a significant excess at all fluxes with respect to the data of the CDFS. We quantified this excess between 15% and 40% with a significance of $2\sigma$. This result is in agreement with that found by other authors (Cappi et al., 2001; Branchesi et al., 2007a). The spatial distribution of point sources in the field of RCS clusters shows a factor of $\sim 6$ over-density at $3\sigma$ confidence level in the inner 150 kpc. The contribution of these point sources to the X-ray emission is limited to few percent in the soft band, showing that the contamination from AGN in the soft band is not severe at least for cluster with $L_X > 10^{44}$ erg s$^{-1}$ up to $z \sim 1$.

Studying the X-ray surface brightness properties of the RCS clusters and evaluating the presence of Cool Core in our sample, we find that high-z optically selected clusters seem to have a lower fraction of CC with respect to high-z clusters detected in the X-ray.

We have performed a strong lensing analysis of the cluster RCS0224-0002. The mass derived from our lensing model is in very good agreement with the one obtained form the X-ray temperature, showing that X-ray data are effective in probing the mass of high-z clusters.

In Chapter 5 I expounded the scientific case of WFXT and I described the main characteristics of the sophisticated WFXT image simulator, to the realization of which I substantially contributed. In this preliminary phase of the mission proposal, simulations are a crucial step not only to support the scientific case of WFXT, but also to test how different technical aspects (e. g. angular resolution, effecttive aree, etc.) influence the overall mission performance.

In this preliminary phase of the mission, the simulations can be improved for the point of view of source population. For instance, a minor contribution to the X-ray sky is given by late-type stars, which are not included in the simulations at present. Beside this, since WFXT is not only aimed at detecting $\sim 10^6$ clusters out to their formation redshift, but it will characterize the physical properties for a good fraction of these clusters, it is important not only focus on simulated images and source detection, but also on spectral simulations of the sources. For example, as shown in Chapter 3, the redshift measurement starting from the X-ray spectrum is a delicate issue in particular as regards the determination of the error. Therefore, also spectral simulation are essential to understand down to which number of net counts spectral properties of the ICM, first of all the redshift, can be measured satisfactorily. In this sense, the ongoing detailed analysis of simulated images will help to understand the balance and the possible compromise to achieve all the goals of the mission, comparing different telescope designs.

Summarizing, X-ray studies of the large scale structure of the Universe are able to constrain both the growth of cosmic structure (and hence cosmology) and the physics of the ICM. The SXCS presented in my work is a new, deep X-ray survey, which is expected to double the number of known X-ray clusters at $z > 1$. The future X-ray survey realized with an *ad hoc* satellite, WFXT, with large Field of View and sharp PSF will address many outstanding cosmological and astrophysical objectives, such as the formation and evolution of clusters of galaxies and associated implications on cosmology and fundamental physics, properties and evolution of AGNs, and cosmic star formation history Giacconi et al. (see 2009). Overall, this Thesis provides a significant contribution to the study of the large scale structure of the Universe and the physics of the ICM, exploiting present and future X-ray facilities.

# A

## THEORETICAL AND TECHNICAL STUFF

In this Appendix I show some theoretical and technical issue adopted in this Thesis. In particular, I will describe the main theoretical characteristics of the Two Dimensional Fitting (TDF) algorithm, how it was derived the $7\sigma$ limit used to separate between point sources and extended sources, how specific XRT ARFs are built to perform the spectral analysis, and details about XRT ECFs.

### A.1 THEORY BEHIND THE TWE DIMENSIONAL FITTING ALGORITHM

The Two Dimensional Fitting (TDF) algorithm is designed to fit at once the surface brightness of all the sources in an X-ray image and the background. It exploits "simulated annealing" technique, based on MCMC and maximum likelihood criterion.

Simulated annealing is a generic probabilistic method for locating a good approximation to the global minimum of a given function in a large search space. The name and inspiration come from annealing used in metallurgy, a technique involving heating and controlled cooling of a material to increase the size of its crystals and reduce their defects.

By analogy with this physical process, each step of the algorithm replaces the current solution by a random "nearby" solution, chosen with a probability that depends on the difference between the corresponding function values and on a global parameter $\tau$ (called the temperature), that is gradually decreased during the process. The slow cooling gives them more chances of finding configurations with lower internal energy than the initial one, so closer to the global minimum.

In TDF the "energy" function to cool is the minus logarithm of the likelihood, which is the probability to get the data from the model. To compute the likelihood the exposure map corrected sum of all the surface brightness profiles of all the sources plus the background is assumed as the model, and the image itself is assumed as the data. Since X-ray images are counts divided into pixels, the likelihood is computed from the Poisson distribution.

In the implementation of TDF the "neighborhood function" to move from one step to the next is chosen simply by randomly changing the value of one parameter between its minumim and maximum allowed values. The sequence of all the steps to reach the minimum is therefore a MCMC.

The temperature is decreased with the square root of the number of steps. In the case of XRT, this is a good compromise between machine time and efficiency in the reconstruction of the global minimum.

In the following Sections I briefly show some theoretical aspects regarding the TDF algorithm.

### A.1.1 *Bayes' Theorem*

In probability theory, Bayes' theorem relates the conditional and marginal probabilities of two random events. It is often used to compute posterior probabilities given observations and prior probability.

Consider two events $A$ and $B$. The Bayes' theorem gives the conditional probability of $A$ given $B$, and reads

$$P(A|B) = \frac{P(B|A)P(A)}{P(B)},$$ (A.1)

where

- $P(A|B)$ is the conditional probability of $A$ given $B$. It is also called the posterior probability because it is derived from or depends upon the specified value of $B$;

- $P(B|A)$ is the conditional probability of $B$ given $A$;

- $P(A)$ is the prior probability or marginal probability of $A$. It is "prior" in the sense that it does not take into account any information about $B$;

- $P(B)$ is the prior or marginal probability of $B$, and acts as a normalizing constant.

These formula becomes more interesting when we consider a set of $N$ events (or hypotheses) $\{H_k, 0 \leq k \leq N-1\}$. For each hypothesis you can apply the Bayes' theorem

$$P(H_k|B) = \frac{P(B|H_k)P(H_k)}{P(B)}.$$ (A.2)

If the hypotheses $H_k$ are collectively exhaustive and mutually exclusive, which means that they cover all the event space and that they are incompatible, simply we can write

$$\sum_{k=0}^{N-1} P(H_k) = 1,$$ (A.3)

that can be used to write the marginal probability of $B$ as

$$P(B) = \sum_{k=0}^{N-1} P(B|H_k)P(H_k).$$ (A.4)

Then

$$P(H_k|B) = \frac{P(B|H_k)P(H_k)}{\sum_{k=0}^{N-1} P(B|H_k)P(H_k)} \tag{A.5}$$

which is another formulation of the Bayes' theorem.

In data analysis, usually, the hypothesis is a model, which depends on a parameter vector $\boldsymbol{\theta}$, and the event $B$ are the data, which can be arranged into a vector $\boldsymbol{d}$. In this case the Bayes' theorem reads

$$P(\boldsymbol{\theta}|\boldsymbol{d}, H) = \frac{P(\boldsymbol{d}|\boldsymbol{\theta}, H)P(\boldsymbol{\theta}|H)}{P(\boldsymbol{d}|H)}, \tag{A.6}$$

where

- $P(\boldsymbol{\theta}|H)$ is the prior probability function, which should encode any prior knowledge we have of the parameters $\boldsymbol{\theta}$ of the model $H$;

- $P(\boldsymbol{\theta}|\boldsymbol{d}, H)$ is the posterior probability function of the parameters $\boldsymbol{\theta}$, which will contain our new or updated knowledge of the parameters $\boldsymbol{\theta}$ after the analysis of the data set $\boldsymbol{d}$;

- $P(\boldsymbol{d}|H)$ is the evidence, or the *a priori* probability of witnessing the new evidence $\boldsymbol{d}$ under all possible choice of the parameters $\boldsymbol{\theta}$;

- $P(\boldsymbol{d}|\boldsymbol{\theta}, H)$ is the likelihood, which is the probability to obtain the observed data set for a specific choice of the parameters $\boldsymbol{\theta}$ of the model $H$.

In a more general formulation we can also take into account different models $H_k$ with different parameters vectors $\boldsymbol{\theta}_k$, but this case is beyond our purpose.

In Equation A.6, the evidence is constant and by definition acts as normalization factor; so the posterior probability depends on the product between the prior and the likelihood. The prior itself is a constant, if we do not have any prior information about the probability function of the parameters, in other words we do not have any *a priori* preference in the choice of the best parameters value (like in a fit). In this last case computing the maximum of the likelihood in the parameter space is the same as computing the best choice of the parameters $\boldsymbol{\theta}$ (i.e. the one with the higher probability). Methods based on this theoretical principles used for fitting a mathematical model to data are called Maximum Likelihood Estimation.

### A.1.2 *Likelihood*

The likelihood is the probability that a real world experiment would generate a specific datum, as a function of the parameters $\boldsymbol{\theta}$ of a mathematical model $H$.

Since X-ray images are counts divided into pixels, the Poissonian distribution is applied. Counts are sampled from the Poisson distribution, and so the best way to assess the quality of model fits is to use the product of individual Poisson probabilities computed in each bin $i$, or the likelihood

$$\mathcal{L} = \prod \frac{M_i^{D_i}}{D_i!} \exp(-M_i) \tag{A.7}$$

where $M_i = S_i + B_i$ is the sum of the source and background model amplitude, and $D_i$ is the number of observed counts in bin $i$.

### A.1.3 *Simulated Annealing*

Consider a nonnegative "energy" function $H(\mathbf{x})$ (where $\mathbf{x} \in \mathcal{H} \subseteq \mathbb{R}^m$) (where $H(\mathbf{x}) > 0$ and $H(\mathbf{x}) < \infty$). Suppose that the objective is to determine $H_* = \min_{\mathbf{x} \in \mathcal{H}} H(\mathbf{x})$, but $\mathcal{H}$ is so large as to make total enumeration impossible. Many stochastic techniques for estimating $H_*$ exist, and among these *simulated annealing* relies on the concepts of acceptance-rejection sampling. Observe that for any two points $\mathbf{x}$ and $\mathbf{y} \in \mathcal{H}$, the ratio between the probability mass function of $\mathbf{x}$ and $\mathbf{y}$

$$\frac{\pi(\mathbf{y})}{\pi(\mathbf{x})} = e^{[H(\mathbf{x}) - H(\mathbf{y})]/\alpha\tau} \tag{A.8}$$

increases as $\tau$ decreases if $H(\mathbf{x}) > H(\mathbf{y})$ and decreases as $\tau$ decreases if $H(\mathbf{x}) < H(\mathbf{y})$. As a consequence, for a fixed $\mathbf{x}$, decreasing $\tau$ amplifies peaks and attenuates troughs in $\pi(\mathbf{y})/\pi(\mathbf{x})$. Simulated annealing exploits this property for estimating $H_*$. In particular, it relies on the Metropolis-Hastings (Metropolis et al., 1953; Hastings, 1970) method to search for the minimum, employing:

- a countably finite–state space $\mathcal{H} = \{\mathbf{x}_i, 0 \leq i \leq v - 1\}$;

- a nominating matrix $\mathbf{R} = \|r_{ij}\|_{i,j=0}^{v-1}$;

- an acceptance function $\alpha_{ij}(\tau) = \min(1, e^{-[H(\mathbf{x}_j) - H(\mathbf{x}_i)]/\tau})$, with $0 \leq i, j \leq v - 1$;

- a nonincreasing *temperature cooling schedule* $\{\tau_l \geq \tau_{l+1}, l \geq 1\}$, with $\lim_{k \to \infty} \tau_k = 0$.

The nominating matrix denotes the transition matrix between two states in the space $\mathcal{H}$. Basically on the step $l \geq 1$ it generates a proposed state transition from the state $\mathbf{x}_i$ to the state $\mathbf{x}_j$ sampled form $r_{ij}$, which is the probability mass function to generate $\mathbf{x}_j$ form $\mathbf{x}_i$ in $\mathcal{H}$

Associated with the nominating matrix is a *neighborhood function* $\{N(\mathbf{x}_l), 0 \leq l \leq v - 1\}$ such that for all $i, j \in \{0, 1, \dots, v - 1\}$

$$r_{ij} > 0 \qquad \text{if } \mathbf{x}_j \in N(\mathbf{x}_i) \tag{A.9}$$
$$= 0 \qquad \text{otherwise,} \tag{A.10}$$

which means that the probability to randomly generate $\mathbf{x}_j$ from $\mathbf{x}_i$ is nonzero only for the *neighbor* state of $\mathbf{x}_i$ in $\mathscr{H}$.

The acceptance function determines whether the chain moves to the state $\mathbf{x}_j$, basically whether the $j$ state is accepted, or remains in the state $\mathbf{x}_i$. The nominating matrix combined with the acceptance function gives the porbability to go from one state to another in the random walk.

Summaryzing this is the algorithm scheme.

---

**Simulated Annealing Algorithm Scheme**

- **Purpose:** To estimate $H_* = \min_{\mathbf{x} \in \mathscr{H}}$ where $\mathscr{H} = \{\mathbf{x}, 0 \leq i \leq v - 1\}$.

- **Input:** Initial state $i_0 \in \{0, 1, \ldots, v - 1\}$, nominating matrix $\mathbf{R}$, in-line procedure for evaluating temperature $\tau_l$, $l \geq 1$, and number of steps $k$.

- **Output:** $\hat{H}_{*k}$ as an estimate of $H_*$ and $L$ denoting the point $\mathbf{x}_L \in \mathscr{H}$ at which $\hat{H}_{*k}$ occurs.

- **Method:**
    - $l \leftarrow 1$, $I \leftarrow i_0$, $L \leftarrow I$, $A \leftarrow H(\mathbf{x}_I)$ and $C \leftarrow A$
    - While $l \leq k$
        * Compute $\tau_l$
        * Randomly generate $J$ form the p.m.f. $\{r_{ij}, 0 \leq j \leq v - 1\}$
        * $B \leftarrow H(\mathbf{x}_J)$
        * If $B \leq A$, $A \leftarrow B$, $C \leftarrow B$ and $L \leftarrow J$
        * Else:
            · Randomly generate $U$ in $[0, 1]$
            · If $U \leq \exp[-(B - A)/\tau_l]$, $A \leftarrow B$
        * $l \leftarrow l + 1$
    - $\hat{H}_{*k} \leftarrow C$
    - Return $\hat{H}_{*k}$ and $L$

---

## A.2 ANALYTICAL FORMULA OF THE $7\sigma$ LIMIT CRITERION

From the definition of $r_{B2}$ in Equation 2.11 I create this function:

$$f(x, A, B, C) = A + C \left[ \left( \frac{(x - B)(\beta - 1)}{\pi} \right)^{1/\beta} - 1 \right]^{1/2}, \tag{A.11}$$

where $x = N/(r_c^2 B)$. By definition

$$r_{B2}(x) = f(x, 0, 0, r_c). \tag{A.12}$$

From $f(x, A, B, C)$ I defined $h(x, A, B, C, D)$:

$$h(x, A, B, C, D) = \begin{cases} f(x, A, B, C) & \text{if } f \in R \\ f(x, 0, 0, D) & \text{otherwise.} \end{cases} \tag{A.13}$$

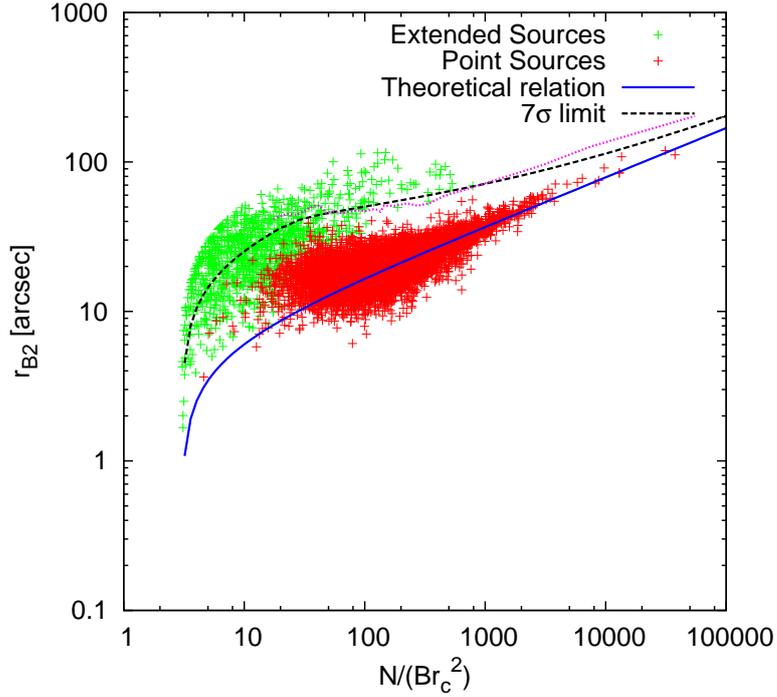

Figure 62: Relation between $r_{B2}$ and the best fit parameters $N/(Br_c^2)$. $N$ is the normalization of the surface brightness profile, $B$ the background counts per square arcsec, and $r_c$ is the core radius. Red points are point sources, green point extended sources. The blue solid line is the "theoretical" relation expected for point sources, the black dashed line is the $7\sigma$ limit of the $r_{B2}$ distribution of point sources, the magenta dotted line is the average plus $7\sigma$ of the $r_{B2}$ distribution for point sources.

Figure 62 is the same of Figure 16, where I also added the magenta dotted lines that represents the average plus $7\sigma$ of the $r_{B2}$ distribution for point sources.

The blue solid line and the black dashed line are obtained with $\beta = 1.5$ and $r_c = 5.5''$. The blue solid line is the "theoretical" relation of $r_{B2}$ for point sources, so $f(x, 0, 0, r_c)$. The black dashed lines mimics the $h(x, A, B, C, D)$, where the values of $A$, $B$, $C$ and $D$ are chosen empirically to reproduce the magenta dotted line at $7\sigma$. One of the possible choice of the parameters is:

- A=35;

- B=19;

- C=5.5;

- D=23.

These functions do not have any particular meaning, they are only useful to define an analytical threshold to distinguish point sources from extended sources.

## A.3 HOMEMADE ROUTINE TO MAKE XRT ARF

The XRT calibration database provides only on-axis ARF, vignetting correction as function of the off-axis angle and of the energy, routines to build exposure maps at fixed energy or non-vignetted exposure maps, but not a routine to build specific ARF at different off-axis angle. I developed this "homemade" routine to compensate for this lack, in particular I demonstrate, starting only from the on-axis ARF, one exposure map at a fixed energy and the non-vignetted exposure map, how it is possible: first to build specific ARF at different off-axis angle in the case of sigle pointing images; second to build specific ARF also in case of multiple pointing images; third to generalize the routine to build ARF for extended sources.

In the following Equations, $x$ and $y$ are the coordinate on the image in pixel, $\theta(x, y)$ is the off-axis angle at the position $(x, y)$, $\nu$ is the energy.

In single pointings XRT images, the vignetting can be modeled as a function of distance from the center of the chip as:

$$V(\nu, \theta(x, y)) = 1 - c(\nu) \times \theta^2(x, y), \tag{A.14}$$

where $\theta$ is the offaxis angle in arcmin and $c$ is a function of the energy $\nu$ of the events of the form:

$$c(E) = K_0 \times K_1^\nu + K_2, \tag{A.15}$$

where $K_0 = 0.000124799$, $K_1 = 1.55219$, $K_2 = 0.00185628$, and the energy is in [keV].

The function $V(\nu, \theta(x, y))$ relates the exposure map, $E(x, y)$, and the non-vignetted exposure map, $E_{novig}(x, y)$,

$$E(\nu, x, y) = V(\nu, \theta(x, y)) \times E_{novig}(x, y). \tag{A.16}$$

The non-vignetted exposure map is basically a flat exposure map equal to the observation exposure time, obtained after removing only bad pixels and bad columns.

The ARF at the off-axis angle $\theta$, $A(\nu, \theta)$, is related to the on-axis ARF, $A(\nu, 0) = A_0(\nu)$, via

$$A(\nu, \theta) = V(\nu, \theta(x, y)) \times A_0(\nu). \tag{A.17}$$

In case of single pointing observation, simply we have

$$\theta(x, y) = \sqrt{(x - x_0)^2 - (y - y_0)^2}, \tag{A.18}$$

where $(x_0, y_0)$ is the aim point coordinate. Then computing the specific ARF at the coordinate $(x, y)$ with Equations A.14, A.17 and A.18 is straightforward.

In case of multiple pointings, the off-axis angle at a pixel coordinate cannot be computed with the simple Equation A.18, but Equations A.16 and A.17 still hold. You can understand this situation with a simple example. Suppose that your image is the sum of two pointings and that at the coordinate $(x, y)$ are associated the off-axis angles $\theta_1(x, y)$ and $\theta_2(x, y)$ in the first and second pointing respectively. Equation A.16 can be applied separately to the two pointings,

$$E_1(\nu, x, y) = V(\nu, \theta_1(x, y)) \times E_{1,novig}(x, y), \tag{A.19}$$

$$E_2(\nu, x, y) = V(\nu, \theta_2(x, y)) \times E_{2,novig}(x, y), \tag{A.20}$$

and the total exposure map and non-vignetted exposure map are given by the sum of the single ones,

$$E(\nu, x, y) = E_1(\nu, x, y) + E_2(\nu, x, y), \tag{A.21}$$

$$E_{novig}(x, y) = E_{1,novig}(x, y) + E_{2,novig}(x, y). \tag{A.22}$$

Now, remembering the definition of $V(\nu, \theta_1(x, y))$ in equation Equation A.14, you have

$$E(\nu, x, y) = \left(1 - c(\nu) \frac{\theta_1^2(x, y)\, E_{1,novig}(x, y) + \theta_2^2(x, y)\, E_{2,novig}(x, y)}{E_{novig}(x, y)}\right) E_{novig}(x, y). \tag{A.23}$$

Equation A.23 is the same as Equation A.16 where the off-axis angle is replaced by an "effective" off-axis angle given by

$$\tilde{\theta}(x, y) = \sqrt{\frac{\sum_{i=1}^{N} \theta_i^2(x, y)\, E_{i,novig}(x, y)}{E_{novig}(x, y)}}, \tag{A.24}$$

where we have generalized to the case of an image obtained by the sum of $N$ pointings. Computing the effective off-axis angle with Equation A.24 is an hard task, since you need to know the off-axis angle and the non-vignetted exposure map of each pointing; whereas it is easy if you invert the Equation A.23 and if you have one exposure map at the fixed energy, $\nu_0$, and the non-vignetted exposure map:

$$\tilde{\theta}(x,y) = \sqrt{\frac{1}{c(\nu_0)} \left(1 - \frac{E(\nu_0, x, y)}{E_{novig}(x, y)}\right)}. \qquad (A.25)$$

Now, with Equations A.14, A.17 and A.25 you can build specific ARF at the position $(x, y)$ also in case of multiple pointing images,

$$A(\nu, x, y) = \left[1 - \frac{c(\nu)}{c(\nu_0)} \left(1 - \frac{E(\nu_0, x, y)}{E_{novig}(x, y)}\right)\right] A_0(\nu). \qquad (A.26)$$

Finally, I want to generalize the routine to build ARF for extended sources. As a matter of fact, Equation A.26 can be used to build ARF only for sources that have all their flux in the single pixel at coordinate $(x, y)$. But extended sources (indeed also point sources) have their flux divided in more than one pixel, so we need to take into account also this effect. This can be done simply by replacing in Equation A.26 $E(\nu, x, y)$ and $E_{novig}(x, y)$ with their values weighted on the normalized profile of the source. This can be easily demonstrate following the same steps that led to derive Equation A.26.

Let define

$$\bar{E}(\nu, x, y) = \int S(x', y') E(\nu, x', y') \, dx' \, dy', \qquad (A.27)$$

$$\bar{E}_{novig}(x, y) = \int S(x', y') E_{novig}(x', y') \, dx' \, dy' \qquad (A.28)$$

as the exposure maps at the coordinate $(x, y)$ weighted for the normalized source profile $S(x, y)$. Multiplying both members of Equation A.23 by $S(x, y)$ and integrating in $dx \, dy$ we obtain

$$\bar{E}(\nu, x, y) = (1 - c(\nu)\, \tilde{\theta}^2)\, \bar{E}_{novig}(x, y), \qquad (A.29)$$

where the effective off-axis angle is redefined as

$$\tilde{\theta}(x, y) = \sqrt{\frac{\sum_{i=1}^{N} \int S(x', y')\, \theta_i^2(x', y')\, E_{i, novig}(x', y')\, dx'\, dy'}{\bar{E}_{novig}(x, y)}}. \qquad (A.30)$$

Therefore Equation A.25 becomes

$$\tilde{\theta}(x, y) = \sqrt{\frac{1}{c(\nu_0)} \left(1 - \frac{\bar{E}(\nu_0, x, y)}{\bar{E}_{novig}(x, y)}\right)}, \qquad (A.31)$$

and finally Equation A.26 becomes

$$A(\nu, x, y) = \left[ 1 - \frac{c(\nu)}{c(\nu_0)} \left( 1 - \frac{\bar{E}(\nu_0, x, y)}{\bar{E}_{novig}(x, y)} \right) \right] A_0(\nu). \tag{A.32}$$

In Chapters 2 and 3, as normalized source profiles I used the normalized best fit profiles obtained by TDF in the soft (0.5-2.0 keV) images. Indeed, this is not completely correct since the source profile itself is function of the energy. But fitting the source profile in 3D space $(\nu, x, y)$ is overly complicated, needing a specific computational effort. At this level, I am not able to quantify the error induced by this approximation, but I noticed in Section 2.9 that the best fit $\beta$ and $r_c$ obtained by TDF in soft and hard images are not so different, suggesting that the correction should not be so severe.

At the end I want to point out which exposure time should be used in *XSpec* to obtain the correct values of the source fluxes during spectral analysis with ARFs calculated with the Equations A.26 and A.32 in case of image obtained by the sum of multiple pointings equations.

In case of one single pointing the situation is quite simple: the exposure time of the image is the one of the pointing and Equation A.17 takes into account for the vignetting correction at the source position $(x, y)$ appropriately scaling the effective area. In case of multiple pointing the nominal exposure time of the image is given by the sum of the exposure time of the pointings, but this is not the "real" or "effective" exposure time for all the pixels in the image, since not all the pixels belong to all the pointings. In particular, $\max(E_{novig}(x, y))$ is the nominal exposure time of the image[1] and $E_{novig}(x, y)$ is the effective exposure time at the position $(x, y)$ that must be used in *XSpec* to measure the correct values of the source fluxes. $\bar{E}(\nu, x, y)$ must be use instead, if we want to take into account also the source profile.

## A.4 DERIVING XRT ENERGY CONVERSION FACTORS

By definition, the Energy Conversion Factor (ECF) is the ratio between the unabsorbed flux in an energy range and the expected count rate in the same energy range, and it is needed to convert the measured count rate into flux. The unabsorbed flux is the flux from a source spectrum as if there were no absorption due to the Galaxy ISM. So the ECF depends on the source spectrum and on the Galactic $N_H$. Extragalactic X-ray sources can be divided into point sources (mostly AGNs and star forming galaxies) and extended sources (diffuse emission from ICM of clusters). I used X-ray thermal emission spectra to compute ECFs for extended source and power law spectra for point soruce. I computed the ECFs in the soft and hard bands, using on-axis ARF and RMF. The plots in Figure 8 show the dependence of average ECFs on the Galactic $N_H$, while the plots in Figures 63 and 64 show the dependence on the different spectral parameters.

---

[1] This is true only in the assumption that at least one pixel belongs to all the pointings.

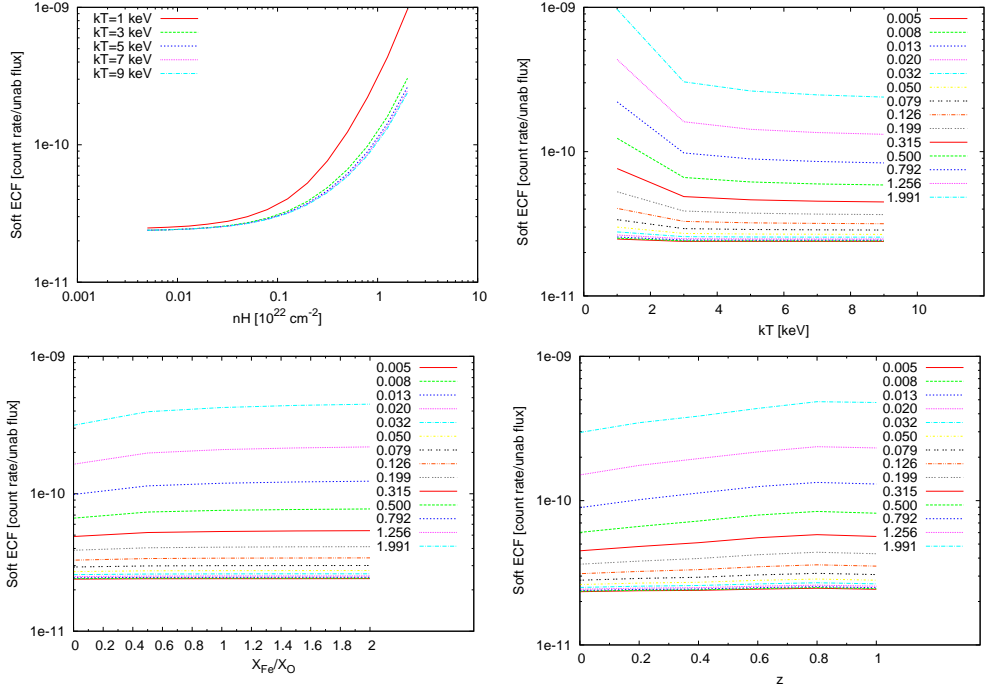

Figure 63: Soft XRT ECFs for extended sources. Top left, average soft ECF as function of the $N_H$ at different temperatures; top right, average soft ECF as function of the temperature; bottom left, average soft ECF as function of the abundance; bottom right, average soft ECF as function of the redshift.

THERMAL BREMSSTRAHLUNG ECF   The spectrum of a X-ray thermal bremsstrahlung emission depends on the redshift of the source and on the temperature and the metallicity of the ICM. I computed ECFs for different values of the parameters.

- $N_H \in [0.005 \times 10^{22} : 2.0 \times 10^{22}]$ cm$^{-2}$

- $kT \in [1.0 : 10.0]$ keV

- $X_{Fe}/X_{\odot} \in [0.0 : 2.0]$

- $z \in [0.0 : 1.0]$

The soft ECF (see Figure 63) has a strong dependence on the galactic $N_H$. For different values of $N_H$ the dependence on redshift and metallicity is negligible with respect to the $N_H$ dependence. For temperature below 2.0 keV the soft ECF increases up to a factor $2-3$ for high values of the $N_H$, otherwise the dependence of the soft ECF on temperature is mild.

The hard ECF (see Figure 64) depends much less on $N_H$, and it increases only for high $N_H$ values. The hard ECF increase slightly with increasing temperature and drecreasing redshift. It is nearly independend on metallicity.

POWER LAW ECF    To describe AGN spectra a single power law with $\gamma = 1.4$ is used. Since there are no free parameters, the ECF depends only on the Galactic $N_H$. As before I computed the ECF for $N_H \in [0.005 \times 10^{22} : 2.0 \times 10^{22}]$ cm$^{-2}$. In the soft band (see Figure 8, left panel) the ECF of AGN has a strong dependence on Galactic $N_H$, and it is very similar to the average ECF of thermal bremsstrahlung sources, especially for $N_H < 0.5 \times 10^{22}$ cm$^{-2}$. In the hard band (see Figure 8, right panel) the dependence on $N_H$ is less severe, and the ECF of AGN is on average 10% larger than that of thermal bremsstrahlung sources.

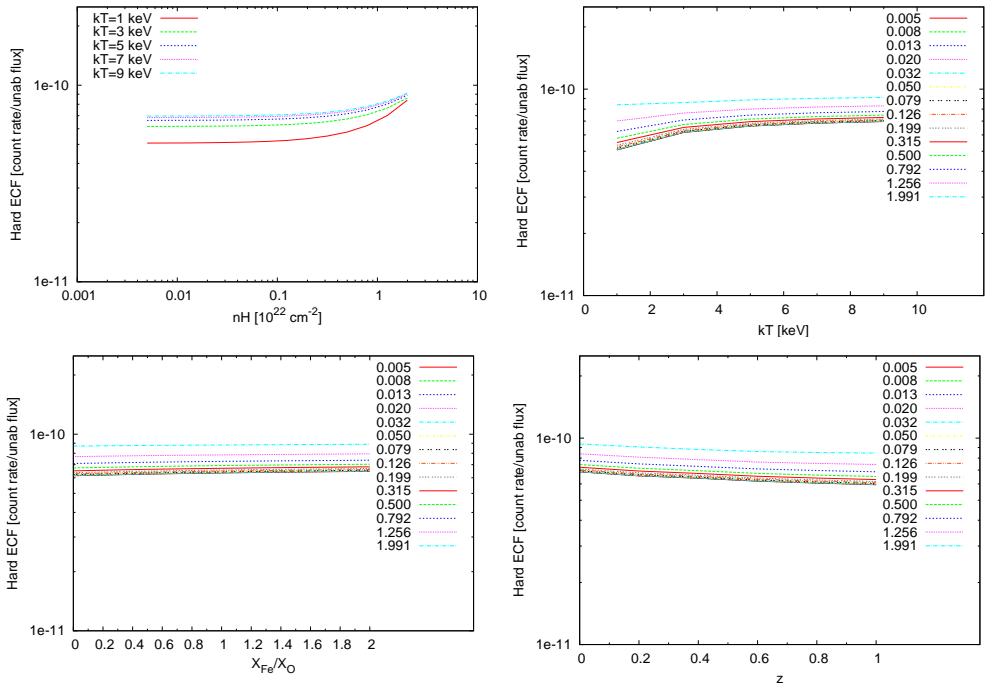

Figure 64: Hard XRT ECFs for extended sources. Top left, average hard ECF as function of the $N_H$ at different temperatures; top right, average hard ECF as function of the temperature; bottom left, average hard ECF as function of the abundance; bottom right, average hard ECF as function of the redshift.

# B

## NOTES ON THE CLUSTERS OF THE SXCS BRIGHT SAMPLE

In this Appendix I present the results on the detailed spectral analysis of the sources in the SXCS bright sample. In particular, I show for each cluster:

- the unfolded X-ray spectrum;

- X-ray color image[1] with regions of detected sources;

- optical DSS image (or SDSS when available) with X-ray countours overlapped;

- redshift histogram of SDSS galaxies close to the X-ray emissione (when SDSS data are available);

- tables with the most relevant cluster propertis (temperature, redshift, gas mass, total mass, luminosity, etc.;

- a brief note about the cluster;

- several marginalized C-stat of the spectral fit: two for the redshift (one zoomed on the minimum and one wide the whole redshift range), one for the temperature and one for the abundance;

- countour plots in the abundance *versus* redshift and temperature *versus* redshift planes;

- scaling relations ($L_X - T_X$, $M - Y_X$ and $M - T$) and gas fraction of the whole sample where the specific cluster is marked with a black cross.

---

1 Cyan corresponds to the energy band 0.5-1.0 keV, magenta to 1.0-2.0 keV and yellow to 2.0-7.0 keV

## B.1 SWJ0217-5014

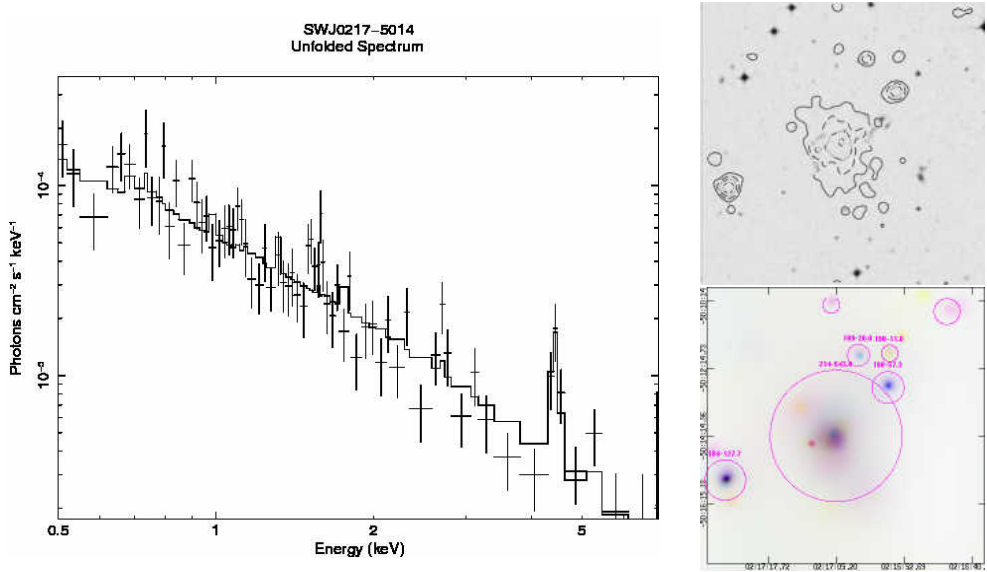

| Name | GRB | R.A. | Dec | Catalogue | Distance | Published z |
|------|-----|------|-----|-----------|----------|-------------|
| SWJ0217-5014 | GRB050406 | 34.270653 | -50.236660 | 1AXG | 0.787 | - |

| Expmap [s] | Net Counts | SNR | Flux [$10^{-13}$ erg/cm²/s] | $N_H$ [$10^{22}$cm⁻²] | Bkg rate [$10^{-3}$ cts/arcsec²] | $r_{ext}$ [arcsec] |
|------------|-----------|-----|------|-----|----------|-------|
| 133717 | 543±30 | 18.0 | 1.02±0.06 | 0.018 | 5.18 | 116.6 |

| $kT$ [keV] | $z$ | $X_{Fe}/X_{\odot}$ | $r_{ext}$ [kpc] | $r_{500}$ [kpc] | $L_{ext}$ [$10^{44}$ erg/s] | $L_{500}$ [$10^{44}$ erg/s] |
|------------|-----|--------------------|------------------|------------------|------------------------------|------------------------------|
| $6.3^{+0.8}_{-0.7}$ | $0.516^{+0.008}_{-0.008}$ | $1.08^{+0.38}_{-0.28}$ | 730±39 | 936±51 | $4.21^{+0.33}_{-0.33}$ | $4.53^{+0.35}_{-0.36}$ |

| $M_{500}$ [$10^{13} M_{\odot}$] | $M_{gas,500}$ [$10^{13} M_{\odot}$] | $f_{gas,500}$ |
|----------------------------------|--------------------------------------|----------------|
| 41.78±6.76 | 2.53±0.62 | 0.060±0.005 |

As one can see from the contour plot, both the temperature and the redshift are well defined. The redshift has only one minimum between 0.50 and 0.53. However the resulting luminosity is a factor three lower compared with the expected luminosity from Branchesi et al. The aperture photometry gives 15% more soft net counts with respect the surface brightness fit. Morover the expmap correction is not severe and the source is far away from the image edge, so there should not be missing counts. The cluster results somehow underluminous. $Y_X$ and $f_{gas}$ are marginally in agreement with previous works by Arnaud et al. (2007) and Sun et al. (2009).

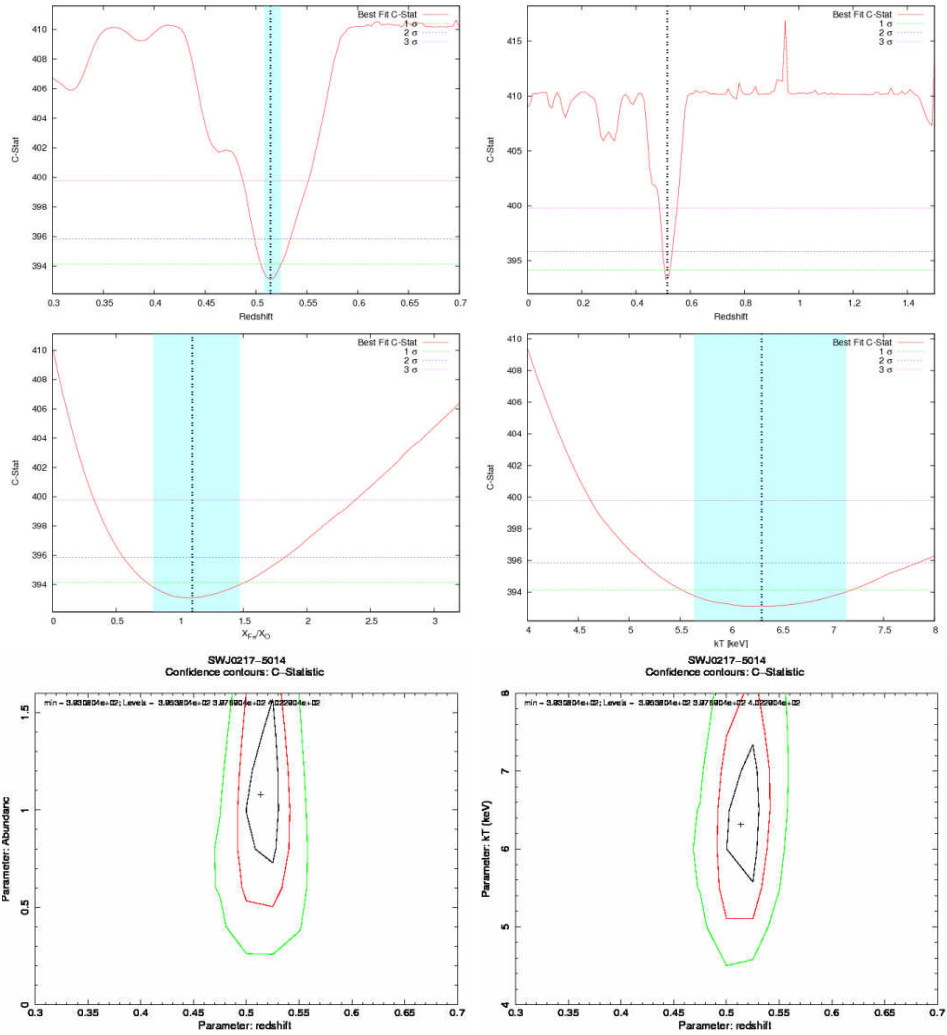

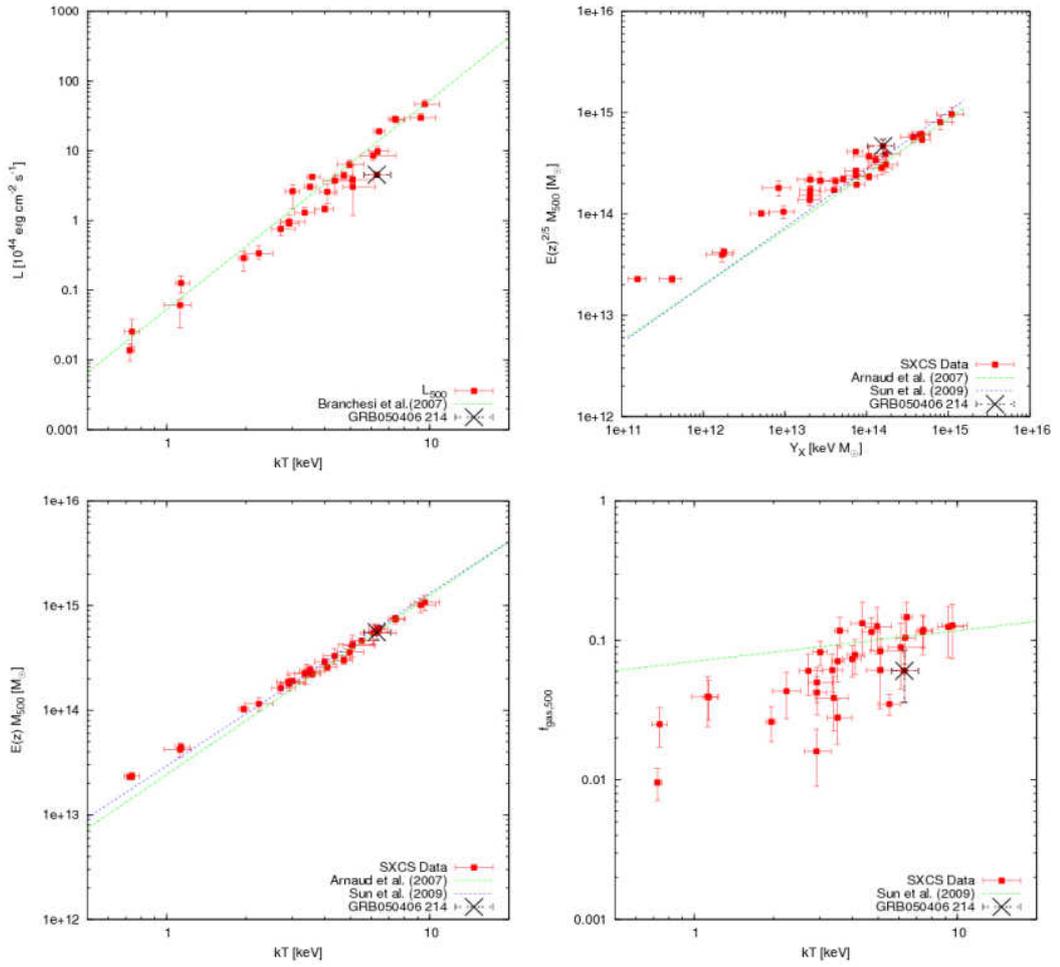

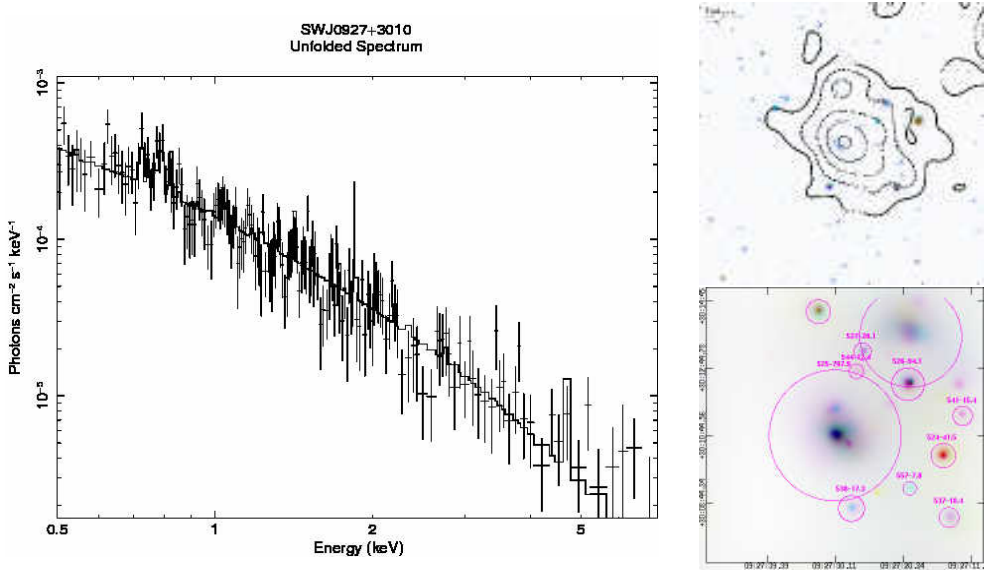

| Name | GRB | R.A. | Dec | Catalogue | Distance | Published $z$ |
|------|-----|------|-----|-----------|----------|------------|
| SWJ0927+3010 | GRB050505* | 141.874588 | 30.179701 | 1AXG | 0.314 | 0.35[7] |

| Expmap [s] | Net Counts | SNR | Flux [$10^{-13}$ erg/cm²/s] | $N_H$ [$10^{22}$cm⁻²] | Bkg rate [$10^{-3}$ cts/arcsec²] | $r_{ext}$ [arcsec] |
|------|------|------|------|------|------|------|
| 85805 | 767±34 | 23.2 | 2.23±0.10 | 0.017 | 7.47 | 116.2 |

| $kT$ [keV] | $z$ | $X_{Fe}/X_\odot$ | $r_{ext}$ [kpc] | $r_{500}$ [kpc] | $L_{ext}$ [$10^{44}$ erg/s] | $L_{500}$ [$10^{44}$ erg/s] |
|------|------|------|------|------|------|------|
| $3.0^{+0.2}_{-0.2}$ | $0.417^{+0.018}_{-0.039}$ | $0.49^{+0.17}_{-0.12}$ | 540±52 | 741±72 | $2.49^{+0.59}_{-1.07}$ | $2.63^{+0.62}_{-1.13}$ |

| $M_{500}$ [$10^{13}M_\odot$] | $M_{gas,500}$ [$10^{13}M_\odot$] | $f_{gas,500}$ |
|------|------|------|
| 16.15±0.59 | 1.33±0.21 | 0.082±0.010 |

The C-stat has two redshift minimum at 0.42 and 0.32 almost indistinguishible. This is mostly due to the low SNR in the spectrum at energies around 5.0 keV, where lies the redshifted K-$a$ iron emission line. From the X-ray contour is visible a substructure 60 arcsec northward form the centrer of X-ray emission, therefore the cluster is not relaxed. Also removing the "clump" the quality of the spectrum and the redshift determination do not improve. However $M-T$, $M-Y_X$ and $f_{gas}$ are in agreement with other observational work. The best redshift seems to be 0.32.

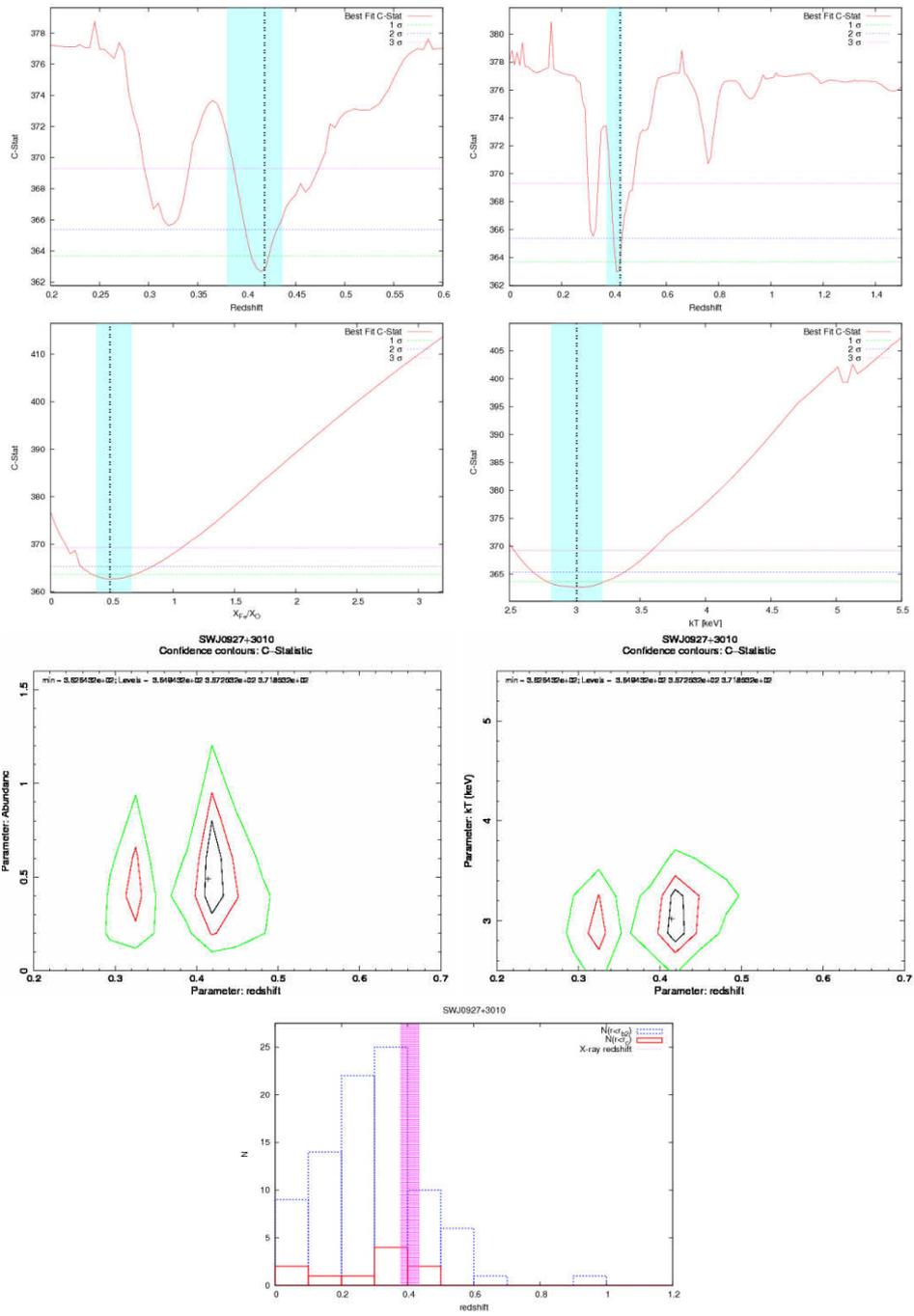

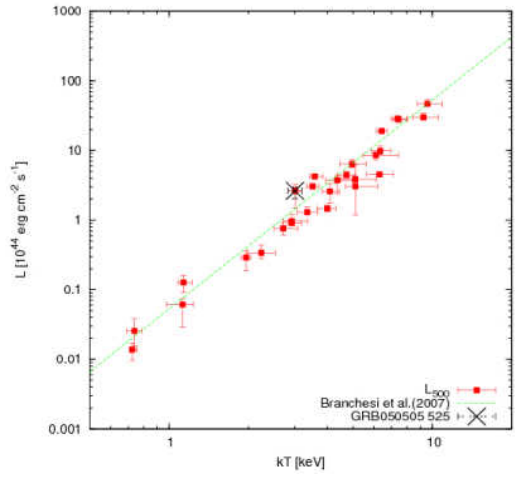
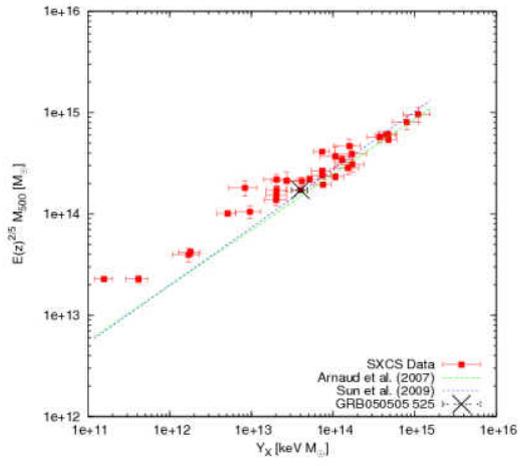

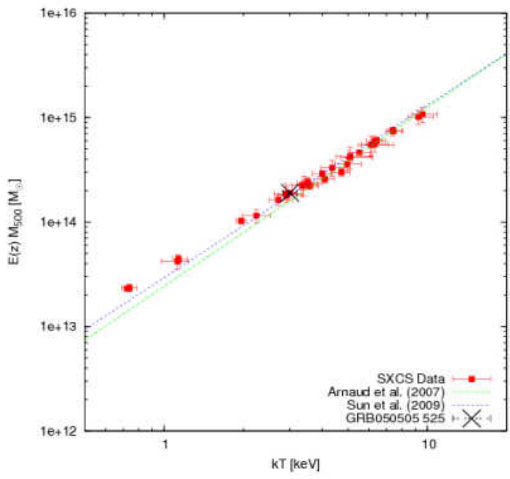
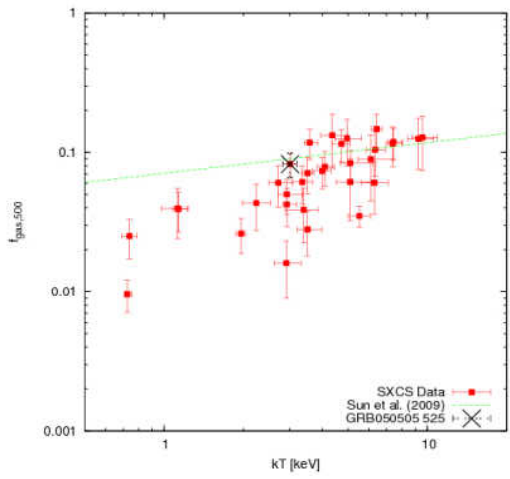

B.3 SWJ0926+3013

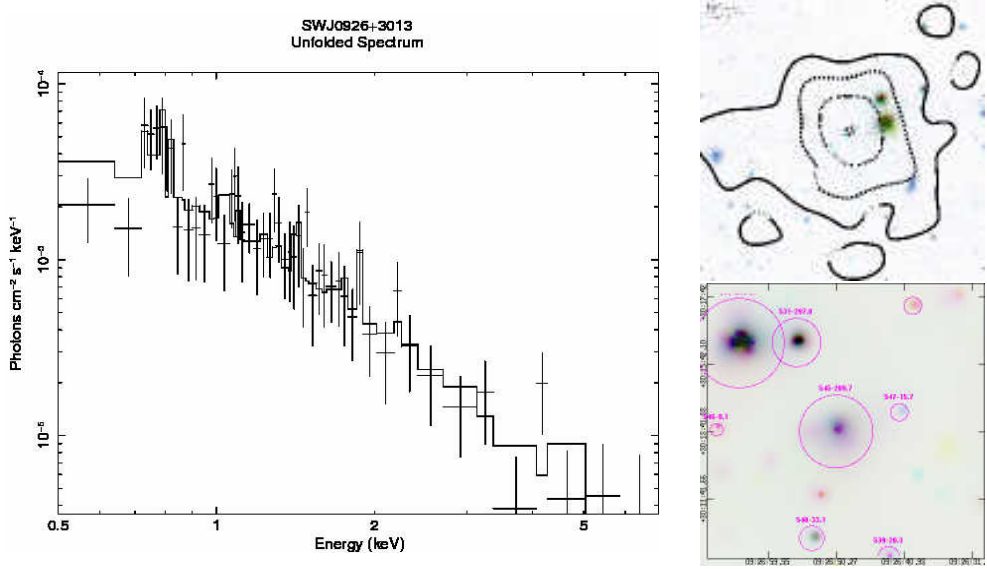

| Name | GRB | R.A. | Dec | Catalogue | Distance | Published $z$ |
|------|-----|------|-----|-----------|----------|---------------|
| SWJ0926+3013 | GRB050505* | 141.708237 | 30.229328 | - | - | - |

| Expmap [s] | Net Counts | SNR | Flux [$10^{-13}$ erg/cm$^2$/s] | $N_H$ [$10^{22}$cm$^{-2}$] | Bkg rate [$10^{-3}$ cts/arcsec$^2$] | r$_{ext}$ [arcsec] |
|------------|-----------|-----|-------------------------------|---------------------------|-------------------------------------|--------------------|
| 164969 | 289±22 | 13.2 | 0.44±0.03 | 0.017 | 7.47 | 65.0 |

| $kT$ [keV] | $z$ | $X_{Fe}/X_\odot$ | $r_{ext}$ [kpc] | $r_{500}$ [kpc] | $L_{ext}$ [$10^{44}$ erg/s] | $L_{500}$ [$10^{44}$ erg/s] |
|------------|-----|------------------|-----------------|-----------------|-----------------------------|-----------------------------|
| $2.9^{+0.4}_{-0.2}$ | $0.412^{+0.058}_{-0.029}$ | $1.21^{+0.83}_{-0.44}$ | 392±42 | 658±71 | $0.84^{+0.20}_{-0.12}$ | $0.96^{+0.23}_{-0.14}$ |

| $M_{500}$ [$10^{13}M_\odot$] | $M_{gas,500}$ [$10^{13}M_\odot$] | $f_{gas,500}$ |
|------------------------------|----------------------------------|----------------|
| 13.83±1.33 | 0.69±0.13 | 0.050±0.005 |

This cluster has only $\sim 300$ net counts, so the $1\sigma$ error on the temperature and on the redshift are $\sim 15-20\%$, even if possibly the redshift error is underestimated. The redshift is determinded for L-shell emission lines. $L-T$ is in good agreement with the best fit by Branchesi et al. (2007), while it seems that $M_{gas}$ is underestimated.

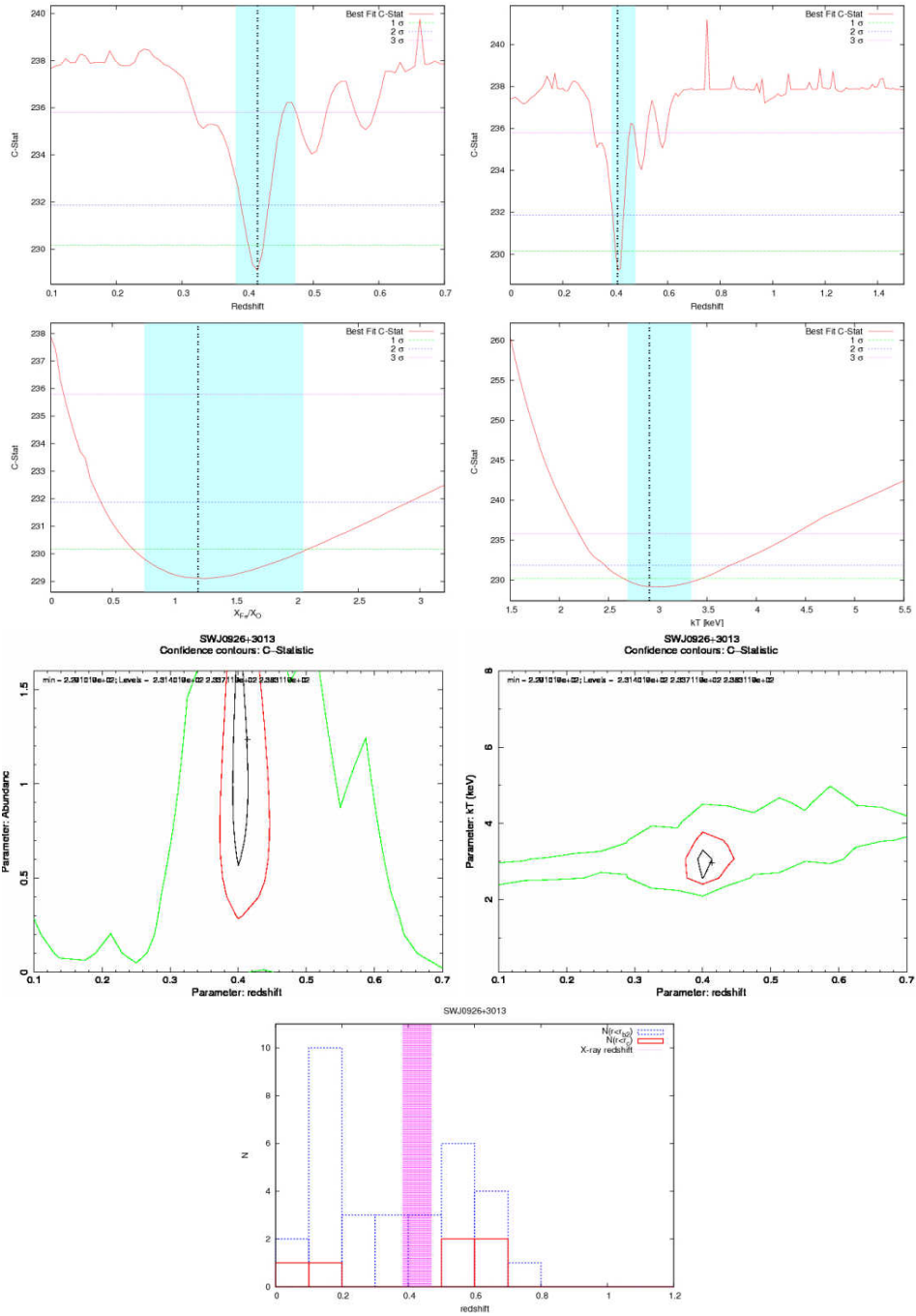

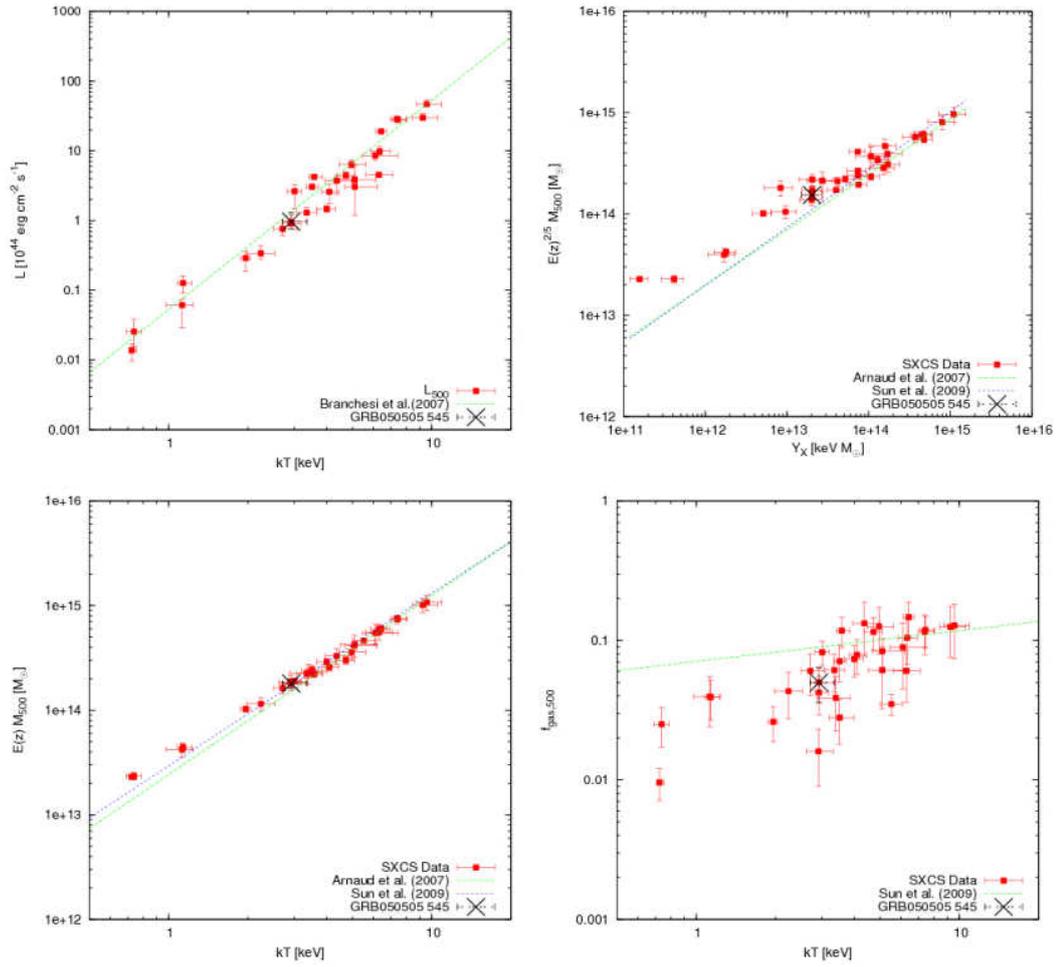

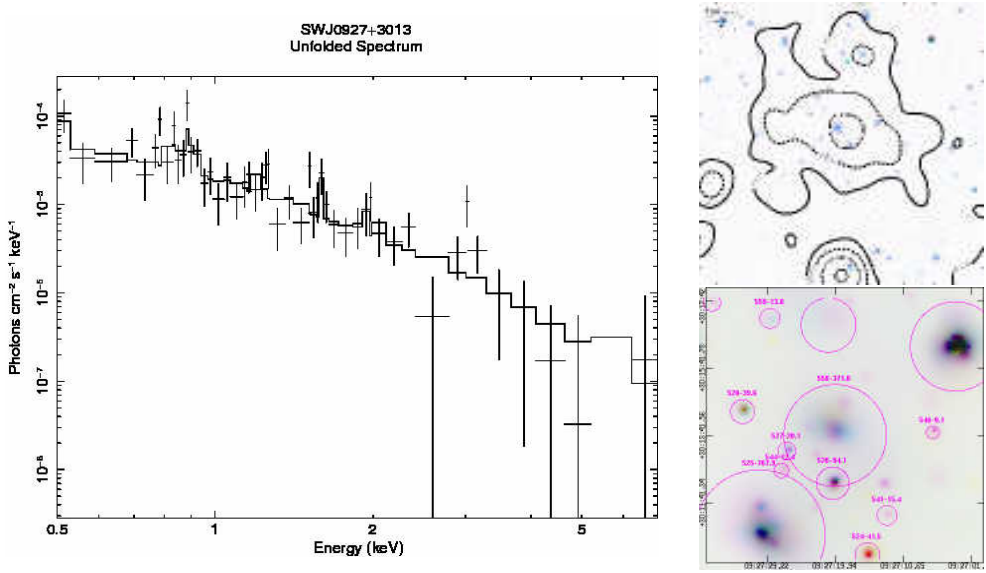

| Name | GRB | R.A. | Dec | Catalogue | Distance | Published $z$ |
|------|-----|------|-----|-----------|----------|---------------|
| SWJ0927+3013 | GRB050505* | 141.831726 | 30.228964 | MaxBCG | 0.186 | 0.30[1,7] |

| Expmap | Net Counts | SNR | Flux | $N_H$ | Bkg rate | $r_{ext}$ |
| [s] | | | [$10^{-13}$ erg/cm$^2$/s] | [$10^{22}$ cm$^{-2}$] | [$10^{-3}$ cts/arcsec$^2$] | [arcsec] |
| 157316 | 370±31 | 13.8 | 0.59±0.05 | 0.017 | 7.47 | 90.3 |

| $kT$ | $z$ | $X_{Fe}/X_\odot$ | $r_{ext}$ | $r_{500}$ | $L_{ext}$ | $L_{500}$ |
| [keV] | | | [kpc] | [kpc] | [$10^{44}$ erg/s] | [$10^{44}$ erg/s] |
| $2.2^{+0.3}_{-0.3}$ | $0.252^{+0.043}_{-0.023}$ | $0.72^{+0.57}_{-0.31}$ | 402±123 | 632±194 | $0.29^{+0.08}_{-0.05}$ | $0.34^{+0.10}_{-0.06}$ |

| $M_{500}$ | $M_{gas,500}$ | $f_{gas,500}$ |
| [$10^{13}M_\odot$] | [$10^{13}M_\odot$] | |
| 9.88±1.41 | 0.43±0.10 | 0.043±0.004 |

The C-stat has two redshift minimum at $\sim$ 0.25 and $\sim$ 1.2 almost indistinguishable. However with the redshift best fit setted to $\sim$ 1.2 the resulting luminosity and total mass are completely in disagreement with scaling laws from previous work, while this is not the case when $z \sim$ 0.25. Moreover the SDSS galaxy distribution is peaked at $z \sim$ 0.25. In any case $M_{gas}$ seems underestimated.

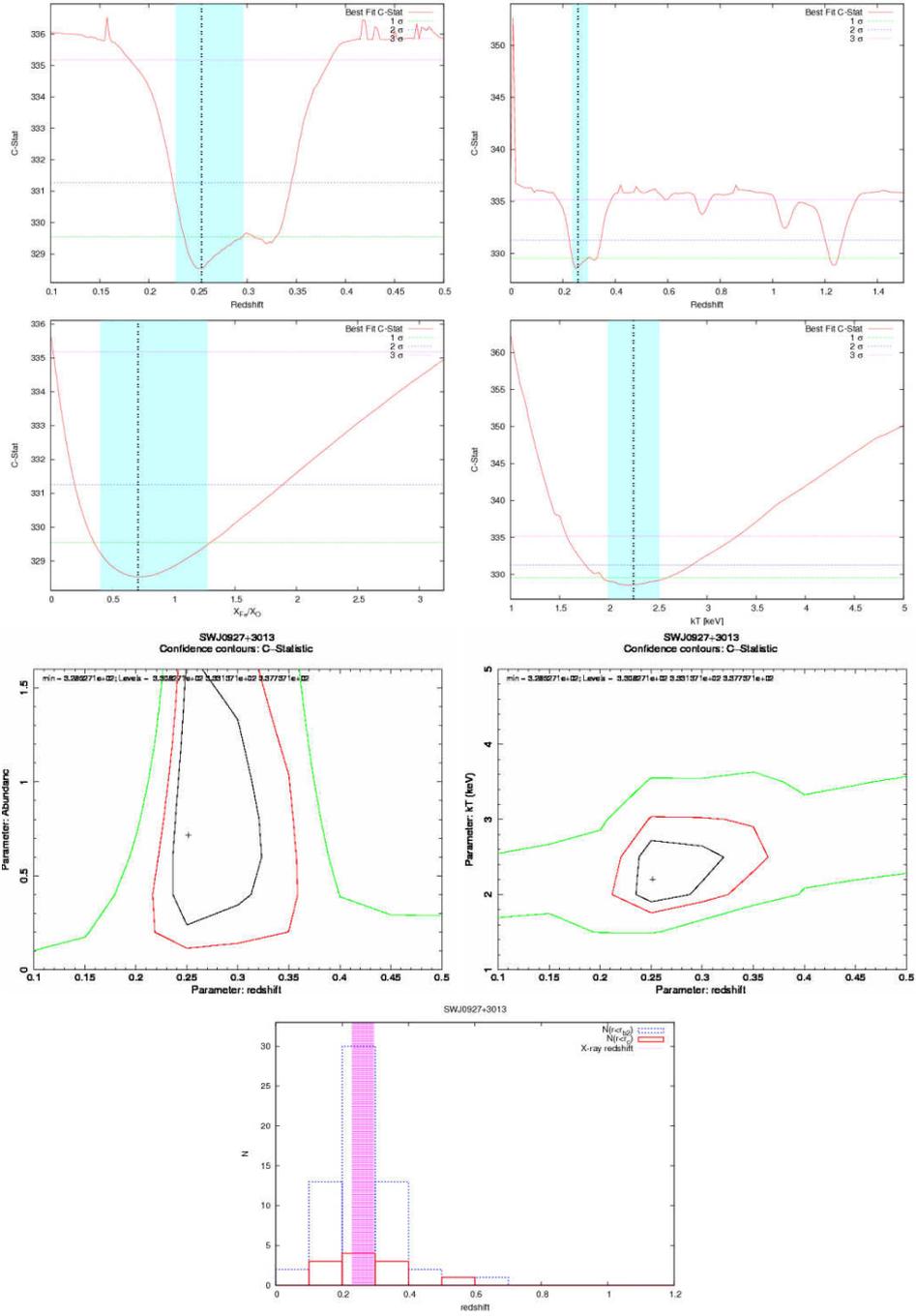

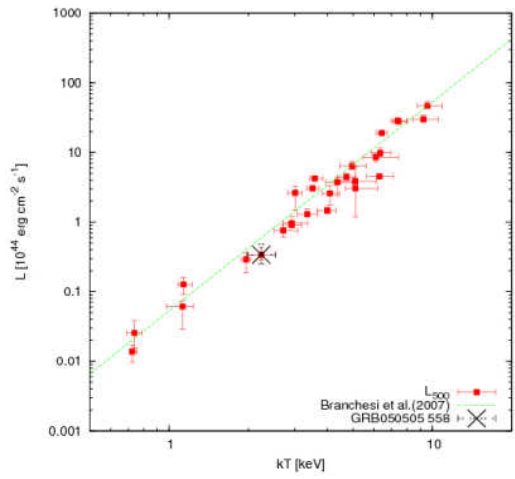
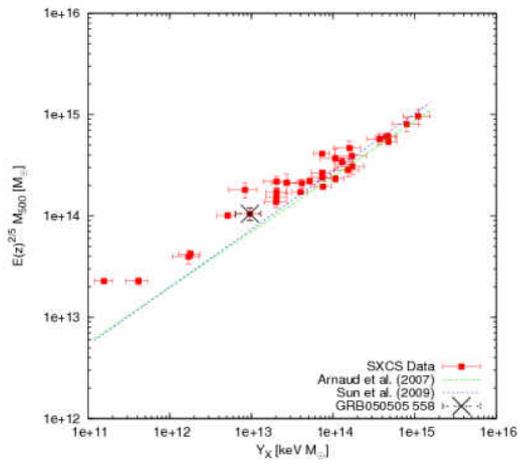
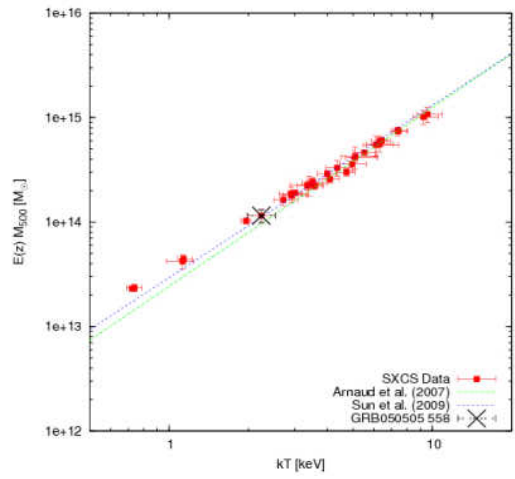
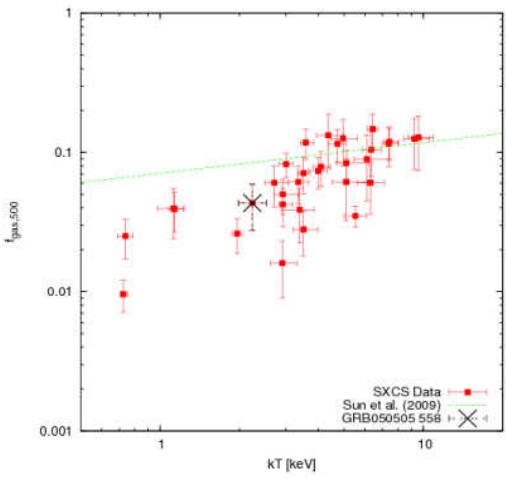

## B.5 SWJ0239-2505

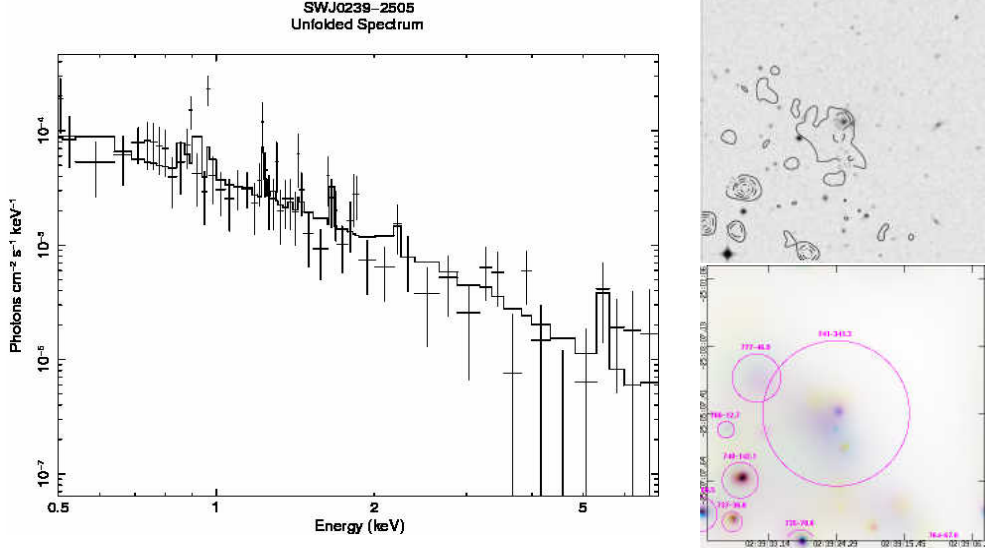

| Name | GRB | R.A. | Dec | Catalogue | Distance | Published $z$ |
|---|---|---|---|---|---|---|
| SWJ0239-2505 | GRB050603 | 39.850075 | -25.084248 | 2dFGRS | 0.084 | 0.174[2] |

| Expmap [s] | Net Counts | SNR | Flux [$10^{-13}$ erg/cm$^2$/s] | $N_H$ [$10^{22}$cm$^{-2}$] | Bkg rate [$10^{-3}$ cts/arcsec$^2$] | $r_{ext}$ [arcsec] |
|---|---|---|---|---|---|---|
| 75421 | 341±27 | 13.5 | 1.15±0.09 | 0.022 | 9.63 | 129.8 |

| $kT$ [keV] | $z$ | $X_{Fe}/X_\odot$ | $r_{ext}$ [kpc] | $r_{500}$ [kpc] | $L_{ext}$ [$10^{44}$ erg/s] | $L_{500}$ [$10^{44}$ erg/s] |
|---|---|---|---|---|---|---|
| $3.4^{+0.5}_{-0.5}$ | $0.201^{+0.005}_{-0.007}$ | $1.19^{+0.70}_{-0.50}$ | 430±24 | 833±46 | $0.27^{+0.04}_{-0.04}$ | $0.33^{+0.04}_{-0.05}$ |

| $M_{500}$ [$10^{13}M_\odot$] | $M_{gas,500}$ [$10^{13}M_\odot$] | $f_{gas,500}$ |
|---|---|---|
| 20.46±4.41 | 0.79±0.16 | 0.040±0.009 |

It seems that the center of this cluster is exactly on the edge, or even outside the image. Most of the flux is not in the image and is not possible to recover it. However in the image there are enogh net counts, ∼ 450, to have a satisfying temperature and redshift measure. It has been removed form the $L - T$ plot because it is out of the FoV of the image and most of its luminosity is undetected, but it seems ok for $M - Y_X$, $M - T$ and $f_{gas}$ plots.

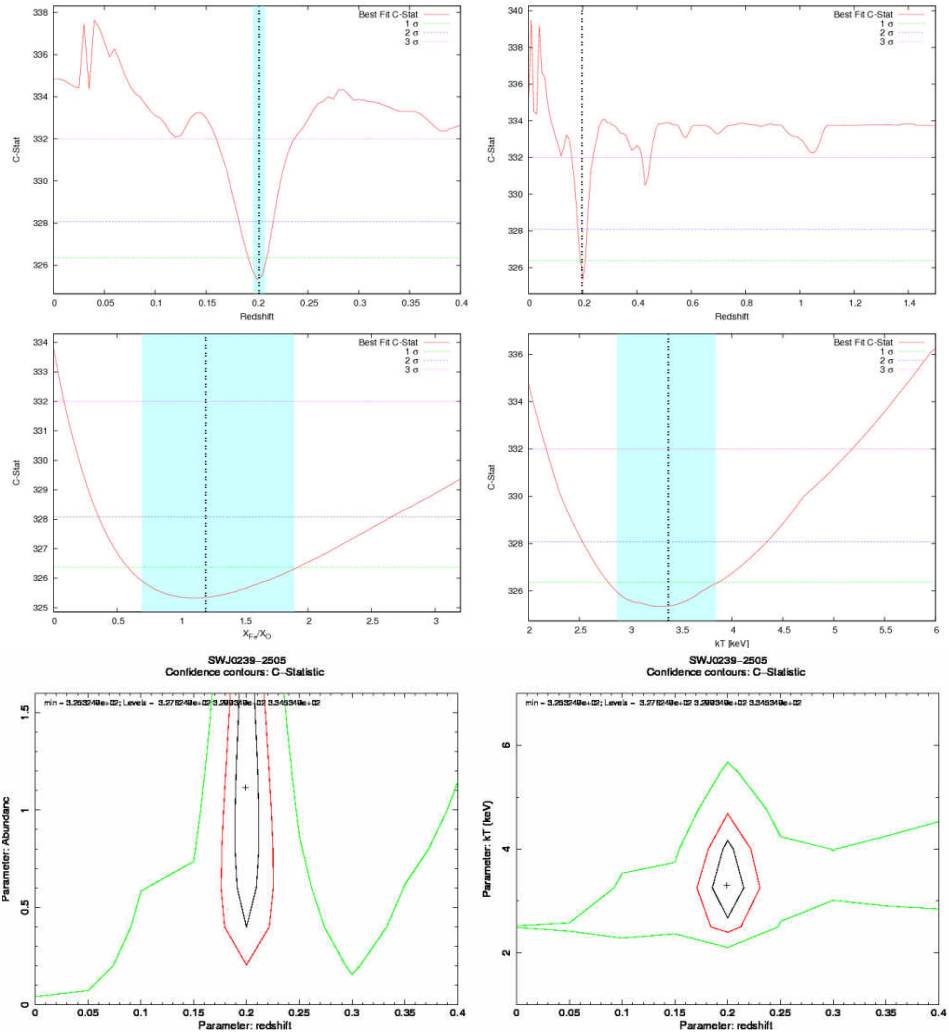

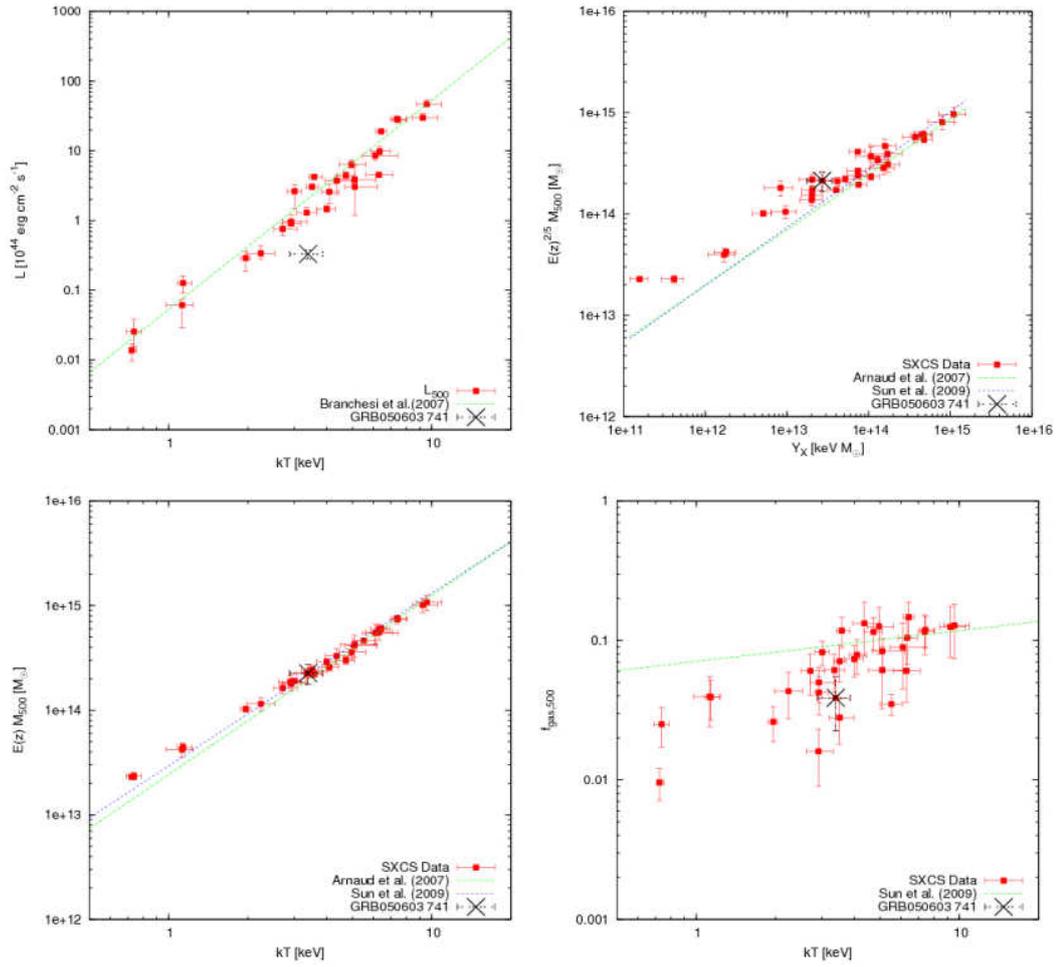

| Name | GRB | R.A. | Dec | Catalogue | Distance | Published $z$ |
|------|-----|------|-----|-----------|----------|---------------|
| SWJ2322+0548 | GRB050803 | 350.702240 | 5.803986 | - | - | - |

| Expmap [s] | Net Counts | SNR | Flux [$10^{-13}$ erg/cm$^2$/s] | $N_H$ [$10^{22}$cm$^{-2}$] | Bkg rate [$10^{-3}$ cts/arcsec$^2$] | $r_{ext}$ [arcsec] |
|-----------|-----------|-----|-------|-----|----------|------|
| 207450 | 1622±55 | 32.9 | 2.15±0.07 | 0.050 | 9.39 | 118.8 |

| $kT$ [keV] | $z$ | $X_{Fe}/X_\odot$ | $r_{ext}$ [kpc] | $r_{500}$ [kpc] | $L_{ext}$ [$10^{44}$ erg/s] | $L_{500}$ [$10^{44}$ erg/s] |
|-----------|-----|------------------|-------|-------|----------|----------|
| $2.9^{+0.3}_{-0.1}$ | $0.227^{+0.003}_{-0.002}$ | $1.09^{+0.25}_{-0.16}$ | 432±18 | 771±33 | $0.82^{+0.04}_{-0.04}$ | $0.91^{+0.05}_{-0.04}$ |

| $M_{500}$ [$10^{13}M_\odot$] | $M_{gas,500}$ [$10^{13}M_\odot$] | $f_{gas,500}$ |
|-----------|-----------|-----------|
| 16.47±1.41 | 0.70±0.15 | 0.042±0.006 |

The redshift is determined with high significance, given the high number of detected emission lines. $L - T$ is consistent with Branchesi et al. (2007), and $M - T$ with Arnaud et al. (2007) and Sun et al. (2009). $M_{gas}$ seems underestimated.

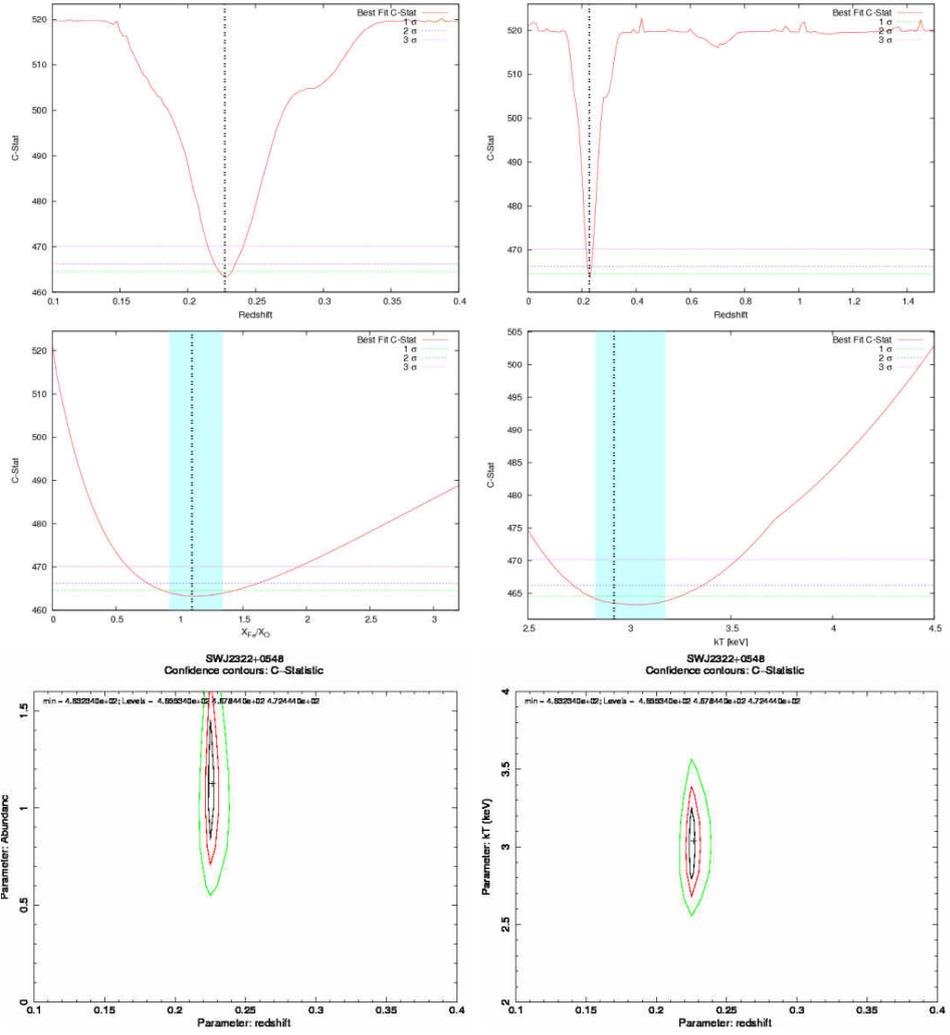

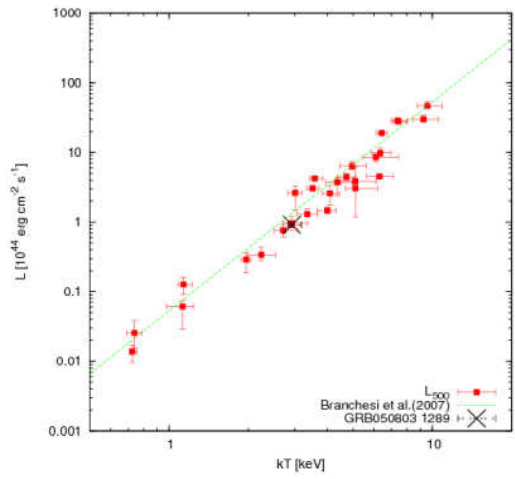

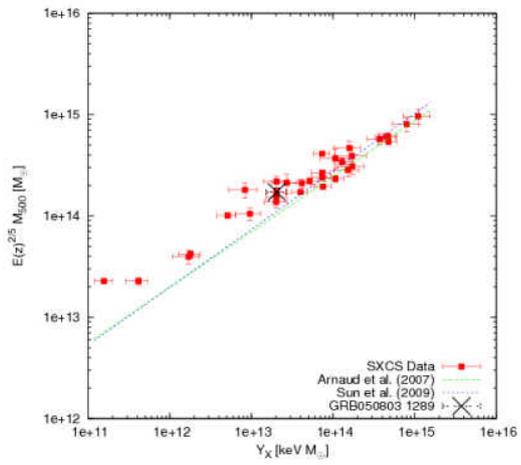

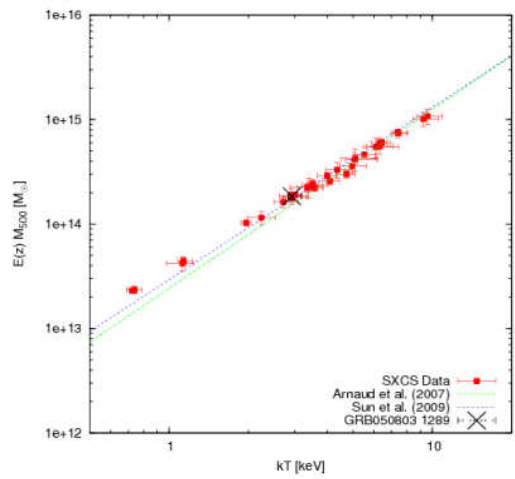

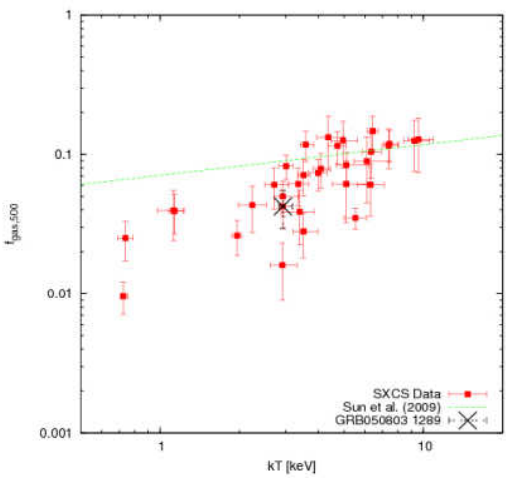

## B.7 SWJ1737+4618

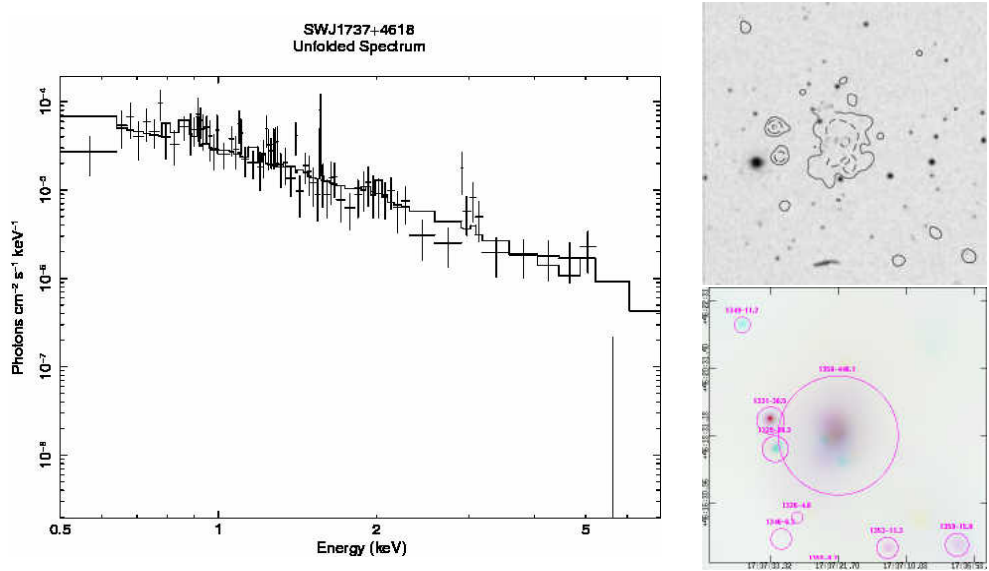

| Name | GRB | R.A. | Dec | Catalogue | Distance | Published $z$ |
|------|-----|------|-----|-----------|----------|---------------|
| SWJ1737+4618 | GRB050814 | 264.339539 | 46.309334 | - | - | - |

| Expmap [s] | Net Counts | SNR | Flux [$10^{-13}$ erg/cm$^2$/s] | $N_H$ [$10^{22}$cm$^{-2}$] | Bkg rate [$10^{-3}$ cts/arcsec$^2$] | $r_{ext}$ [arcsec] |
|------------|-----------|-----|-------------------------------|----------------------------|--------------------------------------|--------------------|
| 137248 | 448±31 | 15.4 | 0.83±0.06 | 0.023 | 6.52 | 106.4 |

| $kT$ [keV] | $z$ | $X_{Fe}/X_{\odot}$ | $r_{ext}$ [kpc] | $r_{500}$ [kpc] | $L_{ext}$ [$10^{44}$ erg/s] | $L_{500}$ [$10^{44}$ erg/s] |
|------------|-----|--------------------|------------------|------------------|------------------------------|------------------------------|
| $3.5^{+0.5}_{-0.3}$ | $0.280^{+0.030}_{-0.032}$ | $0.81^{+0.49}_{-0.29}$ | 416±32 | 825±64 | $0.33^{+0.12}_{-0.12}$ | $0.41^{+0.15}_{-0.15}$ |

| $M_{500}$ [$10^{13} M_{\odot}$] | $M_{gas,500}$ [$10^{13} M_{\odot}$] | $f_{gas,500}$ |
|----------------------------------|--------------------------------------|----------------|
| 20.82±2.70 | 0.58±0.13 | 0.028±0.003 |

The redshift cannot be determied. The main C-stat has four redshift minimum: $\sim 0.21$ and $\sim 0.28$ (which are stronger and indistinguishable), and two relative minimum $\sim 0.6$ and $\sim 1.2$, which cannot be statistically excluded. None of this redshift value provides satisfactory mass and luminosity measurement in agreement with best fit from previous work. However in the X-ray image the diffuse emission is clear and detected with high significance.

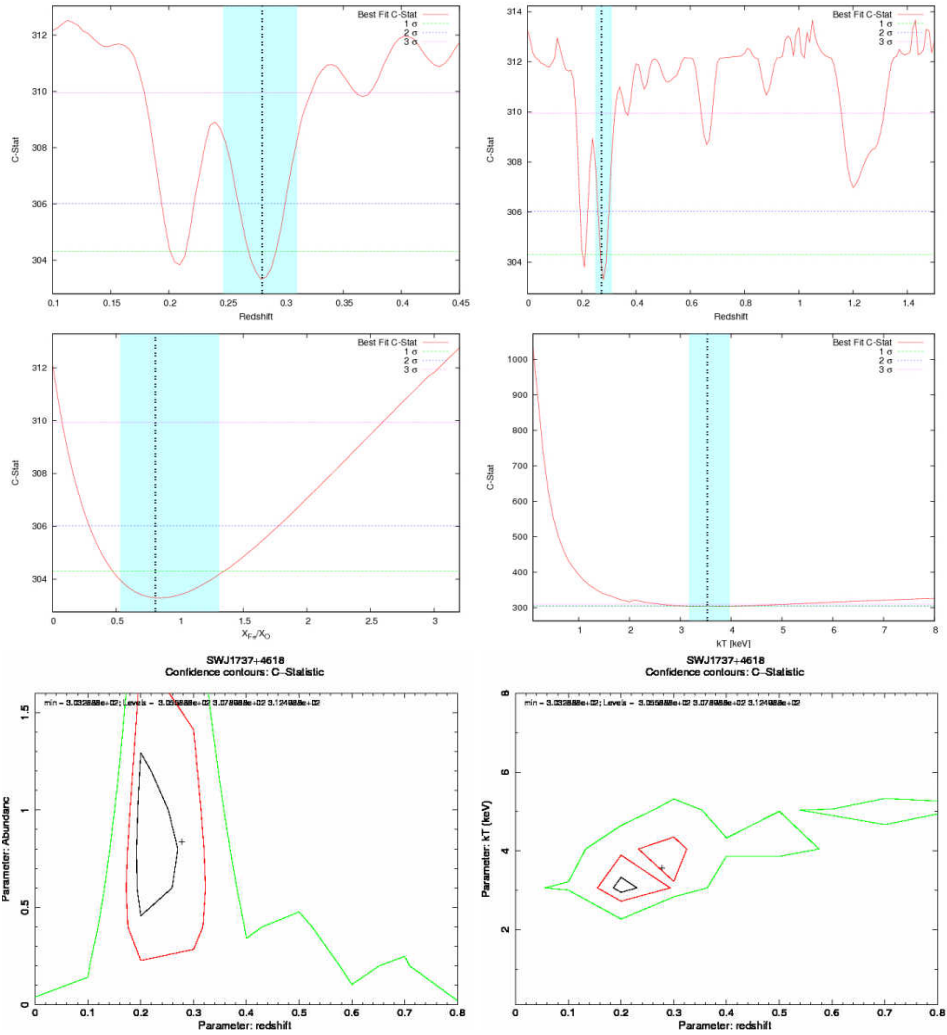

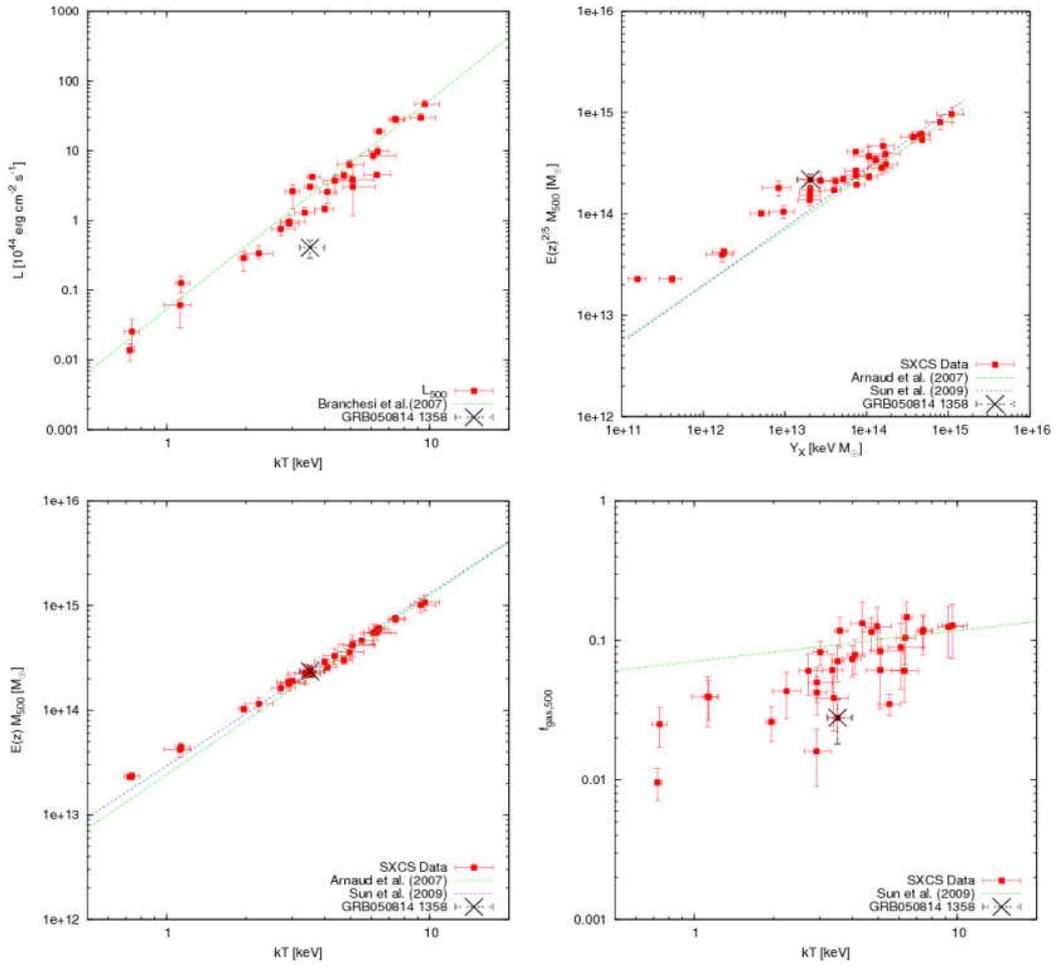

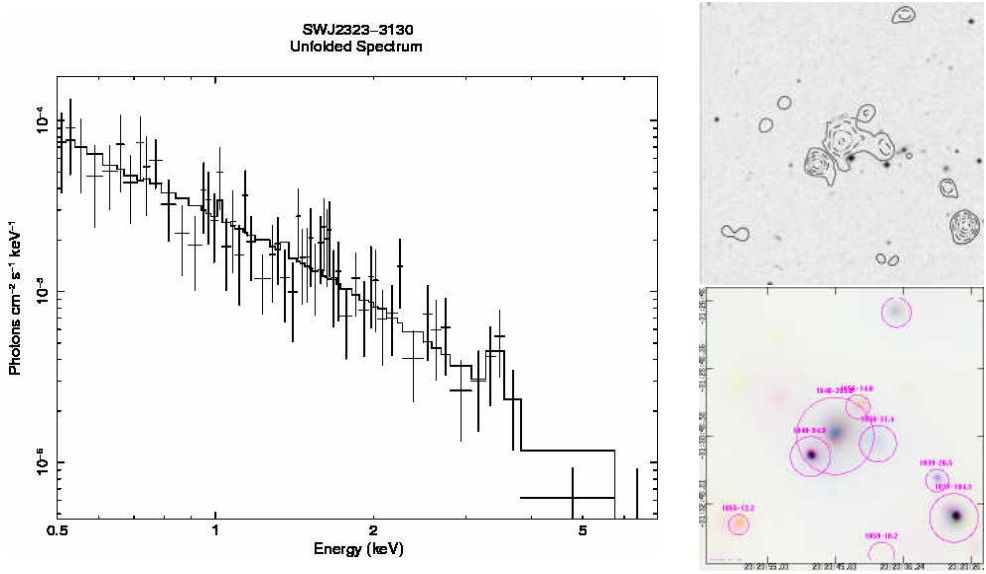

| Name | GRB | R.A. | Dec | Catalogue | Distance | Published $z$ |
|---|---|---|---|---|---|---|
| SWJ2323-3130 | GRB051001 | 350.939148 | -31.512819 | - | - | - |

| Expmap [s] | Net Counts | SNR | Flux [$10^{-13}$ erg/cm$^2$/s] | $N_H$ [$10^{22}$cm$^{-2}$] | Bkg rate [$10^{-3}$ cts/arcsec$^2$] | $r_{ext}$ [arcsec] |
|---|---|---|---|---|---|---|
| 102561 | 204±17 | 11.6 | 0.49±0.04 | 0.011 | 3.76 | 68.9 |

| $kT$ [keV] | $z$ | $X_{Fe}/X_\odot$ | $r_{ext}$ [kpc] | $r_{500}$ [kpc] | $L_{ext}$ [$10^{44}$ erg/s] | $L_{500}$ [$10^{44}$ erg/s] |
|---|---|---|---|---|---|---|
| $6.1^{+1.3}_{-0.7}$ | $0.964^{+0.033}_{-0.037}$ | $0.48^{+0.38}_{-0.24}$ | 548±47 | 708±61 | $8.06^{+1.14}_{-1.20}$ | $8.55^{+1.21}_{-1.27}$ |

| $M_{500}$ [$10^{13}M_\odot$] | $M_{gas,500}$ [$10^{13}M_\odot$] | $f_{gas,500}$ |
|---|---|---|
| 31.36±6.86 | 2.80±0.78 | 0.088±0.005 |

This is the most distant cluster in the SXCS bright sample. According to the X-ray spectral analysis its redshift is $\sim$ 0.96 determined with a significance of $2\sigma$. The scaling relations are in agreement with the best fit values obtained by Branchesi et al. (2007), Arnaud et al. (2007) and Sun et al. (2009).

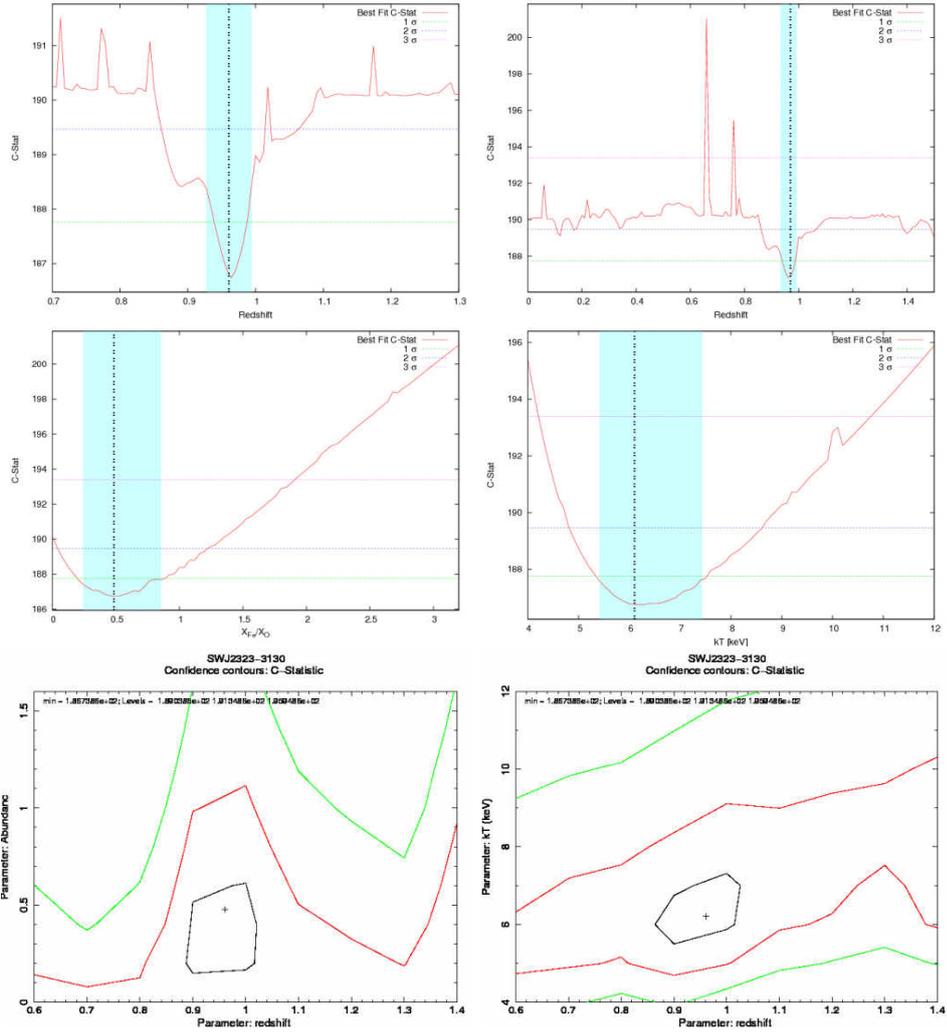

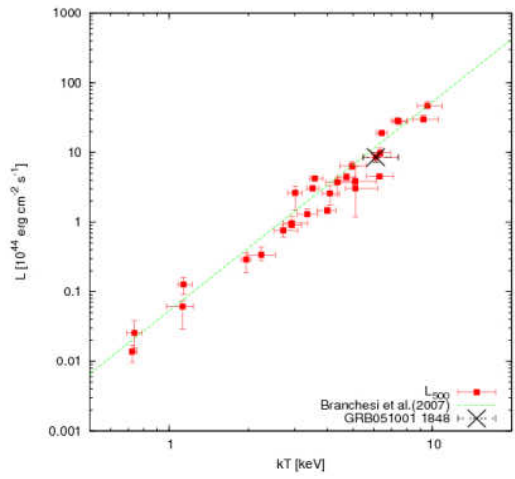

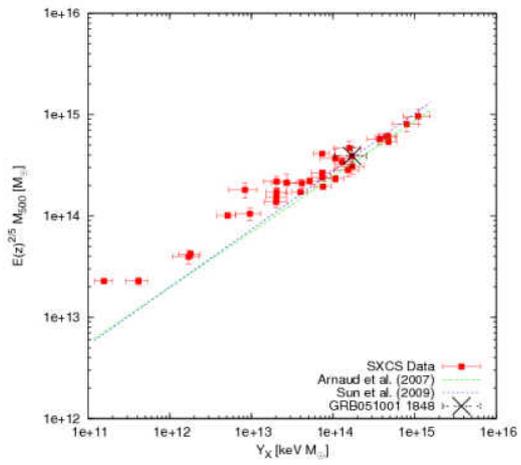

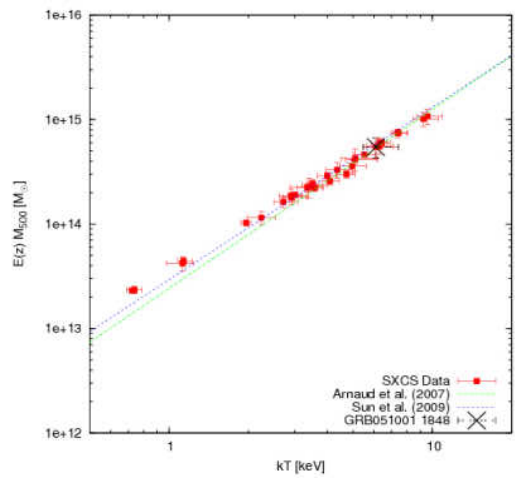

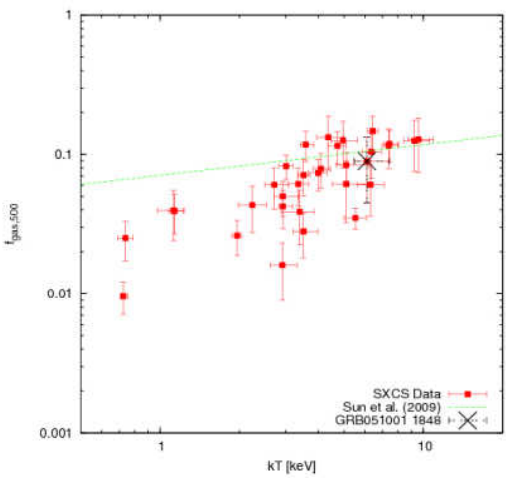

## B.9 SWJ1330+4200

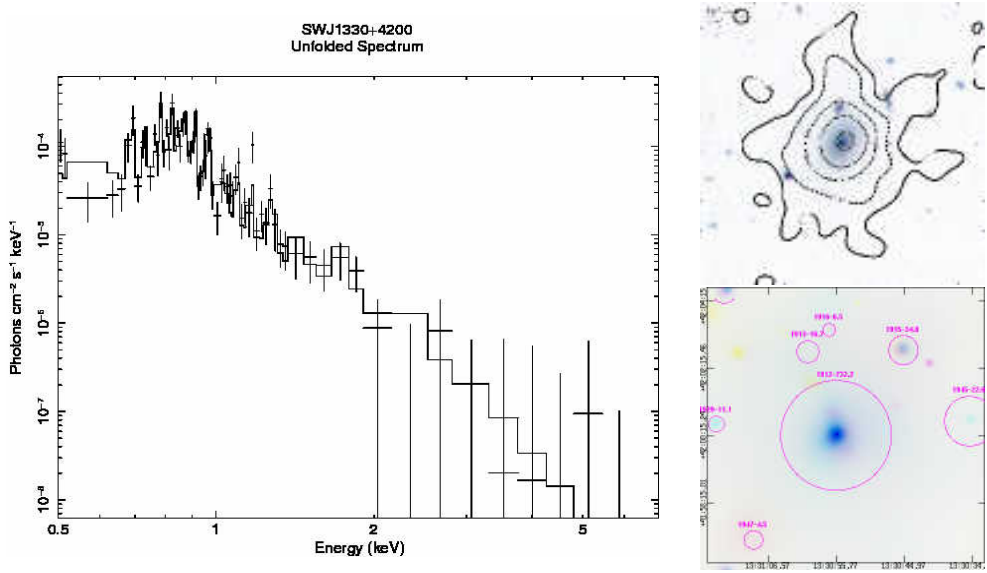

| Name | GRB | R.A. | Dec | Catalogue | Distance | Published $z$ |
|------|-----|------|-----|-----------|----------|---------------|
| SWJ1330+4200 | GRB051008* | 202.731781 | 42.004669 | 2MASX | 0.061 | 0.017 |

| Expmap | Net Counts | SNR | Flux | $N_H$ | Bkg rate | $r_{ext}$ |
| [s] | | | [$10^{-13}$ erg/cm$^2$/s] | [$10^{22}$cm$^{-2}$] | [$10^{-3}$ cts/arcsec$^2$] | [arcsec] |
|--------|-----------|-----|------|-------|----------|-----------|
| 168584 | 732±39 | 22.2 | 1.06±0.06 | 0.010 | 6.70 | 97.8 |

| $kT$ | $z$ | $X_{Fe}/X_\odot$ | $r_{ext}$ | $r_{500}$ | $L_{ext}$ | $L_{500}$ |
| [keV] | | | [kpc] | [kpc] | [$10^{44}$ erg/s] | [$10^{44}$ erg/s] |
|------|-----|------------------|-----------|-----------|-----------|-----------|
| $0.7^{+0.0}_{-0.0}$ | $0.055^{+0.007}_{-0.010}$ | $0.69^{+0.68}_{-0.16}$ | 120±8 | 420±29 | $0.01^{+0.00}_{-0.00}$ | $0.01^{+0.00}_{-0.00}$ |

| $M_{500}$ | $M_{gas,500}$ | $f_{gas,500}$ |
| [$10^{13} M_\odot$] | [$10^{13} M_\odot$] | |
|-----------|---------------|---------------|
| 2.25±0.10 | 0.02±0.00 | 0.010±0.002 |

The three galaxies marked with a "X" in the SDSS image are 2MASXJ13305564+4200176, 2MASXJ13303464+4200406 and 2MASXJ13304379+4156120, and their spectroscopic redshifts are 0.061154, 0.060258 and 0.061124 respectively. They are probably cluster members and their redshift is in agreement with the X-ray redshift. The $L-T$ and the $M-T$ are in agreement with the best fit scaling laws obtained in previous observational works. The $M_{gas}$ is underestimated.

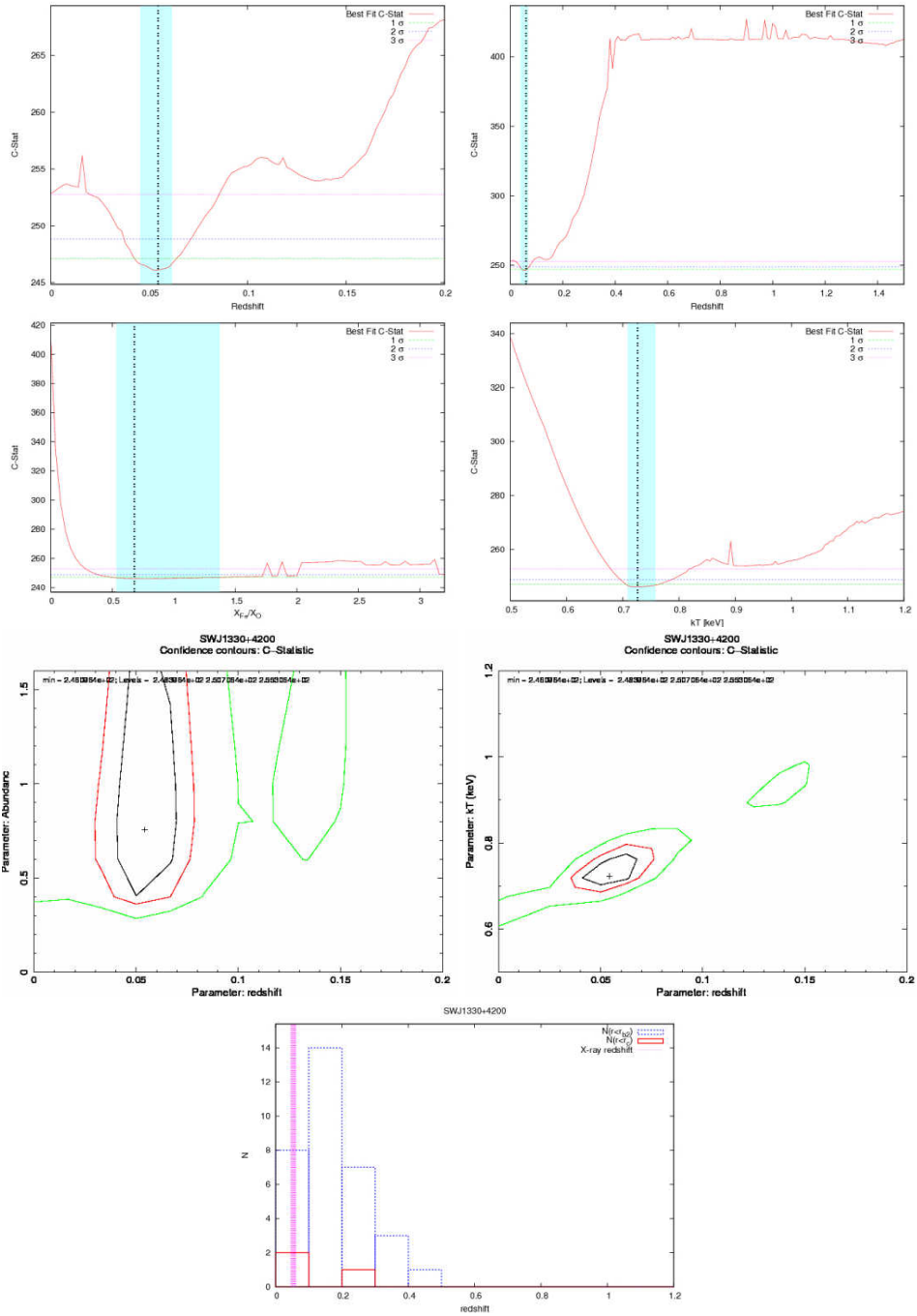

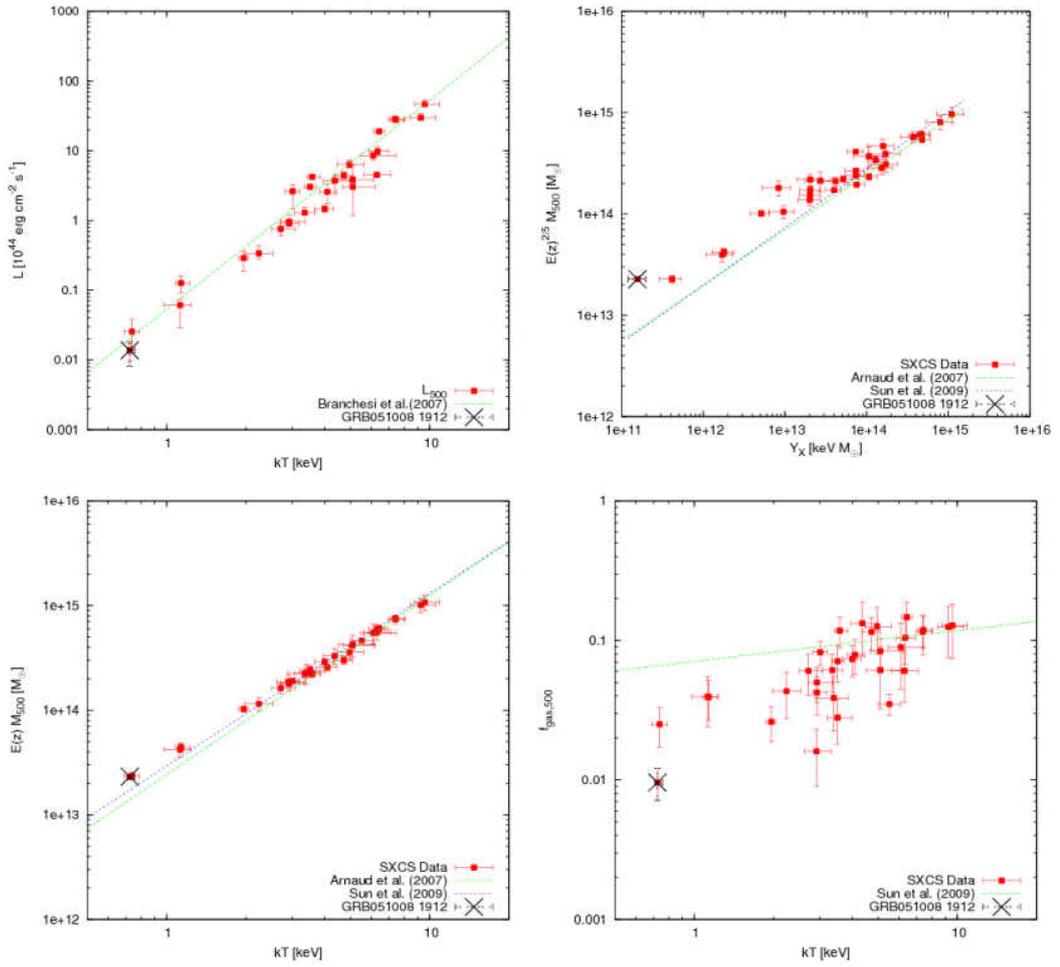

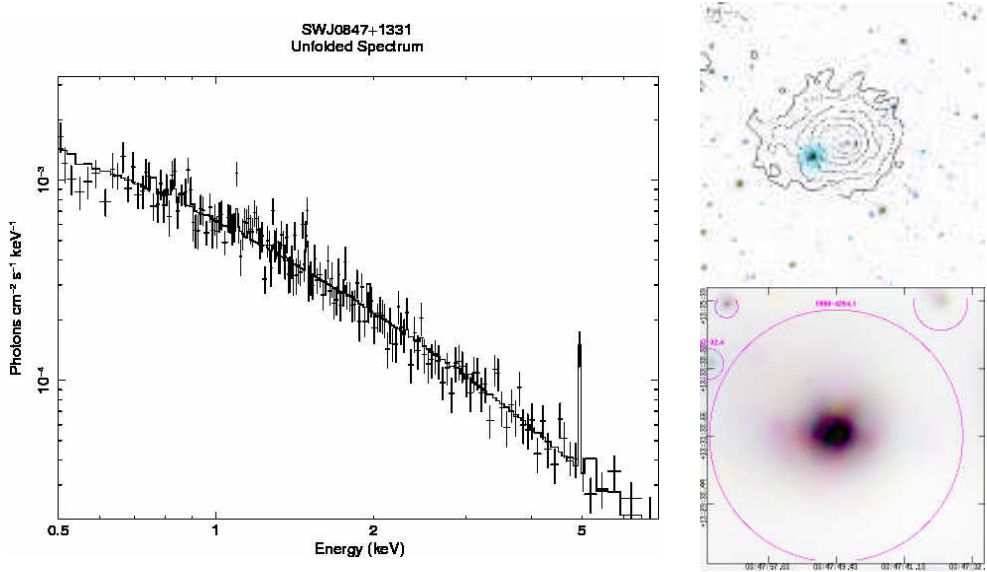

| Name | GRB | R.A. | Dec | Catalogue | Distance | Published $z$ |
|------|-----|------|-----|-----------|----------|---------------|
| SWJ0847+1331 | GRB051016B* | 131.955215 | 13.528370 | - | - | - |

| Expmap [s] | Net Counts | SNR | Flux [$10^{-13}$ erg/cm$^2$/s] | $N_H$ [$10^{22}$cm$^{-2}$] | Bkg rate [$10^{-3}$ cts/arcsec$^2$] | $r_{ext}$ [arcsec] |
|-----------|-----------|-----|------|------|------|------|
| 74419 | 4294±78 | 59.3 | 15.04±0.27 | 0.032 | 8.16 | 223.7 |

| $kT$ [keV] | $z$ | $X_{Fe}/X_\odot$ | $r_{ext}$ [kpc] | $r_{500}$ [kpc] | $L_{ext}$ [$10^{44}$ erg/s] | $L_{500}$ [$10^{44}$ erg/s] |
|-----------|-----|-----|------|------|------|------|
| $6.4^{+0.3}_{-0.3}$ | $0.346^{+0.004}_{-0.003}$ | $0.43^{+0.09}_{-0.08}$ | 1096±91 | 1072±89 | $19.08^{+0.74}_{-0.60}$ | $19.03^{+0.74}_{-0.60}$ |

| $M_{500}$ [$10^{13}M_\odot$] | $M_{gas,500}$ [$10^{13}M_\odot$] | $f_{gas,500}$ |
|------|------|------|
| 50.42±3.33 | 7.43±1.56 | 0.146±0.021 |

The K-$\alpha$ emission line is detected with a very high significance, and the spectrum has a very high SNR. Typical errors are < 5%. The scaling relation are all in agreement with previous works. Also the $M_{gas}$ is in agreement with the data from Sun et al. (2009).

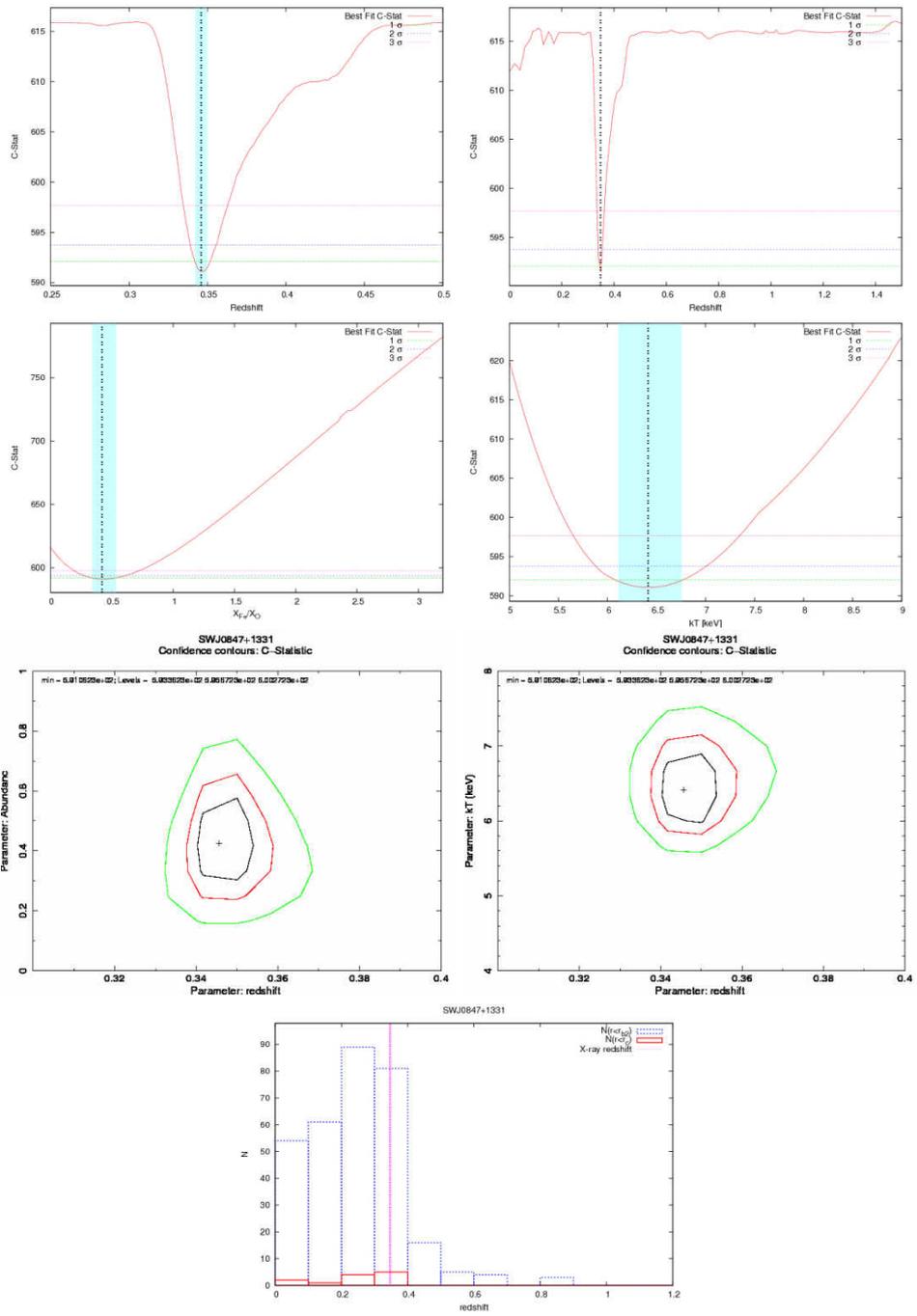

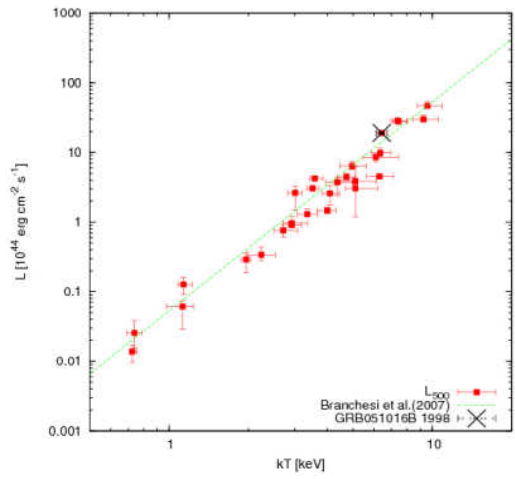

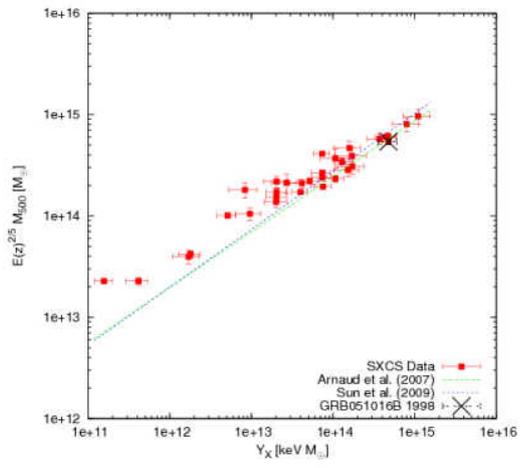

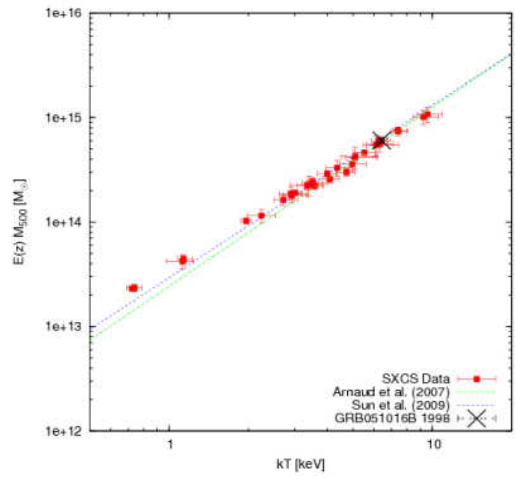

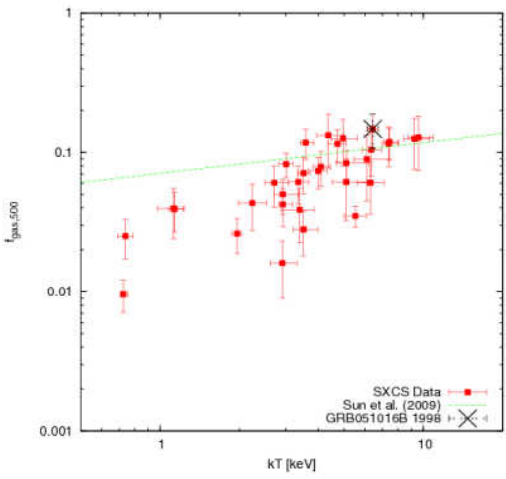

## B.11 SWJ0821+3200

| Name | GRB | R.A. | Dec | Catalogue | Distance | Published $z$ |
|------|-----|------|-----|-----------|----------|---------------|
| SWJ0821+3200 | GRB051227* | 125.307526 | 32.002674 | - | - | - |

| Expmap [s] | Net Counts | SNR | Flux [$10^{-13}$ erg/cm$^2$/s] | $N_H$ [$10^{22}$cm$^{-2}$] | Bkg rate [$10^{-3}$ cts/arcsec$^2$] | $r_{ext}$ [arcsec] |
|------------|------------|-----|-------------------------------|---------------------------|-------------------------------------|--------------------|
| 119159 | 697±40 | 20.6 | 1.53±0.09 | 0.034 | 5.72 | 119.3 |

| $kT$ [keV] | $z$ | $X_{Fe}/X_\odot$ | $r_{ext}$ [kpc] | $r_{500}$ [kpc] | $L_{ext}$ [$10^{44}$ erg/s] | $L_{500}$ [$10^{44}$ erg/s] |
|------------|-----|------------------|-----------------|-----------------|-----------------------------|-----------------------------|
| $5.0^{+0.7}_{-0.5}$ | $0.706^{+0.037}_{-0.019}$ | $0.44^{+0.26}_{-0.19}$ | 855±89 | 727±76 | $6.76^{+1.07}_{-0.73}$ | $6.38^{+1.02}_{-0.69}$ |

| $M_{500}$ [$10^{13}M_\odot$] | $M_{gas,500}$ [$10^{13}M_\odot$] | $f_{gas,500}$ |
|------------------------------|----------------------------------|---------------|
| 24.30±3.49 | 3.07±0.68 | 0.125±0.010 |

The redshift is mostly undetermined. The C-stat has three redshift minimum that are statistically undistinguishable: $\sim 0.3$, $\sim 0.42$ and $\sim 0.7$. Chosing the minimum at $z \sim 0.7$ the resulting luminosity, total mass and gas mass are in agreement with the best fit of the scaling laws found by Brancehsi et al. (2007), Arnaud et al. (2007) and Sun et al. (2009). The choice of the higher redshift is also motivated by the fact that in the SDSS image are not visible any bright galaxies close to the X-ray emission position. Furthermore, the redshift histogram of the SDSS galaxies close to the X-ray emission position has a peak between 0.6 and 0.7.

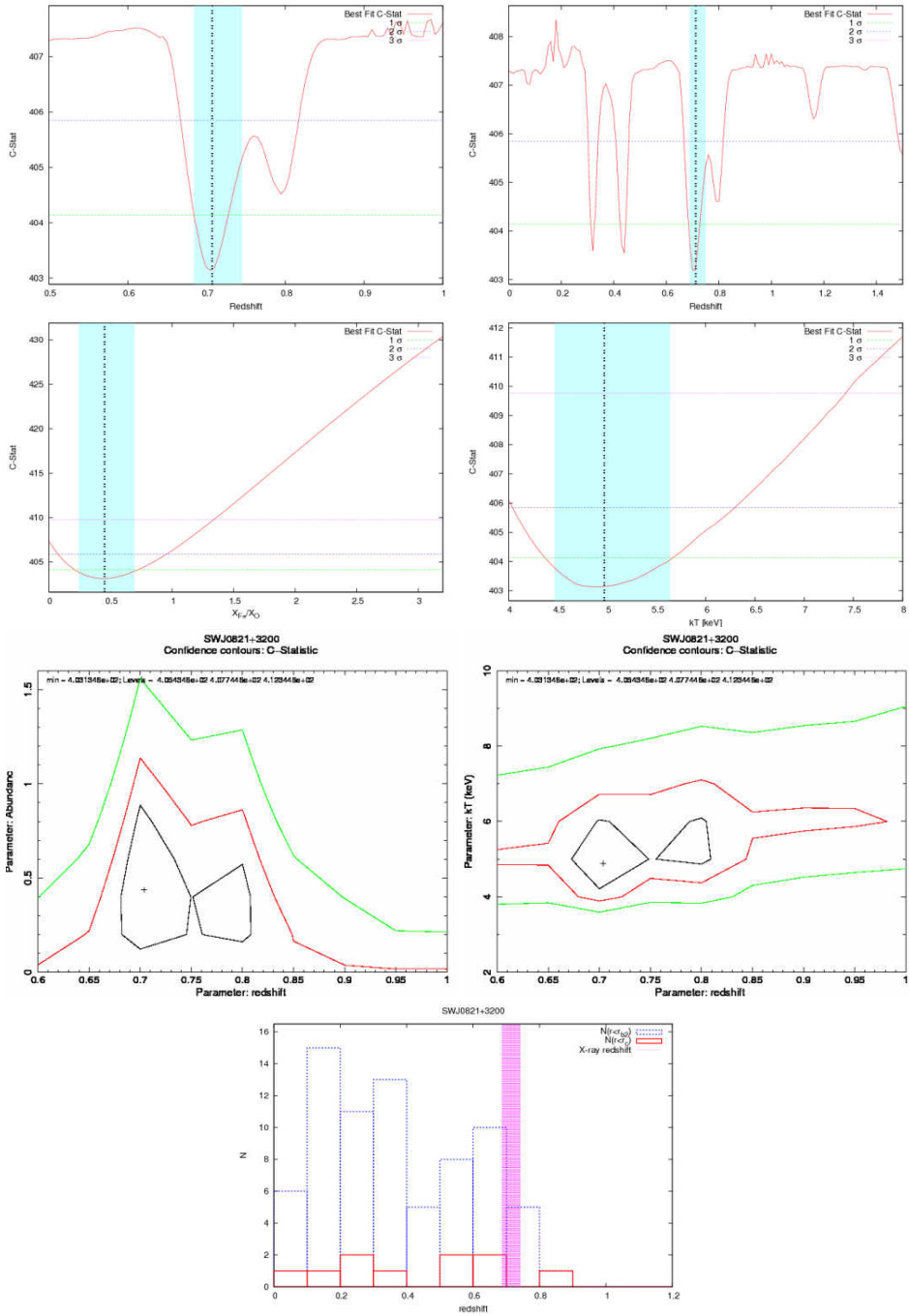

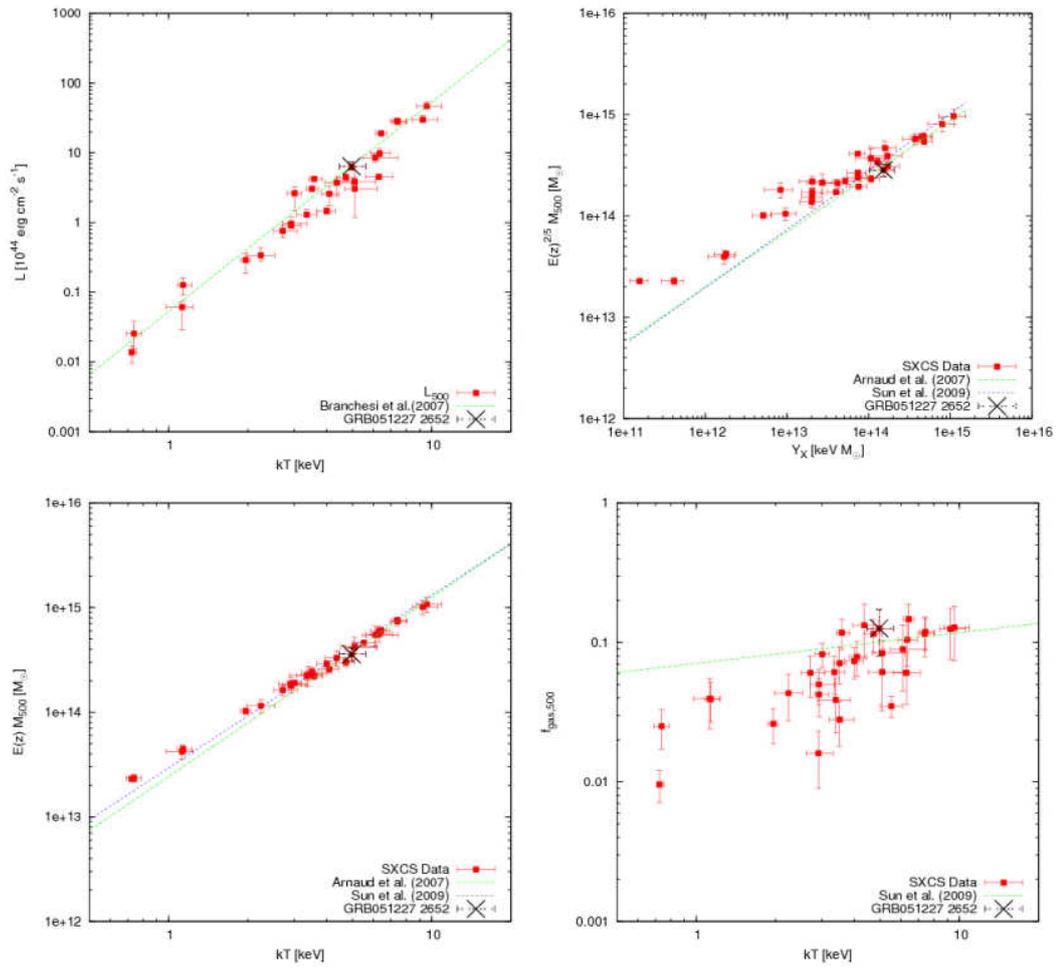

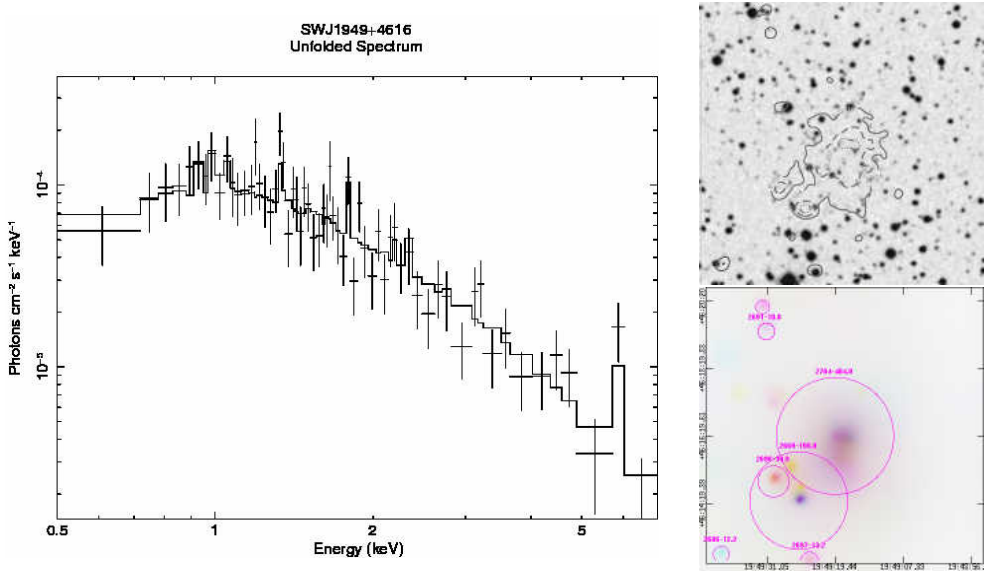

| Name | GRB | R.A. | Dec | Catalogue | Distance | Published $z$ |
|---|---|---|---|---|---|---|
| SWJ1949+4616 | GRB060105 | 297.330414 | 46.272861 | - | - | - |

| Expmap [s] | Net Counts | SNR | Flux [$10^{-13}$ erg/cm$^2$/s] | $N_H$ [$10^{22}$cm$^{-2}$] | Bkg rate [$10^{-3}$ cts/arcsec$^2$] | $r_{ext}$ [arcsec] |
|---|---|---|---|---|---|---|
| 117420 | 404±26 | 14.6 | 1.23±0.08 | 0.148 | 6.05 | 103.7 |

| $kT$ [keV] | $z$ | $X_{Fe}/X_\odot$ | $r_{ext}$ [kpc] | $r_{500}$ [kpc] | $L_{ext}$ [$10^{44}$ erg/s] | $L_{500}$ [$10^{44}$ erg/s] |
|---|---|---|---|---|---|---|
| $3.3^{+0.3}_{-0.3}$ | $0.113^{+0.031}_{-0.005}$ | $0.61^{+0.29}_{-0.19}$ | 375±39 | 837±88 | $1.06^{+0.21}_{-0.05}$ | $1.30^{+0.25}_{-0.06}$ |

| $M_{500}$ [$10^{13} M_\odot$] | $M_{gas,500}$ [$10^{13} M_\odot$] | $f_{gas,500}$ |
|---|---|---|
| 20.25±2.01 | 1.24±0.26 | 0.061±0.007 |

The K-$\alpha$ iron emission line is clear. However the redshift determination is is noisy. $L - T$ and $M - T$ are in agreement with the best fit of the two scaling laws as obtained by previous works on X-ray selected clusters. Instead the $M_{gas}$ is underestimated.

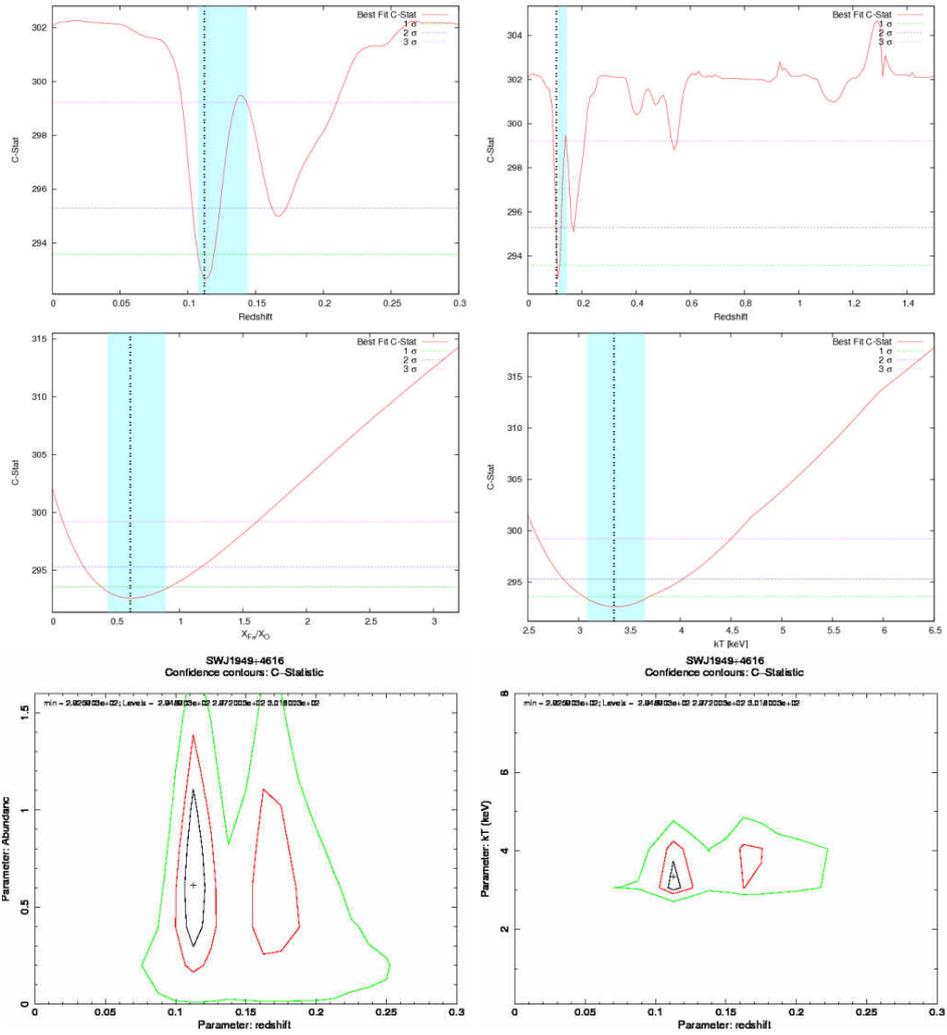

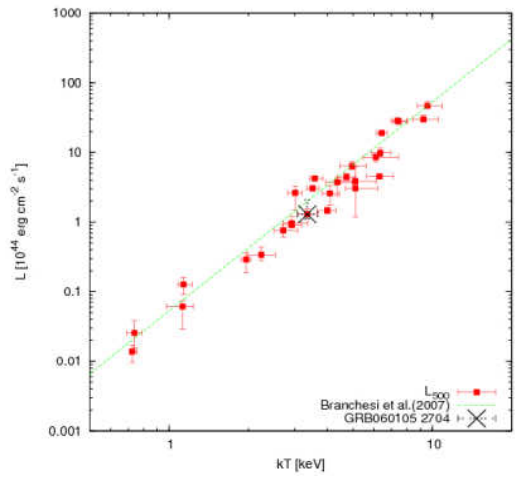

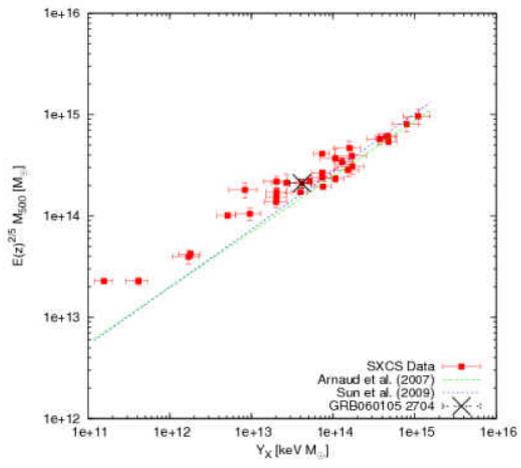

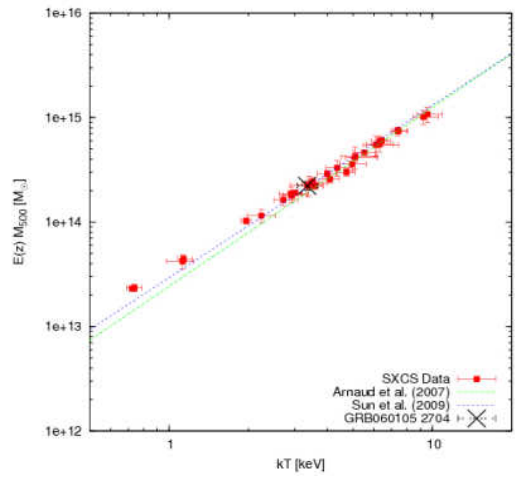

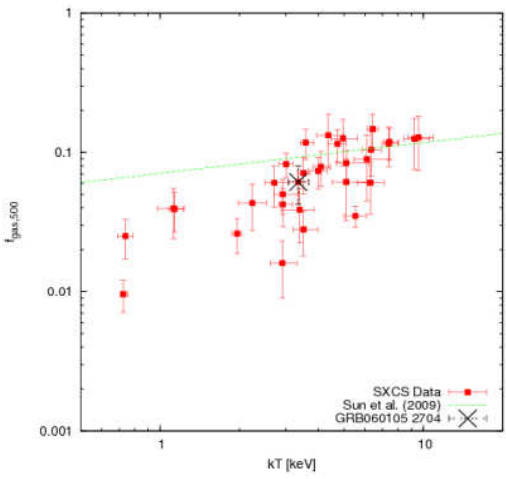

B.13 SWJ1406+2743

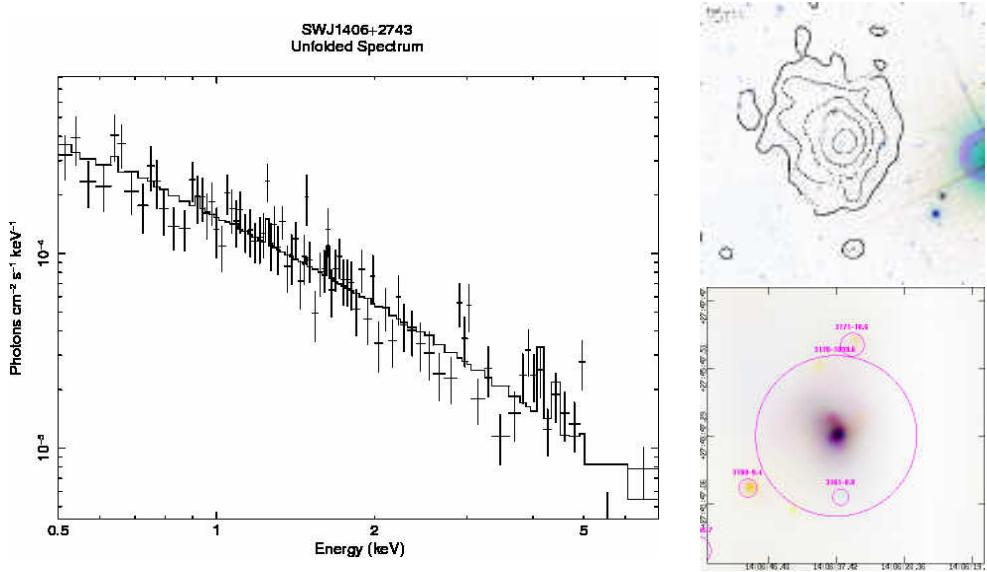

| Name | GRB | R.A. | Dec | Catalogue | Distance | Published $z$ |
|------|-----|------|-----|-----------|----------|---------------|
| SWJ1406+2743 | GRB060204B* | 211.654831 | 27.730812 | WGA | 1.045 | - |

| Expmap | Net Counts | SNR | Flux | $N_H$ | Bkg rate | $r_{ext}$ |
| [s] | | | [$10^{-13}$ erg/cm²/s] | [$10^{22}$cm⁻²] | [$10^{-3}$ cts/arcsec²] | [arcsec] |
|------|-----------|-----|------|-------|----------|-----------|
| 80171 | 1039±42 | 27.5 | 3.23±0.13 | 0.017 | 3.81 | 142.5 |

| $kT$ | $z$ | $X_{Fe}/X_\odot$ | $r_{ext}$ | $r_{500}$ | $L_{ext}$ | $L_{500}$ |
| [keV] | | | [kpc] | [kpc] | [$10^{44}$ erg/s] | [$10^{44}$ erg/s] |
|-------|-----|------------------|-----------|-----------|-----------|-----------|
| $9.2^{+1.3}_{-0.8}$ | $0.600^{+0.057}_{-0.008}$ | $0.74^{+0.31}_{-0.20}$ | 674±70 | 1033±108 | $26.98^{+4.11}_{-1.10}$ | $29.84^{+4.54}_{-1.22}$ |

| $M_{500}$ | $M_{gas,500}$ | $f_{gas,500}$ |
| [$10^{13}M_\odot$] | [$10^{13}M_\odot$] | |
|-----------|---------------|---------------|
| 69.00±10.87 | 8.64±2.08 | 0.124±0.010 |

The redshift is determinad with a significance of $\sim 2\sigma$. The scaling laws are in agreement with Branchesi et al. (2007), Arnaud et al. (2007) and Sun et al. (2009). Also the $M_{gas}$ is in agreement with previous works.

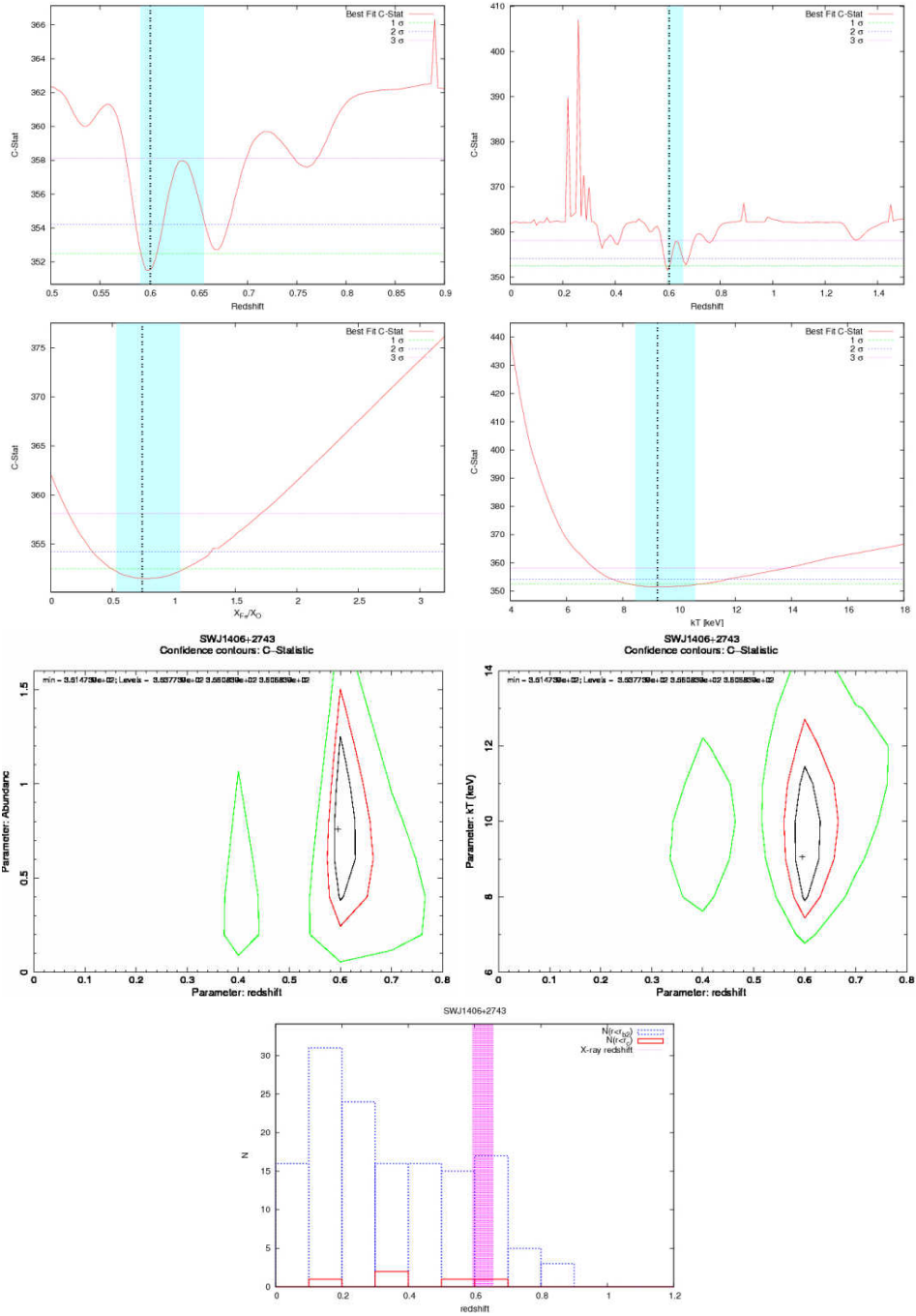

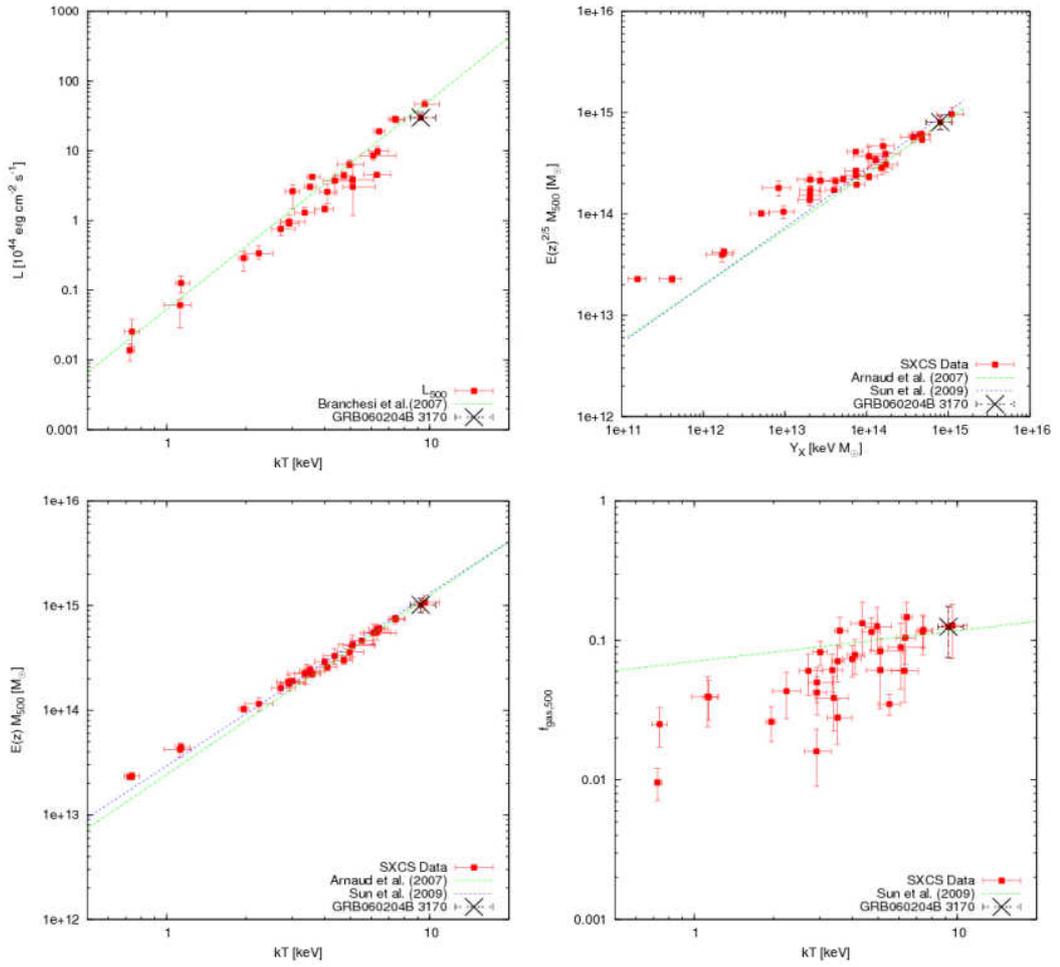

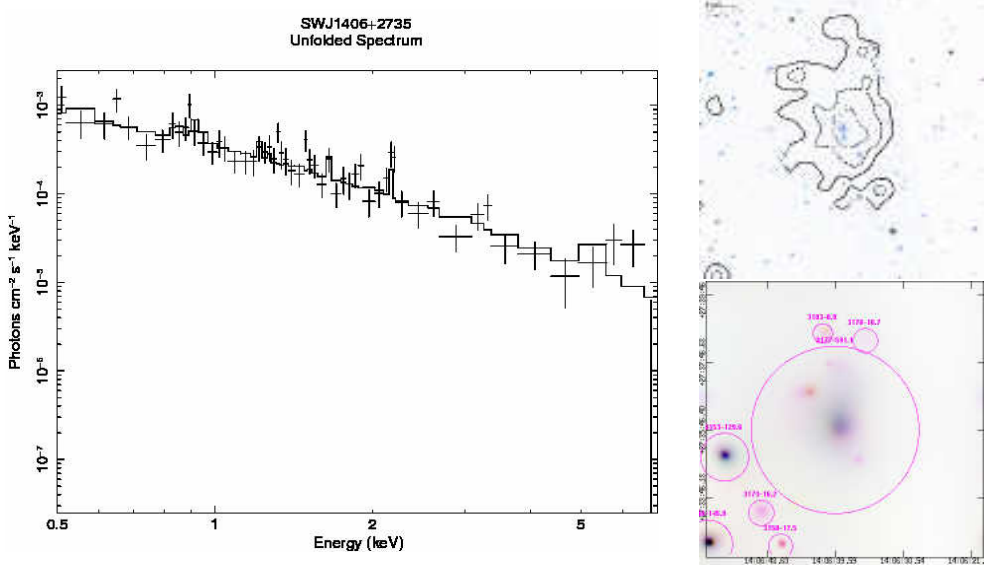

| Name | GRB | R.A. | Dec | Catalogue | Distance | Published $z$ |
|---|---|---|---|---|---|---|
| SWJ1406+2735 | GRB060204B* | 211.664337 | 27.597130 | MaxBCG | 0.161 | $0.25^{1,7}$ |

| Expmap [s] | Net Counts | SNR | Flux [$10^{-13}$ erg/cm²/s] | $N_H$ [$10^{22}$cm⁻²] | Bkg rate [$10^{-3}$ cts/arcsec²] | $r_{ext}$ [arcsec] |
|---|---|---|---|---|---|---|
| 78534 | 591±37 | 18.7 | 1.87±0.12 | 0.017 | 3.81 | 148.4 |

| $kT$ [keV] | $z$ | $X_{Fe}/X_\odot$ | $r_{ext}$ [kpc] | $r_{500}$ [kpc] | $L_{ext}$ [$10^{44}$ erg/s] | $L_{500}$ [$10^{44}$ erg/s] |
|---|---|---|---|---|---|---|
| $4.4^{+0.6}_{-0.4}$ | $0.216^{+0.012}_{-0.032}$ | $1.05^{+0.53}_{-0.36}$ | 520±117 | 939±212 | $3.21^{+0.55}_{-1.14}$ | $3.72^{+0.64}_{-1.32}$ |

| $M_{500}$ [$10^{13}M_\odot$] | $M_{gas,500}$ [$10^{13}M_\odot$] | $f_{gas,500}$ |
|---|---|---|
| 29.65±5.06 | 3.94±0.98 | 0.131±0.010 |

The redshift histogram of SDSS galaxies at the cluster position peaks at the X-ray redshift. The scaling laws are in agreement with Branchesi et al. (2007), Arnaud et al. (2007) and Sun et al. (2009). Also the $M_{gas}$ is in agreement with previous works.

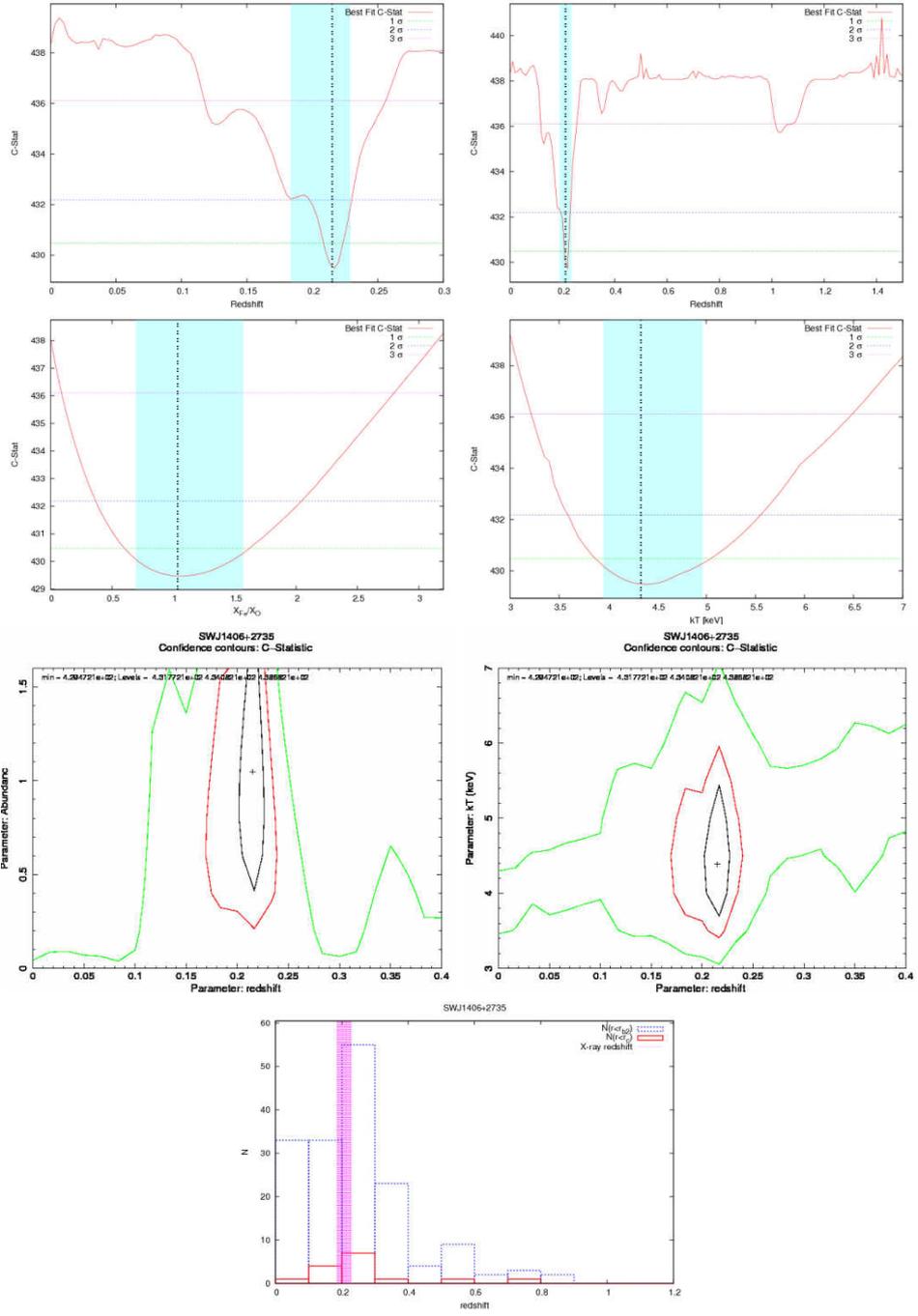

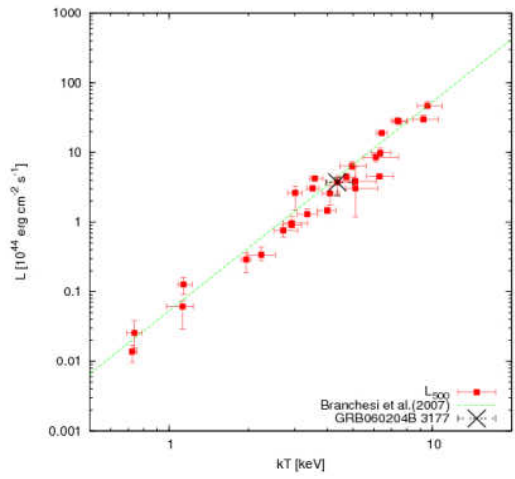
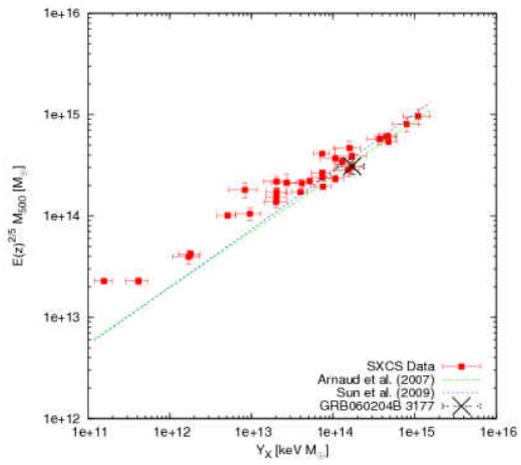

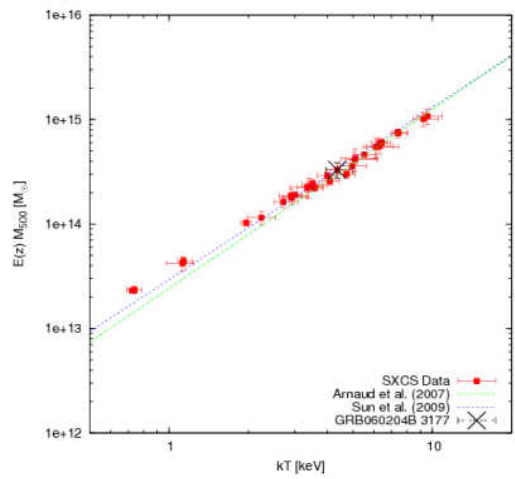
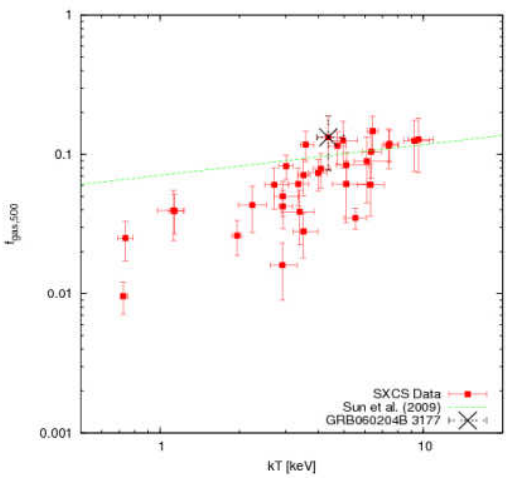

## B.15 SWJ1145+5953

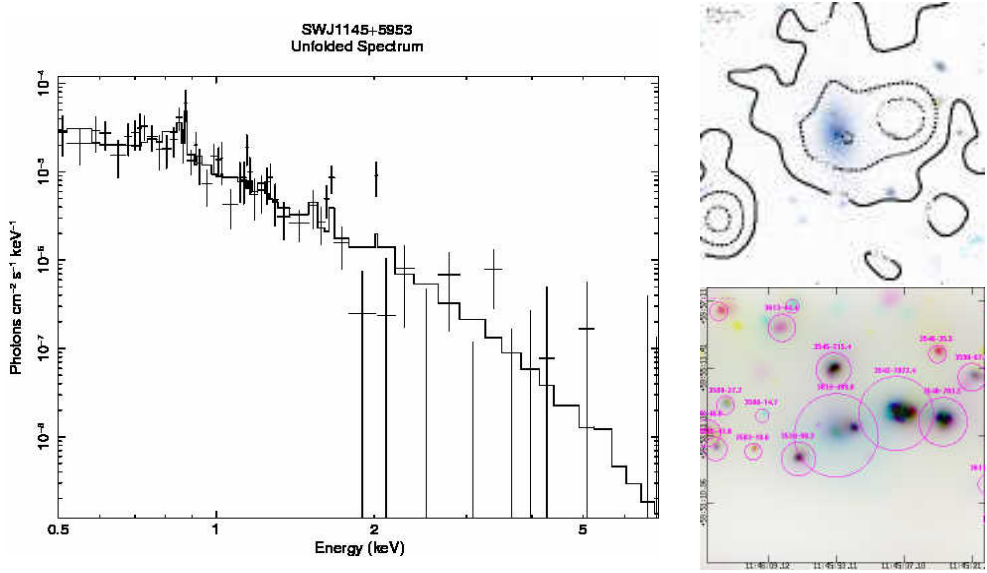

| Name | GRB | R.A. | Dec | Catalogue | Distance | Published $z$ |
|------|-----|------|-----|-----------|----------|---------------|
| SWJ1145+5953 | GRB060319* | 176.469711 | 59.887722 | 2MASX | 0.080 | 0.1463 |

| Expmap [s] | Net Counts | SNR | Flux [$10^{-13}$ erg/cm²/s] | $N_H$ [$10^{22}$cm⁻²] | Bkg rate [$10^{-3}$ cts/arcsec²] | $r_{ext}$ [arcsec] |
|-----------|-----------|-----|------|------|------|------|
| 479992 | 499±37 | 15.6 | 0.26±0.02 | 0.015 | 19.48 | 74.8 |

| $kT$ [keV] | $z$ | $X_{Fe}/X_\odot$ | $r_{ext}$ [kpc] | $r_{500}$ [kpc] | $L_{ext}$ [$10^{44}$ erg/s] | $L_{500}$ [$10^{44}$ erg/s] |
|-----------|-----|------|------|------|------|------|
| $1.1^{+0.1}_{-0.2}$ | $0.200^{+0.014}_{-0.045}$ | $0.18^{+0.11}_{-0.06}$ | 246±49 | 477±96 | $0.05^{+0.01}_{-0.03}$ | $0.06^{+0.01}_{-0.03}$ |

| $M_{500}$ [$10^{13}M_\odot$] | $M_{gas,500}$ [$10^{13}M_\odot$] | $f_{gas,500}$ |
|------|------|------|
| 3.81±0.58 | 0.15±0.04 | 0.039±0.003 |

The spectrum is quite noisy at energies greater than 2.0 keV. The temperature is low ∼ 1.0 keV and the redshift is determined from L-shell emission lines. The redshift of 2MASXJ11455301+5953194 is 0.146273, in agreement with the X-ray redsift. $L - T$ and $M - T$ are in agreement with previuos works by Branchesi et al. (2007), Arnaud et al. (2007) and Sun et al. (2009). The $M_{gas}$ is underestimated.

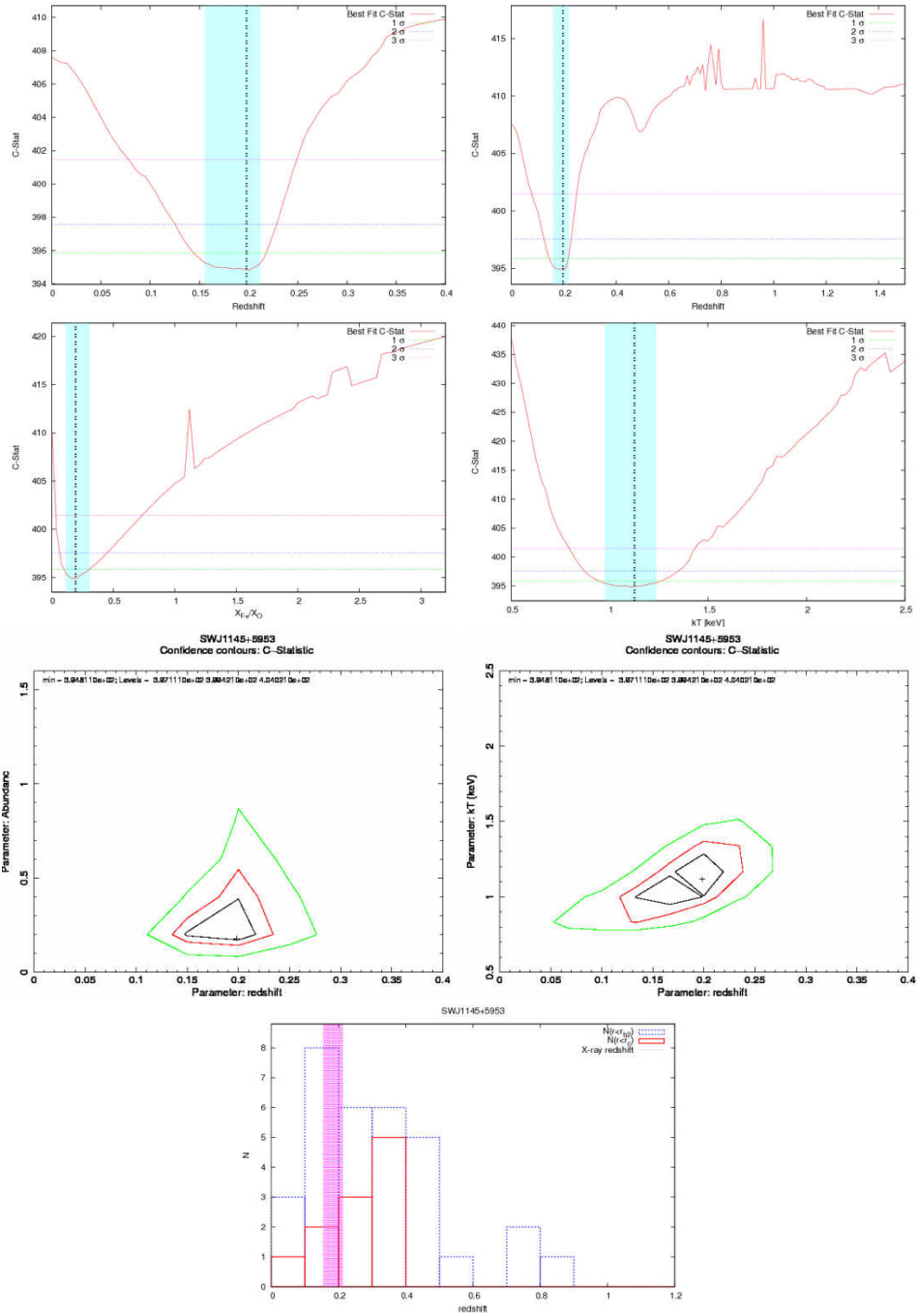

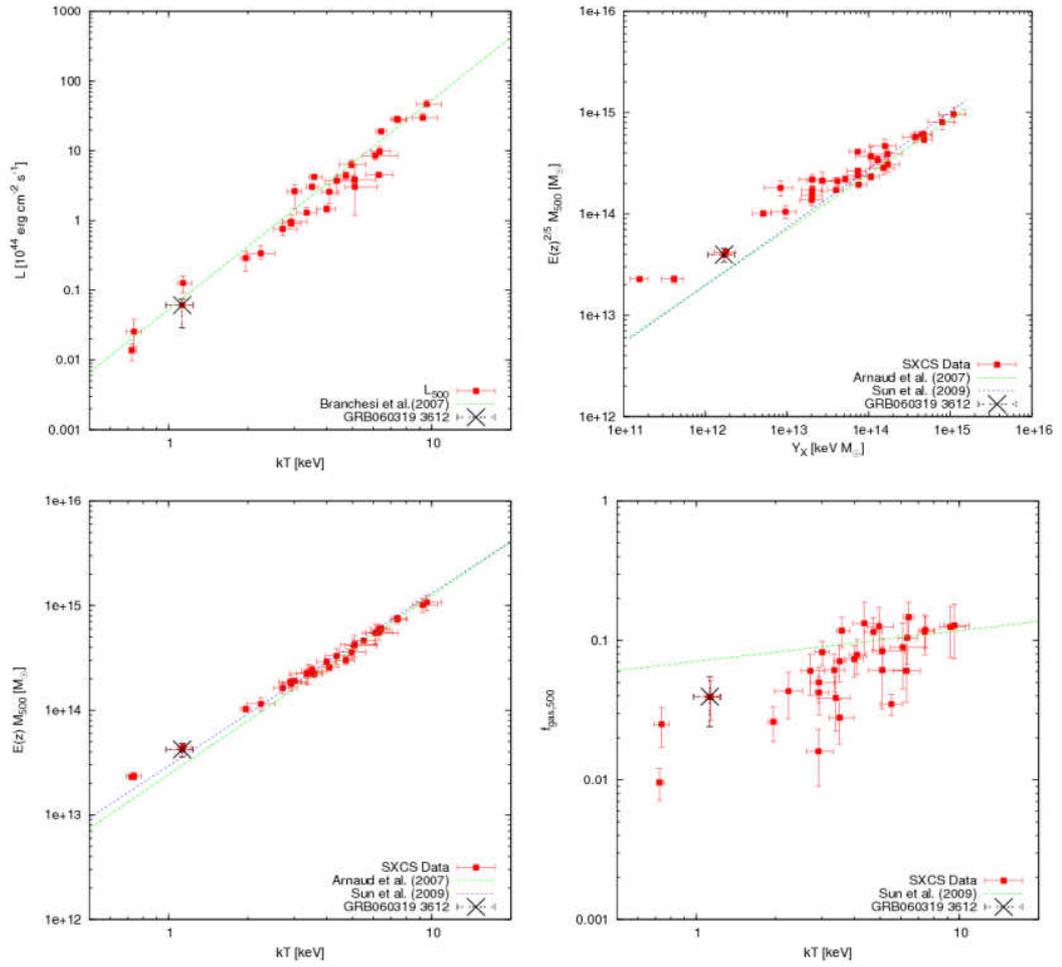

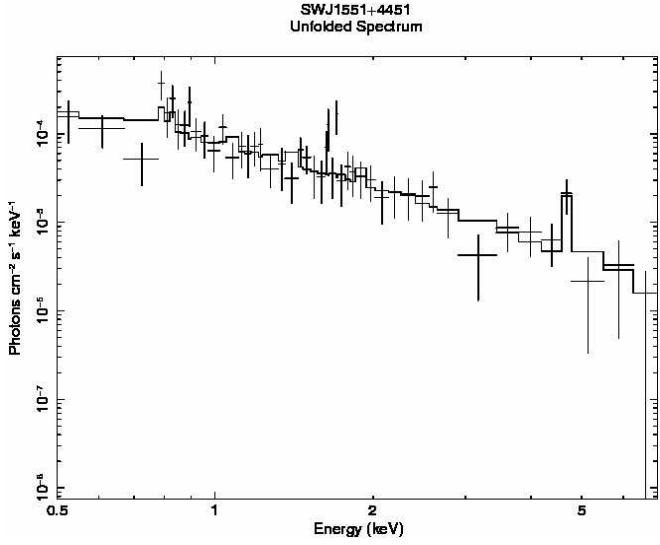

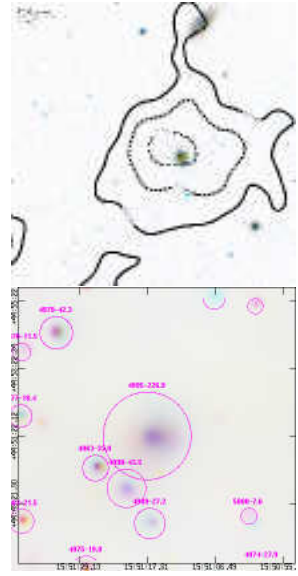

| Name | GRB | R.A. | Dec | Catalogue | Distance | Published $z$ |
|------|-----|------|-----|-----------|----------|---------------|
| SWJ1551+4451 | GRB060904A* | 237.823303 | 44.857132 | - | - | - |

| Expmap [s] | Net Counts | SNR | Flux [$10^{-13}$ erg/cm$^2$/s] | $N_H$ [$10^{22}$cm$^{-2}$] | Bkg rate [$10^{-3}$ cts/arcsec$^2$] | $r_{ext}$ [arcsec] |
|------------|------------|-----|-------------------------------|---------------------------|-------------------------------------|--------------------|
| 147518 | 226±20 | 11.2 | 0.38±0.03 | 0.013 | 5.06 | 78.2 |

| $kT$ [keV] | $z$ | $X_{Fe}/X_\odot$ | $r_{ext}$ [kpc] | $r_{500}$ [kpc] | $L_{ext}$ [$10^{44}$ erg/s] | $L_{500}$ [$10^{44}$ erg/s] |
|------------|-----|------------------|-----------------|-----------------|------------------------------|------------------------------|
| $5.1^{+1.0}_{-0.6}$ | $0.401^{+0.005}_{-0.006}$ | $1.83^{+1.20}_{-0.66}$ | 420±56 | 917±123 | $3.13^{+0.34}_{-0.35}$ | $3.84^{+0.41}_{-0.43}$ |

| $M_{500}$ [$10^{13}M_\odot$] | $M_{gas,500}$ [$10^{13}M_\odot$] | $f_{gas,500}$ |
|------------------------------|----------------------------------|---------------|
| 34.19±8.15 | 2.09±0.49 | 0.063±0.016 |

The redshift is determined very well since the K-$\alpha$ emission line is detected with high significance. $L-T$ and $M-T$ are in agreement with previous works. $M_{gas}$ is underestimated.

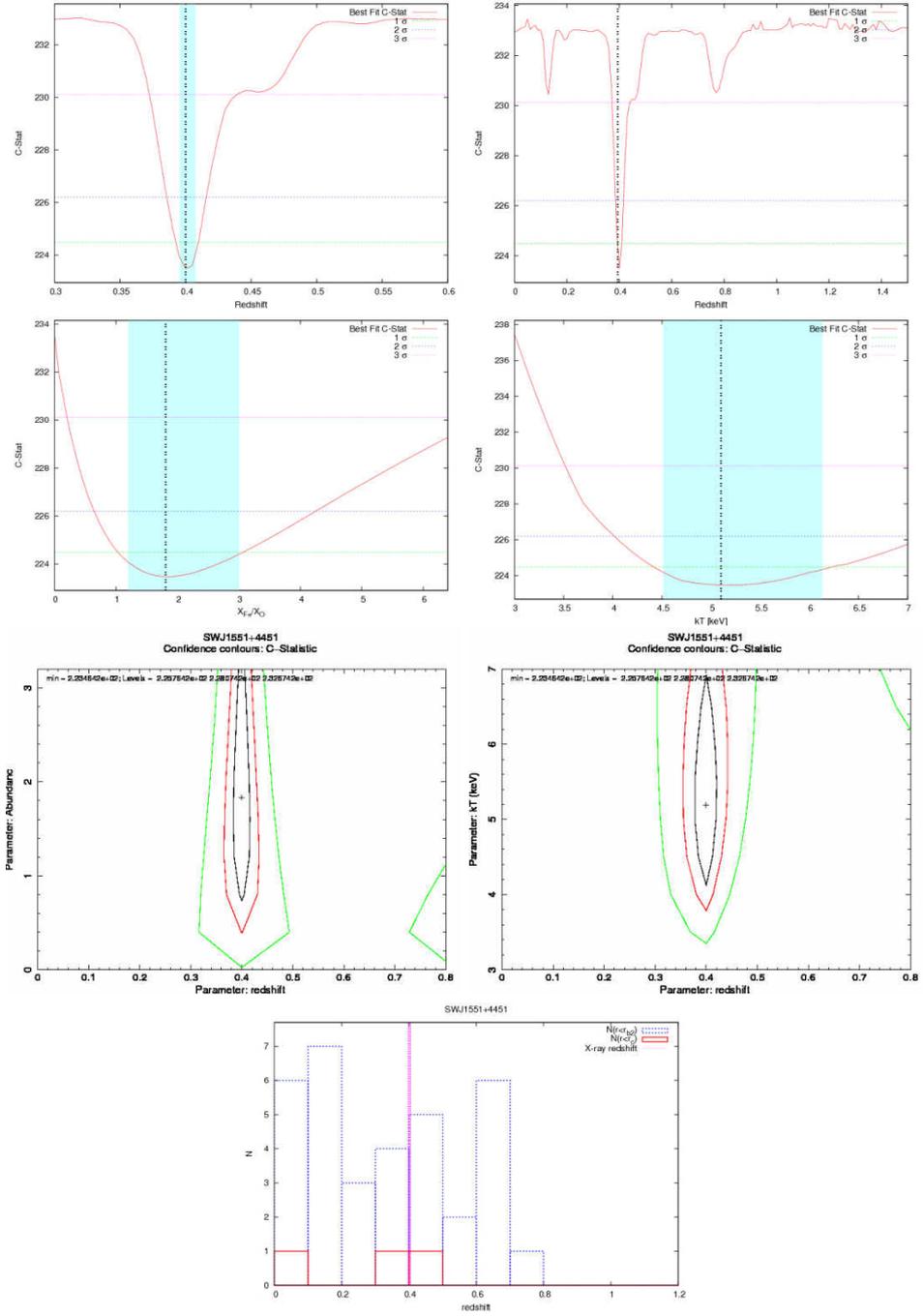

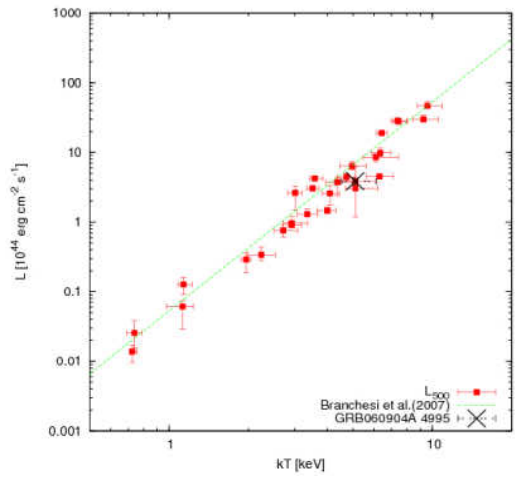
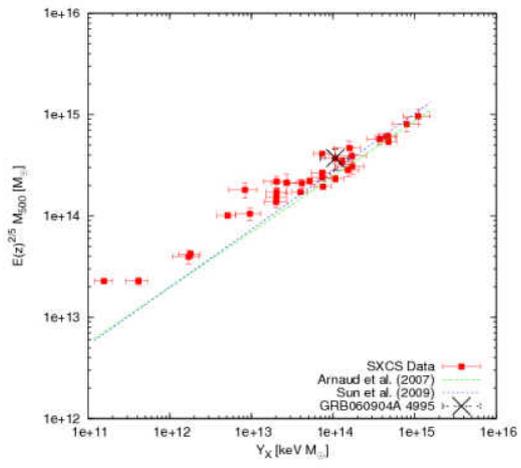

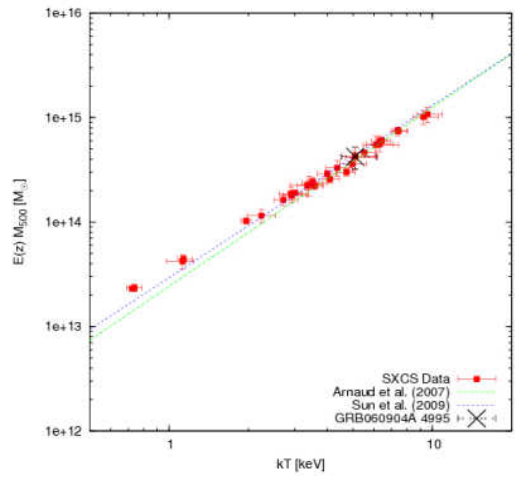
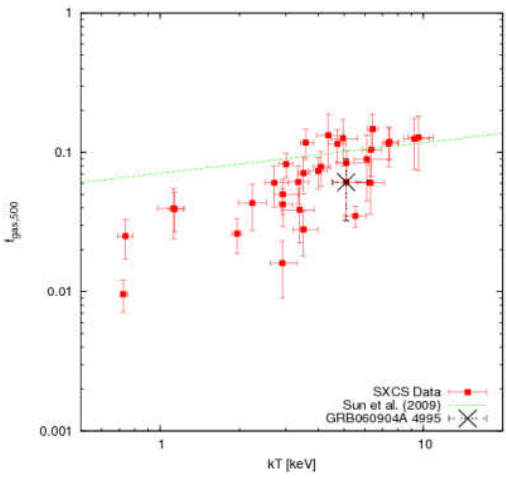

## B.17 SWJ0352-0043

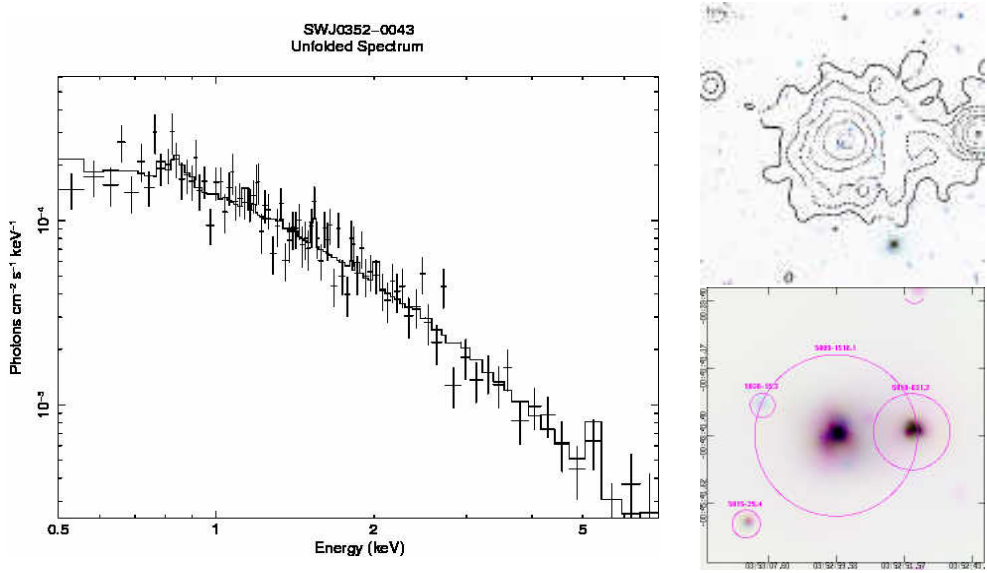

| Name | GRB | R.A. | Dec | Catalogue | Distance | Published $z$ |
|------|-----|------|-----|-----------|----------|---------------|
| SWJ0352-0043 | GRB060904B* | 58.247223 | -0.727059 | SDSS | 0.015 | 0.335 |

| Expmap [s] | Net Counts | SNR | Flux [$10^{-13}$ erg/cm$^2$/s] | $N_H$ [$10^{22}$cm$^{-2}$] | Bkg rate [$10^{-3}$ cts/arcsec$^2$] | $r_{ext}$ [arcsec] |
|------------|-----------|-----|--------|------|----------|--------|
| 106949 | 1516±47 | 34.0 | 4.65±0.15 | 0.114 | 4.07 | 144.1 |

| $kT$ [keV] | $z$ | $X_{Fe}/X_\odot$ | $r_{ext}$ [kpc] | $r_{500}$ [kpc] | $L_{ext}$ [$10^{44}$ erg/s] | $L_{500}$ [$10^{44}$ erg/s] |
|------------|-----|------------------|-----------------|-----------------|-----------------------------|------------------------------|
| $3.5^{+0.2}_{-0.2}$ | $0.311^{+0.008}_{-0.008}$ | $0.37^{+0.11}_{-0.09}$ | 657±60 | 807±74 | $2.96^{+0.23}_{-0.24}$ | $3.05^{+0.24}_{-0.24}$ |

| $M_{500}$ [$10^{13}M_\odot$] | $M_{gas,500}$ [$10^{13}M_\odot$] | $f_{gas,500}$ |
|------------------------------|----------------------------------|---------------|
| 20.74±1.50 | 1.47±0.32 | 0.070±0.010 |

The X-ray redshift is very well determined and in good agreement with the redshift histogram of SDSS galaxies. The scaling relation between luminosity, total mass and temperature are in agreement with the best fit found for X-ray selected clusters by Branchesi et al. (2007), Arnaud et al. (2007) and Sun et al. (2009). Also the $M_{gas}$ is in agreement.

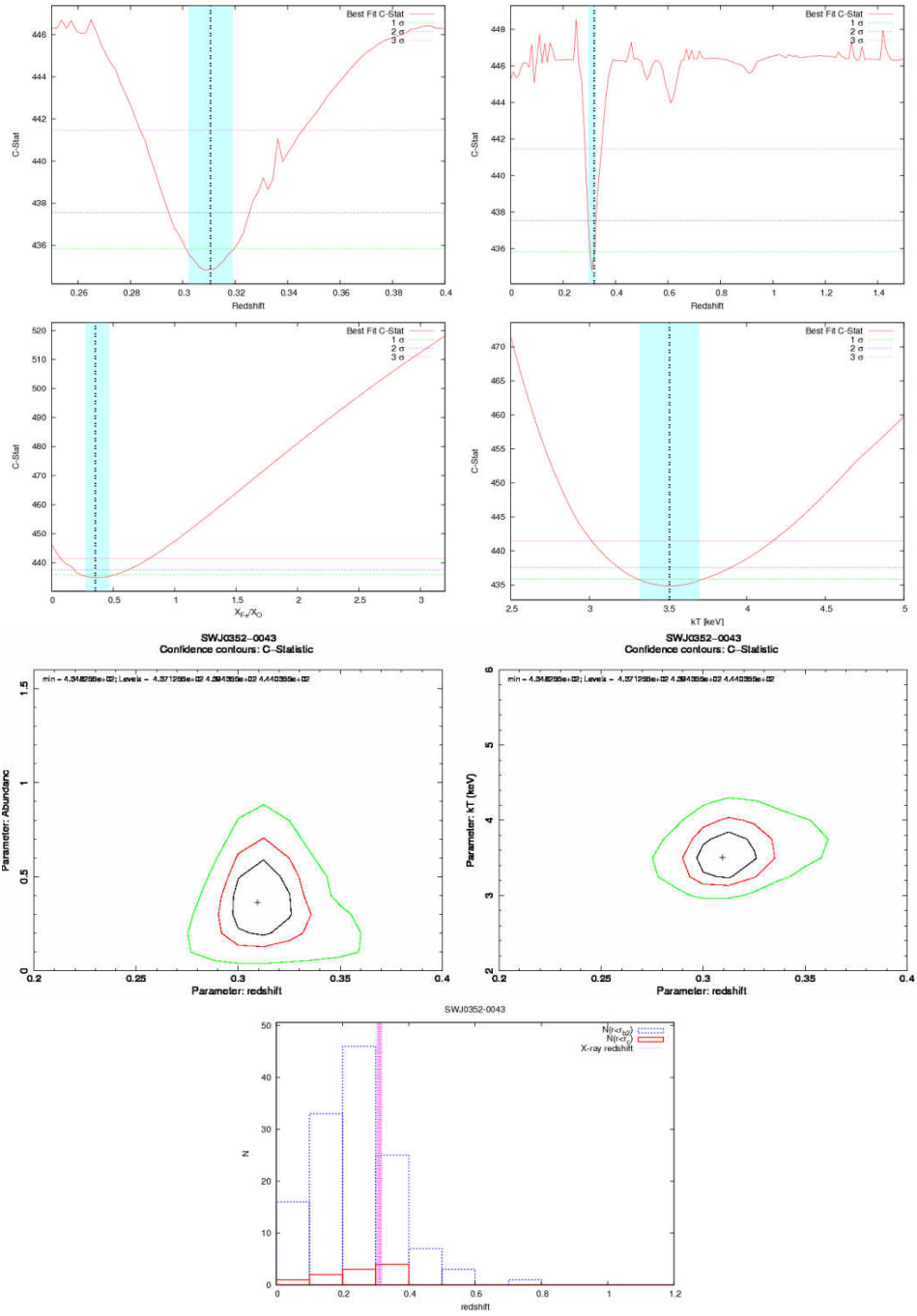

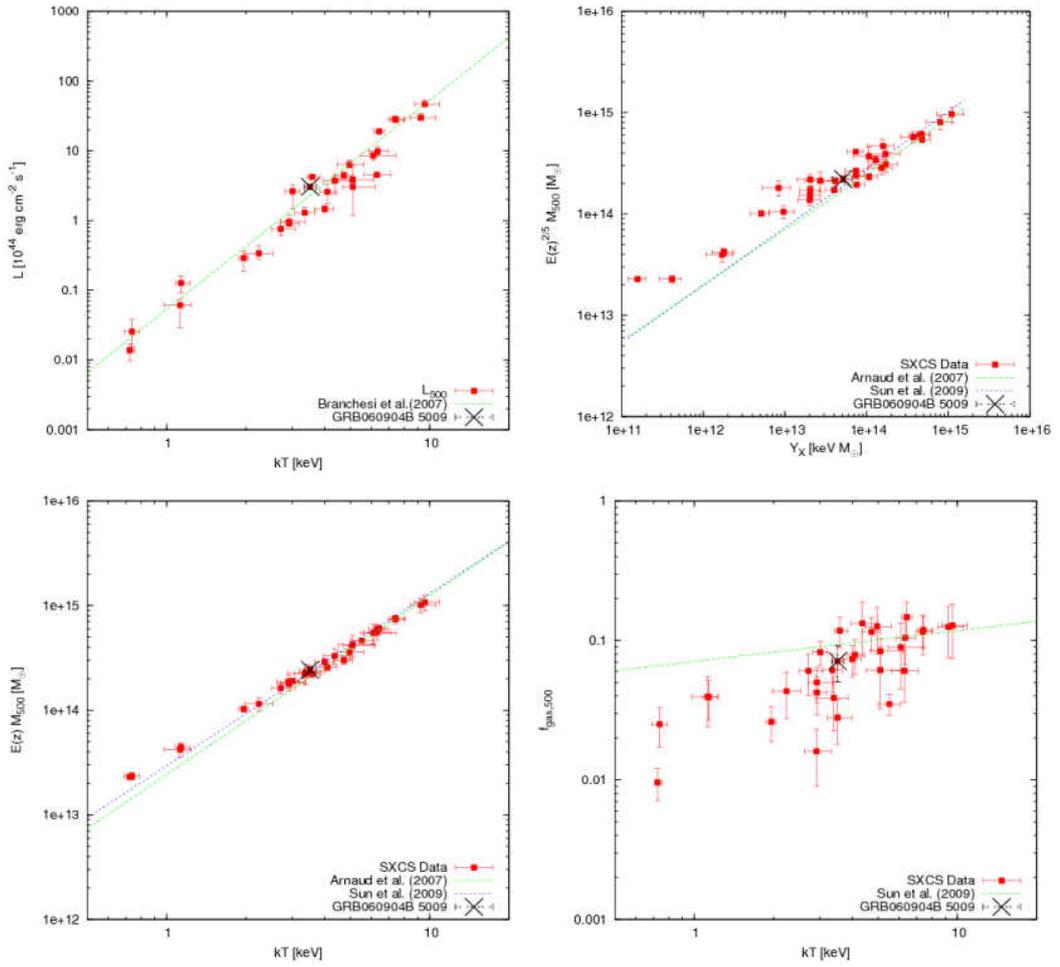

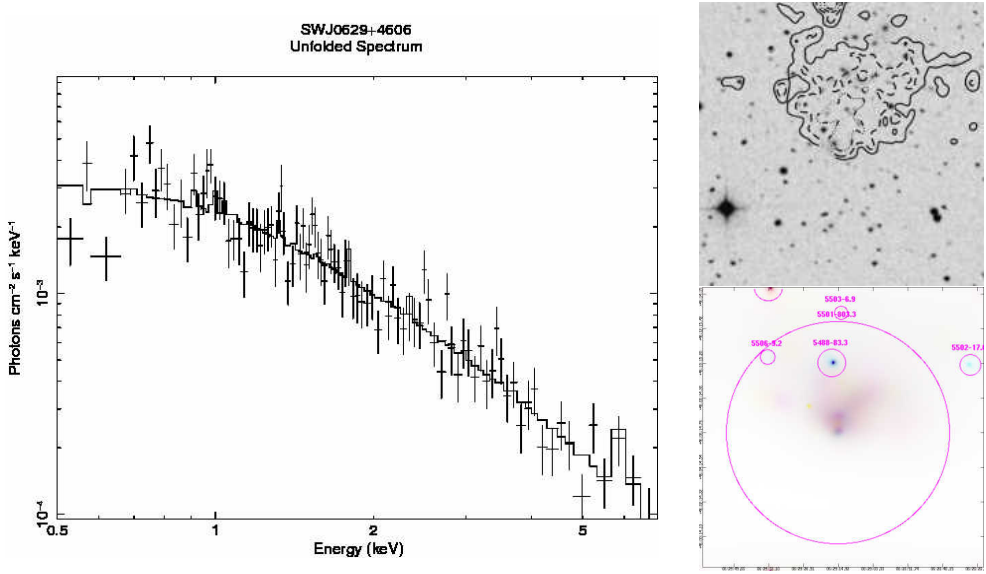

| Name | GRB | R.A. | Dec | Catalogue | Distance | Published $z$ |
|------|-----|------|-----|-----------|----------|---------------|
| SWJ0629+4607 | GRB061028 | 97.311806 | 46.120747 | CIZA ZwCl | 0.148 | 0.13 |

| Expmap | Net Counts | SNR | Flux | $N_H$ | Bkg rate | $r_{ext}$ |
|--------|-----------|-----|------|-------|----------|-----------|
| [s] | | | [$10^{-13}$ erg/cm$^2$/s] | [$10^{22}$cm$^{-2}$] | [$10^{-3}$ cts/arcsec$^2$] | [arcsec] |
| 7220 | 803±29 | 25.3 | 38.09±1.38 | 0.131 | 1.55 | 379.6 |

| $kT$ | $z$ | $X_{Fe}/X_\odot$ | $r_{ext}$ | $r_{500}$ | $L_{ext}$ | $L_{500}$ |
|------|-----|------------------|-----------|-----------|-----------|-----------|
| [keV] | | | [kpc] | [kpc] | [$10^{44}$ erg/s] | [$10^{44}$ erg/s] |
| $6.3^{+0.6}_{-0.5}$ | $0.113^{+0.008}_{-0.005}$ | $0.53^{+0.22}_{-0.16}$ | 393±11 | 1209±36 | $8.23^{+1.34}_{-0.97}$ | $9.86^{+1.61}_{-1.16}$ |

| $M_{500}$ | $M_{gas,500}$ | $f_{gas,500}$ |
|-----------|---------------|---------------|
| [$10^{13}M_\odot$] | [$10^{13}M_\odot$] | |
| 56.59±6.81 | 5.91±1.36 | 0.103±0.012 |

The redshift well determined and it is in agreement with ZwCl. The scaling laws and also the $M_{gas}$ are in agreement with previous works on X-ray selected clusters.

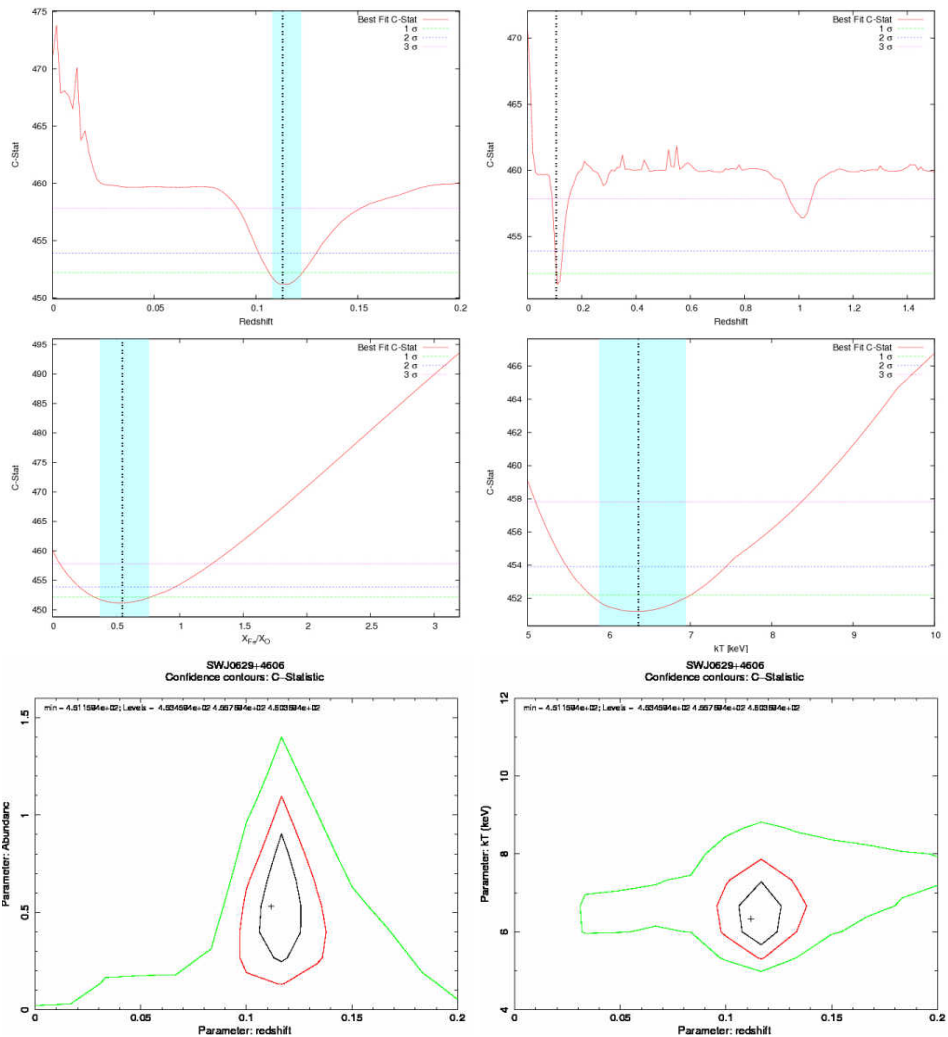

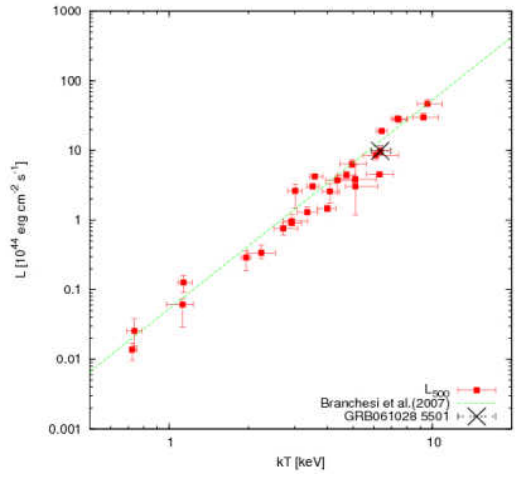
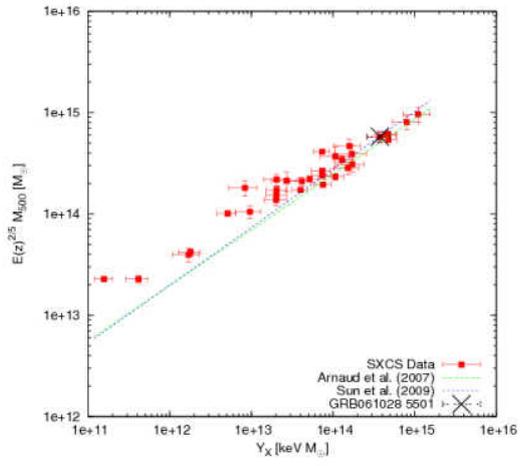
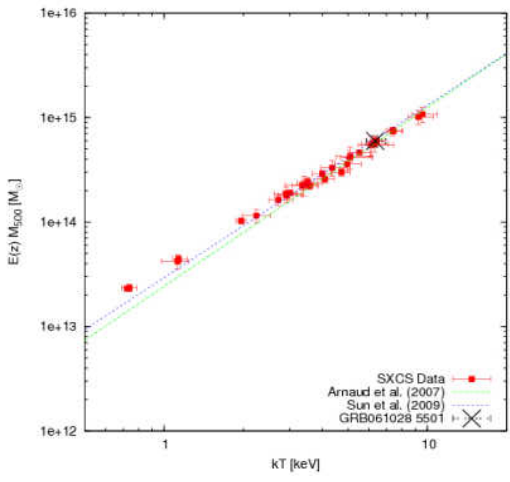
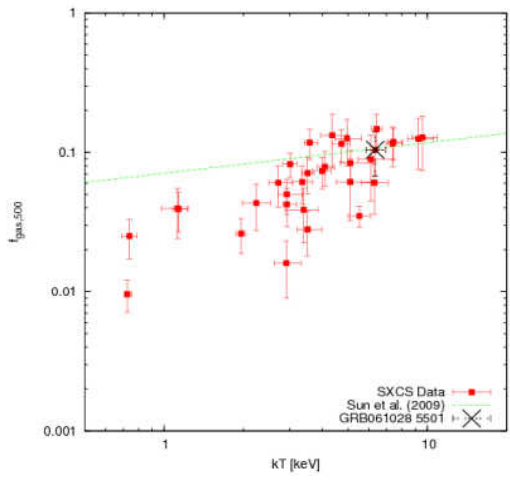

B.19 SWJ0948-1316

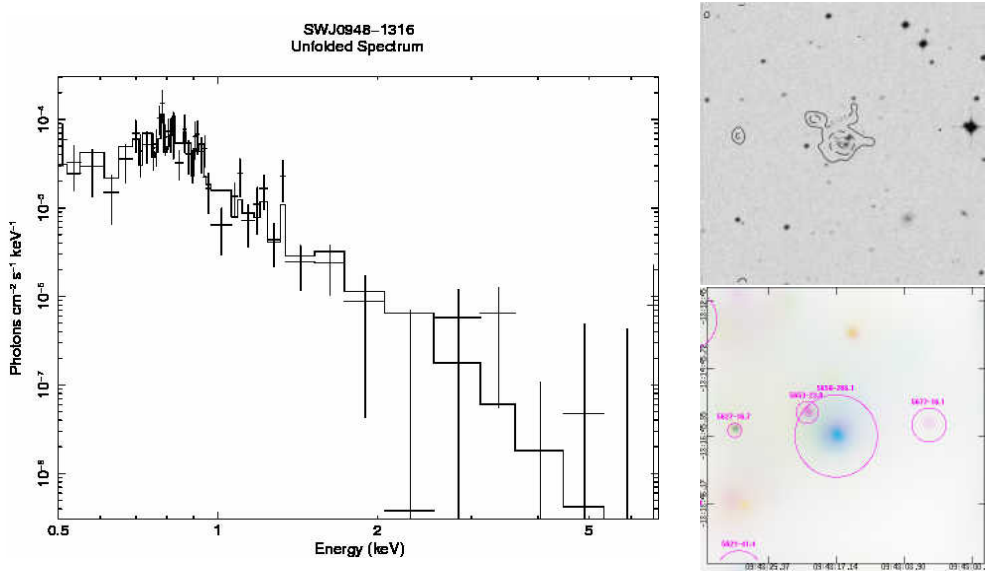

| Name | GRB | R.A. | Dec | Catalogue | Distance | Published $z$ |
|------|-----|------|-----|-----------|----------|---------------|
| SWJ0948-1316 | GRB061121 | 147.070740 | -13.278816 | - | - | - |

| Expmap [s] | Net Counts | SNR | Flux [$10^{-13}$ erg/cm$^2$/s] | $N_H$ [$10^{22}$cm$^{-2}$] | Bkg rate [$10^{-3}$ cts/arcsec$^2$] | $r_{ext}$ [arcsec] |
|------------|------------|-----|------|-----|-----|------|
| 160740 | 266±26 | 12.2 | 0.44±0.04 | 0.040 | 10.01 | 73.7 |

| $kT$ [keV] | $z$ | $X_{Fe}/X_\odot$ | $r_{ext}$ [kpc] | $r_{500}$ [kpc] | $L_{ext}$ [$10^{44}$ erg/s] | $L_{500}$ [$10^{44}$ erg/s] |
|------------|-----|------------------|-----------------|-----------------|------|------|
| $0.7^{+0.0}_{-0.0}$ | $0.108^{+0.022}_{-0.017}$ | $0.36^{+0.46}_{-0.11}$ | 147±49 | 412±139 | $0.02^{+0.01}_{-0.01}$ | $0.03^{+0.01}_{-0.01}$ |

| $M_{500}$ [$10^{13}M_\odot$] | $M_{gas,500}$ [$10^{13}M_\odot$] | $f_{gas,500}$ |
|------------------------------|----------------------------------|---------------|
| 2.23±0.19 | 0.06±0.01 | 0.025±0.004 |

The temperature is low < 1.0 keV and the redshift is determined from the L-shell emission lines. The scaling laws are in agreement with previous works about X-ray selected clusters. The $M_{gas}$ is underestimated.

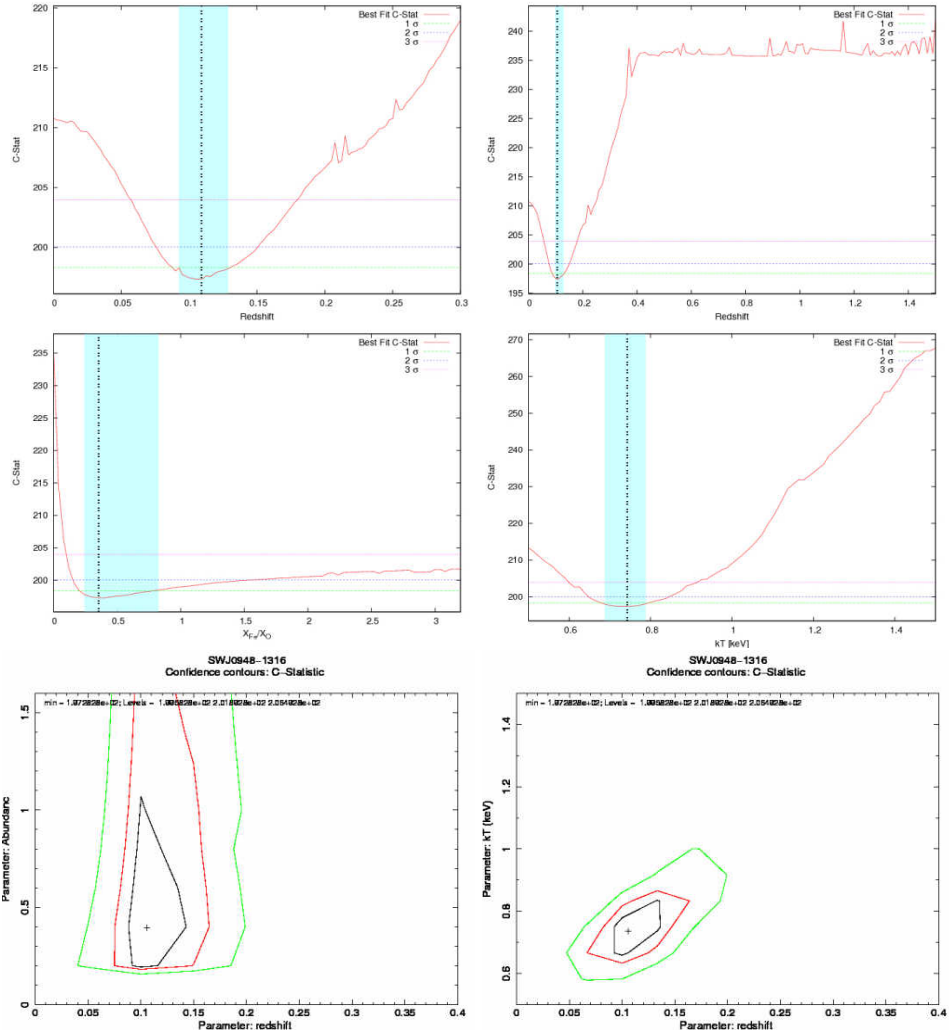

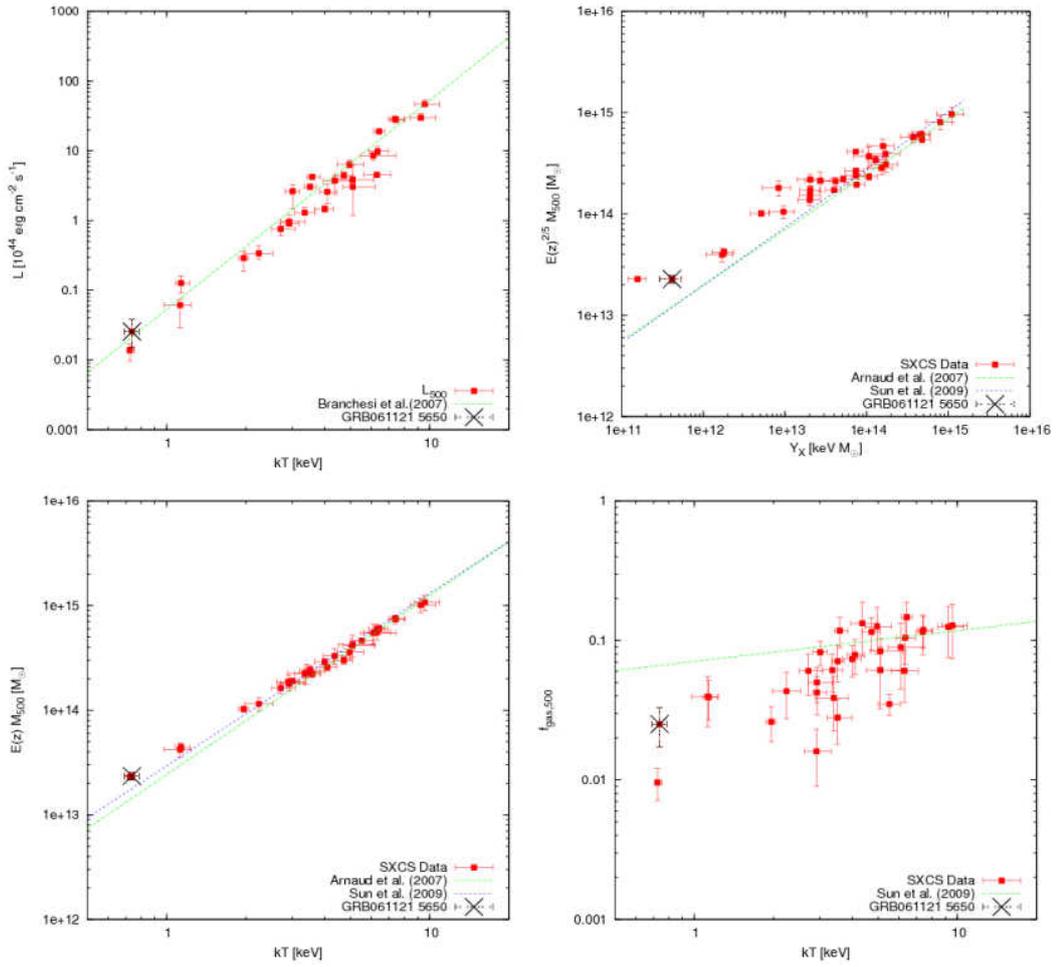

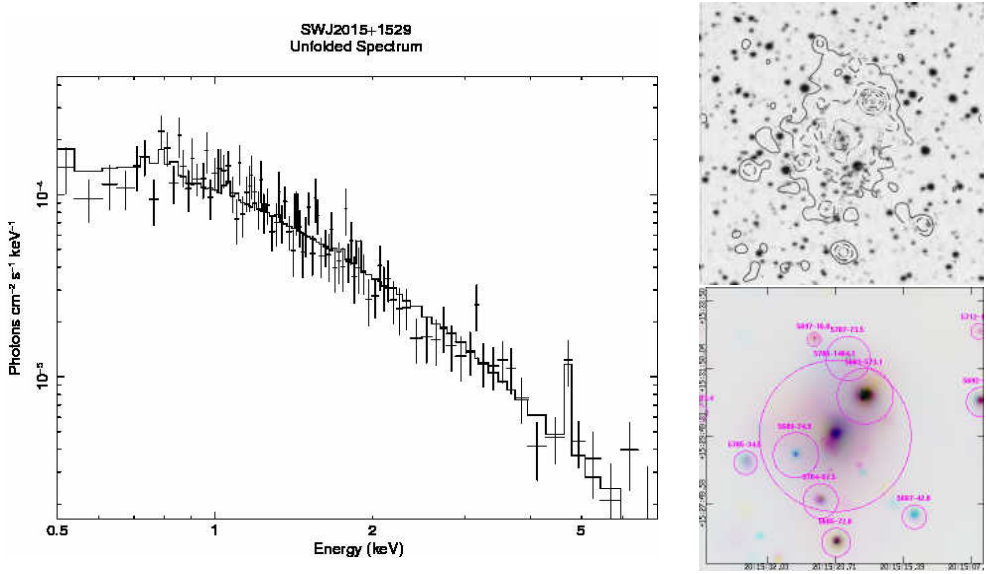

| Name | GRB | R.A. | Dec | Catalogue | Distance | Published $z$ |
|------|-----|------|-----|-----------|----------|---------------|
| SWJ2015+1529 | GRB061122 | 303.848236 | 15.497838 | - | - | - |

| Expmap [s] | Net Counts | SNR | Flux [$10^{-13}$ erg/cm²/s] | $N_H$ [$10^{22}$cm⁻²] | Bkg rate [$10^{-3}$ cts/arcsec²] | $r_{ext}$ [arcsec] |
|------------|------------|-----|------|------|------|------|
| 169258 | 1484±56 | 28.9 | 2.95±0.11 | 0.125 | 10.42 | 134.6 |

| $kT$ [keV] | $z$ | $X_{Fe}/X_\odot$ | $r_{ext}$ [kpc] | $r_{500}$ [kpc] | $L_{ext}$ [$10^{44}$ erg/s] | $L_{500}$ [$10^{44}$ erg/s] |
|------------|-----|------------------|--------|--------|------|------|
| $3.6^{+0.3}_{-0.2}$ | $0.403^{+0.007}_{-0.008}$ | $0.32^{+0.12}_{-0.08}$ | 726±309 | 743±316 | $4.19^{+0.28}_{-0.30}$ | $4.22^{+0.28}_{-0.30}$ |

| $M_{500}$ [$10^{13}M_\odot$] | $M_{gas,500}$ [$10^{13}M_\odot$] | $f_{gas,500}$ |
|------------|------------|------------|
| 17.91±1.10 | 2.11±0.39 | 0.117±0.014 |

The K-$\alpha$ emission line is quite strong and therefore the redshift is well determined. The relation between luminosity, total mass, and gas mass are in agreement with the best fit found by other autors for X-ray selected clusters.

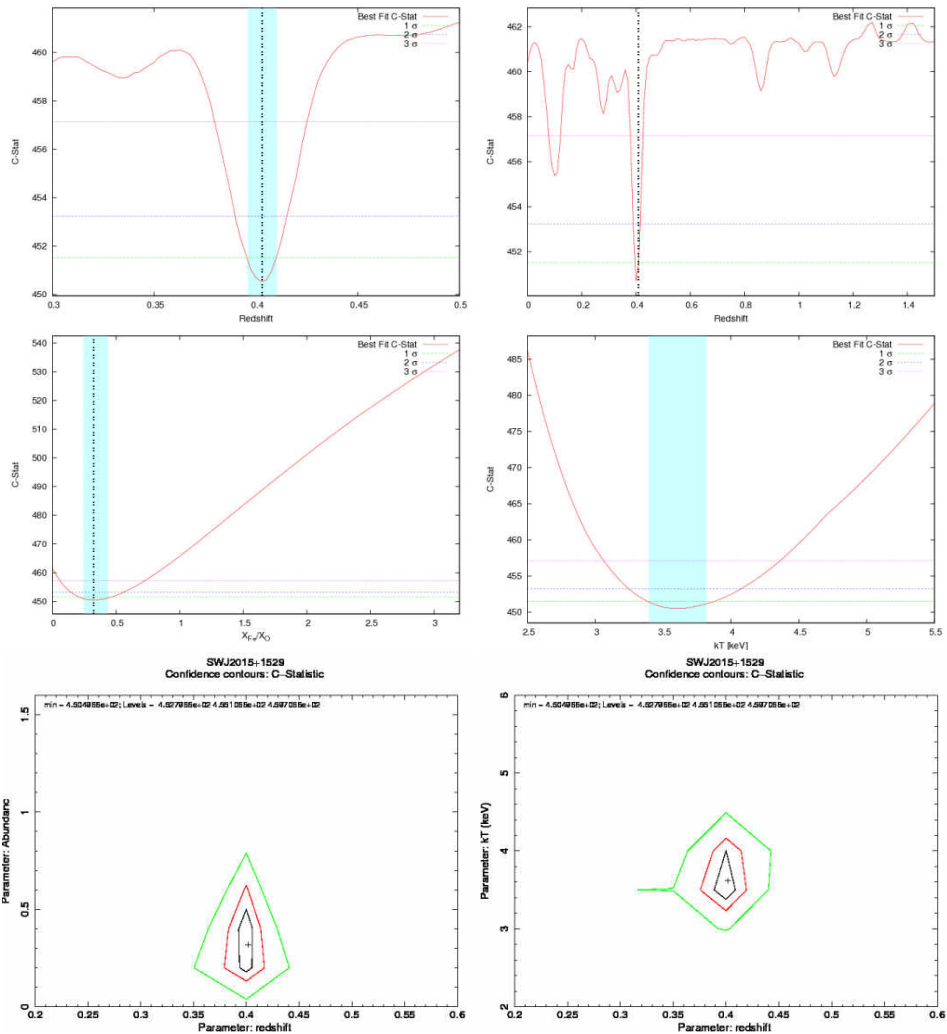

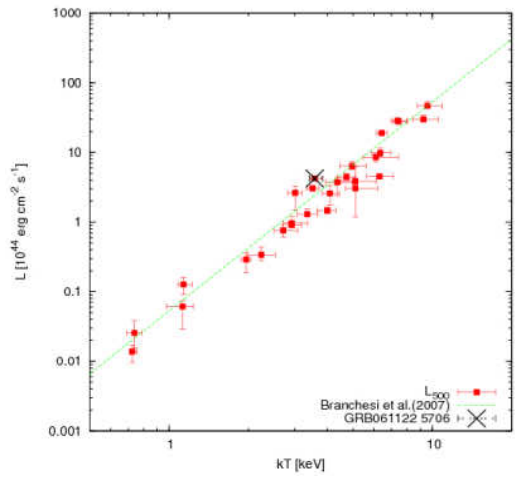
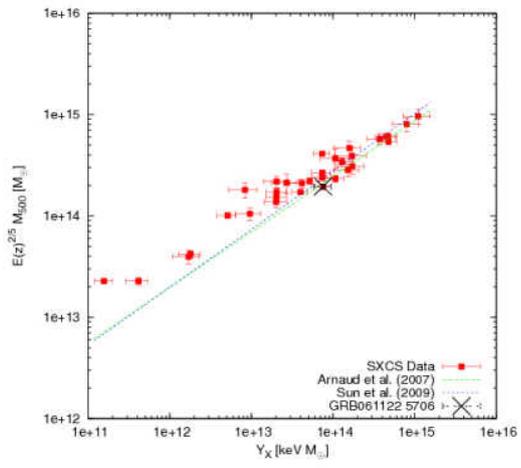
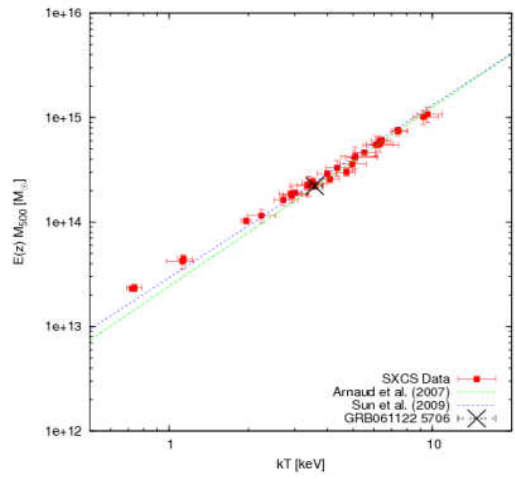
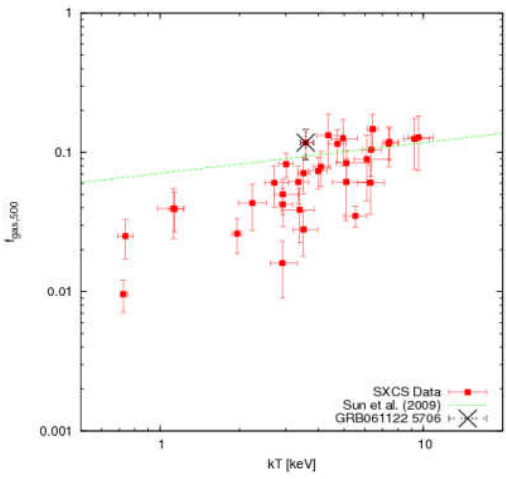

B.21 SWJ0003-5255

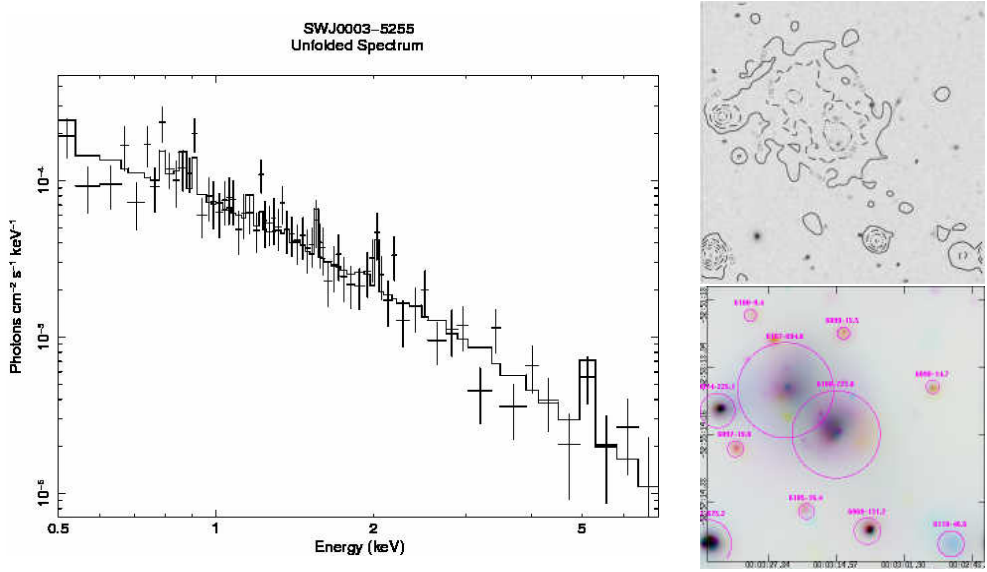

| Name | GRB | R.A. | Dec | Catalogue | Distance | Published z |
|------|-----|------|-----|-----------|----------|-------------|
| SWJ0003-5255 | GRB070110 | 0.808732 | -52.920033 | - | - | - |

| Expmap [s] | Net Counts | SNR | Flux [$10^{-13}$ erg/cm²/s] | $N_H$ [$10^{22}$cm$^{-2}$] | Bkg rate [$10^{-3}$ cts/arcsec²] | $r_{ext}$ [arcsec] |
|------------|-----------|-----|------|------|------|------|
| 316209 | 721±37 | 19.1 | 0.57±0.03 | 0.016 | 18.99 | 77.8 |

| $kT$ [keV] | z | $X_{Fe}/X_\odot$ | $r_{ext}$ [kpc] | $r_{500}$ [kpc] | $L_{ext}$ [$10^{44}$ erg/s] | $L_{500}$ [$10^{44}$ erg/s] |
|------------|---|------|------|------|------|------|
| $4.0^{+0.3}_{-0.3}$ | $0.281^{+0.007}_{-0.007}$ | $0.98^{+0.37}_{-0.27}$ | 330±18 | 870±47 | $1.10^{+0.11}_{-0.11}$ | $1.47^{+0.14}_{-0.14}$ |

| $M_{500}$ [$10^{13}M_\odot$] | $M_{gas,500}$ [$10^{13}M_\odot$] | $f_{gas,500}$ |
|------|------|------|
| 25.18±1.87 | 1.85±0.33 | 0.073±0.008 |

The K-$\alpha$ emission line is clear and allows the determination of redshift. It is a factor $\sim 2$ under-luminous with respect to the $L - T$ relation by Branchesi et al. (2007). This is probably due to the underestimation of the net counts and the difficult reconstruction of the net counts because of the disturbing presence of SWJ0003-5253.

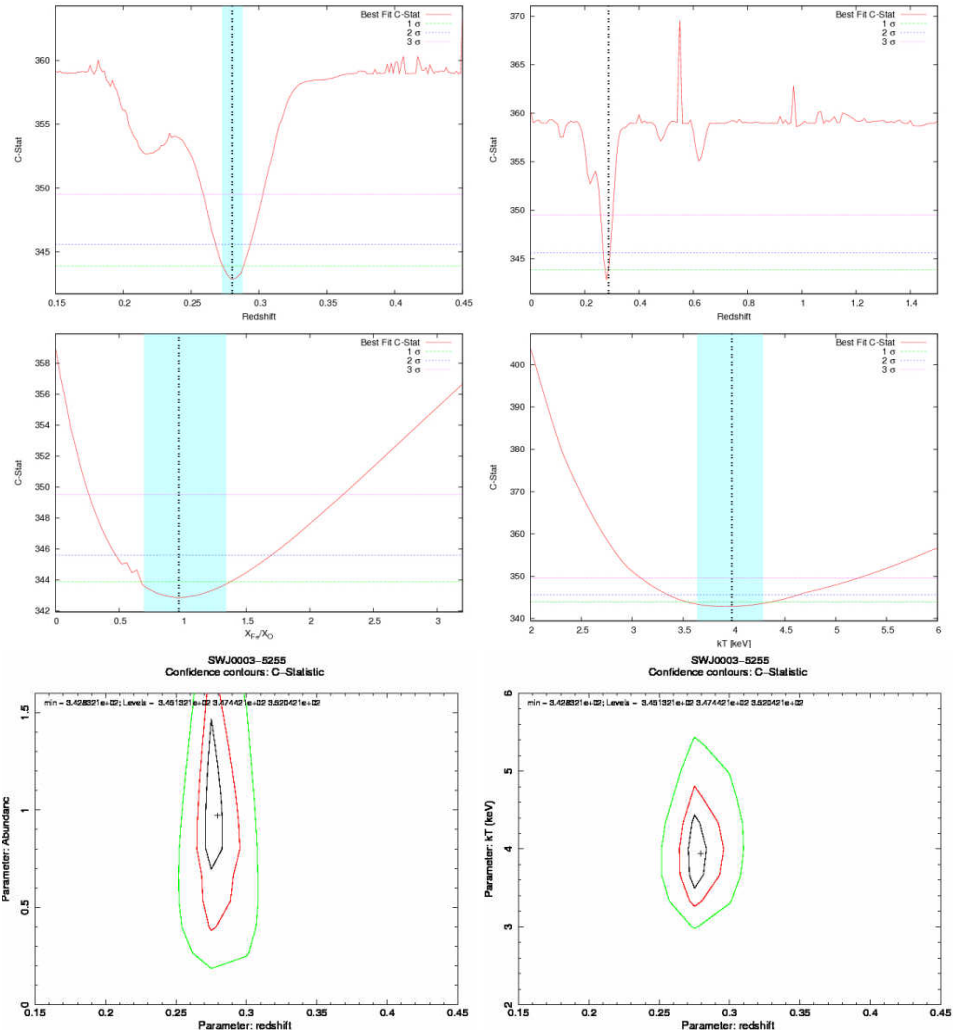

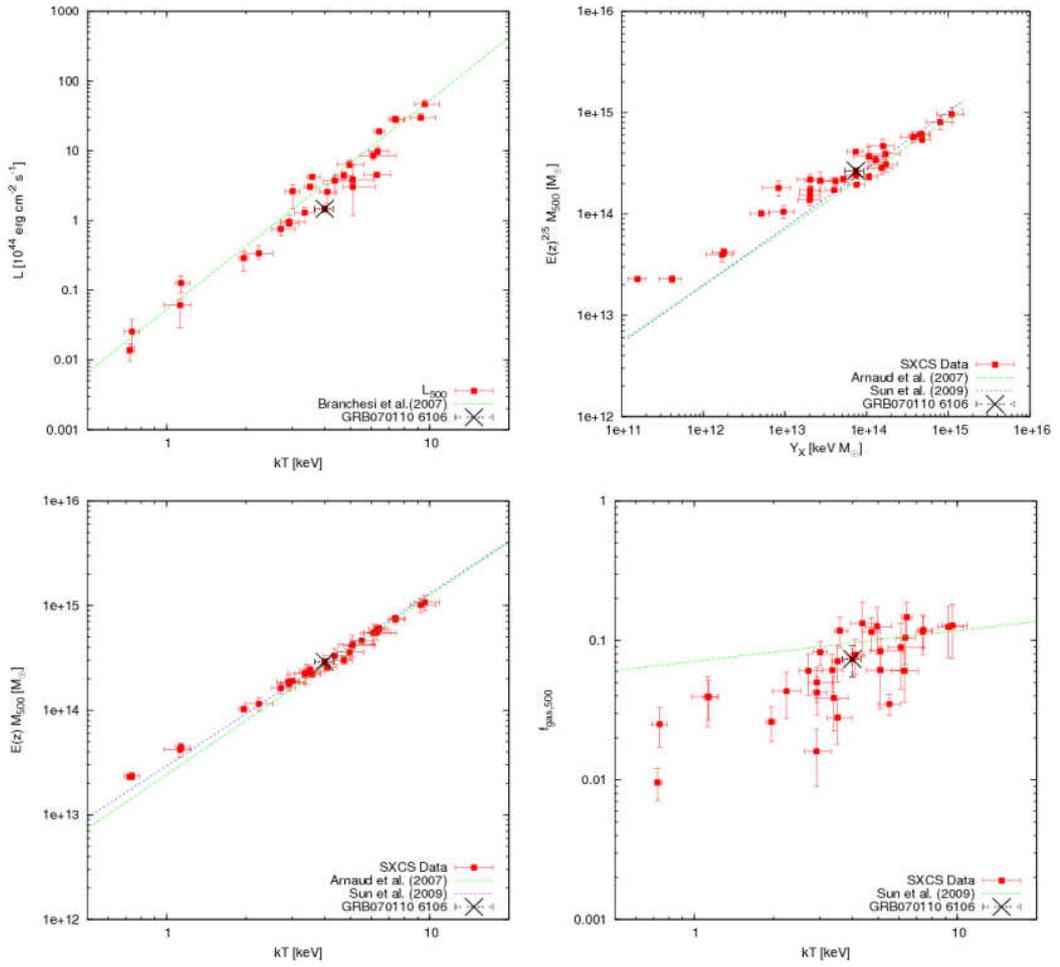

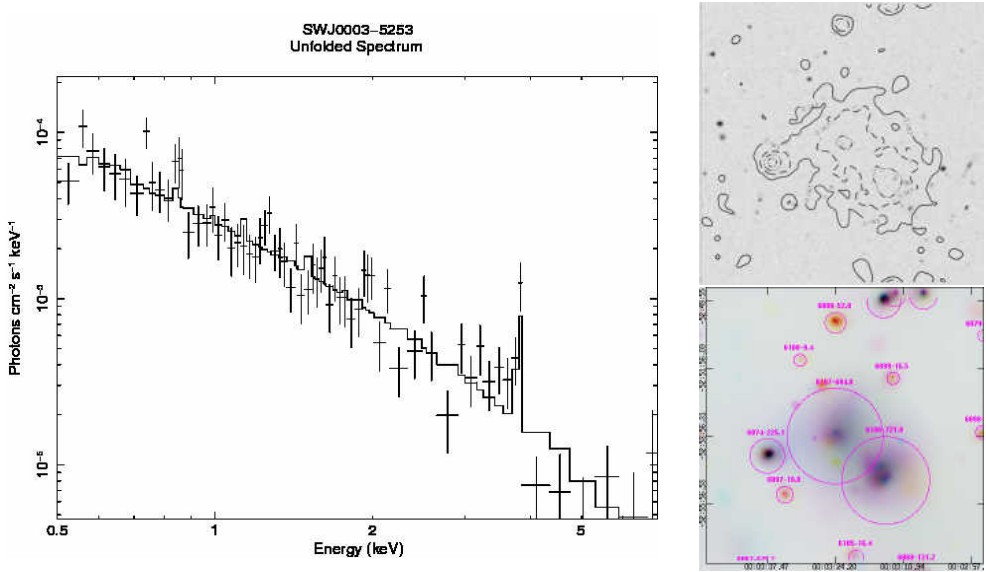

| Name | GRB | R.A. | Dec | Catalogue | Distance | Published $z$ |
|------|-----|------|-----|-----------|----------|---------------|
| SWJ0003-5253 | GRB070110 | 0.849941 | -52.898319 | - | - | - |

| Expmap [s] | Net Counts | SNR | Flux [$10^{-13}$ erg/cm²/s] | $N_H$ [$10^{22}$cm$^{-2}$] | Bkg rate [$10^{-3}$ cts/arcsec²] | r$_{ext}$ [arcsec] |
|------------|-----------|-----|------|------|-----------|-------|
| 303740 | 694±35 | 18.0 | 0.57±0.03 | 0.016 | 18.99 | 84.8 |

| $kT$ [keV] | $z$ | X$_{Fe}$/X$_\odot$ | $r_{ext}$ [kpc] | $r_{500}$ [kpc] | $L_{ext}$ [$10^{44}$ erg/s] | $L_{500}$ [$10^{44}$ erg/s] |
|------|-----|------|------|------|------|------|
| $4.7^{+0.4}_{-0.4}$ | $0.756^{+0.036}_{-0.009}$ | $0.50^{+0.20}_{-0.16}$ | 624±38 | 666±40 | $4.26^{+0.60}_{-0.29}$ | $4.42^{+0.62}_{-0.30}$ |

| $M_{500}$ [$10^{13}M_\odot$] | $M_{gas,500}$ [$10^{13}M_\odot$] | $f_{gas,500}$ |
|------|------|------|
| 19.67±1.52 | 2.27±0.41 | 0.114±0.012 |

The spectrum is moderately disturbed by the presence of SWJ0003-5255, however the K-$\alpha$ iron emission line and the redshift determination is clear. The scaling laws are in agreement with previous works. $M_{gas}$ is in agreement with previous works.

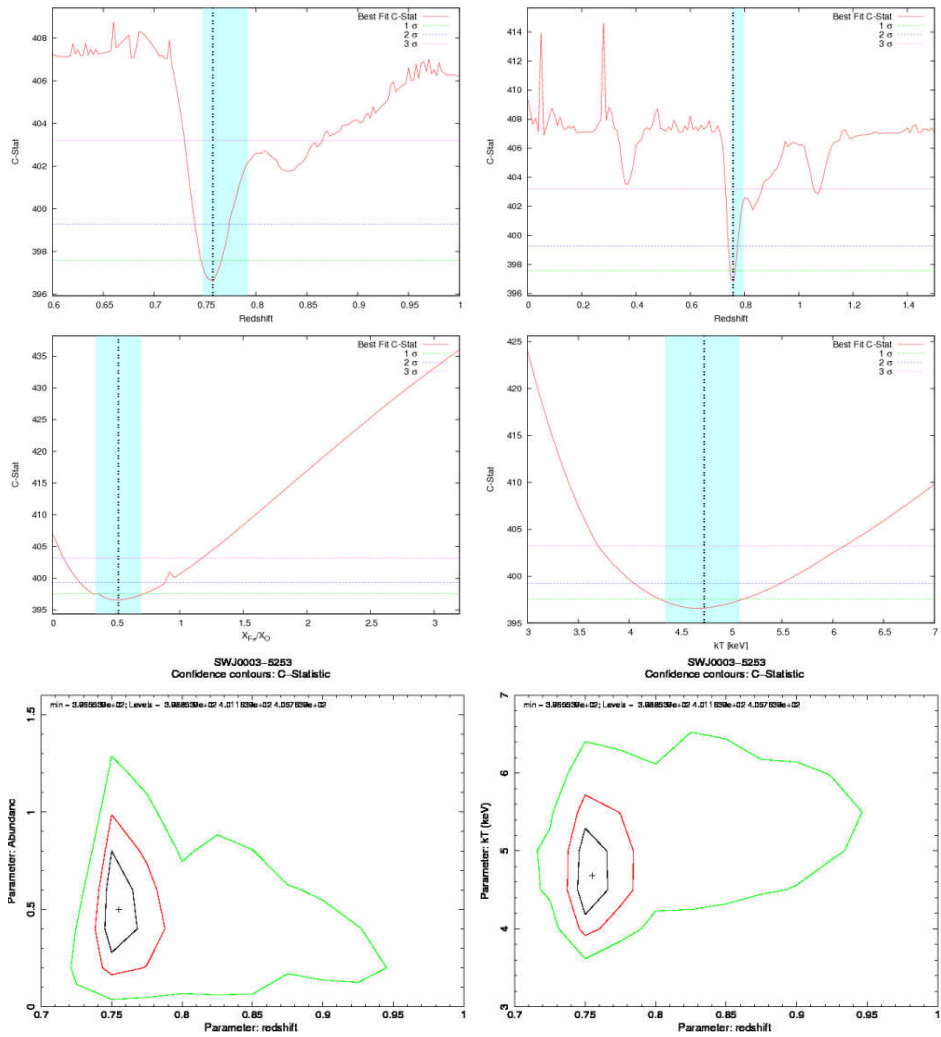

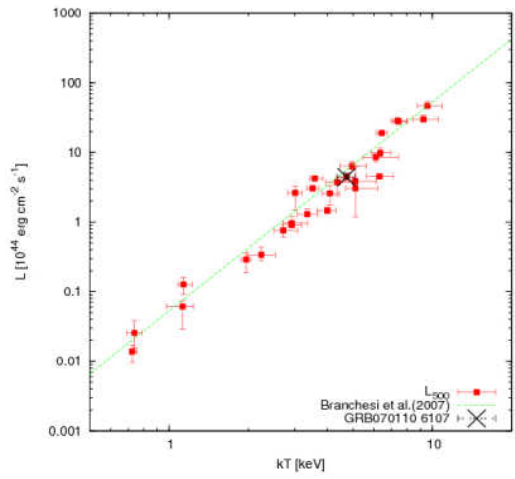
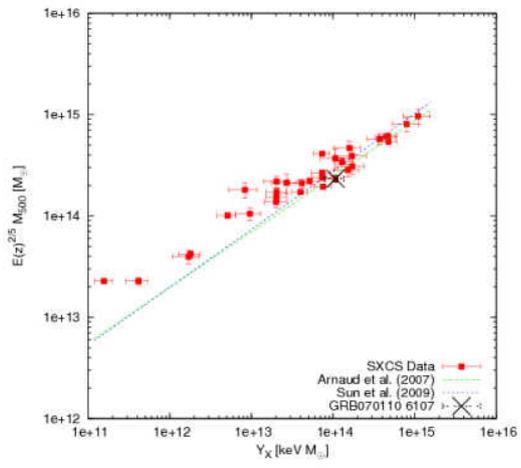
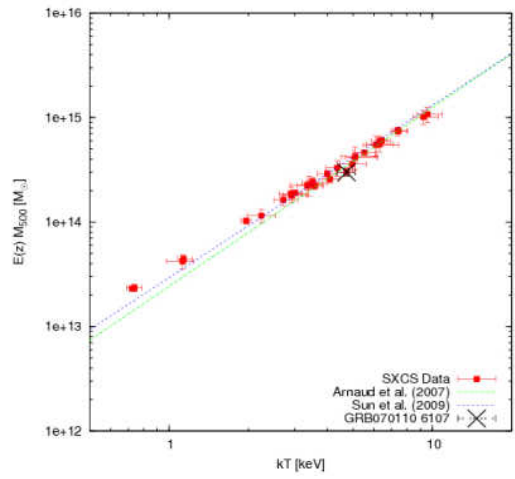
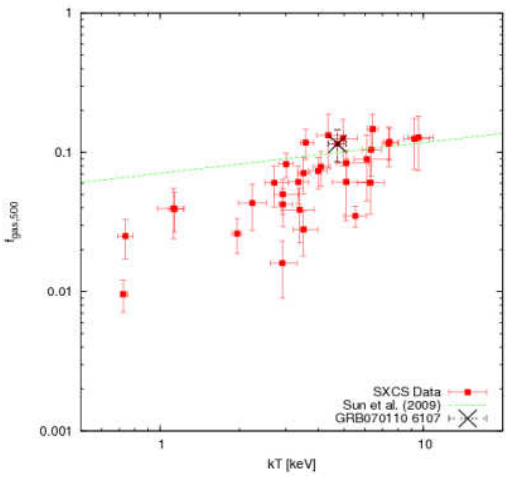

## B.23 SWJ1935+0214

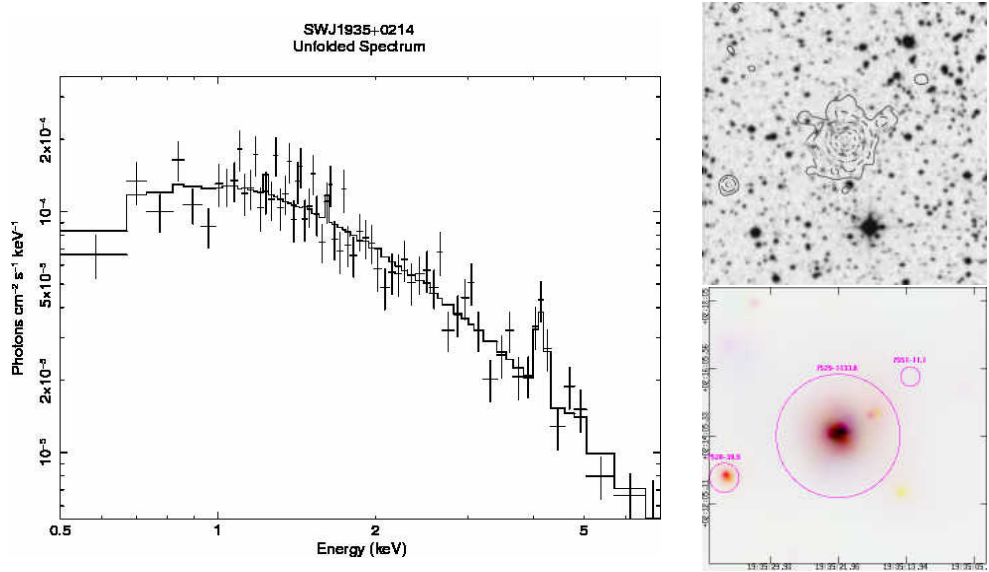

| Name | GRB | R.A. | Dec | Catalogue | Distance | Published $z$ |
|------|-----|------|-----|-----------|----------|-------------|
| SWJ1935+0214 | GRB070917 | 293.840546 | 2.235848 | - | - | - |

| Expmap [s] | Net Counts | SNR | Flux [$10^{-13}$ erg/cm$^2$/s] | $N_H$ [$10^{22}$cm$^{-2}$] | Bkg rate [$10^{-3}$ cts/arcsec$^2$] | $r_{ext}$ [arcsec] |
|------|-----|-----|------|------|------|------|
| 90758 | 1133±41 | 29.4 | 4.71±0.17 | 0.169 | 5.42 | 109.4 |

| $kT$ [keV] | $z$ | $X_{Fe}/X_\odot$ | $r_{ext}$ [kpc] | $r_{500}$ [kpc] | $L_{ext}$ [$10^{44}$ erg/s] | $L_{500}$ [$10^{44}$ erg/s] |
|------|-----|------|------|------|------|------|
| $7.4^{+0.6}_{-0.4}$ | $0.634^{+0.010}_{-0.008}$ | $0.49^{+0.14}_{-0.10}$ | 749±26 | 969±34 | $27.73^{+1.66}_{-1.46}$ | $28.70^{+1.71}_{-1.52}$ |

| $M_{500}$ [$10^{13}M_\odot$] | $M_{gas,500}$ [$10^{13}M_\odot$] | $f_{gas,500}$ |
|------|------|------|
| 52.36±4.92 | 6.06±1.36 | 0.114±0.015 |

The redshift determination is good. Very good agreement with the best fit of the scaling laws found by Branchesi et al. (2007), Aranud et al. (2007) and Sun et al. (2009). Also $M_{gas}$ is in agreement with previous works.

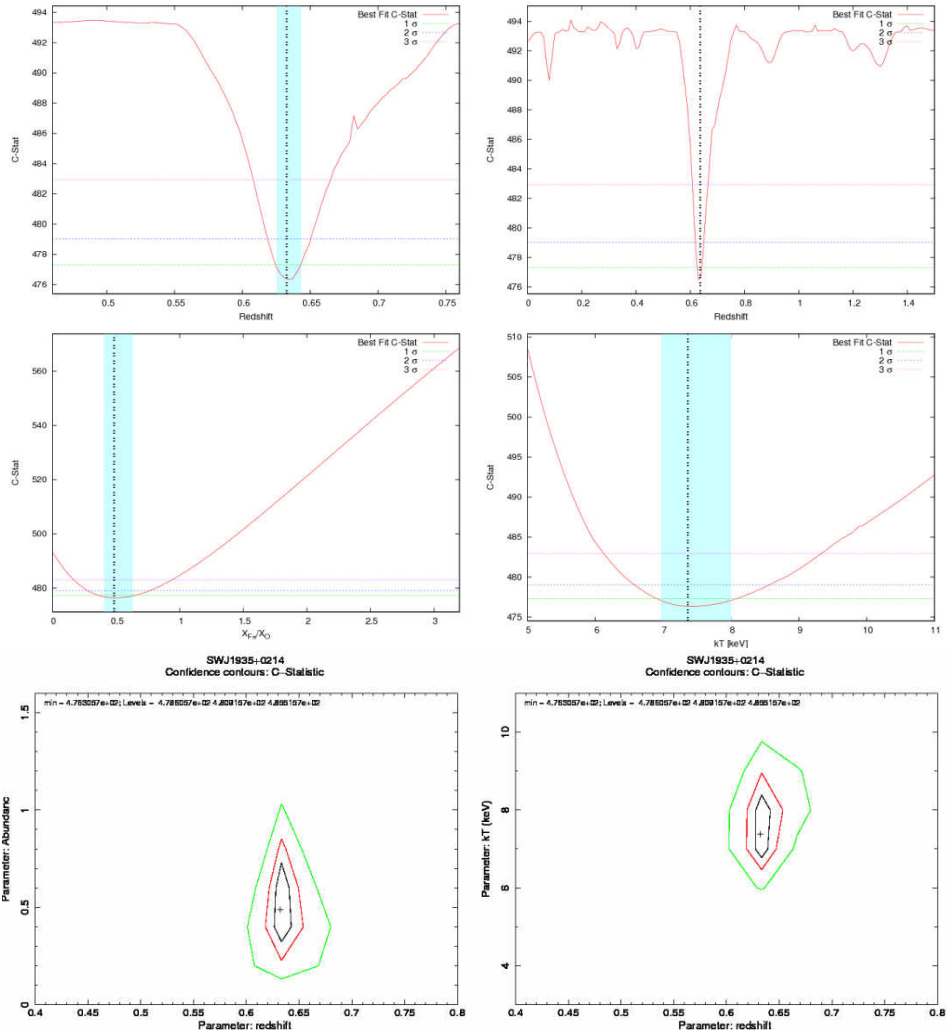

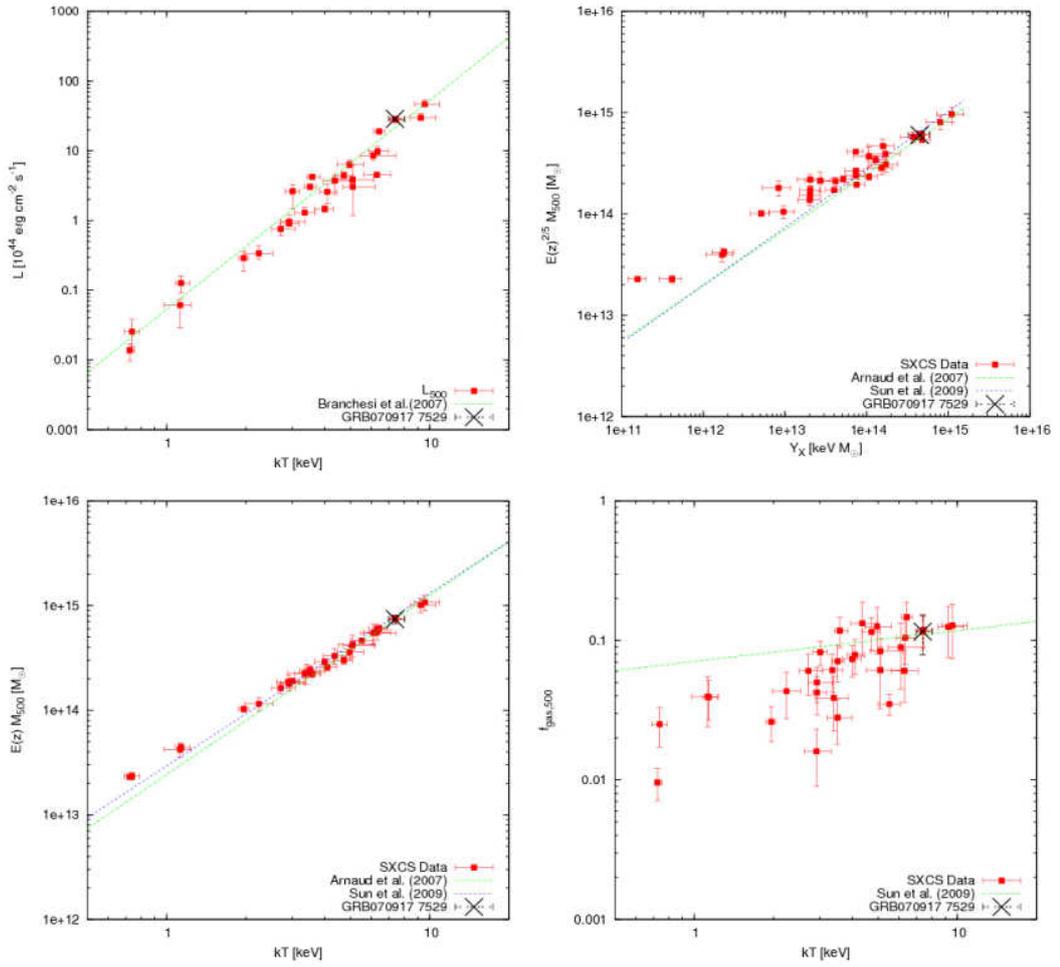

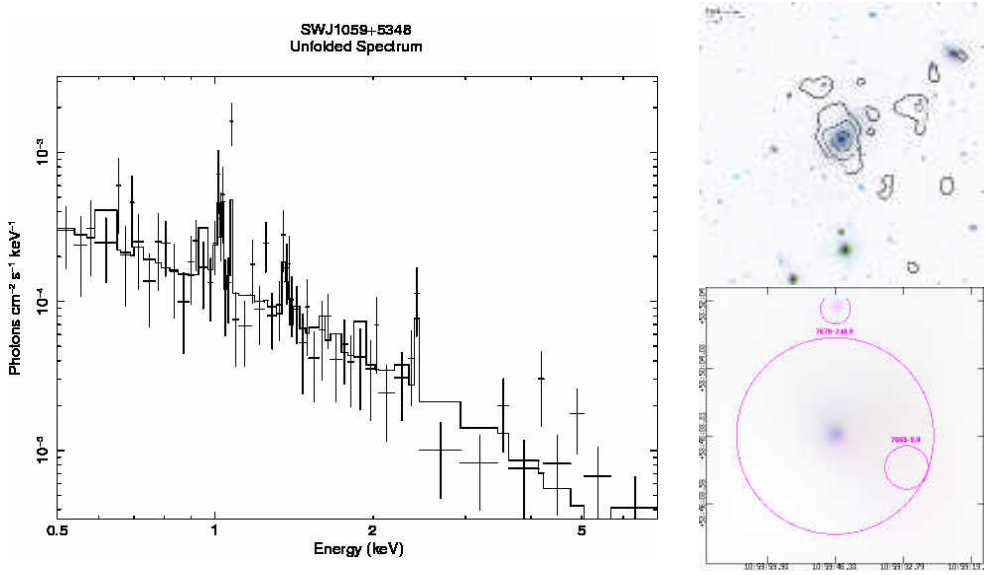

| Name | GRB | R.A. | Dec | Catalogue | Distance | Published $z$ |
|------|-----|------|-----|-----------|----------|---------------|
| SWJ1059+5348 | GRB071018* | 164.941711 | 53.802010 | SDSS-GCl | 0.098 | 0.072[5] |

| Expmap | Net Counts | SNR | Flux | $N_H$ | Bkg rate | $r_{ext}$ |
| [s] | | | [$10^{-13}$ erg/cm$^2$/s] | [$10^{22}$ cm$^{-2}$] | [$10^{-3}$ cts/arcsec$^2$] | [arcsec] |
|--------|------------|-----|------|-------|----------|-----------|
| 22491 | 248±25 | 11.9 | 2.69±0.27 | 0.008 | 1.26 | 174.4 |

| $kT$ | $z$ | $X_{Fe}/X_\odot$ | $r_{ext}$ | $r_{500}$ | $L_{ext}$ | $L_{500}$ |
| [keV] | | | [kpc] | [kpc] | [$10^{44}$ erg/s] | [$10^{44}$ erg/s] |
|------|-----|------------------|-----------|-----------|-----------|-----------|
| $2.9^{+0.4}_{-0.3}$ | $0.086^{+0.009}_{-0.009}$ | $1.52^{+0.88}_{-0.50}$ | 282±145 | 828±426 | $0.15^{+0.05}_{-0.05}$ | $0.20^{+0.06}_{-0.06}$ |

| $M_{500}$ | $M_{gas,500}$ | $f_{gas,500}$ |
| [$10^{13} M_\odot$] | [$10^{13} M_\odot$] | |
|-----------|---------------|---------------|
| 17.80±3.03 | 0.29±0.08 | 0.016±0.002 |

This cluster falls on the edge of the X-ray image and a considerable part of its flux is lost. In fact, it is well below the best fit $L - T$ obtained by Branchesi et al.(2007). However, $\sim 250$ net counts are enough to measure the redshift with high significance and also temperature and iron abundance with error < 20%.

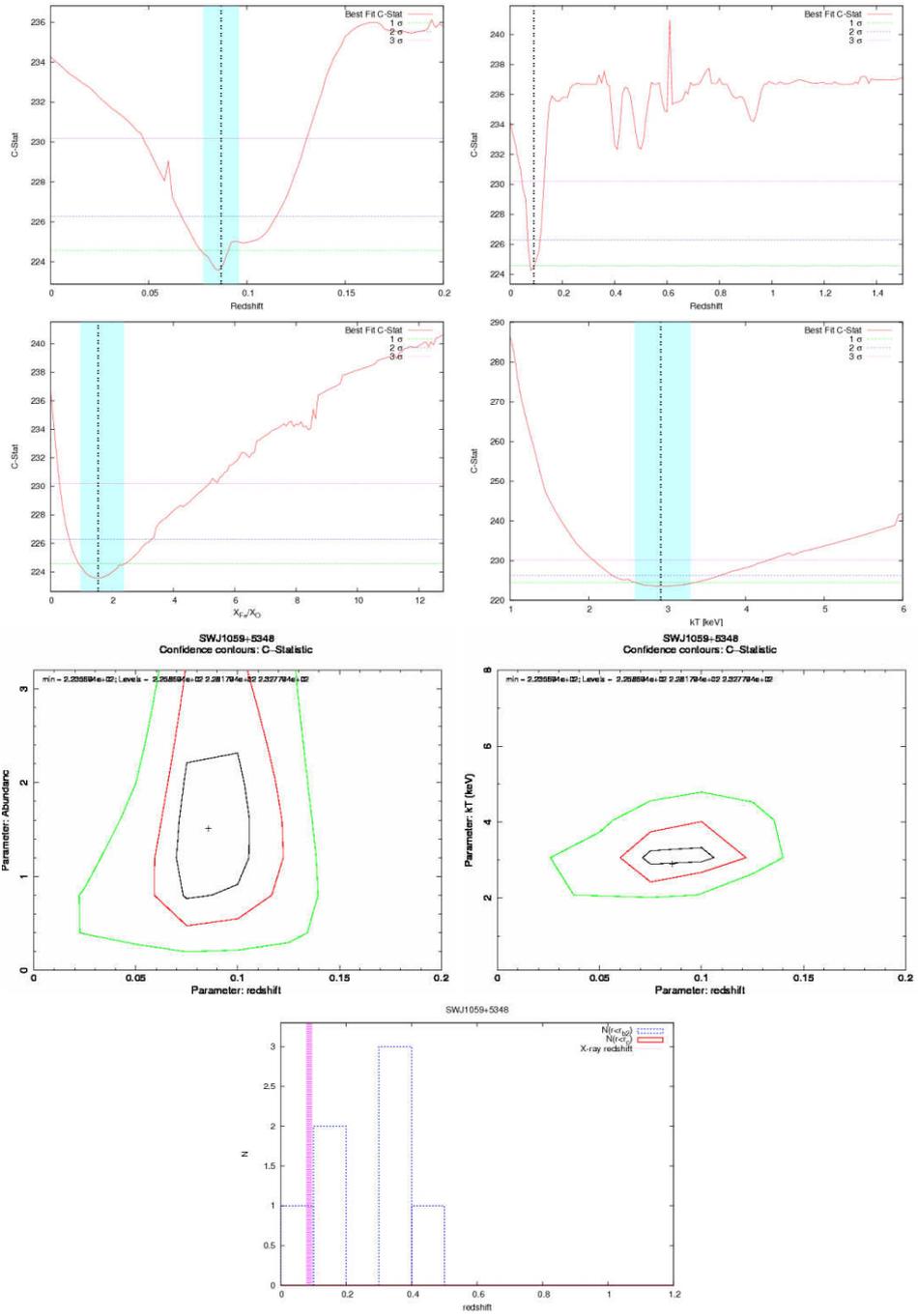

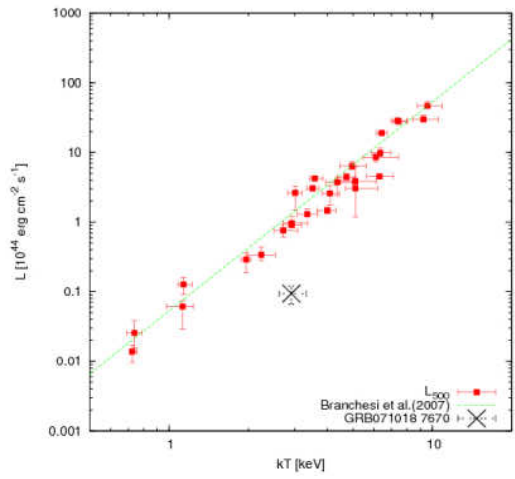
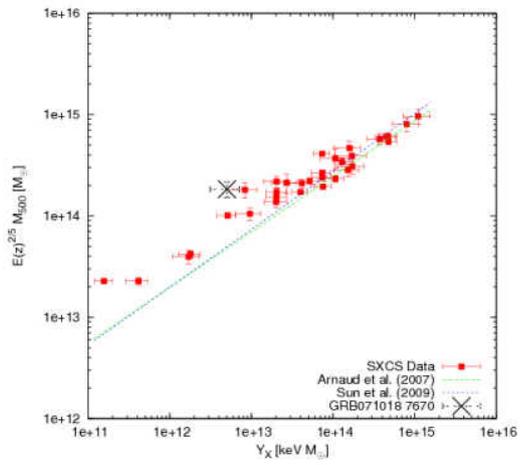
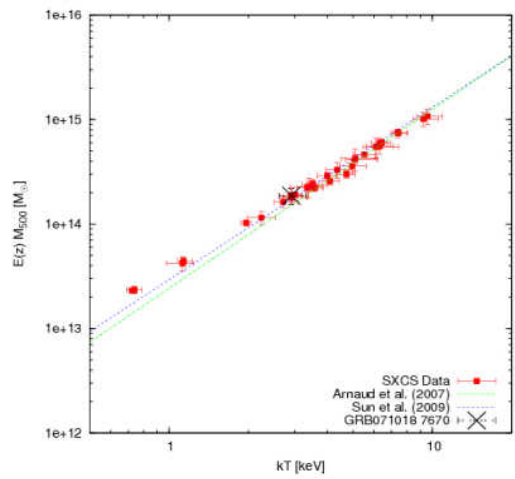
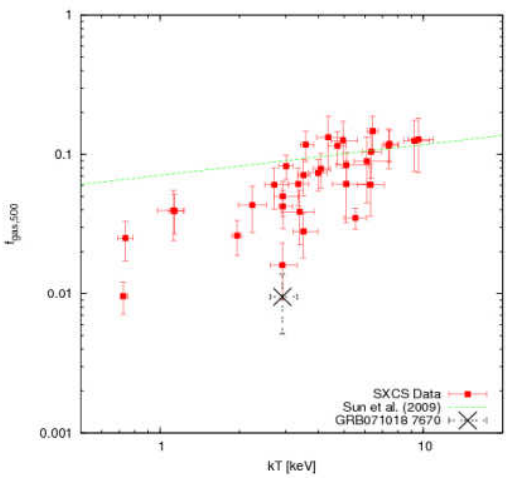

B.25 SWJ2336-3136

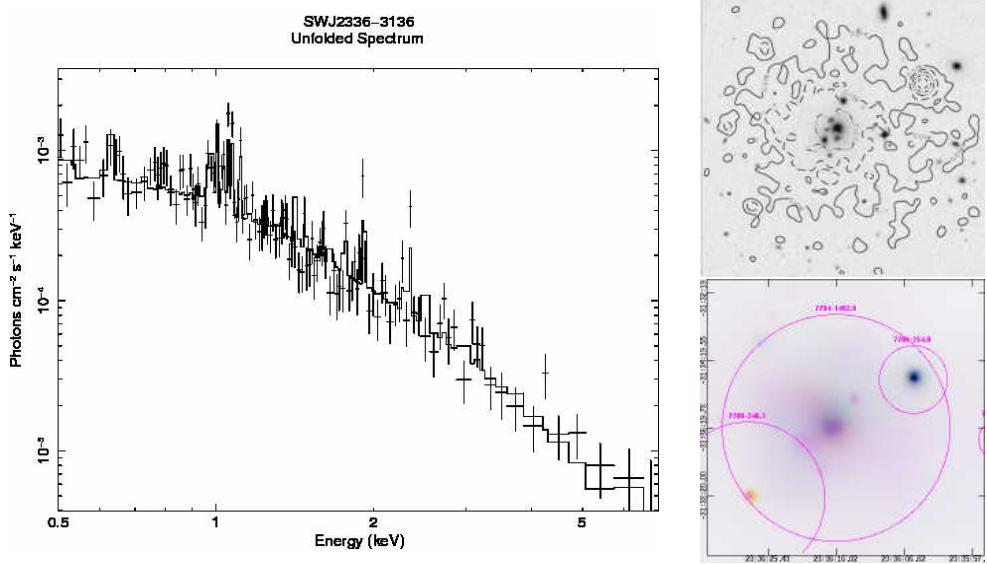

| Name | GRB | R.A. | Dec | Catalogue | Distance | Published $z$ |
|------|-----|------|-----|-----------|----------|---------------|
| SWJ2336-3136 | GRB071028B | 354.066040 | -31.604403 | ABELL | 0.358 | 0.0625[6] |

| Expmap [s] | Net Counts | SNR | Flux [$10^{-13}$ erg/cm$^2$/s] | $N_H$ [$10^{22}$cm$^{-2}$] | Bkg rate [$10^{-3}$ cts/arcsec$^2$] | $r_{ext}$ [arcsec] |
|------------|-----------|-----|------|------|------|------|
| 43924 | 1492±48 | 31.2 | 8.35±0.27 | 0.012 | 3.44 | 201.0 |

| $kT$ [keV] | $z$ | $X_{Fe}/X_\odot$ | $r_{ext}$ [kpc] | $r_{500}$ [kpc] | $L_{ext}$ [$10^{44}$ erg/s] | $L_{500}$ [$10^{44}$ erg/s] |
|-----------|-----|------|------|------|------|------|
| $2.0^{+0.1}_{-0.1}$ | $0.049^{+0.009}_{-0.013}$ | $0.46^{+0.09}_{-0.06}$ | 240±14 | 691±42 | $0.25^{+0.06}_{-0.09}$ | $0.29^{+0.07}_{-0.10}$ |

| $M_{500}$ [$10^{13}M_\odot$] | $M_{gas,500}$ [$10^{13}M_\odot$] | $f_{gas,500}$ |
|------|------|------|
| 9.99±0.63 | 0.26±0.06 | 0.026±0.004 |

The redshift is determined from the L-shell emission lines, and it is in agreement within $1\sigma$ with the literature redshift. The $L-T$ and the $M-T$ are in agreement with previous works. The $M_{gas}$ is underestimated.

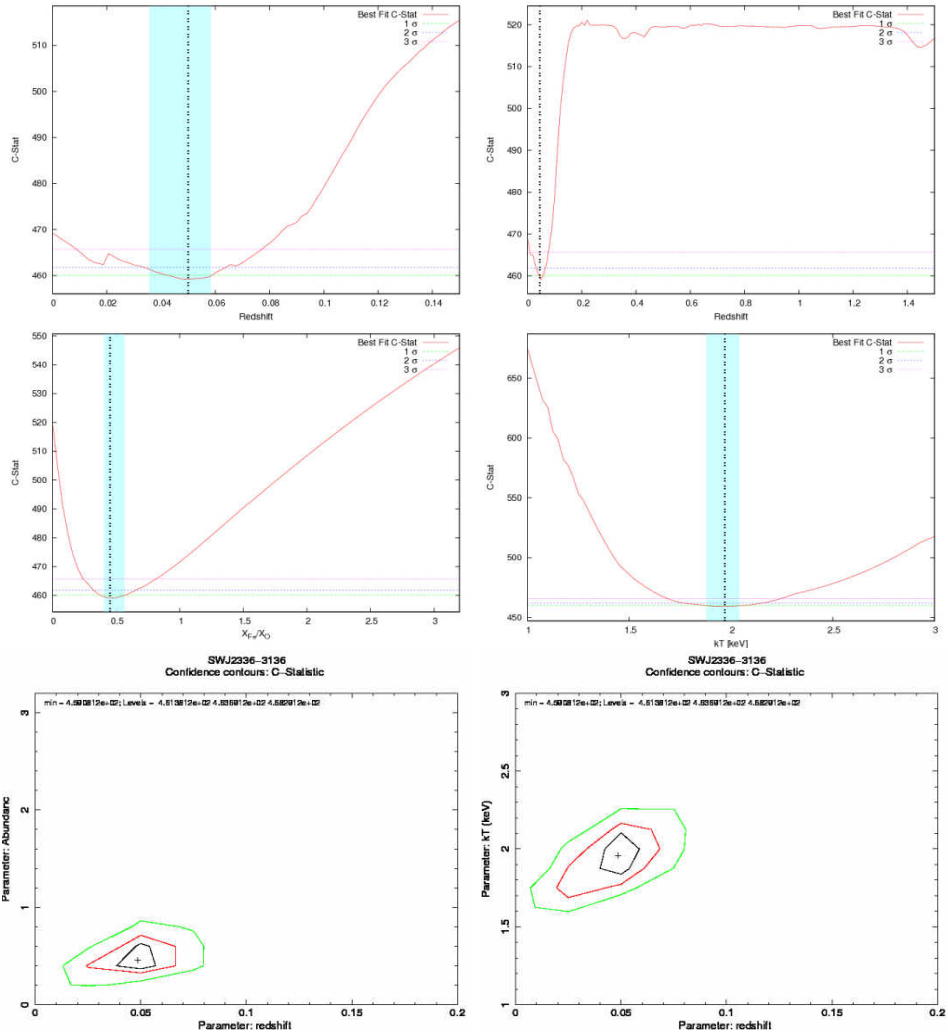

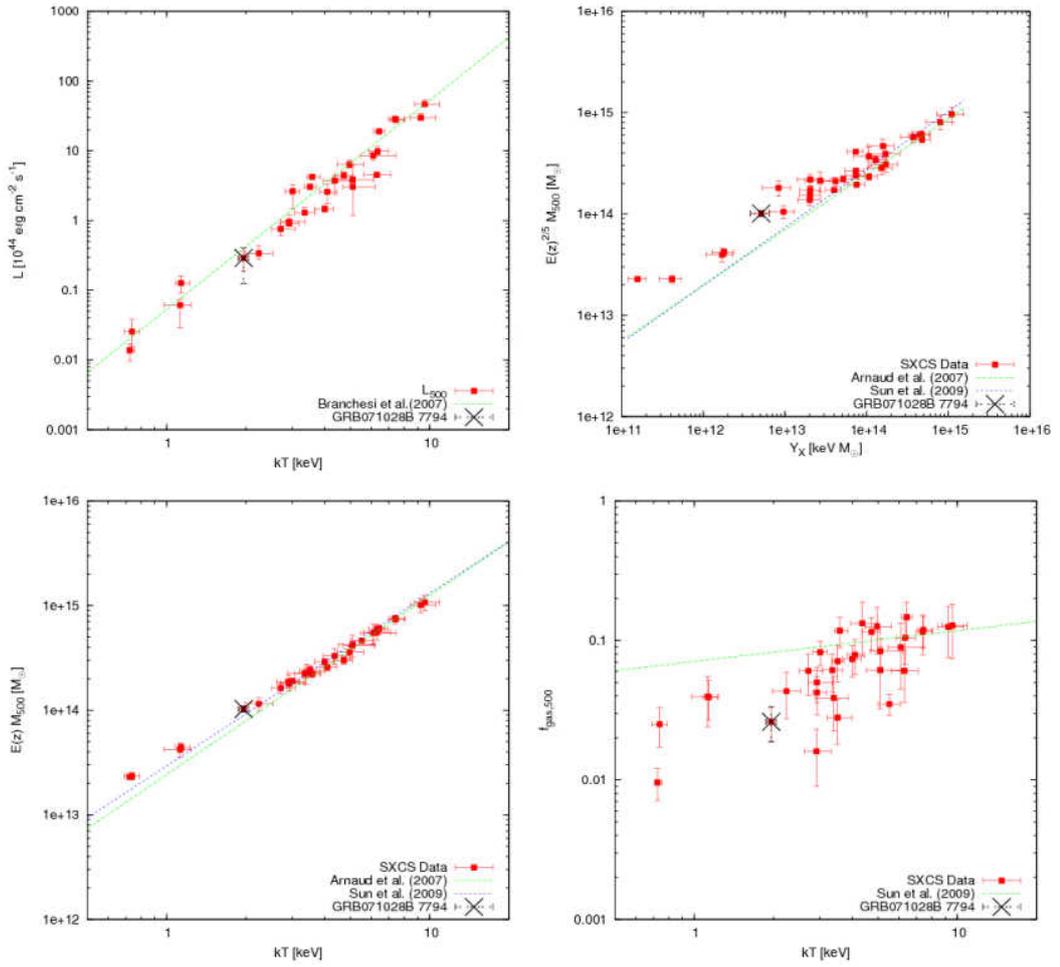

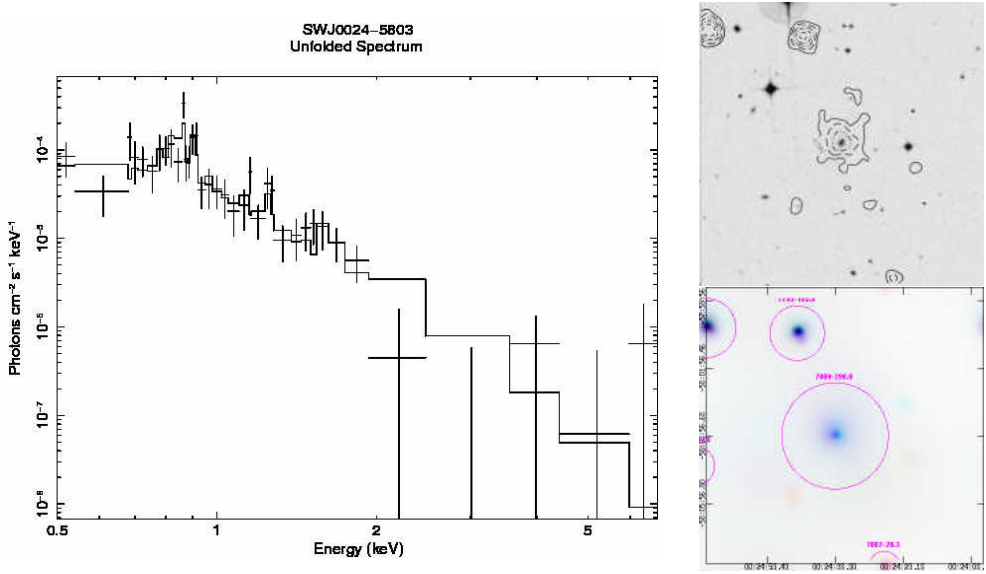

| Name | GRB | R.A. | Dec | Catalogue | Distance | Published $z$ |
|------|-----|------|-----|-----------|----------|---------------|
| SWJ0024-5803 | GRB071031 | 6.157714 | -58.064888 | - | - | - |

| Expmap [s] | Net Counts | SNR | Flux [$10^{-13}$ erg/cm$^2$/s] | $N_H$ [$10^{22}$cm$^{-2}$] | Bkg rate [$10^{-3}$ cts/arcsec$^2$] | r$_{ext}$ [arcsec] |
|------------|-----------|-----|-------|-------|----------|-------|
| 76010 | 298±27 | 13.6 | 0.97±0.09 | 0.012 | 3.84 | 94.1 |

| $kT$ [keV] | $z$ | X$_{Fe}$/X$_\odot$ | $r_{ext}$ [kpc] | $r_{500}$ [kpc] | $L_{ext}$ [$10^{44}$ erg/s] | $L_{500}$ [$10^{44}$ erg/s] |
|------------|-----|----------|-------|-------|----------|----------|
| $1.1^{+0.1}_{-0.1}$ | $0.178^{+0.016}_{-0.016}$ | $0.50^{+0.36}_{-0.16}$ | 282±79 | 491±138 | $0.11^{+0.03}_{-0.03}$ | $0.13^{+0.03}_{-0.04}$ |

| $M_{500}$ [$10^{13}M_\odot$] | $M_{gas,500}$ [$10^{13}M_\odot$] | $f_{gas,500}$ |
|------------|------------|------------|
| 4.04±0.33 | 0.16±0.04 | 0.039±0.006 |

The redshift is determined from L-shell emission lines. The $L-T$ and $M-T$ are in agreement with best fit obtained by other authors. The $M_{gas}$ is underestimated.

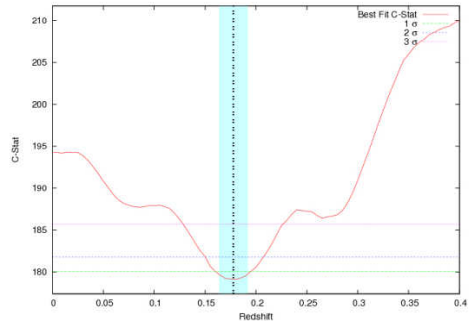
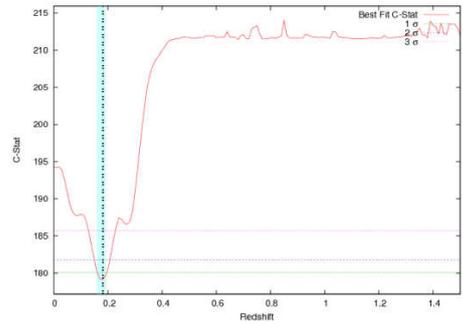

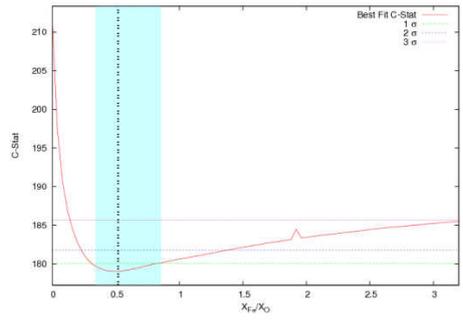
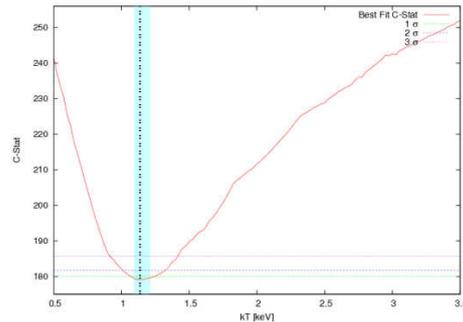

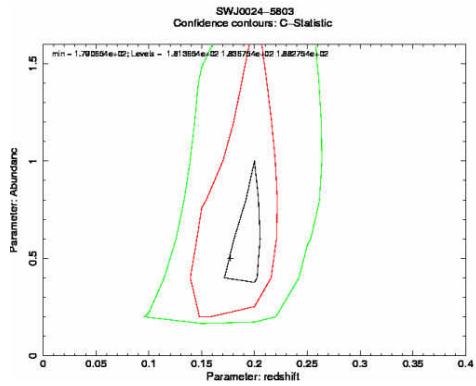
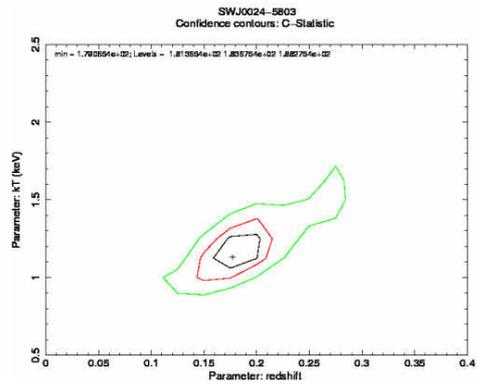

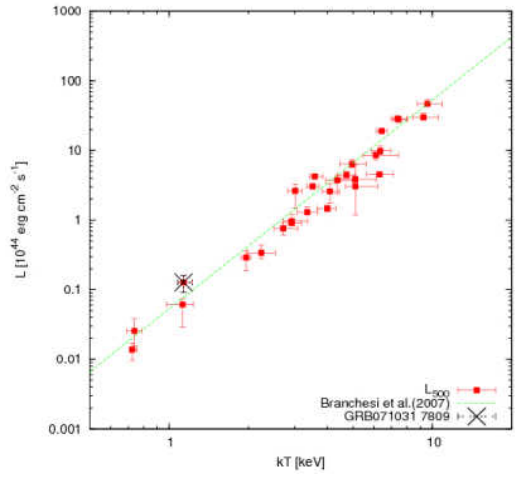
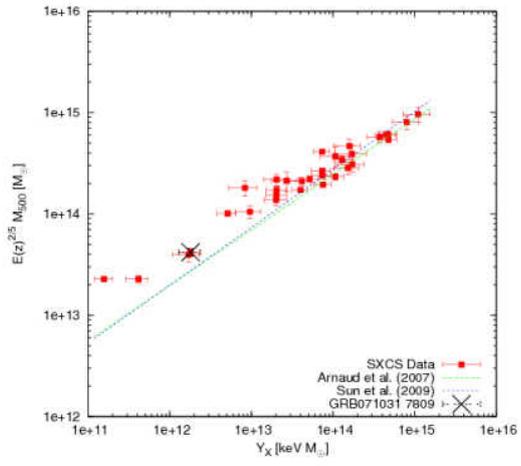

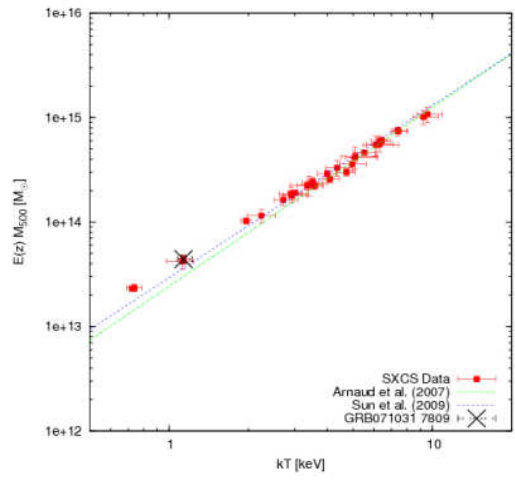
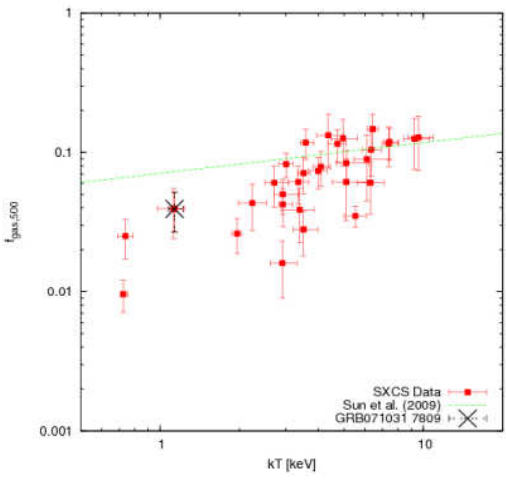

## B.27 SWJ1432+3617

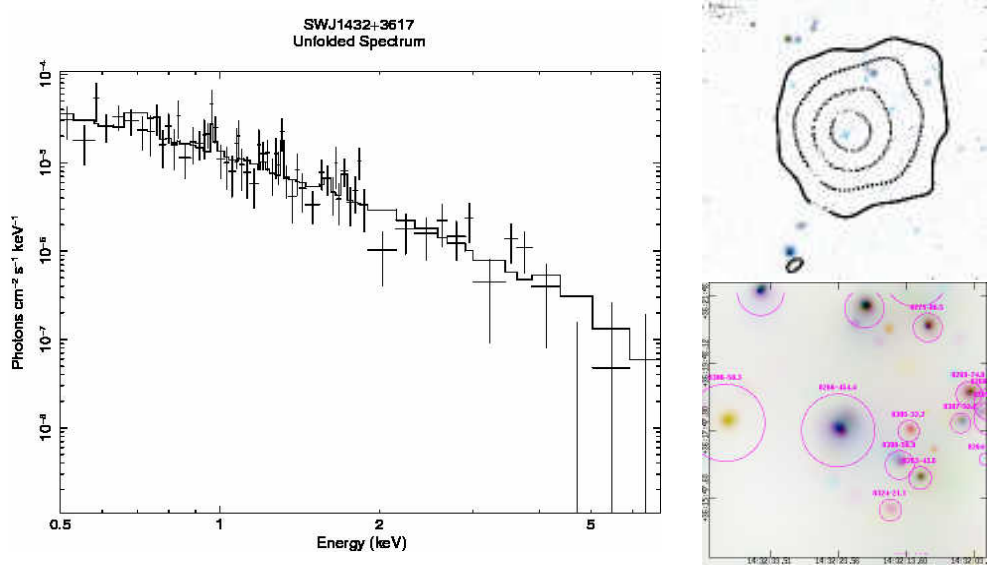

| Name | GRB | R.A. | Dec | Catalogue | Distance | Published $z$ |
|------|-----|------|-----|-----------|----------|---------------|
| SWJ1432+3617 | GRB080319B* | 218.097107 | 36.297855 | - | - | - |

| Expmap [s] | Net Counts | SNR | Flux [$10^{-13}$ erg/cm²/s] | $N_H$ [$10^{22}$cm⁻²] | Bkg rate [$10^{-3}$ cts/arcsec²] | $r_{ext}$ [arcsec] |
|------------|------------|-----|------|-------|----------|-------|
| 333299 | 454±32 | 16.9 | 0.33±0.02 | 0.011 | 13.65 | 64.8 |

| $kT$ [keV] | $z$ | $X_{Fe}/X_\odot$ | $r_{ext}$ [kpc] | $r_{500}$ [kpc] | $L_{ext}$ [$10^{44}$ erg/s] | $L_{500}$ [$10^{44}$ erg/s] |
|------------|-----|-----------------|-------|-------|-------|-------|
| $2.7^{+0.4}_{-0.2}$ | $0.527^{+0.038}_{-0.032}$ | $0.67^{+0.45}_{-0.24}$ | 407±57 | 620±86 | $0.69^{+0.14}_{-0.13}$ | $0.76^{+0.16}_{-0.15}$ |

| $M_{500}$ [$10^{13}M_\odot$] | $M_{gas,500}$ [$10^{13}M_\odot$] | $f_{gas,500}$ |
|------------|------------|------------|
| 12.23±1.45 | 0.74±0.16 | 0.060±0.006 |

The C-stat has four redshift minimum: 0.1, 0.5, 0.9 and 1.2. The minimum at $z \sim 0.5$ is the absolute minimum, but the other can only be excluded with a small significance of $\sim 2\sigma$. Assuming $z \sim 0.5$ as the true redshift, the cluster results 20% less luminous with respect to the $L - T$ best fit by Branchesi et al. (2007). The $M - T$ is in agreement with Arnaud et al. (2007) and Sun et al. (2009). The $M_{gas}$ results slightly underestimated.

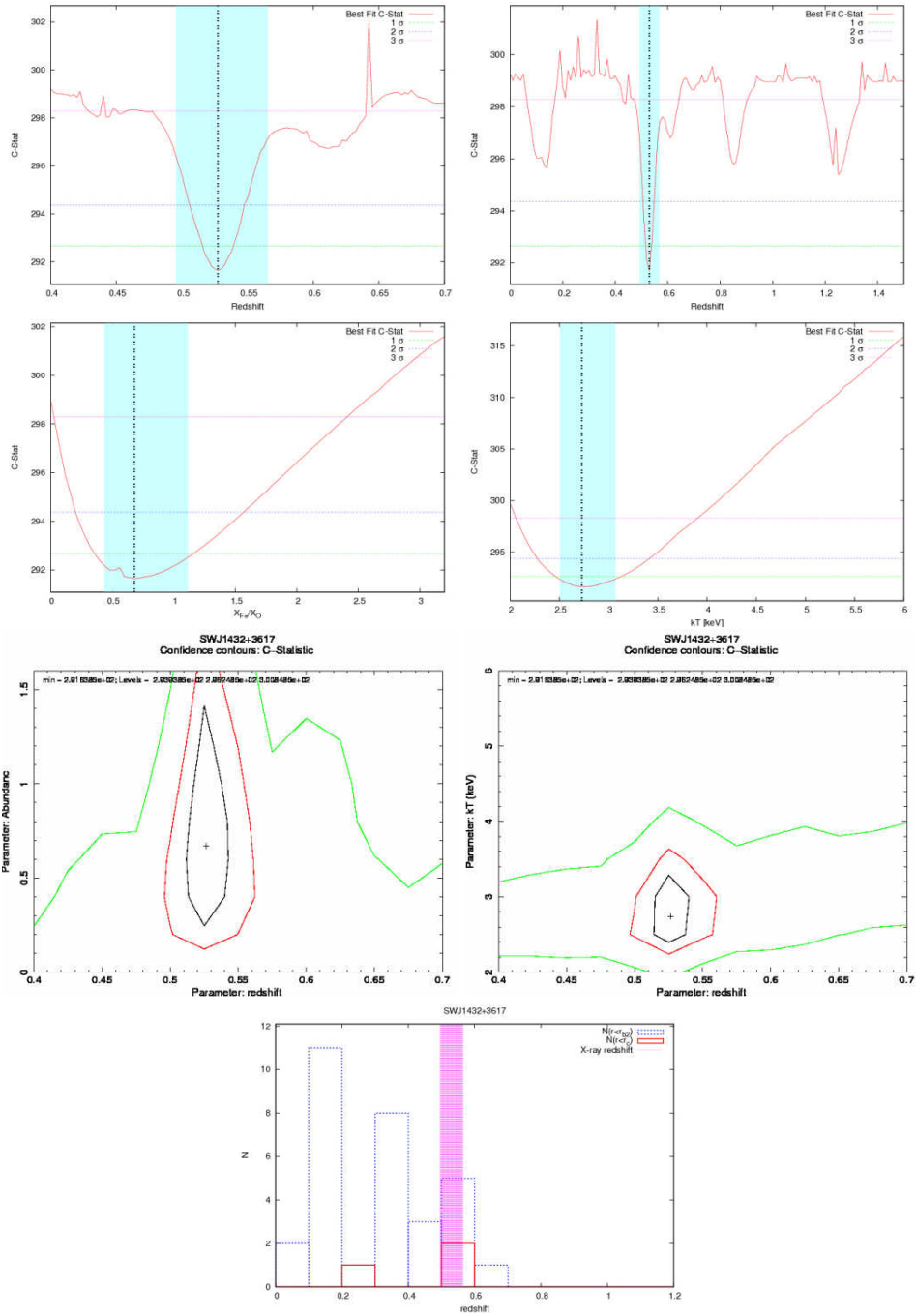

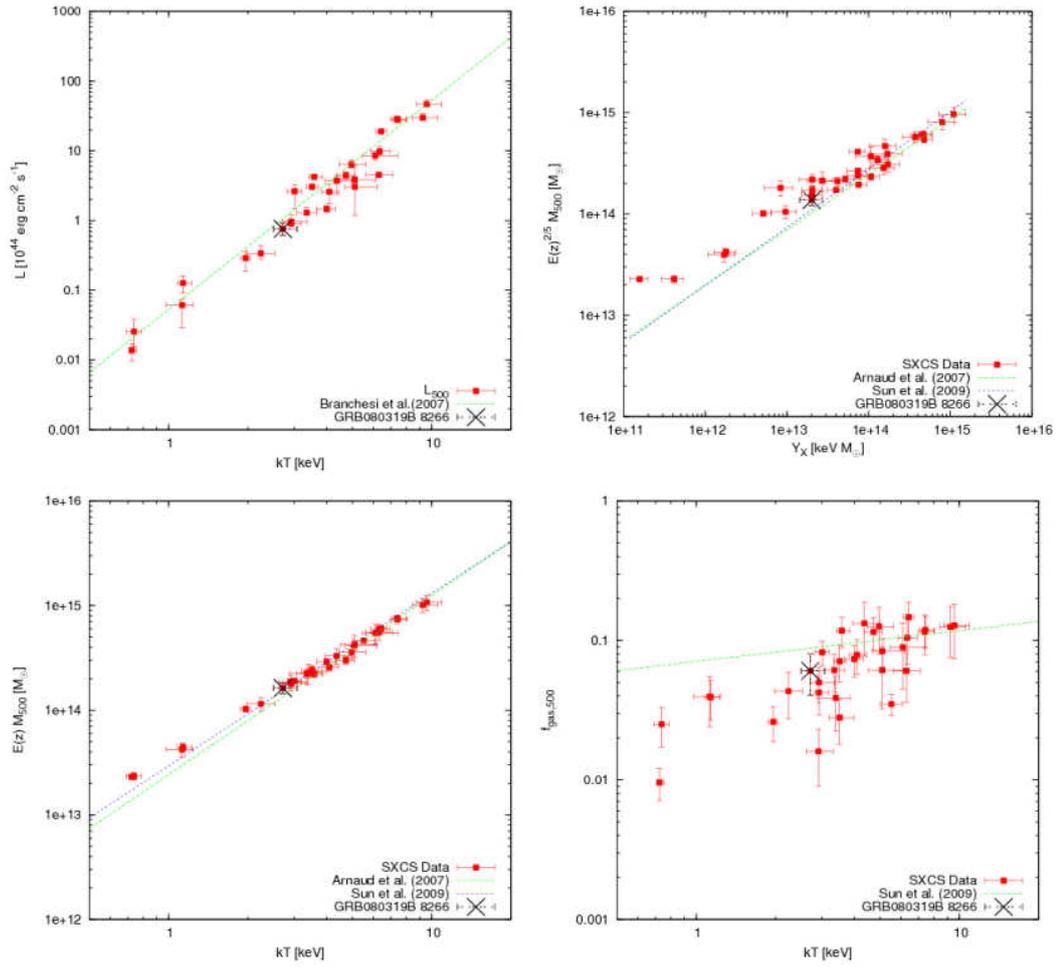

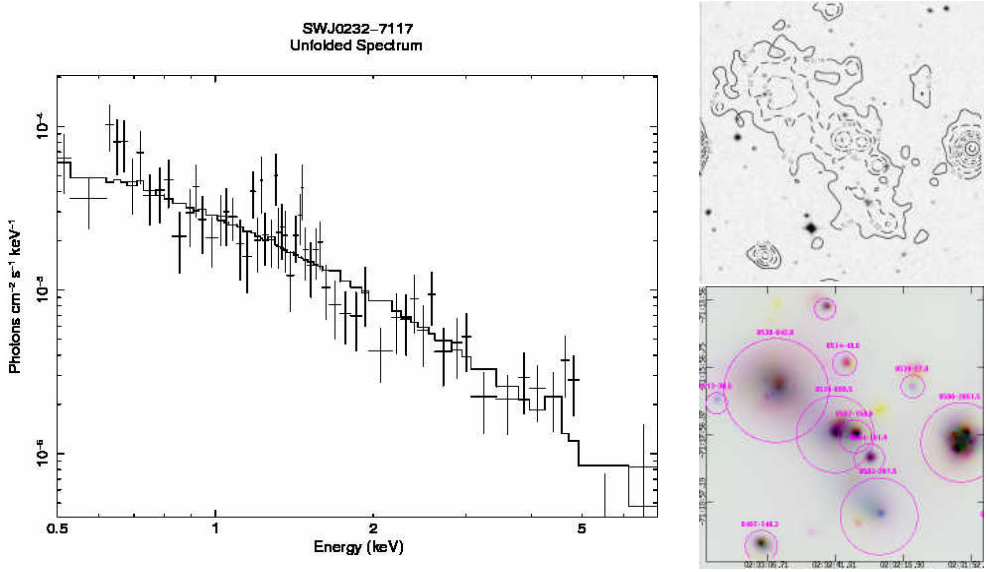

| Name | GRB | R.A. | Dec | Catalogue | Distance | Published $z$ |
|------|-----|------|-----|-----------|----------|---------------|
| SWJ0232-7117 | GRB080411 | 38.172527 | -71.298302 | - | - | - |

| Expmap [s] | Net Counts | SNR | Flux [$10^{-13}$ erg/cm²/s] | $N_H$ [$10^{22}$cm$^{-2}$] | Bkg rate [$10^{-3}$ cts/arcsec²] | $r_{ext}$ [arcsec] |
|------------|-----------|-----|------|-----|----------|--------|
| 324085 | 699±44 | 21.1 | 0.61±0.04 | 0.058 | 14.59 | 68.8 |

| $kT$ [keV] | $z$ | $X_{Fe}/X_\odot$ | $r_{ext}$ [kpc] | $r_{500}$ [kpc] | $L_{ext}$ [$10^{44}$ erg/s] | $L_{500}$ [$10^{44}$ erg/s] |
|------------|-----|------------------|-----------------|-----------------|------------------------------|------------------------------|
| $5.1^{+1.1}_{-0.4}$ | $0.550^{+0.335}_{-0.193}$ | $0.19^{+0.27}_{-0.14}$ | 459±136 | 815±242 | $2.70^{+2.78}_{-1.65}$ | $3.04^{+3.14}_{-1.86}$ |

| $M_{500}$ [$10^{13}M_\odot$] | $M_{gas,500}$ [$10^{13}M_\odot$] | $f_{gas,500}$ |
|------------------------------|----------------------------------|---------------|
| 30.00±2.95 | 2.52±0.34 | 0.084±0.003 |

The redshift is not determined clearly, however the minimum in the C-stat should fall between $0.2 < z < 0.8$. Forcing the redshift in this range in the spectral fit, the resulting luminosity, total mass and gas mass are in agreement with the best fit of the scaling laws found by other authors for X-ray selected clsuters.

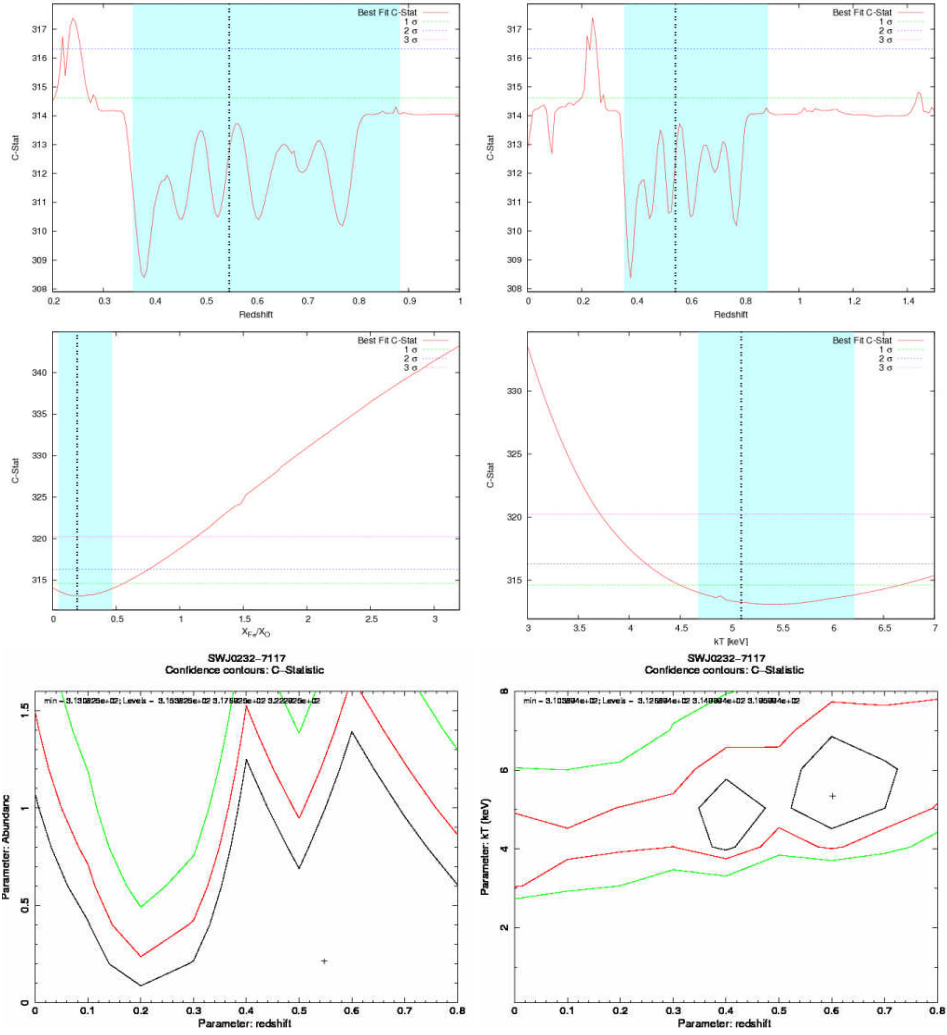

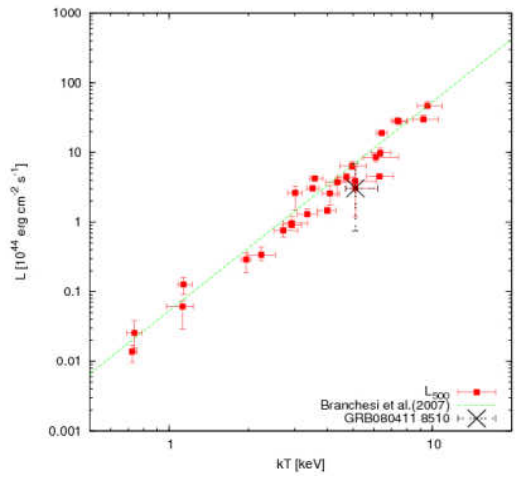

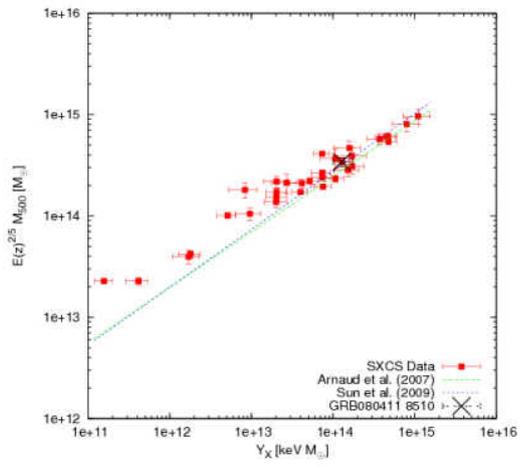

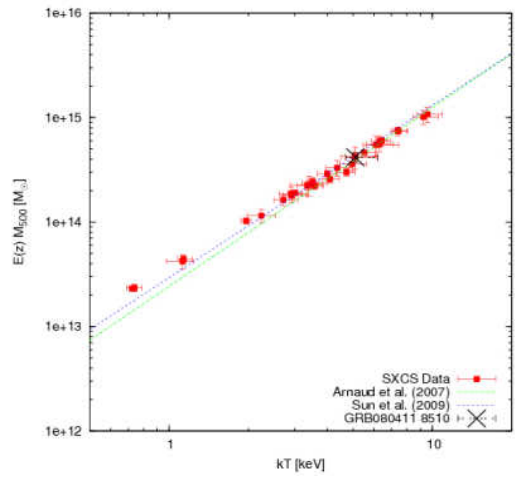

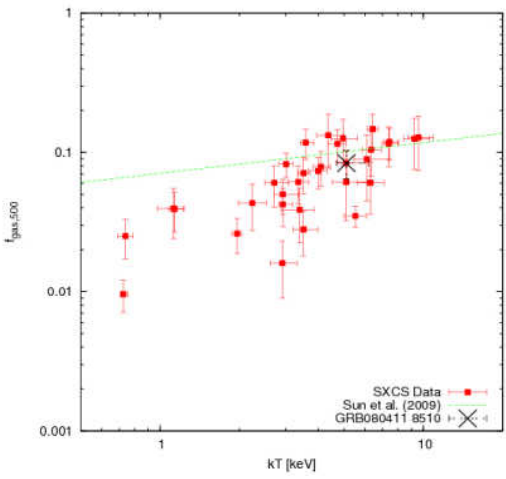

## B.29 SWJ0233-7116

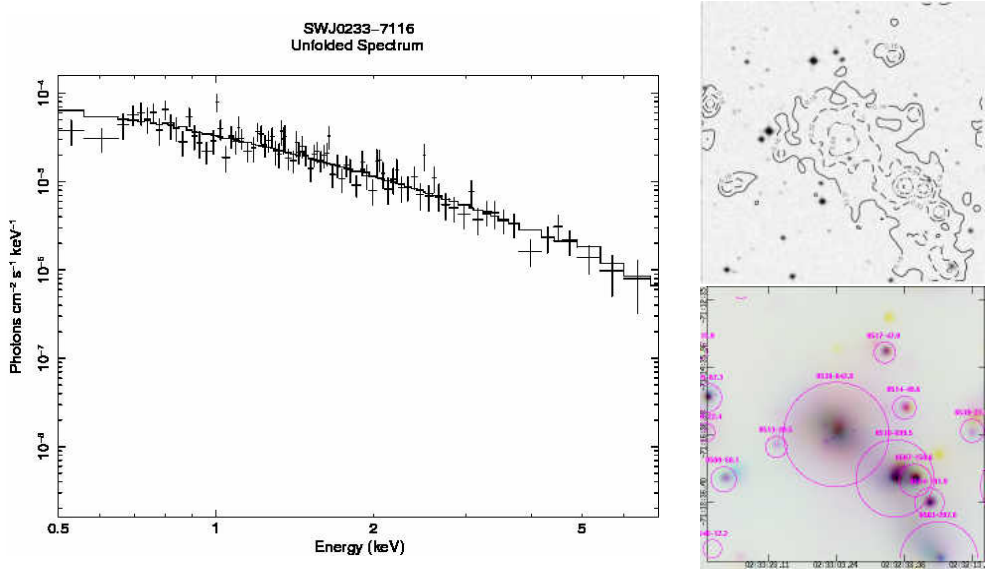

| Name | GRB | R.A. | Dec | Catalogue | Distance | Published $z$ |
|------|-----|------|-----|-----------|----------|---------------|
| SWJ0233-7116 | GRB080411 | 38.262535 | -71.276505 | - | - | - |

| Expmap [s] | Net Counts | SNR | Flux [$10^{-13}$ erg/cm$^2$/s] | $N_H$ [$10^{22}$cm$^{-2}$] | Bkg rate [$10^{-3}$ cts/arcsec$^2$] | $r_{ext}$ [arcsec] |
|------------|-----------|-----|-------------------------------|---------------------------|-------------------------------------|--------------------|
| 324059 | 842±37 | 21.2 | 0.73±0.03 | 0.058 | 14.59 | 92.7 |

| $kT$ [keV] | $z$ | $X_{Fe}/X_\odot$ | $r_{ext}$ [kpc] | $r_{500}$ [kpc] | $L_{ext}$ [$10^{44}$ erg/s] | $L_{500}$ [$10^{44}$ erg/s] |
|------------|-----|------------------|-----------------|-----------------|------------------------------|------------------------------|
| $5.5^{+0.6}_{-0.5}$ | $0.366^{+0.171}_{-0.154}$ | $0.20^{+0.26}_{-0.15}$ | 470±38 | 971±79 | $1.08^{+1.05}_{-0.96}$ | $1.34^{+1.31}_{-1.18}$ |

| $M_{500}$ [$10^{13}M_\odot$] | $M_{gas,500}$ [$10^{13}M_\odot$] | $f_{gas,500}$ |
|------------------------------|-----------------------------------|---------------|
| 38.20±1.69 | 1.34±0.17 | 0.035±0.003 |

The redshift is undetected. Possibly the redshift minimum is between $0.2 < z < 0.5$ but with a small significance $< 1\sigma$. Fitting the spectrum forcing the redshift in this range, the cluster results a factor $\sim 5$ underluminous with respect to the $L - T$ best fit by Branchesi et al. (2007), while the $M - T$ is in agreement with Arnaud et al. (2007) and Sun et al. (2009). The $M_{gas}$ is underestimated.

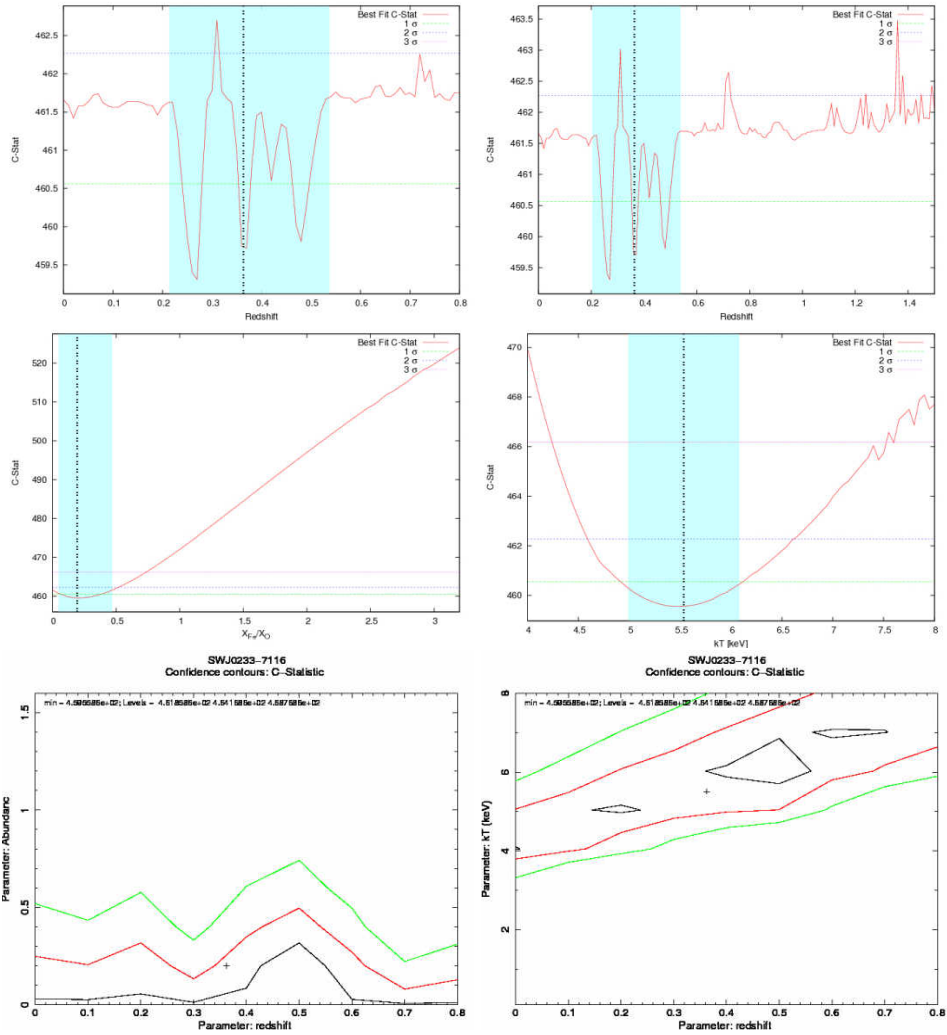

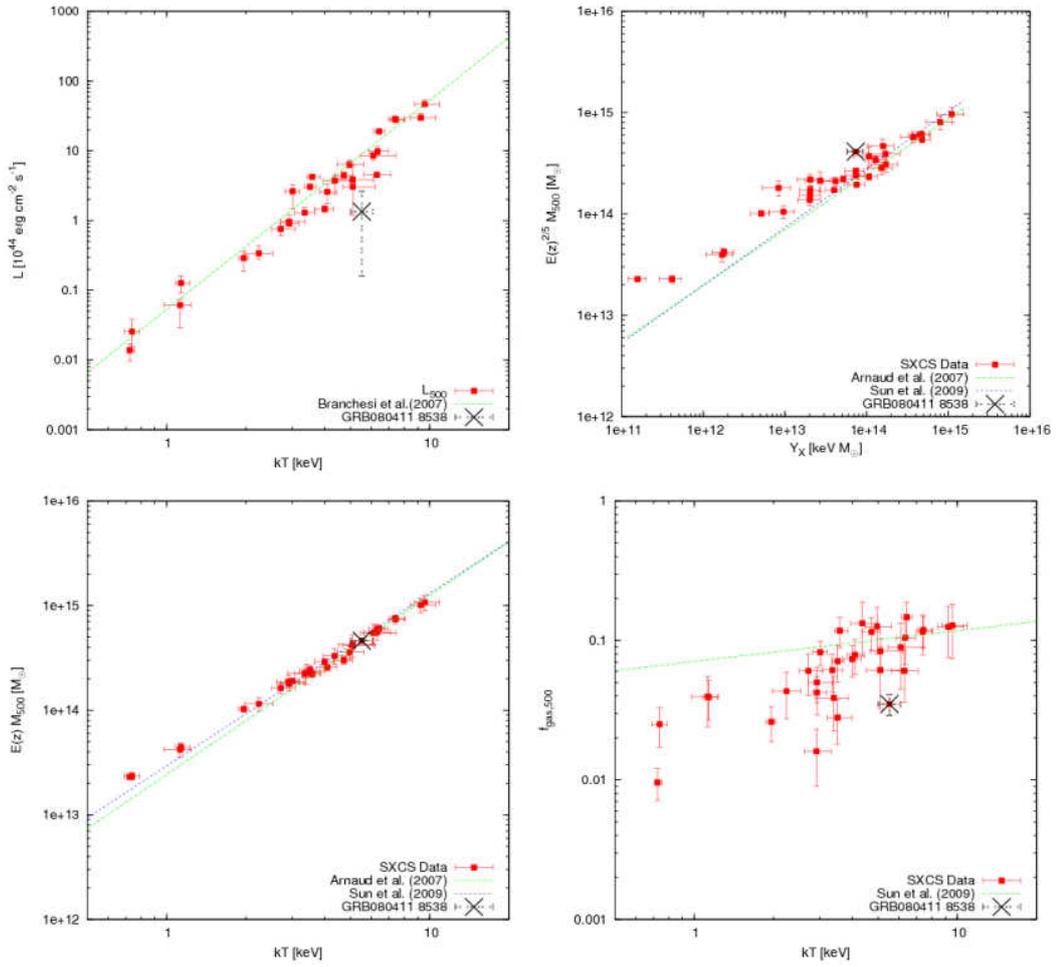

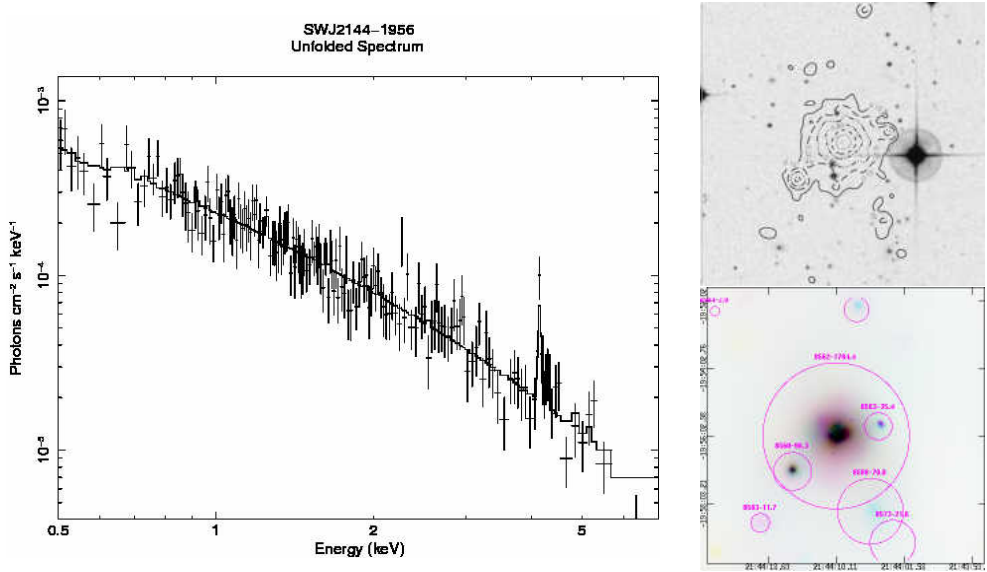

| Name | GRB | R.A. | Dec | Catalogue | Distance | Published $z$ |
|------|-----|------|-----|-----------|----------|---------------|
| SWJ2144-1956 | GRB080413B | 326.041351 | -19.933523 | RXS | 0.537 | - |

| Expmap | Net Counts | SNR | Flux | $N_H$ | Bkg rate | $r_{ext}$ |
| [s] | | | [$10^{-13}$ erg/cm$^2$/s] | [$10^{22}$cm$^{-2}$] | [$10^{-3}$ cts/arcsec$^2$] | [arcsec] |
|------|-----------|-----|------|-------|----------|-----------|
| 78298 | 1784±51 | 37.5 | 5.92±0.17 | 0.031 | 5.62 | 129.4 |

| $kT$ | $z$ | $X_{Fe}/X_\odot$ | $r_{ext}$ | $r_{500}$ | $L_{ext}$ | $L_{500}$ |
| [keV] | | | [kpc] | [kpc] | [$10^{44}$ erg/s] | [$10^{44}$ erg/s] |
|------|-----|---------|-----------|-----------|-----------|-----------|
| $7.4^{+0.6}_{-0.3}$ | $0.610^{+0.008}_{-0.008}$ | $0.50^{+0.13}_{-0.10}$ | 878±39 | 980±43 | $27.38^{+1.36}_{-1.35}$ | $27.76^{+1.38}_{-1.37}$ |

| $M_{500}$ | $M_{gas,500}$ | $f_{gas,500}$ |
| [$10^{13}M_\odot$] | [$10^{13}M_\odot$] | |
|-----------|---------------|---------------|
| 53.40±3.39 | 6.36±1.23 | 0.118±0.015 |

The redshift is well determined and the scaling laws are in agreement with the best fit from previuos works on X-ray selected clusters. Also the $M_{gas}$ is in agreement with previous works.

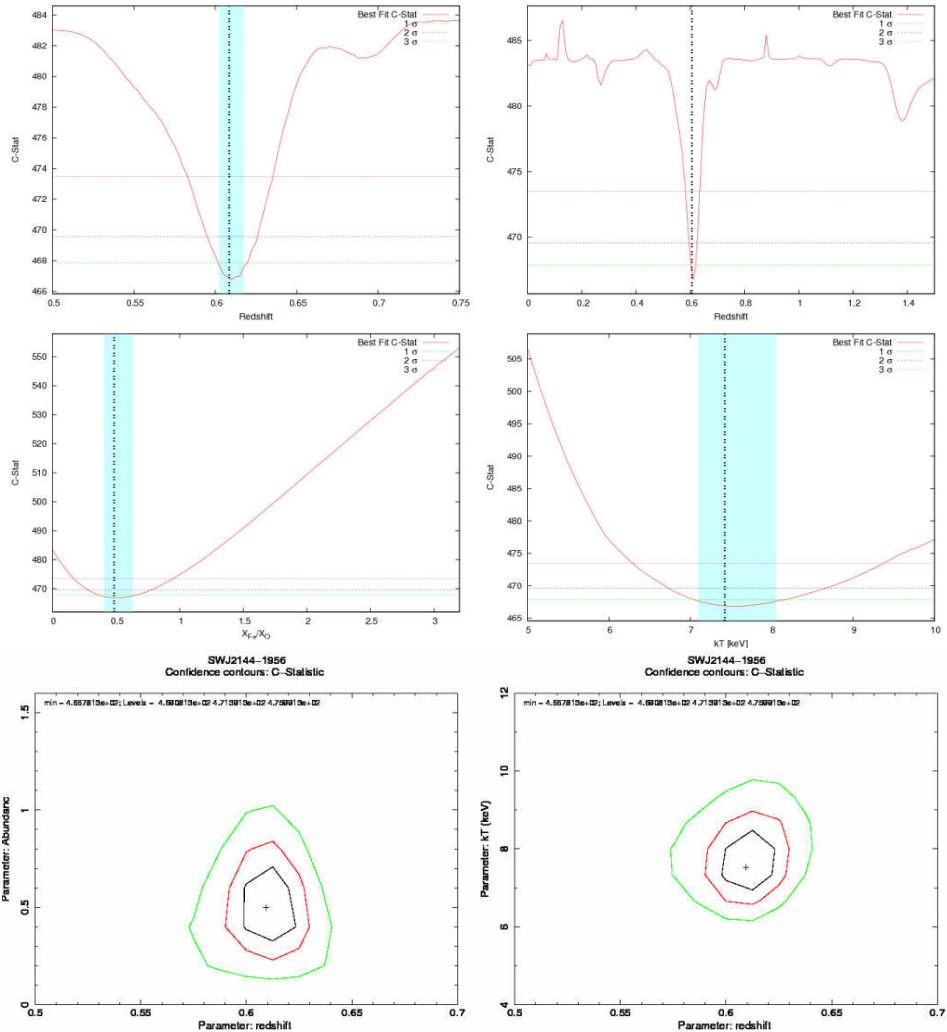

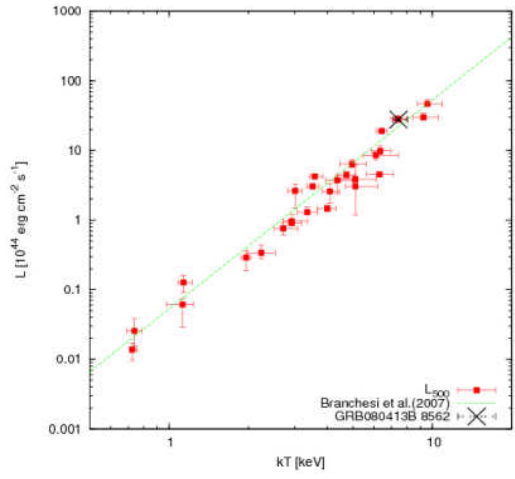
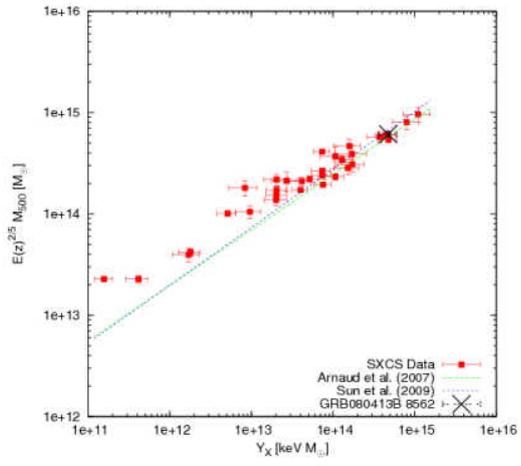

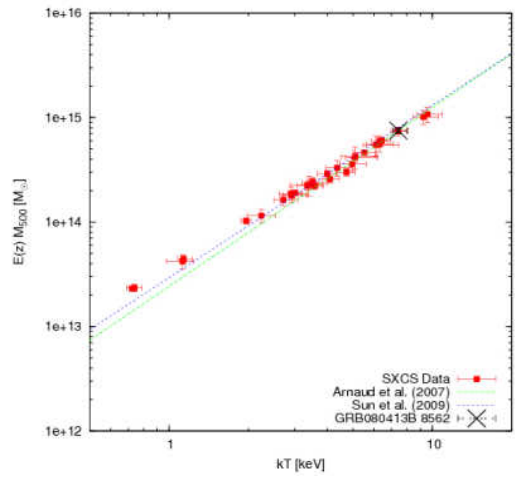
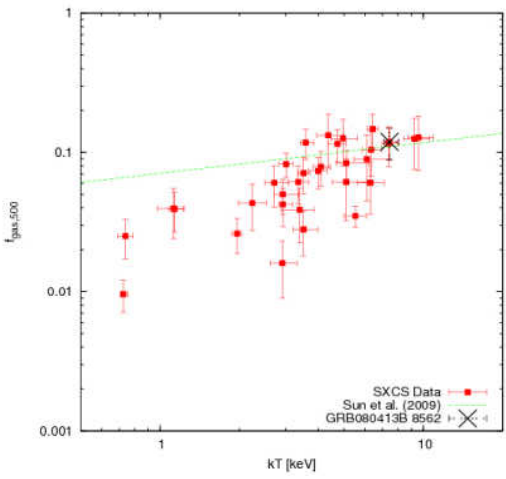

## B.31 SWJ2145-1959

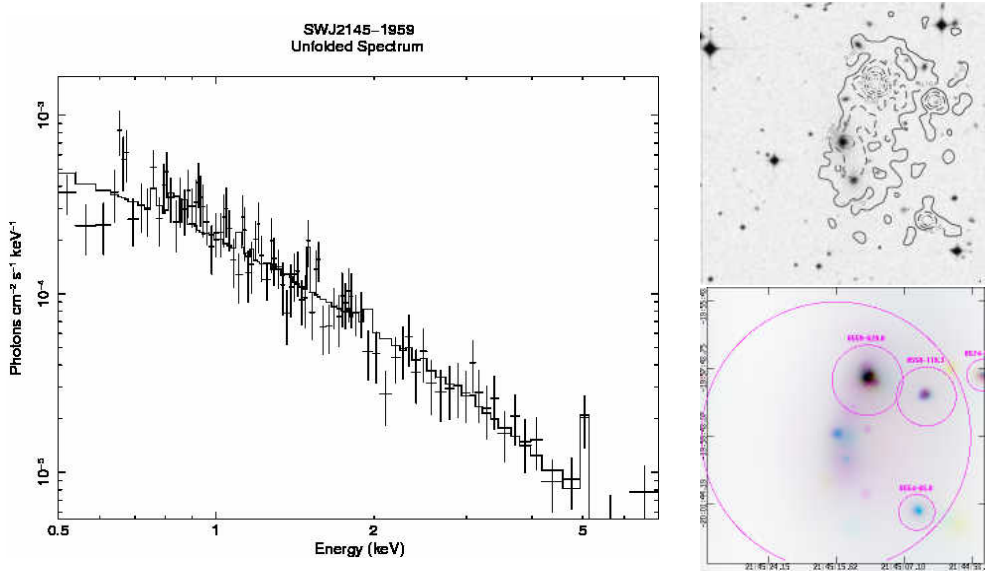

| Name | GRB | R.A. | Dec | Catalogue | Distance | Published $z$ |
|------|-----|------|-----|-----------|----------|--------------|
| SWJ2145-1959 | GRB080413B | 326.314331 | -19.994783 | LOV2003 | 0.018 | 0.0587 |

| Expmap [s] | Net Counts | SNR | Flux [$10^{-13}$ erg/cm$^2$/s] | $N_H$ [$10^{22}$cm$^{-2}$] | Bkg rate [$10^{-3}$ cts/arcsec$^2$] | $r_{ext}$ [arcsec] |
|------------|-----------|-----|------|------|------|------|
| 64304 | 1457±71 | 27.6 | 5.88±0.29 | 0.031 | 5.62 | 238.3 |

| $kT$ [keV] | $z$ | $X_{Fe}/X_\odot$ | $r_{ext}$ [kpc] | $r_{500}$ [kpc] | $L_{ext}$ [$10^{44}$ erg/s] | $L_{500}$ [$10^{44}$ erg/s] |
|-----------|-----|-----------------|-----------------|-----------------|-----------------------------|-----------------------------|
| $4.1^{+0.4}_{-0.3}$ | $0.335^{+0.013}_{-0.014}$ | $0.49^{+0.19}_{-0.16}$ | 519±247 | 862±409 | $1.89^{+0.59}_{-0.61}$ | $2.57^{+0.80}_{-0.84}$ |

| $M_{500}$ [$10^{13}M_\odot$] | $M_{gas,500}$ [$10^{13}M_\odot$] | $f_{gas,500}$ |
|------------------------------|----------------------------------|---------------|
| 22.97±1.91 | 1.81±0.35 | 0.078±0.009 |

There is a foreground elliptical galaxy 2MASXJ21451552-1959406 at redshift 0.058700 that disturb the spectrum and produce a secondary minimum in the C-stat at $z \sim 0.05$. However the cluster redshift $\sim 0.33$ is clear. Scaling laws are in agreement with the best fit found in previous works.

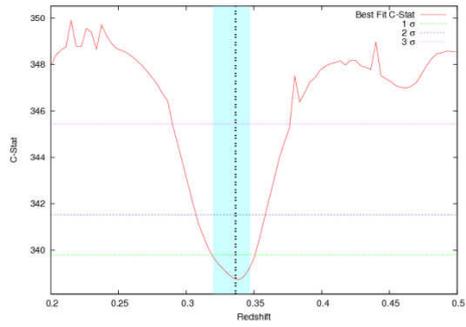
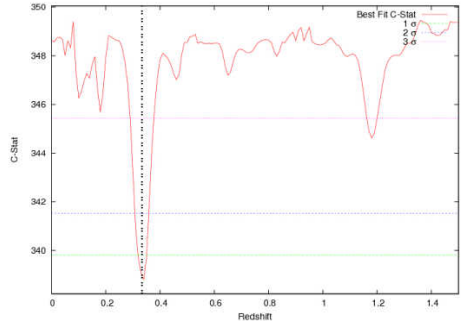

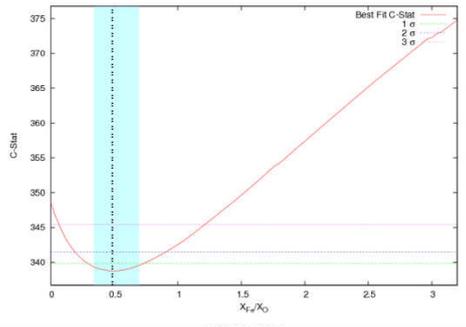
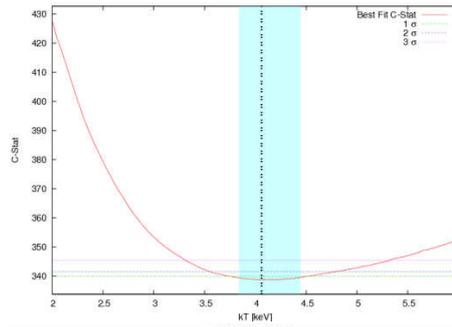

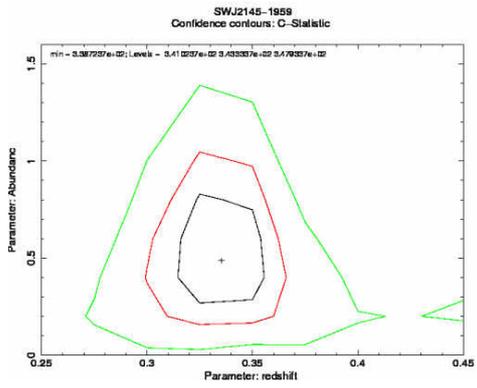
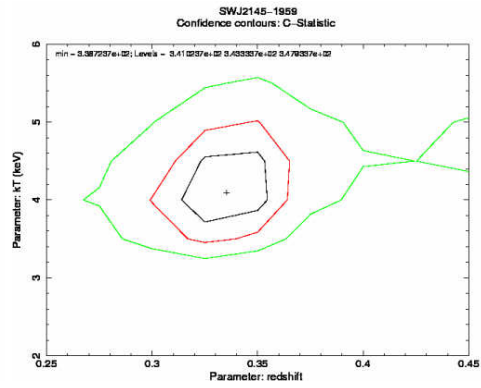

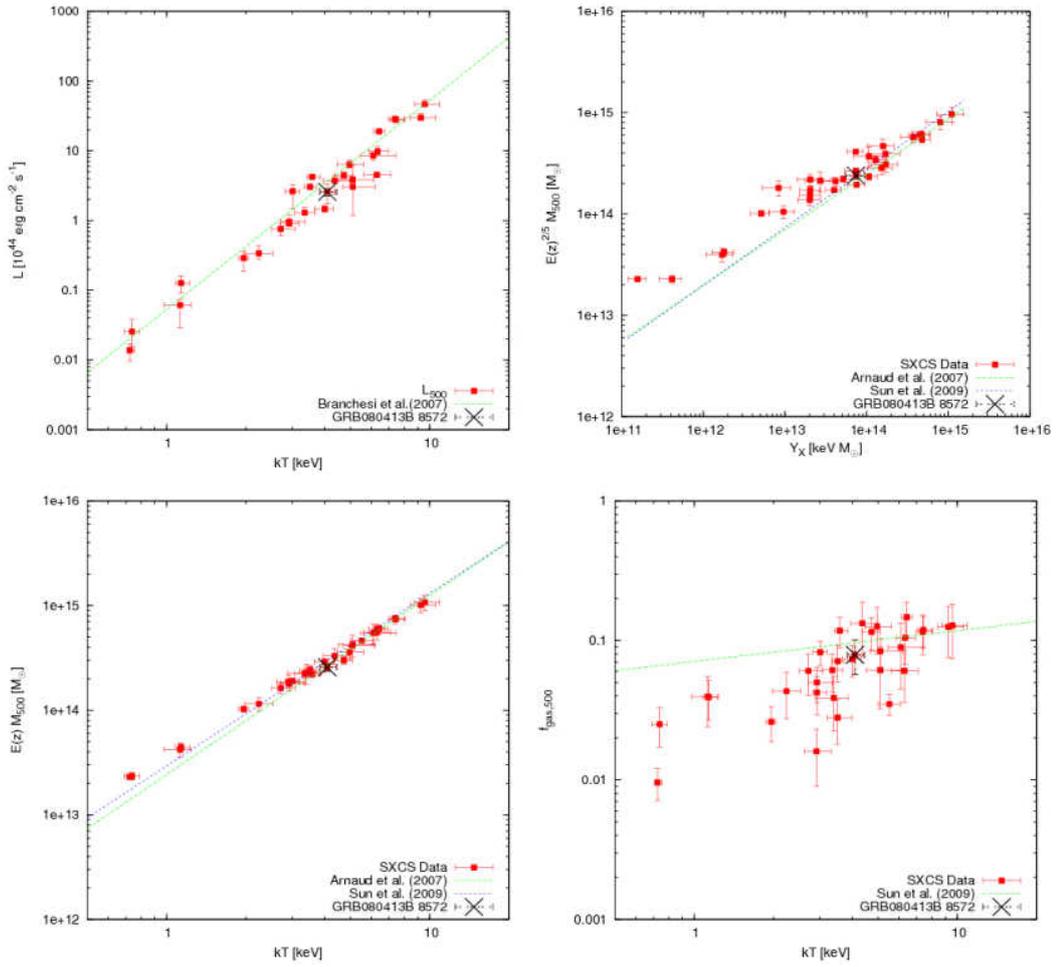

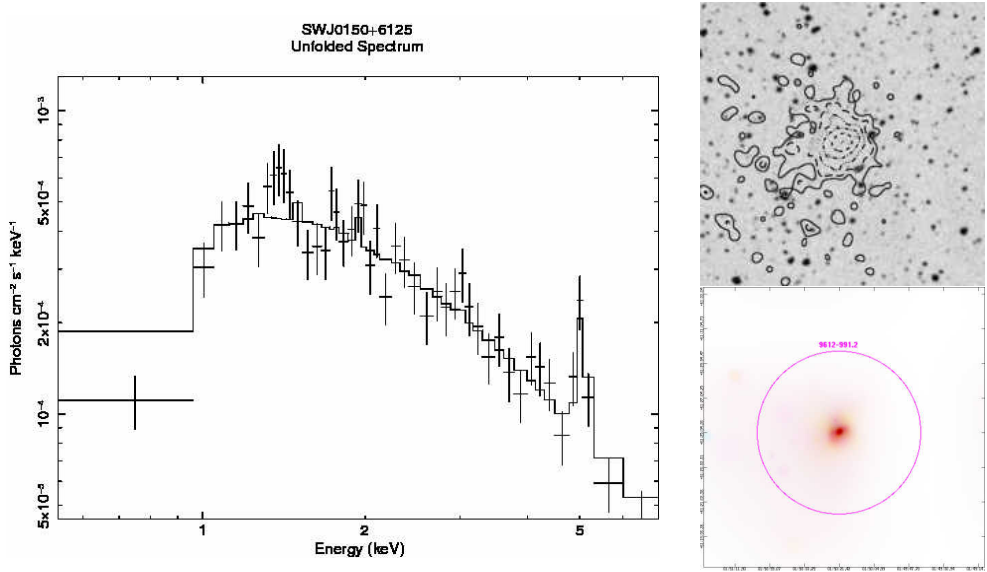

| Name | GRB | R.A. | Dec | Catalogue | Distance | Published $z$ |
|------|-----|------|-----|-----------|----------|-----------|
| SWJ0150+6125 | GRB081024A* | 27.589357 | 61.418316 | WGA | 0.180 | - |

| Expmap | Net Counts | SNR | Flux | $N_H$ | Bkg rate | $r_{ext}$ |
|--------|------------|-----|------|-------|----------|-----------|
| [s] | | | $[10^{-13}$ erg/cm²/s] | $[10^{22}$cm$^{-2}]$ | $[10^{-3}$ cts/arcsec²] | [arcsec] |
| 13989 | 991±41 | 27.2 | 79.73±3.34 | 0.771 | 0.93 | 278.0 |

| $kT$ | $z$ | $X_{Fe}/X_\odot$ | $r_{ext}$ | $r_{500}$ | $L_{ext}$ | $L_{500}$ |
|------|-----|------------------|-----------|-----------|-----------|-----------|
| [keV] | | | [kpc] | [kpc] | $[10^{44}$ erg/s] | $[10^{44}$ erg/s] |
| $9.6^{+1.3}_{-0.8}$ | $0.344^{+0.020}_{-0.005}$ | $0.80^{+0.23}_{-0.19}$ | 919±393 | 1298±556 | $43.27^{+6.45}_{-2.52}$ | $46.67^{+6.96}_{-2.72}$ |

| $M_{500}$ | $M_{gas,500}$ | $f_{gas,500}$ |
|-----------|---------------|---------------|
| $[10^{13}M_\odot]$ | $[10^{13}M_\odot]$ | |
| 89.99±14.85 | 11.53±2.95 | 0.126±0.012 |

There is a very strong Galactic absorption. The redshift is well determined with a strong K-$\alpha$ emission line. The $L-T$, $M-Y_X$ and $M-T$ are in agreement with previous works.

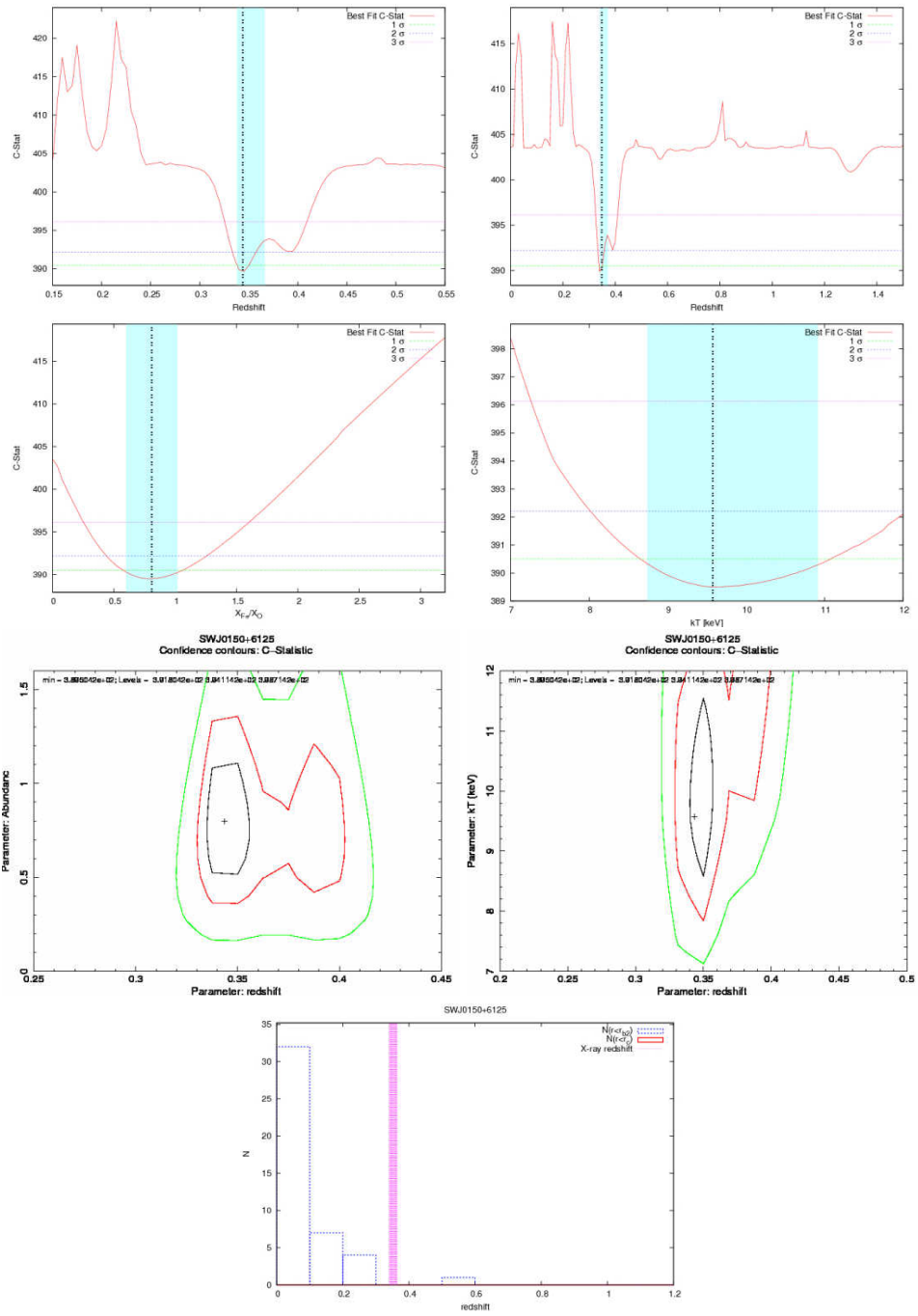

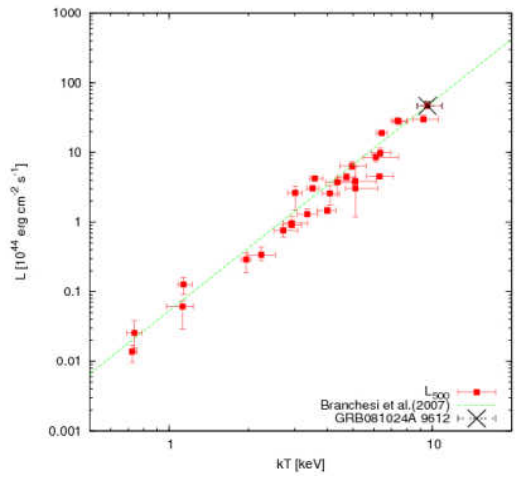
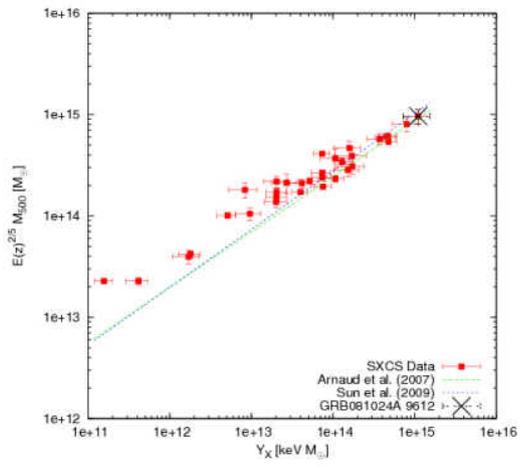
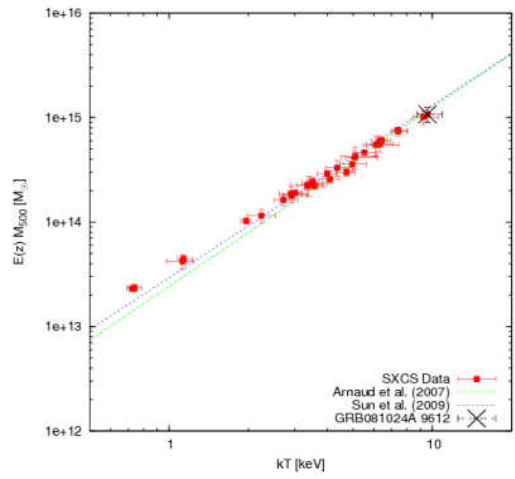
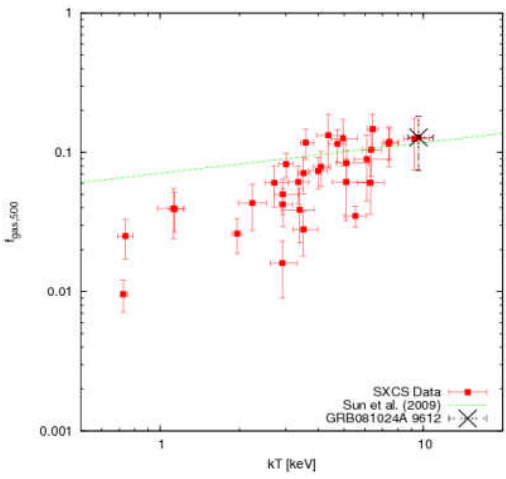